\documentclass[10pt,titlepage]{book}
\usepackage[spanish,english]{babel}
\usepackage[utf8]{inputenc}
\usepackage{fancyhdr}
\usepackage{amsmath}
\usepackage{amstext}
\usepackage{amssymb}
\usepackage{array}
\usepackage[OT1]{fontenc}
\usepackage{textcomp}
\usepackage{wrapfig}
\usepackage{multirow}
\usepackage{enumerate}
\usepackage[nice]{nicefrac}
\usepackage[tight]{units}
\usepackage[dvips]{color,graphicx}
\usepackage[a4paper,hmargin={33mm,27mm},top=25mm,height=225mm,includeheadfoot,asymmetric]{geometry} 
\usepackage{booktabs}
    \makeatletter
    \def\cleardoublepage{\clearpage\if@twoside \ifodd\c@page\else%
        \hbox{}%
        \thispagestyle{empty}%              % Empty header styles
        \newpage%
        \if@twocolumn\hbox{}\newpage\fi\fi\fi}
    \makeatother
\begin{document}
\selectlanguage{english}
% Definitions
\def\Dx{$\Delta x$}
\def\dmp{$\Delta_{mp}$}
\def\dmean{$\bar{\Delta}$}
\def\meandedx{$\langle dE/dx\rangle$}
\def\Meandedx{$\langle \frac{dE}{dx}\rangle$}
\def\ra{\rightarrow}
\newcommand\nutau{{\nu_\tau}}
\newcommand\anutau{\bar\nu_\tau}
\newcommand\numu{{\nu_\mu}}
\newcommand\anumu{\bar\nu_\mu}
\newcommand\nue{{\nu_e}}
\newcommand\anue{\bar\nu_e}
\newcommand\goldens{{\it golden sample }}
\newcommand{\LamO}{\ensuremath{\mathnormal{\Lambda}}}
\newcommand{\lambdac}{\ensuremath{\LamO_{\mathit{c}}^{+}}}
\newcommand{\sigmacp}{\ensuremath{\mathnormal{\Sigma}_{\mathit{c}}^{+}}}
\newcommand{\sigmacpp}{\ensuremath{\mathnormal{\Sigma}_{\mathit{c}}^{++}}}
\providecommand{\openone}{\leavevmode\hbox{\small1\kern-3.8pt\normalsize1}}

\frontmatter

\begin{titlepage}
\begin{center} 
%\makebox[15.0cm][l]{Preliminary}\\[1.1cm]
{\large \textsc{Dpto. de F\'isica Te\'orica y del Cosmos \& CAFPE}}\\
{\normalsize \textsc{Universidad de Granada} }\\[4.cm]
{\huge \textbf{Study Of Accelerator Neutrino Interactions in a Liquid Argon TPC} } \\[10.cm]
\end{center}

\begin{flushleft}
\noindent Memoria presentada por \\[0.2cm]
\quad {\large \textbf{Alberto Mart\'inez de la Ossa Romero.}} \\[.5cm]
%{para optar al grado de }\\[0.2cm]
%\quad {\large \textbf{Doctor en Ciencias F\'isicas}}\\[1.cm]
\noindent Directores: \\[0.2cm]
\quad Dr. \textbf{Francisco del \'Aguila Gim\'enez} y \\[0.1cm]
\quad Dr. \textbf{Antonio Bueno Villar.} \\[0.5cm]
\end{flushleft}

\begin{flushright}
- Enero de 2007 -
\end{flushright}
\end{titlepage}  

\chapter*{}
\thispagestyle{empty}
D. Francisco del \'Aguila Gim\'enez, Catedr\'atico de Universidad, y 
D. Antonio Bueno Villar, Profesor Titular de Universidad, 

\vspace*{0.5cm}
{\textbf{CERTIFICAN:} que la presente memoria, 
\textsc{Study Of Accelerator Neutrino Interactions In A Liquid Argon TPC}, 
ha sido realizada por D. Alberto Mart\'inez de la Ossa Romero bajo su direcci\'on 
en el Dpto. de F\'isica Te\'orica y del Cosmos, as\'i como que
\'este ha disfrutado de estancias en el extranjero por un periodo superior
a tres meses, tanto en el CERN (Suiza) como en el Instituto Superior T\'ecnico 
de Lisboa (Portugal).

\vspace*{4.cm} 
\begin{flushright} 
Granada, 24 de Enero de 2005

\vspace*{6.cm} 
Fdo: \ \ Francisco del \'Aguila Gim\'enez \ \ \ \ \ \ \ \ \ \ \ \ \ \ \ Antonio Bueno Villar 
\end{flushright}

\include{thanks}

\tableofcontents

\mainmatter

\typeout{}
\chapter*{Introduction}
\markboth{Introduction}{Introduction}
\addcontentsline{toc}{chapter}{Introduction}

In the thirties, W. Pauli put forward the idea of a new neutral particle that appeared 
together with the electron in the final state of nuclear beta decay. 
In its origin, this was ``a desperate solution'' to preserve the idea of energy 
conservation in physical processes. Apparently energy disappeared in beta decay 
reactions without leaving any trace. In fact, that was the case since energy was 
carried away by the invisible \emph{neutrino}. More than twenty years elapsed, 
till C. Cowan and F. Reines observed for the first time electron antineutrinos 
coming from a nuclear reactor. Despite the multiple applications neutrinos had in 
Particle Physics through these years, we still know very little about these elusive 
particles. 
Almost a decade ago, we had strong experimental evidence that atmospheric 
neutrinos do \emph{oscillate} (they change their leptonic \emph{flavour}) in their path to 
the detector. A similar mechanism is at the heart of the so-known solar neutrino 
problem. This set of  results was beautifully confirmed through experiments done
with man-made neutrinos (those coming from reactors and accelerators). 
We are now convinced that neutrinos are massive particles, however with oscillation 
experiments, we cannot tell anything about absolute neutrino masses. 
There are other experiments that carefully measure the kinematics of reactions 
were neutrinos are involved (like the tritium beta decay) that are able to 
restrict the absolute mass of the electron neutrino below 2 eV. 
Although massive, neutrinos have a very tiny mass compared to their corresponding 
charged leptons. 

The evidence for neutrino masses gave rise to a true revolution 
in Particle Physics, since this was the first solid clue about the existence of 
new Physics beyond the Standard Model. Within this model, the neutrino was 
exclusively left-handed and therefore could not be given a Dirac mass term. 
Experimentally we have never observed right-handed neutrinos, but there is no 
fundamental principle that forbids those neutrino states. 
In fact, many extensions of the Standard Model do contemplate their existence. 

After a brief introduction of the Standard Model in Chapter~\ref{section:theory}, 
we will see how to build mass terms for neutrinos. We will deal with fundamental 
questions 
like the Dirac or Majorana nature of neutrinos, the hierarchy problem, the 
\emph{see-saw} mechanism or with the possible existence of heavy neutrinos at the 
TeV scale. With rather generality, we introduce the neutrino mixing matrix 
that allows us to describe the results obtained in neutrino oscillation experiments. 
Since many theoretical models propose the existence of heavy neutrinos, 
weakly mixed with the light ones, we perform a phenomenological study in 
Sec.~\ref{sec:HeavyNu} on the precision that next-generation lepton colliders will 
have in establishing limits to such an admixture and on the heavy neutrino masses. 
This work is thoroughly described in Ref.~\cite{paper1}. 

The rest of the thesis is devoted to assess the potentialities of Liquid 
Argon (LAr) TPCs for the study of neutrino properties. Nowadays there is a 
tremendous experimental effort going on in order to precisely measure the 
mixing angles, neutrino masses and the existence of a possible CP violation 
in the leptonic sector. In this context, LAr TPCs will allow us to study 
neutrino interactions with matter with unprecedented precisions, since the 
detector is fully sensitive and acts as a fine-grained track detector, 
recording very precisely the deposited energies of particles that propagate 
through argon, making thus possible  an accurate reconstruction of a large 
variety of events. Moreover, this device records events with an imaging quality 
similar to those of ancient bubble chambers (see Fig.~\ref{fig:FirstNu2}), 
but with the additional advantage of providing fully-automatized three dimensional 
reconstruction of the events. 

\begin{figure}[ht]
\begin{center}
 \begin{tabular}{c c}
  \includegraphics[width=7cm]{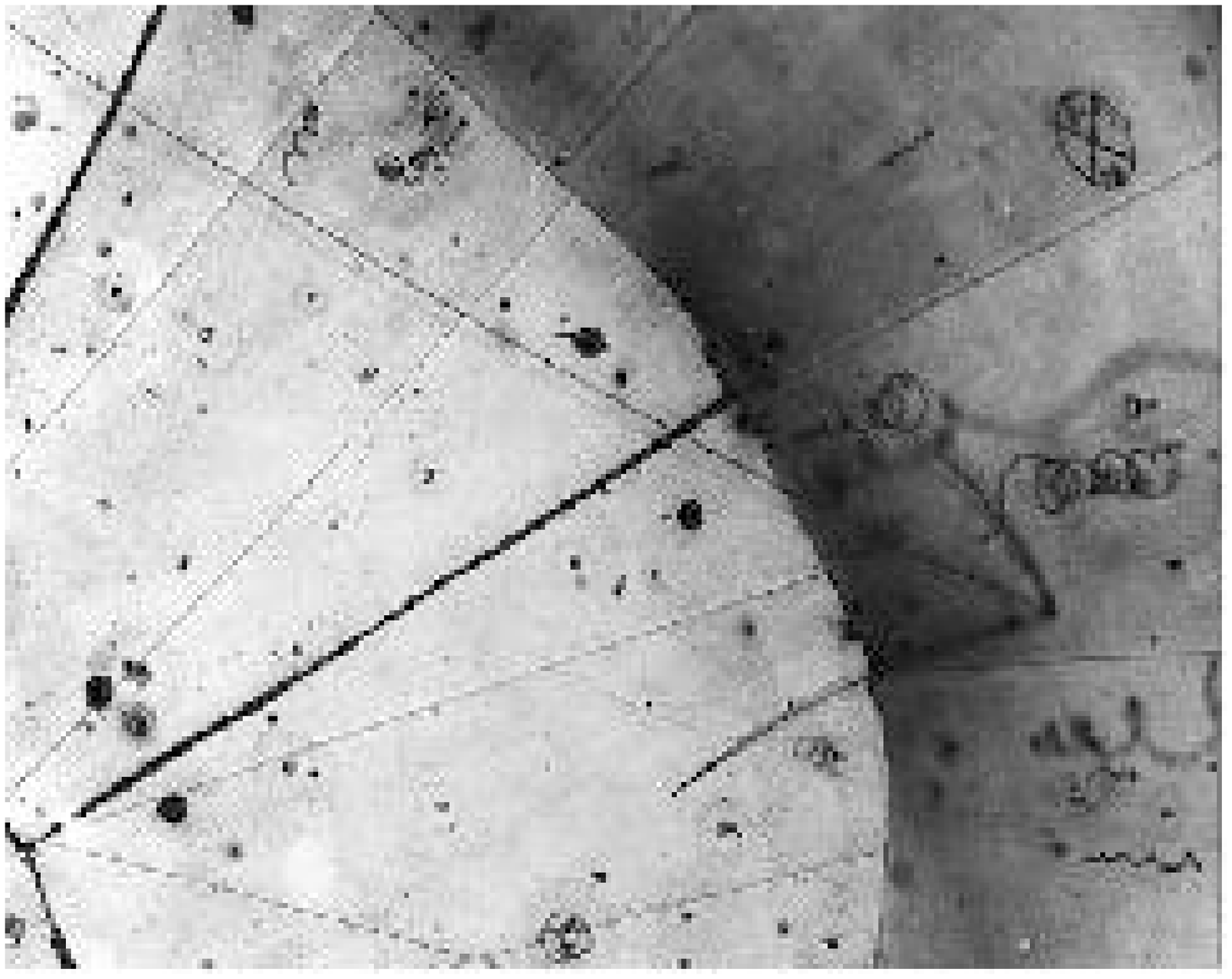} &
  \includegraphics[width=7.2cm]{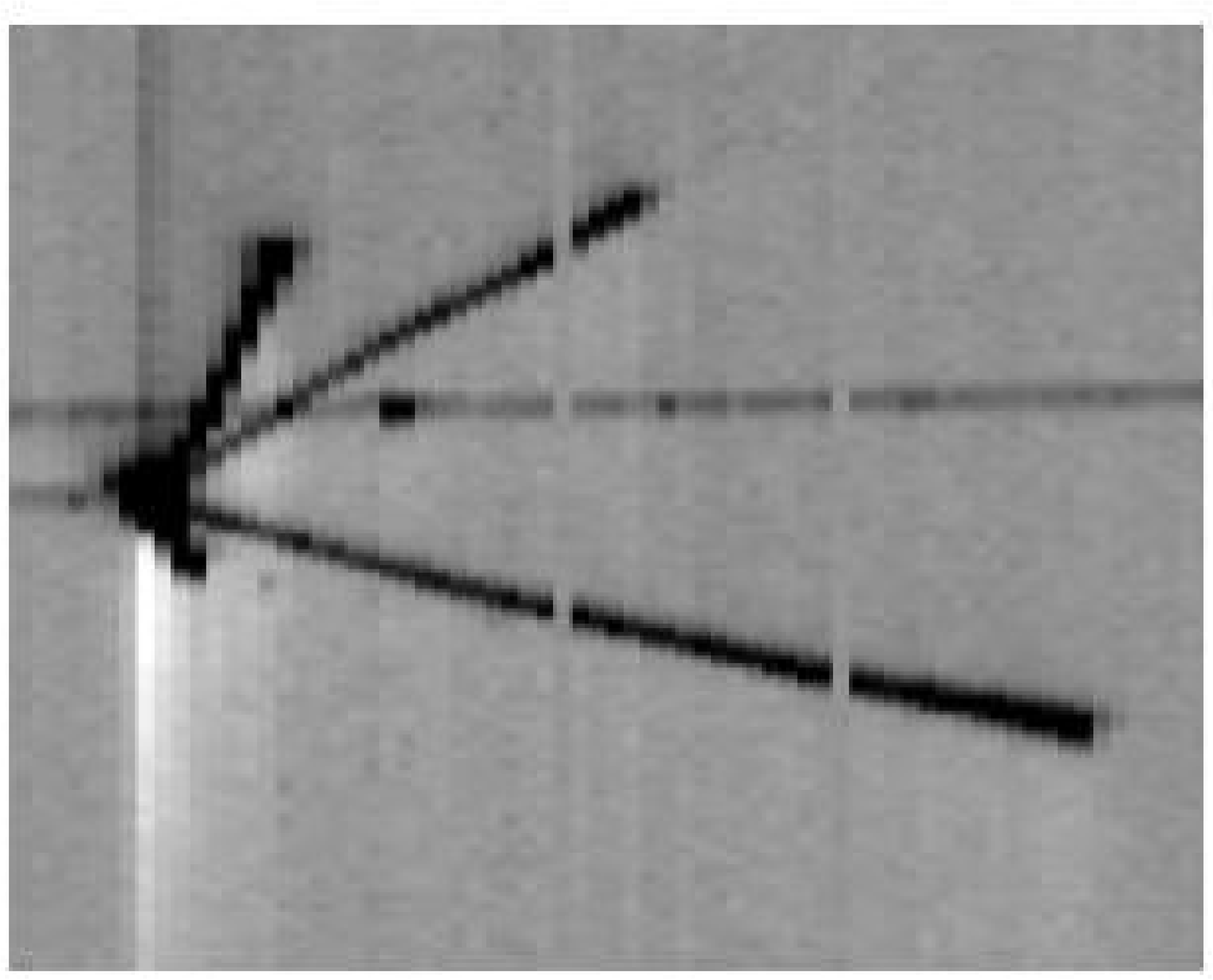}
 \end{tabular}
\caption[The world's first neutrino observation in a bubble chamber]
{(Left) First observation of a neutrino hitting a proton in a bubble chamber. 
The collision occurred at the point where three tracks emanate on the right of the 
photograph. (Right) Another neutrino-nucleon interaction, this time measured
by one of the read-out planes of a Liquid Argon TPC.}
\label{fig:FirstNu2}
\end{center}
\end{figure}

\vspace{-0.4cm}
The LAr TPC technology is discussed in Chapter~\ref{chap:Larintro}, where we 
discuss all the 
relevant technical details. One of the major challenges we faced during 
this work was the development of a set of tools and procedures that allowed 
us to perform a three dimensional and calorimetric reconstruction of  the 
recorded neutrino interactions. The goal was to exploit the detector capabilities 
in order to obtain an extremely accurate reconstruction of the neutrino event 
kinematics. To this purpose, we have worked with a small LAr TPC prototype having 
an active volume of 50 liters (see Chapter~\ref{chap:50Ldetector} for further details). 
In Chapter~\ref{chap:recon} we discuss the reconstruction tools. Thanks to them, 
in Chapter~\ref{chap:50Lexperiment}
we have carried out a careful study of quasi-elastic muon neutrino interactions 
recorded in our detector. Those multi-GeV neutrinos were produced in the CERN 
WANF neutrino facility. Despite the small size of our prototype, we have 
carried out a full study of the $\nu_\mu n \rightarrow \mu^- p$ quasi-elastic 
reactions. Those reactions were extensively studied during the sixties and the 
seventies in experiments with bubble chambers, however we still lack a precise 
modeling of the behavior of nuclear matter after a neutrino interacts with it. 
A systematic study of neutrino interactions with LAr TPCs will allow to improve 
sensibly the Monte-Carlo simulations of nuclear effects. The study of quasi-elastic 
interactions, published in Ref.~\cite{50Lqe}, has been completed with the measurement 
of the charged-current quasi-elastic total cross section for multi-GeV neutrinos. 
This measurement, together with a careful evaluation of the associated systematic 
errors is described in Sec.~\ref{sec:xsection} and has given rise to the article 
described in Ref.~\cite{50Lxsec}. 

The thesis also includes, in Chapter~\ref{chap:tausearch}, a study of the 
capabilities a LAr TPC would have when searching for direct $\nu_\tau$ 
appearance in $\nu_\mu - \nu_\tau$ oscillations by means of kinematical criteria. 
We carefully analyze the performance of several statistical techniques for pattern 
recognition, applied to look for very few $\nu_\tau$ events in a 
beam mostly composed of  $\nu_\mu$. 
The study has been published in Ref.~\cite{oscsearch}.

\selectlanguage{spanish}
\typeout{}
\chapter*{Introducción}
\markboth{Introducción}{Introducción}
\addcontentsline{toc}{chapter}{Introducción}

Corrían los años 30 cuando por vez primera W.~Pauli propuso la existencia 
de una nueva partícula neutra que acompañaba al electrón en los procesos 
de desintegración beta. En su origen, esto no fue más que una ``solución 
desesperada'' por salvaguardar el sagrado principio de la conservación de la energía: 
parecía ser entonces que en determinados procesos de desintegración nuclear 
la energía desaparecía, se esfumaba sin dejar rastro. Efectivamente así
ocurría, se la llevaba el invisible \emph{neutrino}. 
Así las cosas, todavía tuvieron que pasar más de 20 años hasta que C. Cowan
y F. Reines consiguieran \emph{observar} por primera vez neutrinos procedentes
de un reactor nuclear. A pesar del tiempo transcurrido desde su descubrimiento, 
hoy en día los neutrinos continuan siendo un misterio. 
Hace aproximadamente una década se produjo la confirmación experimental de que los
neutrinos procedentes del sol y la atmósfera terrestre \emph{oscilan} o \emph{cambian}
de \emph{sabor} leptónico en el transcurso de su camino hacia el detector. 
En los años sucesivos, la existencia de estas oscilaciones pudo ser corroborada 
por experimentos de última generación realizados con neutrinos producidos 
artificialmente en reactores y aceleradores.
Aunque los experimentos de oscilaciones implican necesariamente que deben existir 
distintos estados de neutrino cuya diferencia de masas sea finita, 
nada dicen acerca del valor absoluto de las mismas.
Existen, sin embargo, otros experimentos basados en el estudio cinemático de reacciones de
neutrinos (como la desintegración beta del tritio) que restringen la escala de 
masas por debajo de 2 eV.
La conclusión de estos resultados experimentales es clara: 
Los neutrinos tienen, aunque muy pequeña, una masa no nula.

El descubrimiento de que los neutrinos son masivos supuso una auténtica revolución 
en la Física de Partículas,
ya que  representaba la primera evidencia sólida acerca de la existencia de nueva 
Física más allá del Modelo Estándar.
En él, el neutrino es una partícula sin las componentes quirales 
diestras necesarias para dotar de masa a cualquier fermión de Dirac. 
Esto último se infiere del hecho de que, dentro de los límites de error experimentales, 
nunca se han observardo neutrinos con helicidad positiva,
aunque en realidad no existe ningún principio físico fundamental que lo prohíba.
De hecho, existen numerosas hipótesis que dan lugar a modelos teóricos 
que extienden el Modelo Estándar e incorporan la posibilidad de tener neutrinos 
con masas distintas de cero.

Tras un breve repaso del Modelo Estándar, en el Capítulo~\ref{section:theory}
veremos cómo se pueden construir términos de masa para los neutrinos.
Cuestiones como el carácter de Dirac o de Majorana de los neutrinos, el problema 
de las jerarquías y el mecanismo del \emph{balancín}, o la posible existencia de 
neutrinos pesados a la escala del TeV, son tratadas en este Capítulo. 
De una forma efectiva y completamente general introducimos la matriz de mezcla de los neutrinos,
que en el marco del Modelo Estándar minimal con neutrinos masivos 
y conjuntamente con las masas de éstos, permite describir los fenómenos
de oscilaciones de neutrinos.
%Desde el punto de vista de la experimentación y la fenomenología, 
%el cononcimiento de los parámetros de esta matriz define completamente el problema de la masa y 
%la naturaleza de los neutrinos. 
Puesto que muchos de los modelos teóricos proponen en sus soluciones la existencia 
de neutrinos pesados, débilmente mezclados con los lígeros,
en la Sección~\ref{sec:HeavyNu}, realizamos un estudio fenomenológico
sobre la precisión con la que los colisonadores de leptones de próxima generación
podrán establecer los límites de esa mezcla y de las masas de los neutrinos pesados.
Este trabajo se detalla en la publicación de la Ref.~\cite{paper1}.

%De una forma totalmente general, podemos suponer la existencia de componentes quirales 
%diestras, desacopladas de la interacción electrodébil, que se mezclan con los neutrinos 
%para dar origen a sus masas. La forma en que estos se mezclan viene completamente parametrizada
%por la así llamada \emph{matriz de mezcla de los neutrinos}. 
%Muchos de los modelos teóricos proponen en sus soluciones la existencia de neutrinos 
%pesados, débilmente mezclados con los lígeros.
%En la Sección~\ref{sec:HeavyNu}, realizamos un estudio fenomenológico
%sobre las capacidades que, colisonadores de leptones de próxima generación, 
%tendrían en la exploración de los límites de la mezcla y las masas de los 
%neutrinos pesados. Este trabajo ha sido publicado en la Ref.~\cite{paper1}.

El resto de la tesis está dedicado a mostrar el potencial que poseen las cámaras de proyección 
temporal (TPC) de Argón Líquido (LAr) para el estudio de las propiedades de los neutrinos.
Hoy en dia se está realizando un enorme esfuerzo experimental para determinar con precisión
las masas de los neutrinos, sus ángulos de mezcla y una posible violación de CP en el 
sistema leptónico. En este contexto, las TPCs de Argón Líquido van a permitir estudiar 
las interacciones de los neutrinos con la materia con precisiones nunca antes alcanzadas.
Todo el volumen sensible de este dispositivo actúa como un detector de trazas 
y de energía depositada con una alta precisión y granularidad, haciendo así
posible una precisa reconstrucción cinemática de una gran variedad de sucesos. 
Además, su dispositivo de lectura es capaz de registrar los sucesos en distintas 
orientaciones con una calidad de visualización similar a la de las antiguas cámaras de 
burbujas (ver Fig.~\ref{fig:FirstNu}), pero con la ventaja adicional de hacer posible 
una reconstrucción tridimensional.
 
\begin{figure}[ht]
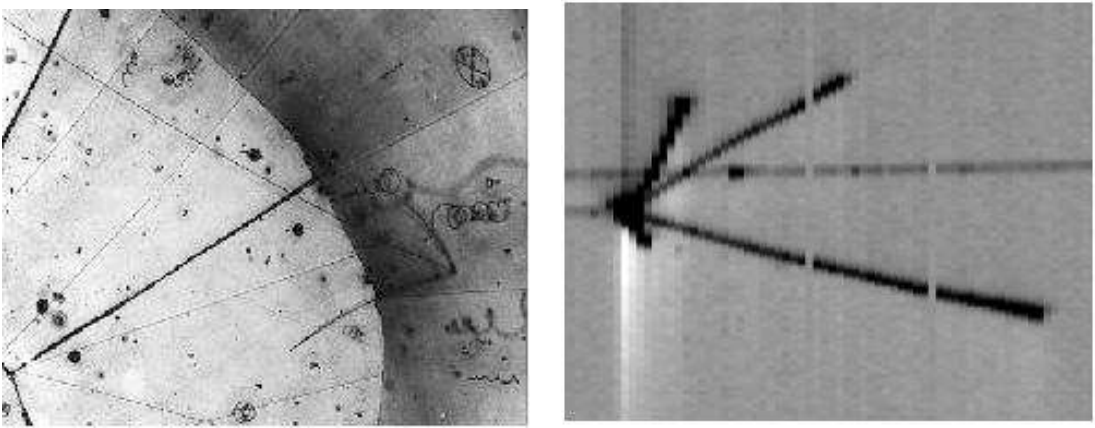

\begin{center}
 \begin{tabular}{c c}
  \includegraphics[width=7cm]{./FirstNu} &
  \includegraphics[width=7.2cm]{./LArNuEvt}
 \end{tabular}
\caption[Primera observación en la historia de un neutrino usando una cámara de 
burbujas]
{(Izquierda) Primera observación de un neutrino golpeando a un protón en una 
cámara de burbujas. La colisión ocurre a la derecha de la imagen, donde de repente, 
surgen tres trazas (un muón, un protón y un pión). 
(Derecha) Otra interacción neutrino-nucleón, esta vez recogida por uno de los planos
de detección de una cámara de deriva de Argón Líquido.}
\label{fig:FirstNu}
\end{center}
\end{figure}

La tecnología del Argón Líquido se presenta en el Capítulo~\ref{chap:Larintro}, 
donde se exponen los principios de la técnica de detección, las capacidades de 
visualización y algunas consideraciones técnicas de interés.
Uno de los mayores retos de nuestra investigación consistía en el desarrollo de 
un conjunto de herramientas y procedimientos que hicieran posible la reconstrucción 
calorimétrica y tridimensional de los sucesos ionizantes que se producen en el 
detector. 
El objetivo, explotar las capacidades del detector para lograr una medida 
sin precedentes de la cinemática de las interacciones de neutrinos.
Para tal efecto, se ha trabajado a fondo con un pequeño prototipo 
de Argon Líquido con unos 50 litros de volumen activo, descrito en detalle en
el Capítulo~\ref{chap:50Ldetector}.
El Capítulo~\ref{chap:recon} está dedicado a la descripición de las herramientas
de reconstrucción, así como a la discusión de su rendimiento y posibilidades.
Gracias a los progresos obtenidos en los algoritmos de reconstrucción, 
en el Capítulo~\ref{chap:50Lexperiment} hemos realizado un 
estudio de las interacciones quasi-elásticas de neutrinos del muón, procedentes
de un acelerador de partículas a energías de varias decenas de GeV, registradas
en el interior de la TPC de 50 litros. Pese a su limitado tamaño, el conjunto
de sucesos recogidos ha permitido un análisis completo de la cinemática de las
interacciones quasi-elasticas del tipo $\nu_\mu n \rightarrow \mu^- p$. 
Aunque este tipo de reacciones fueron extensamente estudiadas en los experimentos 
con cámaras de burbujas en los años 60 y 70, todavía no existe una modelización 
precisa de los mecanismos de interacción del neutrino con la materia nuclear. 
El estudio sistemático de las interacciones de neutrinos mediante los detectores
de Argón Líquido permitiría una mejora substancial de las simulaciones Monte-Carlo,
que tan importantes son siempre en la experimentación con neutrinos.
Este trabajo, publicado en la Ref.~\cite{50Lqe}, ha sido completado posteriormente 
con la medida de la sección eficaz de colisión quasi-elástica, vía corrientes cargadas,
para neutrinos multi-GeV dispersados por núcleos de Argón. Dicha medición, junto 
con la evaluación de sus errores sistemáticos está expuesta en la 
Sección~\ref{sec:xsection}
de esta Tesis, y puede consultarse en la Ref.~\cite{50Lxsec}.

La memoria de la Tesis concluye en el Capítulo~\ref{chap:tausearch}, 
dedicado al estudio de las capacidades que una cámara de Argón Líquido tendría a la 
hora de realizar una búsqueda directa de la oscilación $\nu_\mu$--$\nu_\tau$ mediante 
criterios cinemáticos. Allí también analizamos el rendimiento de distintas técnicas 
estadísticas para el reconocimiento de señales, aplicadas al problema de búsqueda de 
$\nu_\tau$ en un haz compuesto mayoritariamente de $\nu_\mu$. 
Toda la discusión al respecto se encuentra publicada en la Ref.~\cite{oscsearch}.

\selectlanguage{english}

%%%%%%%%%%%%%%%%%%%%%%%%%%%%%%%%%%
\chapter{Neutrino interactions}
%%%%%%%%%%%%%%%%%%%%%%%%%%%%%%%%%%
\label{section:theory}
%%%%%%%%%%%%%%%%%%%%%%%%%%%%%%%%%%%%%%%%%%%%%%%

The main goal of this Thesis is to prove the optimal behaviour 
of Liquid Argon for detection of light neutrinos, with the aim 
of studying neutrino oscillations. Let us then first in this 
Section briefly review what we know about massive neutrinos. 
Thus, after introducing the \emph{Standard Model} (SM) of Particle 
Physics, we revise the two types, Dirac and Majorana, of neutrino 
masses and the \emph{see-saw} mechanism for three generations. 
Then, we derive the oscillation probabilities for flavor transitions 
between light neutrinos, and discuss the limits on the mixing between 
light and heavy neutrinos.

\section{Overview of the Standard Model}
\label{sec:SM}
%%%%%%%%%%%%%%%%%%%%%%%%%%%%%%%%%%%%%%%%%%%%%%%
Our present understanding of the elementary particles and their interactions 
is described by the so called Standard Model 
\cite{PeskinQFT,HalzenQuarksLeptons,Griff,CottSM}. 
It contains: 
\begin{itemize}
\item{\emph{Fermions}} with spin \nicefrac{1}{2}, which are the particles of which 
matter is made.
\item{\emph{Gauge Bosons}} with spin 1, which act as mediators of the different interactions.
\item{\emph{Scalars}} with spin 0, which are responsible of the spontaneous breaking of 
the electroweak symmetry.
\end{itemize} 
The fermions are either \emph{leptons} which are not affected by the strong interactions 
or \emph{quarks} which are the fundamental constituents of \emph{hadrons}, 
the particles interacting via the strong force. Leptons and quarks can be grouped into 
three generations according to their masses as shown in Tab.~\ref{table:fermions}. 
Ordinary matter consists of fermions of the first generation. 
The higher mass fermions of the second and third generation are produced by 
particle accelerators and cosmic rays at higher energies. 
For every fermion there is an associated anti-fermion with the same mass but opposite 
additive quantum numbers. 
\begin{table}[t]
\begin{center}
\begin{tabular}{|l|>{$}c<{$}|>{$}c<{$}|>{$}c<{$}|}
\hline
\multicolumn{4}{|c|}{Quarks}\\ \hline
\text{Generation}&\text{Flavour}&\text{Charge (e)}&\text{Mass (\unit{MeV}/$\unit{c}^2$)}\\ \hline
\text{1st}        & u & +\nicefrac{2}{3} & 1.5-4\\ 
                  & d & -\nicefrac{1}{3} & 4-8\\ \hline
\text{2nd}& c     & +\nicefrac{2}{3} & 1150-1350 \\ 
                  & s & -\nicefrac{1}{3} & 80-130 \\ \hline
\text{3rd}        & t & +\nicefrac{2}{3} & 171400 \pm 2100 \\ 
                  & b & -\nicefrac{1}{3} & 4100-4400 \\ \hline
\multicolumn{4}{c}{}\\ \hline
\multicolumn{4}{|c|}{Leptons}\\ \hline
\text{Generation}&\text{Flavour}&\text{Charge (e)}&\text{Mass (\unit{MeV}/$\unit{c}^2$)}\\ \hline
\text{1st}        & \nu_{e}   &  0 &  < 3\times 10^{-6} \\ 
                  & e^-       & -1 &  0.511  \\ \hline
\text{2nd}        & \nu_{\mu} &  0 & <0.19  \\ 
                  & \mu^-     & -1 & 105.7   \\ \hline
\text{3rd}        & \nu_{\tau}&  0 &  <18.2 \\ 
                  & \tau^-    & -1 & 1777    \\ \hline
\end{tabular}

\caption[The fermions of the Standard Model]{Properties of the SM fermions 
\cite{PDG}. Since quarks do not exist as 
free particles, their masses cannot be measured directly, but have to be derived 
from hadron properties. Therefore, the values of the quark masses depend on 
how they are defined. The \emph{constituent} quark masses are inferred 
from the hadron mass spectra, while the masses that enter as parameters in 
the QCD Lagrangian are called \emph{current} masses. 
The values stated in the Table refer to the latter.}
\label{table:fermions}
\end{center}
\end{table}
\begin{table}[t]
\begin{center}
\begin{tabular}{|l|c|>{$}c<{$}|>{$}c<{$}|c|}
\hline
\text{Interaction}&\text{Gauge}&\text{Relative}&\text{Range (m)}&\text{Participating}\\
                  &\text{boson(s)}&\text{strength}&            &\text{fermions}\\
\hline
strong          & gluons ($g$)       & 1         & 10^{-15} & quarks       \\ 
electromagnetic & photons ($\gamma$) & 10^{-2}   & \infty        & all charged  \\ 
weak            & $W^{\pm}$, $Z^0$        & 10^{-7}   & 10^{-18}      & all     \\ 
gravitational   & [graviton (G)]     & 10^{-39}  & \infty        & all          \\ \hline
\end{tabular}

\caption[Interactions of the Standard Model]{SM interactions and corresponding 
gauge bosons. The relative strengths of the forces 
are roughly given for short distance scales of a few \unit{GeV} \cite{BookPerkins}. 
Since, it has not been observed, the graviton may be considered a hypothetical 
particle for the time being.}
\label{table:interactions}
\end{center}
\end{table}
Fermions interact with each other via the exchange of the gauge bosons listed in 
Tab.~\ref{table:interactions}. The photon is the gauge boson which mediates the 
electromagnetic interactions and couples to all charged particles. 
All fermions interact weakly exchanging the massive gauge bosons $W^{\pm}$ and $Z^0$. 
The strong interaction acts only on quarks and is carried by gluons. 
To complete the list of the known forces, gravity has to be mentioned. 
It affects all particles and is assumed to be carried by the graviton, a spin-2 boson. 
However, since the effects of gravity are negligible when considering 
the interactions of elementary particles, it will not be further discussed.

The interactions of the Standard Model are based on the principle of local gauge 
invariance, which demands the \emph{Lagrangian} of the theory to be invariant under 
transformations of the gauge group 
\begin{equation}
SU(3)_C\times SU(2)_L\times U(1)_Y \; .
\label{eq:SM}
\end{equation}
QCD is governed by the symmetry group $SU(3)_C$ and the electroweak theory by 
$SU(2)_L\times U(1)_Y$. Every fermion in the SM belongs to a representation of the 
gauge group which uniquely determines its dynamics. The different representations
where the fermions lay are summarized in Tab.~\ref{table:particleSM}.
The non-Abelian character of $SU(3)_C$ and $SU(2)_L$ leads to 
self-interactions between the gauge bosons. In the case of QCD, this self-interaction 
accounts for the fact that quarks are confined into hadrons, 
i.\ e., they cannot be observed as individual particles. However, probed at higher 
energies they begin to behave as free particles, a property referred to as 
asymptotic freedom. 
\begin{table}[t]
\setlength\extrarowheight{2pt}
\begin{center}
\begin{tabular}{|>{$}c<{$} >{$}c<{$} >{$}c<{$}||>{$}c<{$}|>{$}c<{$}|>{$}c<{$}|}
\hline
\multicolumn{3}{|c||}{\text{Field (particle)}} & SU(3)_C & SU(2)_L & U(1)_Y \\ \hline \hline
  \begin{pmatrix} \nu_e \\ e \end{pmatrix}_L  
& \begin{pmatrix} \nu_\mu \\ \mu \end{pmatrix}_L 
& \begin{pmatrix} \nu_\tau \\ \tau \end{pmatrix}_L
& 1 & 2 & -\nicefrac{1}{2} \\ \hline
e_R & \mu_R & \tau_R 
& 1 & 1 & -1 \\ \hline
  \begin{pmatrix} u \\ d^\prime \end{pmatrix}_L  
& \begin{pmatrix} c \\ s^\prime \end{pmatrix}_L 
& \begin{pmatrix} t \\ b^\prime \end{pmatrix}_L
& 3 & 2 & \nicefrac{1}{6} \\ \hline
u_R & c_R & t_R 
& \bar{3} & 1 & \nicefrac{2}{3} \\ \hline
d^\prime_R & s^\prime_R & b^\prime_R 
& \bar{3} & 1 & -\nicefrac{1}{3} \\ \hline\hline
\end{tabular}

\caption[Representation assignment of the SM fermions]
{Representation assignment of the SM fermions.}
\label{table:particleSM}
\end{center}
\end{table}

Whereas the gauge bosons of the electromagnetic and the strong interaction are massless,
the theory of the weak interactions must also accommodate the massive $W^{\pm}$ and $Z^0$ 
gauge boons. In order to achieve this, while retaining the gauge invariant 
structure of the theory, the Higgs mechanism was introduced. This, however, requires 
at least one additional physical particle, the Higgs boson, yet to be found.

Although the SM is a mathematically consistent, renormalizable 
field theory, it cannot be the final answer \cite{StandardModelTests:1995}. 
Even if the recently discovered neutrino masses were incorporated, more fundamental 
shortcomings can be noted. For example, the minimal version of the model has more 
than 20 parameters, which, as most physicists believe,  are too many for a 
fundamental theory. Although incorporated in the SM, there is no 
explanation for the fundamental fact of charge quantization or for the 
CP (charge-conjugation and parity) violating interactions. 
There is no profound compelling for the existence of the heavier fermion generations 
either. 
Moreover, there is no prediction of the charged fermion masses which vary by more than 
five orders of magnitude. As accelerators reach higher 
energies and results become more precise answers for some of the current questions 
will be certainly found. 

Further details on the structure of the electroweak interactions can be found 
in Appendix \ref{sec:WeakInt}.

%%%%%%%%%%%%%%%%%%%%%%%%%%%%%%%%%%%%%%%%%%%%%%%%%%%%%%%%%%%%

%\footnotetext{If the masses of the quarks $u$,$d$,$c$,$s$ are assumed to be equal, the strong interaction becomes invariant under $SU(4)$ ``rotations'' of these quark flavors}
%%% Local Variables: 
%%% mode: latex
%%% TeX-master: "thesis"
%%% End: 

\section{Massive neutrinos}
\label{sec:neumass}

Neutrinos introduce a hierarchy problem: why are they so
much lighter than their charged partners? In the SM this hierarchy is 
made natural making the neutrino massless. 
Recent experimental results strongly indicate the existence
of oscillations for neutrinos produced in the
atmosphere~\cite{SUPERK,MACRO,SOUDAN2} and those coming from the
sun~\cite{SNO1,SNO2,SNO3}, which can be interpreted introducing 
finite, non-degenerate neutrino masses. The possibility of considering 
neutrinos as massive particles allows for new phenomena such as
the already mentioned flavor mixing and oscillations, but also
magnetic moments and decays, with very important consequences in
astrophysics, cosmology and particle physics. Moreover, since most
extensions of the SM predict the existence of massive neutrinos, this
should be considered as a potentially rich field to look for physics
beyond the SM.

In this section we briefly review the formulation of neutrino masses, 
extending the minimal SM \cite{BookMoha,BookBranco,NuMix,NuMass}.

\subsection*{Dirac mass}
\label{sec:diracmass}

If as the other SM fermions neutrinos have right-handed counterparts 
in $SU(2)_L$ singlets, the standard Higgs mechanism can generate the 
corresponding Dirac mass term
\begin{equation}
\mathcal{L}^\text{D} = -m_{\text{D}}\,\overline{\nu}\,\nu \; ,
\label{eq:diracmass1}
\end{equation}
where we assume for simplicity only one family to start with. 
In terms of the \emph{chiral} fields $\nu_L$ (left-handed neutrino) 
and $\nu_R$ (right-handed neutrino) 
\begin{equation}
\mathcal{L}^\text{D} = -m_{\text{D}} \left( \overline{\nu}_L \, \nu_R 
                                          + \overline{\nu}_R \,\nu_L \right) 
\; ,
\label{eq:diracmass2}
\end{equation}
mixing fields of opposite chirality. 
$ m_{\text{D}} = y \, v / \sqrt{2}$ 
where $y$ is a dimensionless Yukawa coupling and $v / \sqrt{2}$ is the 
vacuum expectation value of the Higgs field. 
This mass term is invariant under the global transformation 
$\nu \to e^{i\alpha}\, \nu$, which implies the
conservation of a global quantum number, in this case the lepton number $L$.

All experiments to date are consistent with the neutrino being
left-handed and the anti-neutrino being right-handed. Whether the two
other states exist in Nature is at present unknown. They would not take part in 
the weak interactions and for this reason are named \emph{sterile}.
The main theoretical problem associated with this mechanism is the smallness 
of the neutrino mass compared to the mass of the charged leptons. Since
particle masses are directly determined by the particle couplings
to the Higgs field, this implies couplings differing by several orders
of magnitude for particles in the same generation. The relation between 
lepton masses of different generations may be thought of as being
``accidental'', but a large mass difference between particles belonging to
the same $SU(2)_L$ doublet is more difficult to argue. 
However, light neutrinos can have also masses in the absence of their 
right-handed counterparts.

\subsection*{Majorana mass}

In 1937 Majorana \cite{MAJORANA} discovered that a massive neutral fermion 
as a neutrino can be described by a spinor $\nu$ with only two independent 
components, imposing the so-called Majorana condition 
\begin{equation}
\nu = \nu^c \; ,
\label{eq:majcond}
\end{equation}
where
$ \nu^c = \mathcal{C} \overline{\nu}^{T} = \mathcal{C} {\gamma^0}^T \nu^* $
is the operation of charge conjugation, and $\mathcal{C}$ 
the charge-conjugation matrix (see \cite{BookBranco}).
The Majorana condition \eqref{eq:majcond} can be conveniently expressed in
terms of its left-handed and right-handed components
\begin{equation}
\begin{split}
\nu_R = \nu_L^c \; , & \qquad  \nu_L = \nu_R^c \; ,
\end{split}
\label{eq:majcond2}
\end{equation}
Thus, the right-handed component $\nu_R$
of the Majorana neutrino field $\nu$ is not independent but obtained from the 
left-handed component $\nu_L$ through charge conjugation\footnote{We assume 
the notation $\nu_L^c \equiv (\nu_L)^c$, which is a 
right-handed field, contrary to $(\nu^c)_L$, which is a left-handed field.}, 
and the Majorana field can be written 
\begin{equation}
\nu = \nu_L + \nu_L^c \; .
\label{eq:maj}
\end{equation}
Using the constraint \eqref{eq:majcond2} in the mass term \eqref{eq:diracmass2},
we obtain the Majorana mass term
\begin{equation}
\mathcal{L}^{\text{M}}_\text{L} = - \frac{1}{2} \, m_{\text{M}}
\left( \overline{\nu_L^c} \, \nu_L + \overline{\nu_L} \, \nu_L^c\right) \; ,
\label{eq:majmass}
\end{equation}
where we have inserted the factor $1/2$ to properly normalized the 
kinetic term in the Lagrangian.
Hence, a mass term for purely left-handed particles (and
right-handed anti-particles) can be constructed \eqref{eq:majmass}. 
Thus, in contrast to the Dirac case a neutrino mass term is possible even without
additional (unobserved) right-handed neutrinos. It requires just one
helicity state of the particle and the opposite helicity state of the
anti-particle.

The Majorana mass term is not invariant under global transformations. 
Indeed, a lepton number violation of $\Delta L = \pm 2$ is produced when 
considering a Majorana mass term. In fact, a renormalizable theory which
includes a Majorana mass term for $\nu_L$ can not be
constructed using the Higgs sector of the SM. In general, 
a Majorana mass for the neutrino would be interpreted as a hint for
new physics at higher energies, whereas a Dirac mass could be accommodated
adding to the SM the right-handed neutrino states but requiring very 
small neutrino Yukawa couplings.

\subsection*{General mass terms and the see-saw mechanism}
\label{sec:genneumassterm}

The same reasoning as for $\nu _L$ can be applied to a hypothetical 
(in principle) and independent right-handed Majorana neutrino, which can be
equivalently written $n = n_R + n_R^c$ 
\footnote{Right-handed sterile neutrinos will be denoted by $n$ 
in what follows, and by $N$ when discussing collider limits.}
with a mass term  
\begin{equation}
\mathcal{L}^{\text{M}}_\text{R} = - \frac{1}{2} \, m_{\text{R}}
\left( \overline{n_R} \, n_R^c + \overline{ n_R^c} \, n_R\right) \; .
\label{eq:majmassneu2}
\end{equation}
The right-handed Majorana neutrino does not need to be the ``partner''
of the (ordinary) left-handed neutrino, but could be, instead, an
independent heavier state.
In general, if both exist and are independent, in addition to the Dirac 
mass term \eqref{eq:diracmass2} also the Majorana mass terms for $\nu_L$ 
and $n_R$, in \eqref{eq:majmass} and \eqref{eq:majmassneu2} respectively, are allowed.
%
%\begin{equation}
%\mathcal{L}^{\text{M}}_L = - \frac{1}{2} \, m_L
%\left( \overline{\nu_L^c} \, \nu_L + \overline{\nu_L} \, \nu_L^c \right)
%\,,
%\qquad
%\mathcal{L}^{\text{M}}_R = - \frac{1}{2} \, m_R
%\left( \overline{n_R^c} \, n_R + \overline{n_R} \, n_R^c \right)
%\,.
%\end{equation}
%
Then, the general mass term
\begin{equation}
\mathcal{L}^{\text{Mass}} =
\mathcal{L}^{\text{D}} + \mathcal{L}^{\text{M}}_L + \mathcal{L}^{\text{M}}_R
\label{0091}
\end{equation}
can be written as
\begin{equation}
\mathcal{L}^{\text{Mass}} = - \frac{1}{2}
\begin{pmatrix} \overline{\nu_L^c} & \overline{n_R} \end{pmatrix}
\begin{pmatrix} m_L & m_{\text{D}} \\ m_{\text{D}} & m_R \end{pmatrix}
\begin{pmatrix} \nu_L \\ n_R^c \end{pmatrix} + \text{H.c.}
%\, = -\frac{1}{2} \, \overline{N_L^c} \, M \, N_L + \text{H.c.}
\label{eq:genmassterm}
\end{equation}
%
%where, for convenience, we introduce matrix notation 
%
%\begin{equation}
%M = \begin{pmatrix} m_L & m_{\text{D}} \\ m_{\text{D}} & m_R \end{pmatrix}
%\,, \qquad N_L = \begin{pmatrix} \nu_L \\ n_R^c \end{pmatrix} \,.
%\label{eq:numat}
%\end{equation}
%
It is clear that the chiral fields $\nu_L$ and $n_R$
do not have a definite mass, since they are coupled by the Dirac mass term.
In order to find the fields with definite masses, it is necessary to diagonalize
the mass matrix in Eq.~\eqref{eq:genmassterm}. 
This is done by a unitary transformation 
\begin{equation}
U^T \, M \, U = \begin{pmatrix} m_1 & 0 \\ 0 & m_2 \end{pmatrix} \; ,
\end{equation}
with $m_k$, $k=1,2$, real and positive and $U$ the unitary mixing matrix 
relating the $\nu_L$ and $n_R$ fields with the mass eigenstates $\nu_{1\,L}$ and 
$\nu_{2\,L}$. 
This diagonalization leads to the following mixing matrix and eigenvalues 
\begin{equation}
\begin{split}
U = \begin{pmatrix} i\cos\theta & \sin\theta \\ -i\sin\theta & \cos\theta \end{pmatrix}
\,,\qquad \tan 2\theta = \frac{2m_D}{m_R - m_L} \,, \\ \\
m_{1,2} = \frac{1}{2} \sqrt{4m_D^2+(m_R-m_L)^2} \mp \frac{m_R+m_L}{2} \; .
\label{eq:diag}
\end{split}
\end{equation}
The corresponding mass eigenstates are given by 
\begin{equation}
\begin{pmatrix} \nu_{1\,L} \\ \nu_{2\,L} \end{pmatrix} =
\begin{pmatrix} i\cos\theta & -i\sin\theta \\\sin\theta  & \cos\theta \end{pmatrix}\,
\begin{pmatrix} \nu_L \\ n_R^c \end{pmatrix} \; .
\end{equation}
An important fact to note is that the diagonalized mass term 
\begin{equation}
\mathcal{L}^{\text{Mass}} = 
\frac{1}{2} \sum_{k=1,2} m_k \, \overline{\nu_{kL}^c} \, \nu_{kL} + \text{H.c.} 
\end{equation}
is a sum of Majorana masses for the Majorana fields 
\begin{equation}
\nu_k = \nu_{kL} + \nu_{kL}^c \; , \qquad  k=1,2 \,.
\end{equation}

There are two interesting physical limits. In the first
case, $m_D \gg m_L, m_R$. Then, $\nu_1$ and $\nu_2$ become
almost degenerate with a mass $m_{1,2} \simeq m_D$ and mixing angle
$\theta \simeq 45^\circ$. Thus, the two Majorana eigenstates form a
\emph{pseudo-Dirac} neutrino state, and the associated mass term in 
the Lagrangian conserves lepton number to a good approximation. 
$\nu_L$ and $\nu_R$ are the active and sterile components (in the SM),
respectively, and are maximally mixed. However, as we have already
mentioned (Sec.\ref{sec:diracmass}), this solution does not explain 
why neutrino masses are so small.

The other more interesting case is the see-saw limit~\cite{SEESAW},
obtained for $m_R \gg m_D ,  m_L$, in which case the masses for $\nu_1$
and $\nu_2$ are 
\begin{equation}
m_1 \simeq \frac{m_D^2}{m_R} - m_L , \quad \quad \quad 
m_2 \simeq m_R \; ,
\end{equation}
differing enormously from each other. If $m_L = 0$, $m_1$ increases with
decreasing $m_2$ for a fixed value of $m_D$. Furthermore, the mixing
angle $\theta$ approaches zero, so that $\nu_1$ and $\nu_2$ are
completely decoupled. Usually, $m_D$ is assumed to be of the order of
the quark and charged lepton masses, and $m_R$ is a large mass of the
order of some unification scale beyond the SM. This forces the mass of
the ordinary neutrino to be small, and postulates a very heavy
right-handed neutrino, therefore decoupled from low energy physics. 
This see-saw mechanism is by far the most attractive 
explanation of the smallness of the ordinary neutrino masses.

\subsection{Neutrino mixing}
\label{sec:NuMix}

The hypothesis of neutrino mixing was first proposed by
B. Pontecorvo for the $\nu_e \to \nu^c_e$
transitions~\cite{PONTECORVO}, in analogy with $K^0 \to \bar{K}^0$. 
After the discovery of the muon neutrino~\cite{LEDERMAN}, 
neutrino mixing between different neutrino flavors was 
introduced~\cite{MNS}. This phenomenon appears
because the neutrino flavor eigenstates
$\nu_\alpha$, $\alpha = e, \mu, \tau$, which take part in weak
interactions, are generally quantum-mechanical superpositions of
different neutrino mass eigenstates (see Sec.~\ref{sec:CCint}), 
which describe the propagation of neutrinos in space-time.

From a large variety of experimental data (see \cite{StandardModelTests:1995}), 
it is well known that there are just three neutrinos that participate in the weak 
interactions.
Hence, let us consider three left-handed chiral fields $\nu_{e L}$, $\nu_{\mu L}$, 
$\nu_{\tau L}$ that describe the three active flavor neutrinos and three corresponding 
right-handed chiral fields ${n_1}_R$, ${n_2}_R$, ${n_3}_R$ that describe 
three sterile neutrinos~\footnote{Let us remark, however, that the number of sterile 
neutrinos is not constrained by terrestrial experimental data, 
because they cannot be detected, and could well be different from three.},
which are blind to weak interactions. The corresponding 
mass term is given by Eq.~\eqref{eq:genmassterm} with
\begin{equation}
\mathcal{L}^{\text{Mass}} = - \frac{1}{2}
\, \begin{pmatrix} \overline{\nu_L^c}\, & \overline{n_R} \end{pmatrix}
\, \begin{pmatrix} M_L\, & M_D^T \\ M_D\, & M_R \end{pmatrix} 
\, \begin{pmatrix} \nu_L \\ n_R^c \end{pmatrix} + \text{H.c.} \; ,
\label{eq:massterm}
\end{equation}
where $M_L$, $M_R$ and $M_D$ are $3\times 3$ matrices and 
\begin{equation}
\nu_L = \begin{pmatrix} \nu_{e L}\\ \nu_{\mu L} \\ \nu_{\tau L} \end{pmatrix}\,
\qquad\text{and}\qquad
n_R = \begin{pmatrix} n_{1 R}\\ n_{2 R} \\ n_{3 R} \end{pmatrix}\, .
\end{equation}
The $6\times 6$ matrix $\mathcal{M}$ can be diagonalized by a unitary matrix 
$\mathcal{U}$, analogous to the one in Eq.~\eqref{eq:diag}.
The mass term is then written as
\begin{equation}
\mathcal{L}^{\text{Mass}} = - \frac{1}{2} \sum_{k=1}^{6} m_k
\,\overline{\nu_{kL}^c} \, \nu_{kL} + \text{H.c.} \; ,
\end{equation}
which is a sum of Majorana mass terms for the Majorana neutrino fields.
The mixing relations between mass and weak interaction eigenstates read 
\begin{equation}
\nu_{{\alpha}L} = \sum_{k=1}^{6} \mathcal{U}_{{\alpha}k} \, \nu_{kL} ,  
\qquad \alpha=e,\mu,\tau\,,\qquad
{n_a}_R^c=\sum_{k=1}^{6} \mathcal{U}_{ak} \, \nu_{kL} , \qquad a=1, 2, 3 \; .
\label{sec:mixing}
\end{equation}
This mass term allows the implementation of the two former limits. 
In the case of $M_{L,R} = 0$ we have three light Dirac neutrinos, and 
when the $M_R$ eigenvalues are much larger than the $M_L$ and $M_D$ 
entries, the see-saw mechanism is at work.  
This case is, however, considerably more complicated than the 
one generation case, briefly discussed above. 
The see-saw mechanism naturally gives a neutrino mass scenario with 
a light and a heavy neutrino sector, which are to a very good 
approximation decoupled. 

In the first case ($M_{L,R} = 0$) the mixing matrix $\,\mathcal{U}$ can be obtained
in two steps. One diagonalizes $M_D$ first, and afterwards splits the Dirac neutrinos
in their respective mass eigenstates (as explained above for one flavour).
The first step is fulfilled by a $\,\mathcal{U}$ matrix block-diagonal, 
with the (unitary) $3\times 3$ upper-left part $U$ describing the 
charged current interactions\footnote{When we refer to the block-diagonal structure
of $\,\mathcal{U}$ is this case, we do to this first step.}.
In the second case, with the eigenvalues $M_R$ much larger than $M_L$ and $M_D$, 
the matrix is only to a very good approximation block-diagonal. 
Although in neutrino oscillations, which is the main concern of this work, it 
can be in practice also taken to be exactly block-diagonal in the 
Majorana case. 
In Section \ref{sec:HeavyNu} we will briefly review the 
present as well as the expected limits 
on the small off-diagonal entries giving the mixing between 
light and heavy neutrinos.

For low-energy phenomenology, the upper-left submatrix $U$ is 
sufficient to describe the effective mixing of the active flavor neutrinos
\footnote{This matrix is analogous to the Cabibbo-Kobayashi-Maskawa matrix for quarks
(see Appendix~\ref{sec:quarkCC}), and it is usually called Maki-Nakagawa-Sakata (MSN) 
matrix \cite{MNS}. Hence, it diagonalizes the neutrino mass matrix 
in the current eigenstate basis where the charged leptons have a well-defined mass.}.
This scenario, called ``three-neutrino mixing'',
can accommodate the experimental evidence for neutrino oscillations
in solar and atmospheric neutrino experiments~\footnote{In definite models with very 
heavy right-handed neutrinos the baryon asymmetry observed in the universe can 
originate from the interactions of these neutrinos, and then from an initial 
lepton asymmetry~\cite{Fukugita:1986hr} 
(see for example Refs.~\cite{Branco:2002kt,Pilaftsis:2003gt,GonzalezFelipe:2003fi,Hambye:2004jf,Boubekeur:2004ez,Abada:2004wn}
for specific models and Ref.~\cite{Buchmuller:2004nz} for a review).}.

Taking into account the experimental constraints from current oscillation 
experiments, the best-fit value for the mixing matrix $U$ is~\cite{NuMatGiunti}:
\begin{equation}
U_{\mathrm{bf}}
\simeq
\left( \begin{matrix}
-0.83 & 0.56 &  0.00 \\
 0.40 & 0.59 &  0.71 \\
 0.40 & 0.59 & -0.71
\end{matrix} \right) \; .
\label{eq:NuMixExp}
\end{equation}
In terms of confidence limits at $90\%$ ($3\sigma$), the range of variation 
on the magnitude of the corresponding matrix elements are~\cite{NuMatConcha}:
\begin{equation}
|U|=\begin{pmatrix} 
(0.73)\,0.79-0.86\,(0.88)& 
(0.47)\,0.50-0.61\,(0.67)& 
0-0.16\, (0.23)\\
(0.17)\,0.24-0.52\,(0.57)&
(0.37)\,0.44-0.69\,(0.73)& 
(0.56)\,0.63-0.79\,(0.84)\\
(0.20)\,0.26-0.52\,(0.58)&
(0.40)\,0.47-0.71\,(0.75)& 
(0.54)\,0.60-0.77\,(0.82)
\end{pmatrix} \; .
\label{eq:NuMixExpLim}
\end{equation}
This mixing matrix, with all elements except $U_{e3}$ large,
is called ``bilarge''.
It is very different from the quark mixing matrix,
which has small off-diagonal entries.
Such a difference may be an important
piece of information for our understanding
of the physics beyond the SM,
which presumably involves some sort of quark-lepton unification.
 
\section{Neutrino oscillations}
Due to the different propagation of the neutrino mass eigenstates, the flavour 
content of the neutrino propagating state changes with time. Since neutrinos
are detected through charged current processes, which are sensitive to the
neutrino flavor, this oscillating behaviour may be observable.

\subsection*{Oscillations in Vacuum}

Neglecting the small mixing with the heavy sector, the active neutrinos 
relevant to oscillation experiments can be written 
\begin{equation}
\nu_{{\alpha}L} = \sum_{i=1}^{3} U_{{\alpha}i} \, \nu_{iL} , 
\qquad \alpha=e,\mu,\tau \; ,
\label{eq:3nmix}
\end{equation}
where $\nu_{1L}$, $\nu_{2L}$, $\nu_{3L}$ are the left-handed components of
the light neutrino mass eigenstates.
Let us assume that a flavour eigenstate $|\nu_\alpha\rangle$ was produced at $x=0$
and $t=0$, with 
\begin{equation}
|\nu_\alpha\rangle = \sum_{i=1}^3 U_{\alpha i}^*\,|\nu_i\rangle \; .
\end{equation}
The space-time dependence of the free mass eigenstates
$|\nu_i(t)\rangle$ of momentum $\vec{p}$ and energy $E_i = \sqrt{p^2+m_i^2}$
reads 
\begin{equation}
|\nu_i(t)\rangle = e^{i(\vec{p} \vec{x} - E_i t)}\,|\nu_i\rangle \; .
\end{equation}
It follows from Eq.~\eqref{eq:3nmix} that the space-time evolution
of the state $|\nu_\alpha\rangle$ is then given by 
\begin{equation}
|\nu_\alpha(t)\rangle = e^{i\vec{p}\vec{x}}\,\sum_{i=1}^3 U_{\alpha i}^*\,e^{-i E_i t} 
|\nu_i\rangle \; .
\label{eq:oscillation}
\end{equation}
If the masses $m_i$ are not equal, the three terms of
the sum in Eq.~\eqref{eq:oscillation} get out of phase and the
state $|\nu_\alpha(t)\rangle$ acquires components $|\nu_\beta\rangle$ with $\beta \ne
\alpha$. The probability with which the neutrino produced as
$|\nu_\alpha\rangle$ is converted into $|\nu_\beta\rangle$ in a different point of
space-time is given by 
\begin{eqnarray}
P_{\nu_\alpha \to \nu_\beta}  & = &|\langle\nu_\beta\,|\nu_\alpha(t)\rangle|^2 
= \left| \sum_i U^*_{\alpha i}\, e^{-i(E_i t-\vec{p}\vec{x})}\,U_{\beta i}\right|^2 
\label{eq:3families}\\ 
& \simeq & \delta_{\alpha \beta} - 4\sum_{i>j}\, U_{\alpha i} U_{\beta i}
U_{\alpha j} U_{\beta j}\, \sin^2 \left(\Delta m^2_{ij} \frac{L}{4 E}\right) 
\; , \nonumber
% & = & \textrm{Re}\, \sum_{i,j} U_{\alpha i} U^*_{\alpha j} U^*_{\beta
%i} U_{\beta j}\, e^{-i\frac{\Deltam_{ij}^2}{2E}x} \nonumber
\end{eqnarray}
where $L$ is the distance between the production and detection points,
$E=p$ the neutrino energy for $m_i = 0$, and $\Delta m^2_{ij} \equiv m^2_i-m^2_j$. 
In the last line we have assumed that neutrinos are highly relativistic 
($m_i \ll p$ and $L \equiv x=t$), and that the mixing matrix is real. 
We see that in order to have a non-vanishing oscillation
probability, it is necessary that at least two neutrino mass eigenstates are
non-degenerate, and that the matrix $U$ has non-vanishing off-diagonal
elements.

This probability satisfies some important relations. First of all, it
becomes evident that neutrino oscillation experiments are not directly
sensitive to the scale of neutrino masses, but to the difference of their
squares. CPT invariance implies $P_{\nu_\alpha \to \nu_\beta} =
P_{\nu^c_\beta \to \nu^c_\alpha}$. For two (or more) flavour mixing,
we also have $P_{\nu_\alpha \to \nu_\beta} = P_{\nu_\beta \to \nu_\alpha}$~\cite{CABIBBO}
if CP is conserved. 
These relations are valid when neutrinos propagate in vacuum. When
interactions with matter affect the neutrino propagation they may no
longer hold because the medium itself is in general not symmetric under
CP and CPT. Finally, the unitarity of $U$ guarantees that the total 
probability adds to 1,
\begin{equation}
P_{\nu_\alpha \to \nu_\alpha} = 1 - \sum_{\beta \ne \alpha}
P_{\nu_\alpha \to \nu_\beta} \; .
\end{equation}
The neutrino oscillation phenomena are described 
by five free parameters. Only two out of the three $\Delta m^2_{ij}$
appearing in Eq.~\eqref{eq:3families} are independent,
since $\Delta m^2_{21} + \Delta m^2_{32} = \Delta m^2_{31}$. On the
other hand, if we ignore the possibility of CP violation, the matrix $U$ 
is real and can be parameterized by three independent mixing angles,
$\theta_{12}$, $\theta_{13}$ and $\theta_{23}$.

For simplicity, it is customary to consider only the case of mixing
between two neutrino families. In such a case, the mixing matrix
$U$ involves only one real mixing angle $\theta$, and the 
oscillation probability simplifies to 
\begin{equation}
P_{\nu_\alpha \to \nu_\beta}  = \left| \delta_{\alpha \beta} -
\sin^2(2\theta) \sin^2 \left(\Delta m^2 \frac{L}{4E}\right)\right| \; .
\end{equation}
Hence, the oscillation amplitude is given by $\sin^2(2\theta)$ and the
oscillation length ($\lambda$) by
\begin{equation}
\lambda = \frac{4\pi\, E}{\Delta m^2} = \frac{2.48\, E
\textrm{ [GeV]}}{\Delta m^2 \textrm{ [eV}^2]} \textrm{ [km]} \; .
\end{equation}
Thus, in order to observe neutrino oscillations, $\lambda$ must be
less or of the order of the source-detector distance. This can be
rewritten as $\Delta m^2 \ge E/L$. Therefore, to search for small
squared mass differences, the distance $L$ should be large and/or the
energy of the neutrinos small.

Back to the general three flavour mixing framework, the observed 
hierarchy $\Delta m^2_{21} \ll \Delta m^2_{31} \simeq \Delta m^2_{32}$ 
makes the oscillation phenomena to decouple and the two flavour
mixing model to be a good approximation in limited regions. Each
oscillation probability can be written as the sum of a ``short'' and a
``long'' component 
\begin{equation}
P_{\nu_\alpha \to \nu_\beta} = P_{\nu_\alpha \to \nu_\beta}^{\rm short} +
P_{\nu_\alpha \to \nu_\beta}^{\rm long} \; ,
\end{equation}
with
\begin{eqnarray}
P_{\nu_\alpha \to \nu_\beta}^{\rm short} &\simeq& 4 U^2_{\alpha 3}
U^2_{\beta 3} \sin^2(\Delta m^2_{31} \frac{L}{4 E}) \; , \nonumber \\
P_{\nu_\alpha \to \nu_\beta}^{\rm long} &=& -4 U_{\alpha 1} U_{\beta 1}
U_{\alpha 2} U_{\beta 2} \sin^2(\Delta m^2_{21} \frac{L}{4 E}) \; . 
\nonumber 
\end{eqnarray}
If we assume the mixing matrix to be nearly diagonal, $\cos
\theta_{ij} \gg \sin \theta_{ij}$, then $P_{\nu_e \to \nu_\tau}$ and
$P_{\nu_\mu \to \nu_\tau}$ are dominated by the short component
and are sensitive to $ \Delta m^2_{31} \simeq \Delta m^2_{32}$, 
whereas $P_{\nu_\mu \to \nu_e}$ is
dominated by the long component and is sensitive to $\Delta m^2_{21}$. 
A neutrino oscillation experiment can be sensitive to either the short or
to both the short and the long components by selecting the
appropriate $L/E$ range. 
Tab.~\ref{table:ranges} shows the typical
energies, base-lines and minimum detectable $\Delta m^2$ for the usual
neutrino sources, namely, nuclear reactors, accelerators, cosmic ray
collisions at the atmosphere and the sun.
\begin{table}[!ht]
\begin{center}
\begin{tabular}{|c|c|c|c|c|c|}\cline{2-6}
\multicolumn{1}{c|}{}& \multirow{2}{*}{\textbf{Reactor}} & \multicolumn{2}{|c|}{\textbf{Accelerator}} & \multirow{2}{*}{\textbf{Atmospheric}} & \multirow{2}{*}{\textbf{Solar}} \\\cline{3-4}
\multicolumn{1}{c|}{}&                                 & \emph{Short base-line} & \emph{Long base-line}  &                              &                        \\\hline
\hline
$E$ (MeV) & $\leq$ 10 & 30 -- 10$^5$ &  30 -- 10$^5$ & 10$^3$ & $\leq$ 14 \\
$L$ (m)  & 10$^1$--10$^2$ & 10$^2$--10$^3$ &  10$^4$--10$^7$ & 10$^4$--10$^7$ & 10$^{11}$ \\
$\Delta m^2$ (eV$^2$) & 10$^{-2}$ & 10$^{-1}$ & 10$^{-4}$ & 10$^{-4}$ & 10$^{-11}$\\
\hline
\end{tabular}
\caption
[Features of the different usual neutrino sources]{Neutrino energies,
baseline distances and minimum testable mass square difference for
reactor, accelerator, atmospheric and solar neutrinos.}
\label{table:ranges}
\end{center}
\end{table}

\subsection*{Oscillations in Matter} 

When neutrinos propagate in matter, a subtle but important effect
takes place which alters the way in which neutrinos oscillate into
one another. In constrast with the oscillation in vacuum, where 
the time evolution of the neutrino state is given by the 
different propagation of the mass eigenstates, the presence of 
matter affects differently the neutrino flavor eigenstates. 
While all neutrino species have the same neutral 
interactions with matter, the $\nu_e$ weak eigenstate 
has a slightly different index of refraction 
than the $\nu_{\mu,\tau}$ weak eigenstates, 
due to its charged current interaction with the electrons in 
the medium.  
This different index of refraction for $\nu_e$ alters the time 
evolution of the system from that in vacuum.
This phenomenon is known as the MSW effect~\cite{MikSir,Wolf}.
However, it is of little relevance to the experimental discussion in 
the following Chapters.

\section{Limits on heavy neutrino masses and mixings}
\label{sec:HeavyNu}

The neutrino mixing matrix $\mathcal{U}$ in Sec.~\ref{sec:NuMix} 
can be safely assumed to be block-diagonal in neutrino oscillations, at least 
given the present experimental precision. For a discussion of possible deviations 
from unitarity of the upper-left block $U$, which might be 
eventually observable in neutrino propagation experiments, 
see for instance~\cite{kagan,bekman}. 
The small mixing between light and heavy neutrinos can be also probed 
in other experiments, in particular at large colliders \cite{Corfu}. 
In the following, and for completeness, we first discuss present limits and 
afterwards the expected bounds at the International Linear Collider (ILC) 
\cite{paper1,paper2}, 
because lepton colliders are more sensitive to heavy neutrino signals. 
Let us first introduce a more convenient notation for the $6\times6$ mixing 
matrix $\mathcal{U}$,   
\begin{equation}
\mathcal{U} = \left( \! \begin{array}{cc}
U & V \\ V' & U'
\end{array} \! \right) \,,
\label{eq:mixmassext}
\end{equation}
where $U$ ($U'$) describes the mixing between the light (heavy)
neutrinos and $V$ ($V'$) parameterises the light-heavy (heavy-light) neutrino
mixing. For $V = 0$ the matrix $U$ is the usual $3 \times 3$ unitary 
MNS matrix.

\subsection*{Indirect limits}

The most stringent constraints on light-heavy neutrino mixing result from 
tree-level contributions to processes involving neutrinos as external states 
like $\pi \to \ell \bar \nu$ and $Z \to \nu \bar \nu$, and from new one-loop
contributions to processes with only external charged leptons like $\mu \to e
\gamma$ and $Z \to \ell \bar \ell '$ 
\cite{kagan,LL,pil1,NRT,pil2,bernabeu,illana,BrancoSantiago}.
These processes constrain the quantities~\footnote{In the following 
in this Section the subindices $\ell$ and $i$ stand for $\nu_{\ell}$ 
and $\nu_{i}$, respectively, and label the corresponding current and mass 
eigenstates. The equality below is a reflection of the unitarity of $\mathcal{U}$.}
\begin{equation}
\Omega_{\ell \ell'} \equiv \delta_{\ell \ell'} - \sum_{i=1}^3 U_{\ell i}
U_{\ell' i}^* = \sum_{i=1}^3 V_{\ell i} V_{\ell' i}^* \,,
\end{equation}
because in the former case we must sum over the external light neutrinos (which
are not distinguished) and in the latter the sum is over the heavy neutrinos 
running in the loop. 
The first type of processes in particular tests universality. A global fit to
experimental data gives \cite{kagan,bekman}
\begin{equation}
\Omega_{ee} \leq 0.0054 \,, \quad \Omega_{\mu \mu} \leq 0.0096 \,, \quad
\Omega_{\tau \tau} \leq 0.016 \; ,
\label{eps1}
\end{equation}
with a 90\% confidence level (CL). These limits do not depend on the heavy
neutrino masses and are model-independent to a large extent. They imply that
heavy neutrino mixing with the known charged leptons is very small,
$\sum_i |V_{\ell i}|^2 \leq 0.0054$, 0.0096, 0.016 for $\ell = e,\mu,\tau$,
respectively. The bound on $\Omega_{ee}$ alone does not guarantee
that  neutrinoless double beta decay is within experimental limits for the
range of heavy neutrino masses we are interested in (larger or of the order of 
the electroweak scale) \cite{M}.

The second type of processes, involving flavour changing
neutral currents (FCNC), get new contributions only at the one loop level 
when the SM is extended only  with neutrino singlets, as in our case. These
contributions, and hence the bounds, depend on the heavy neutrino masses. 
In the limit $m_{N_i} \gg M_W$, they imply \cite{bernabeu}
\begin{equation}
|\Omega_{e \mu}| \leq 0.0001 \,, \quad |\Omega_{e \tau}| \leq 0.01 \,, \quad
|\Omega_{\mu \tau}| \leq 0.01 \,.
\label{eps2}
\end{equation}
Except in the case of the first two families, for which experimental constraints
on lepton flavour violation are rather stringent, these limits are of a similar
size as for the diagonal elements. An important difference, however, is that
(partial) cancellations may operate among heavy neutrino contributions. There
can be cancellations with other new physics contributions as well. 
We are interested in determining the ILC discovery potential and the limits on
neutrino masses and mixings which could be eventually established. Then,
we must allow for the
largest possible neutrino mixing and FCNC, although they require cancellations
or fine-tuning.

\subsection*{Direct limits}

In order to discuss the ILC limits we concentrate on the lightest heavy 
neutrino $N_1$, and follow the analysis in \cite{paper1}. 
We omit the subindex 1 and denote the 
corresponding mixing matrix elements $V_{\ell N}$.    
There are two interesting scenarios:
({\em i\/}) the heavy neutrino only mixes with the electron; ({\em ii\/}) it
mixes with $e$ and either $\mu$, $\tau$, or both. A third less interesting
possibility is that
the heavy neutrino does not mix with the electron. We discuss these three cases
in turn.
In the first two cases the heavy neutrino couples directly to the initial state 
$e^+ e^- \rightarrow N \nu$, and one must search for it in its main decay mode 
$e^+ e^- \rightarrow \ell W \nu$ with $W \rightarrow j j$. 
In Appendix \ref{sec:NuMassCalc} we provide 
a more detailed discussion of this process. Here, it is enough to note that 
one has to look for a peak in the $\ell jj$ mass distribution. 
In Fig.~\ref{fig:mnmw} (a) we plot the $ejj$ invariant mass $m_{ejj}$ for 
$V_{e N} = 0.073$. The center of mass energy is taken to be 500 GeV. 
The solid line corresponds to the SM plus a
300 GeV Majorana neutrino, being the dotted line the SM
prediction. The width of the peak is due
to energy smearing applied in our Monte Carlo and not to the intrinsic 
neutrino width $\Gamma_N = 0.14$~GeV. We must emphasize that the total cross 
sections and the derived limits do not depend on the nature of the heavy 
neutrino, but the angular distributions do depend on its Majorana or Dirac 
mass character.
\begin{figure}[!ht]
\begin{center}
\vspace{5mm}
\begin{tabular}{cc}
\includegraphics[width=7.2cm]{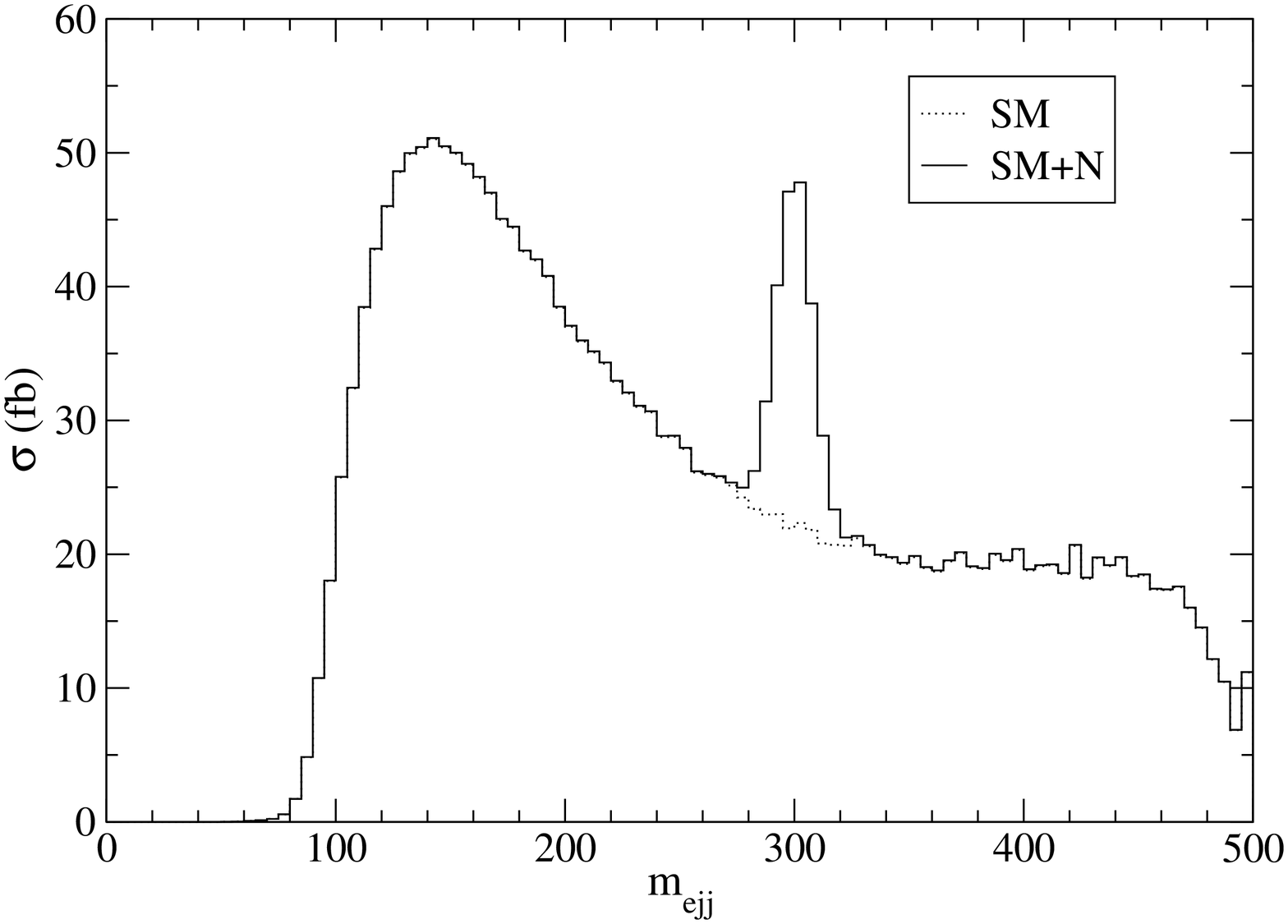} &
\includegraphics[width=7.2cm]{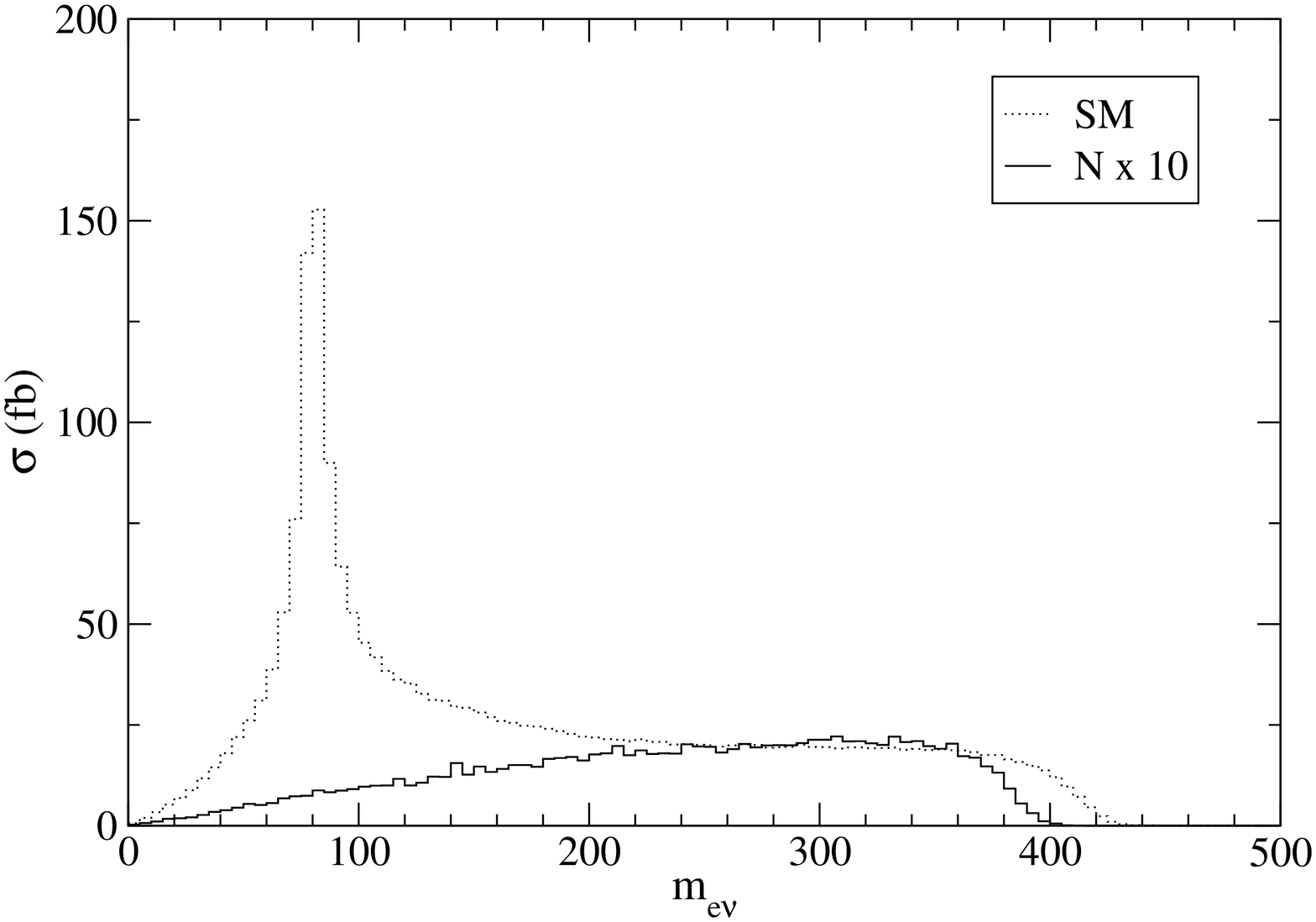} \\
(a) & (b)
\end{tabular}
\caption[$ejj$ and $e\nu$ invariant mass distributions at ILC]
{Kinematical distributions of the $ejj$ invariant mass (a)
and the $e\nu$ invariant mass (b).}
\label{fig:mnmw}
\end{center}
\end{figure}
In Fig.~\ref{fig:mnmw} (b) we show the $W \rightarrow e \nu$ peak 
which characterizes the main SM background $e^+ e^- \rightarrow W^+ W^-$. 

With convenient cuts on these two distributions in the different leptonic 
channels $\ell j j \nu$, $\ell = e, \mu , \tau$, one can derive the 
corresponding combined limits using the signal reconstruction method 
explained in \cite{paper2}. 
The integrated luminosity is assumed to be 345 fb$^{-1}$,
as expected after one year of running.
In Fig.~\ref{fig:limits-ILC} we plot 
the combined limits on $V_{eN}$ and $V_{\mu N}$ or $V_{\tau N}$ for 
the same heavy neutrino mass of 300 GeV.
At the ILC the sensitivities in the
muon and electron channels are similar, and both are better than in
$\tau W \nu$
production. This can be clearly observed in both plots: a $\mu N W$ coupling has
little effect on the limits on $V_{eN}$, but a coupling with the tau decreases
the sensitivity, because the decays in the tau channel are harder to observe.
The direct limit on $V_{eN}$, $V_{\mu N}$ obtained here improves the indirect
one (the solid line in Fig.~\ref{fig:limits-ILC}) only for $V_{\mu N} \lesssim 0.01$. 
However, as it has already been remarked, the latter is not general and can be 
evaded with cancellations among heavy neutrino contributions \cite{paper1}.
\begin{figure}[!ht]
\vspace{0.5cm}
\begin{center}
\begin{tabular}{cc}
\includegraphics[width=7.2cm]{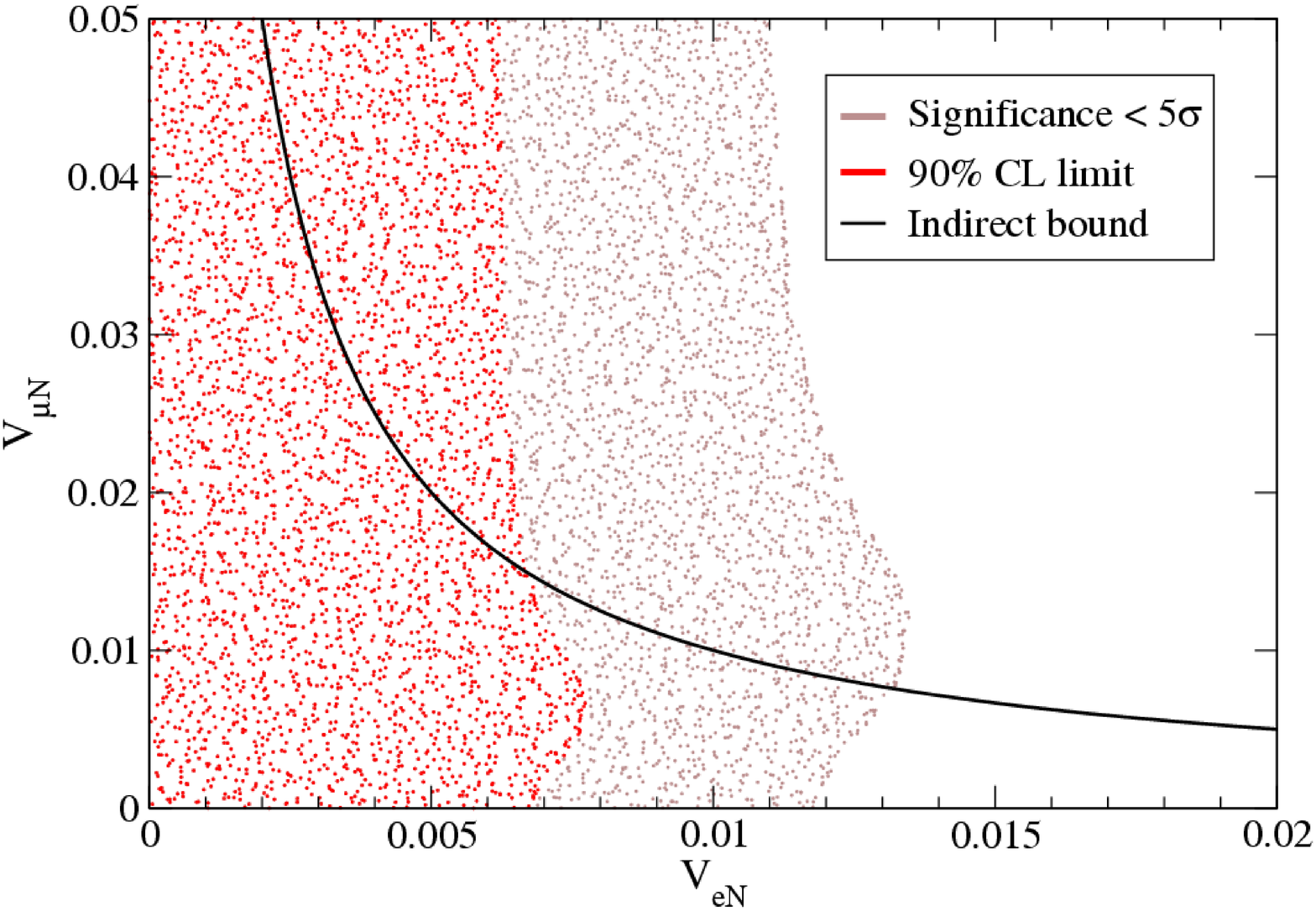} &
\includegraphics[width=7.2cm]{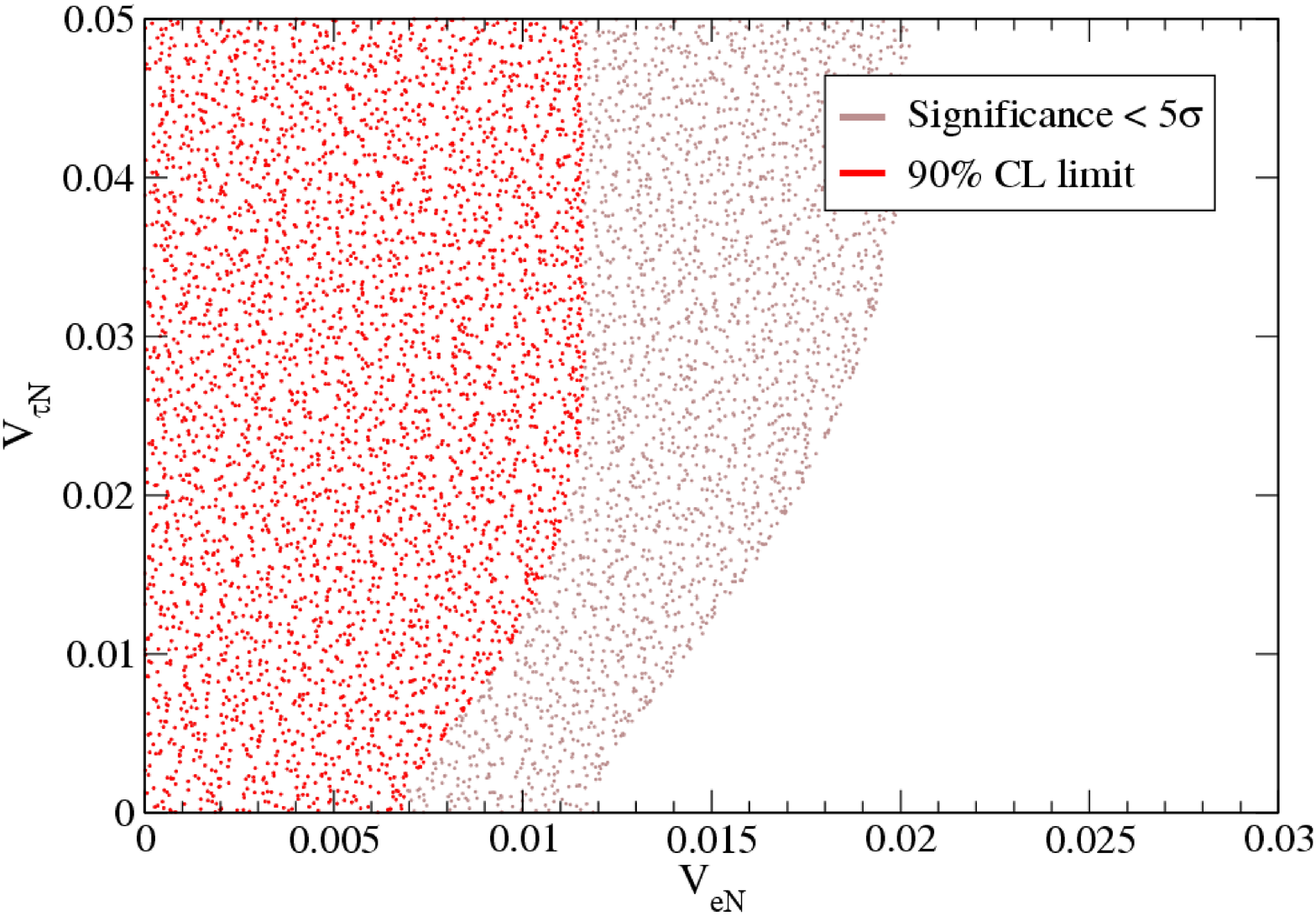} \\
(a) & (b)
\end{tabular}
\caption[Combined limits on $V_{eN, \mu N, \tau N}$ for $m_N = 300$ GeV at ILC]
{Combined limits obtained at ILC on: $V_{eN}$ and $V_{\mu N}$, for
$V_{\tau N} = 0$ (a); $V_{eN}$ and $V_{\tau N}$, for $V_{\mu N} = 0$ (b). The
red areas represent the 90\% CL limits if no signal is observed. The white areas
extend up to present bounds, and correspond to the region where a combined
statistical significance of $5\sigma$ or larger is achieved. The indirect limit
from $\mu - e$ LFV processes is also shown.}
\label{fig:limits-ILC}
\end{center}
\end{figure}

These limits are rather independent of the heavy neutrino mass. 
The dependence of the total $e^\mp W^\pm \nu$ cross section on $m_N$
can be seen in Fig.~\ref{fig:massdep-ILC} (a), for $V_{eN} = 0.073$, $V_{\mu N}
= V_{\tau N} = 0$. For a heavier $N$
the cross sections are smaller and thus the limits on $V_{eN}$ are worse.
However, up to $m_N = 400$ GeV this is compensated by the fact that the SM
background also decreases for larger $m_{ejj}$. The limits on $V_{eN}$ are
shown in Fig.~\ref{fig:massdep-ILC} (b) as a function of $m_N$, assuming
that the heavy neutrino only mixes with the electron.
\begin{figure}[!ht]
\vspace{0.4cm}
\begin{center}
\begin{tabular}{cc}
\includegraphics[width=7.2cm]{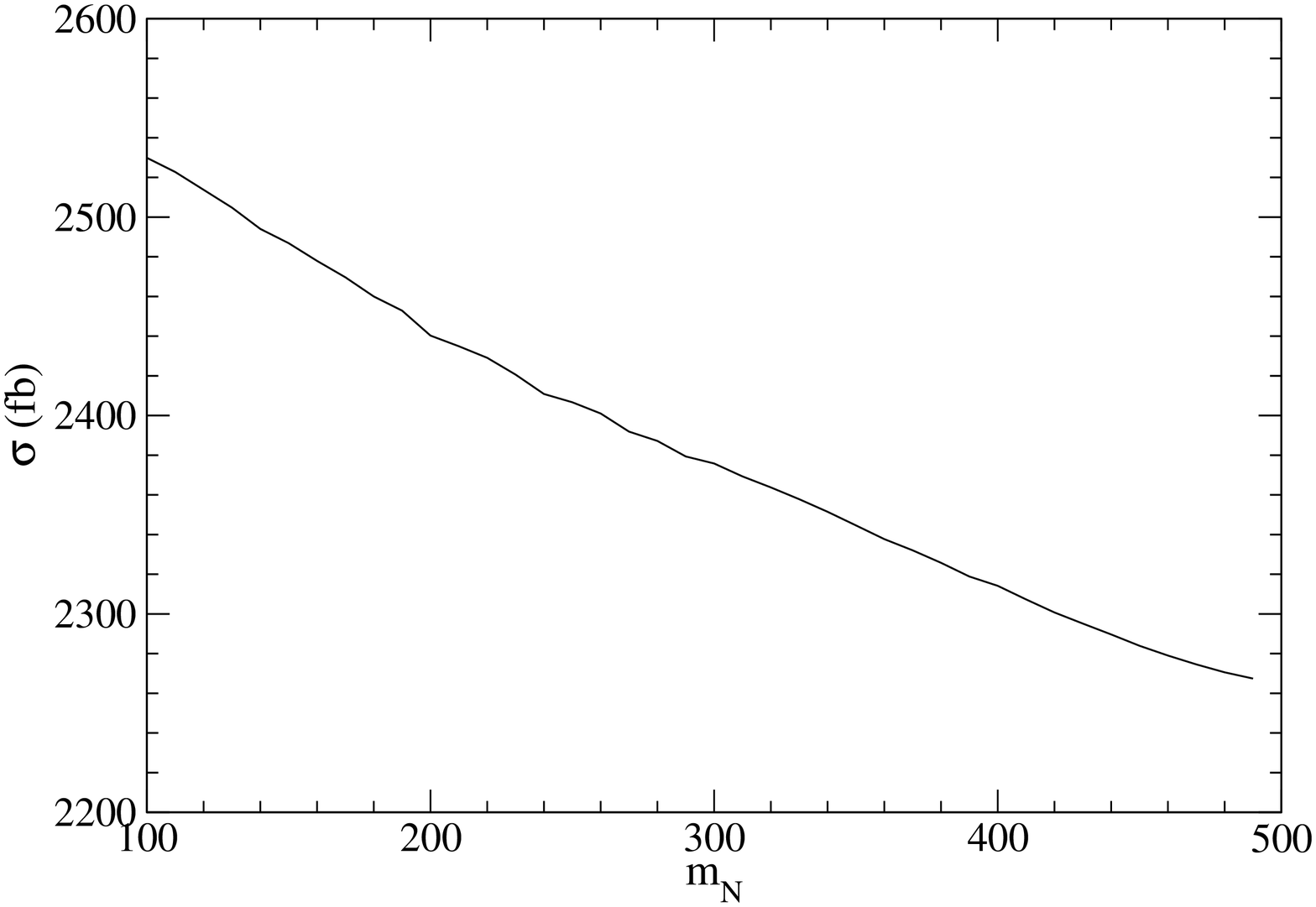} &
\includegraphics[width=7.2cm]{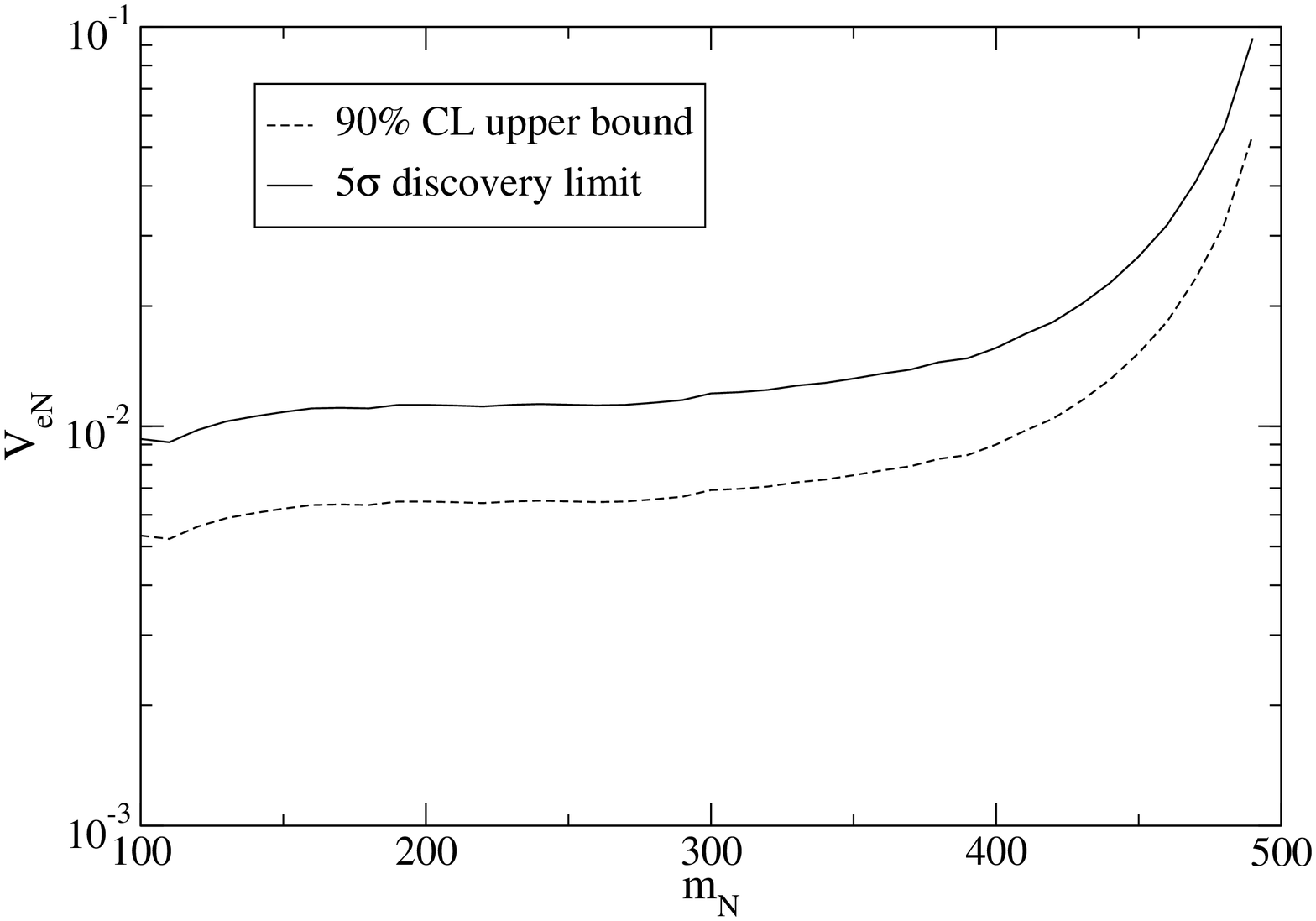} \\
(a) & (b)
\end{tabular}
\caption[Cross section for $e^+ e^- \to e^\mp W^\pm \nu$ 
and discovery limits on $V_{eN}$ at ILC]
{(a) Cross section for $e^+ e^- \to e^\mp W^\pm \nu$ at ILC for
$V_{eN} = 0.073$ and different values of $m_N$.
(b) Dependence of the discovery and upper limits on $V_{eN}$ on the
heavy neutrino mass. Both plots assume mixing only with the electron.}
\label{fig:massdep-ILC}
\end{center}
\end{figure}

A heavy neutrino signal in the $\mu$ or $\tau$ channels is observable 
only if the neutrino also mixes with the electron. 
We now quantify this statement. 
We consider a heavy neutrino coupling only to the
muon, with $V_{\mu N} = 0.098$, or only to the tau, with
$V_{\tau N} = 0.13$. The beam polarisations $P_{e^-}=0.8$,
$P_{e^+}=-0.6$, opposite to the previous ones, are used to enhance the signal
and reduce the SM background. The cross sections for the SM and SM plus a heavy
neutrino are shown in Tab.~\ref{tab:cs3} for these two cases. For $N$ mixing
only with the muon, the statistical significance of the signal is
$S/\sqrt B = 1.85$  for one year of running, and 7.3 years would be necessary
to observe a $5 \sigma$ deviation. If the heavy neutrino only
mixes with the $\tau$, the statistical significance is only
$S/\sqrt B = 0.76$, in which case a luminosity 43 times larger is required to
achieve a $5\sigma$ evidence.
\
\begin{table}[!ht]
\begin{center}
\begin{tabular}{|l|c|c|}
\cline{2-3}
\multicolumn{1}{c|}{}& $N-\mu$ & $N-\tau$ \\\hline
SM        & 1.10 & 0.389 \\
SM + $N$  & 1.20 & 0.414 \\\hline
\end{tabular}
\caption[$e^+ e^- \to \ell^\mp W^\pm \nu$ cross sections for 
a heavy neutrino coupling only to $\ell$]
{$e^+ e^- \to \ell^\mp W^\pm \nu$ cross sections (in fb) for 
a heavy neutrino coupling only to the muon (first column, $\ell=\mu$)
or coupling only to the tau (second column, $\ell=\tau$).}
%and convenient kinematical cuts.}
\label{tab:cs3}
\end{center}
\end{table}

\chapter{The Liquid Argon Time Projection Chamber}% and the ICARUS detector}
\label{chap:Larintro}

During the 1960s and 1970s, bubble chamber detectors~\cite{BUBBLE-I,BUBBLE-II,BUBBLE-III} 
led to the discovery of many new elementary particles. Filled with a
superheated liquid, the bubble chamber creates a track of small
bubbles when a charged particle crosses the chamber, locally bringing
the liquid to boil. Bubble chambers produce high spatial resolution
pictures of ionizing events, and allow a precise investigation of the
nuclear processes occurring in the medium. However, in recent times,
the use of bubble chambers has been superseded by electronic
detectors, since bubble chambers are difficult to trigger, and
experiments with high statistics are not really practicable because of
the time consuming analysis of bubble chamber pictures. The attempt to
merge the superb imaging capabilities of traditional bubble chambers
and the advantages of electronic read-out in a single detector led
C. Rubbia to propose the Liquid Argon (LAr) time projection chamber
(TPC) in 1977~\cite{Rubbia77}. The detector is essentially a cryostat
filled with a liquified noble gas and equipped with an electronic
readout that measures the ionization charge produced by the passage of
charged particles. This detector provides three-dimensional imaging
and, since the ionization charge is proportional to the energy
deposition, also acts as a calorimeter of very fine granularity and
high accuracy. Thus, this device is ideal to study particle
interactions and does not present the problems of traditional bubble
chambers, since the electronic read-out allows the self-triggering and
automatic processing and analysis of the events.

ICARUS is a project, proposed in 1985~\cite{ICARUS-I,ICARUS-II}, for
the installation of a large LAr TPC in the Gran Sasso Laboratory
(LNGS), Italy, for the study of neutrino physics and matter
stability. Several years of R\&D with prototypes of increasing size
have allowed to overcome the major technological problems in the
establishment of this technique.

\section{Principles of the Time Projection Chamber}

The Time Projection Chamber (TPC) is essentially a three dimensional tracking device,
capable of providing imaging of an ionizing particle track and a measurement of its
specific energy loss, {\sl dE/dx}. It consists of a pair of parallel electrodes immersed
in an ionization medium (a gas or a liquid) and connected to a high voltage power supply 
(see Fig.~\ref{fig:SchemaTPC}), producing a homogeneous electric field perpendicular to the electrodes.

\subsection{Detection principle}
%%%%%%%%%%%%%%%%%%%%%%%%%%%%%%%%
\label{sec:detprin}

A particle passing through the gas or the liquid will ionize the medium along the track
and create ion--electron pairs (see fig.~\ref{fig:SchemaTPC}). The applied electric field
suppresses the immediate recombination of the pairs (important mainly in a liquid
medium), pulling the electrons and ions apart. The ionization electrons of the particle
track will drift along the electric field lines to the anode, which is composed of 
an array of position sensitive sensors.

\begin{figure}[!ht]
\begin{center}
\includegraphics[width=9cm]{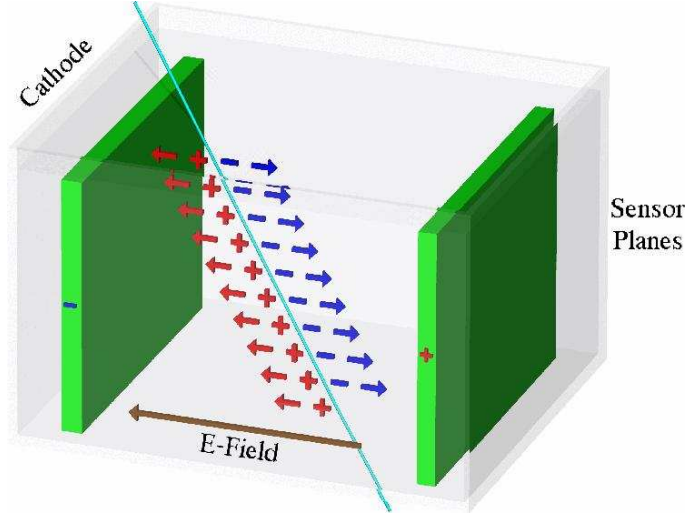} 
\caption[Schematic view of a TPC.]
{\label{fig:SchemaTPC} Schematic view of a TPC. A charged particle is passing through
an ionization medium (e.g. LAr) and the electrons drift along the electric field lines 
to the sensors.}
\end{center}
\end{figure}

\subsubsection{Imaging of events}
\label{sec:imaging}
%%%
The sensors located in the anode of the detector consist of a set of planes of 
multiple parallel wires (hereafter called \emph{wire planes}) (see fig.~\ref{fig:SchemaTPCwires}) . 
We can obtain two-dimensional projections of the ionization point, 
constraining one spatial coordinate from the wire collecting the charge (or a current being
induced) and the other from the drift time. In order to obtain
three-dimensional coordinates, at least two wire planes with different
wire orientations are necessary; therefore, the read-out system must be non-destructive.

\begin{figure}[!ht]
\begin{center}
   \begin{tabular}{ c }
   \includegraphics[width=10cm]{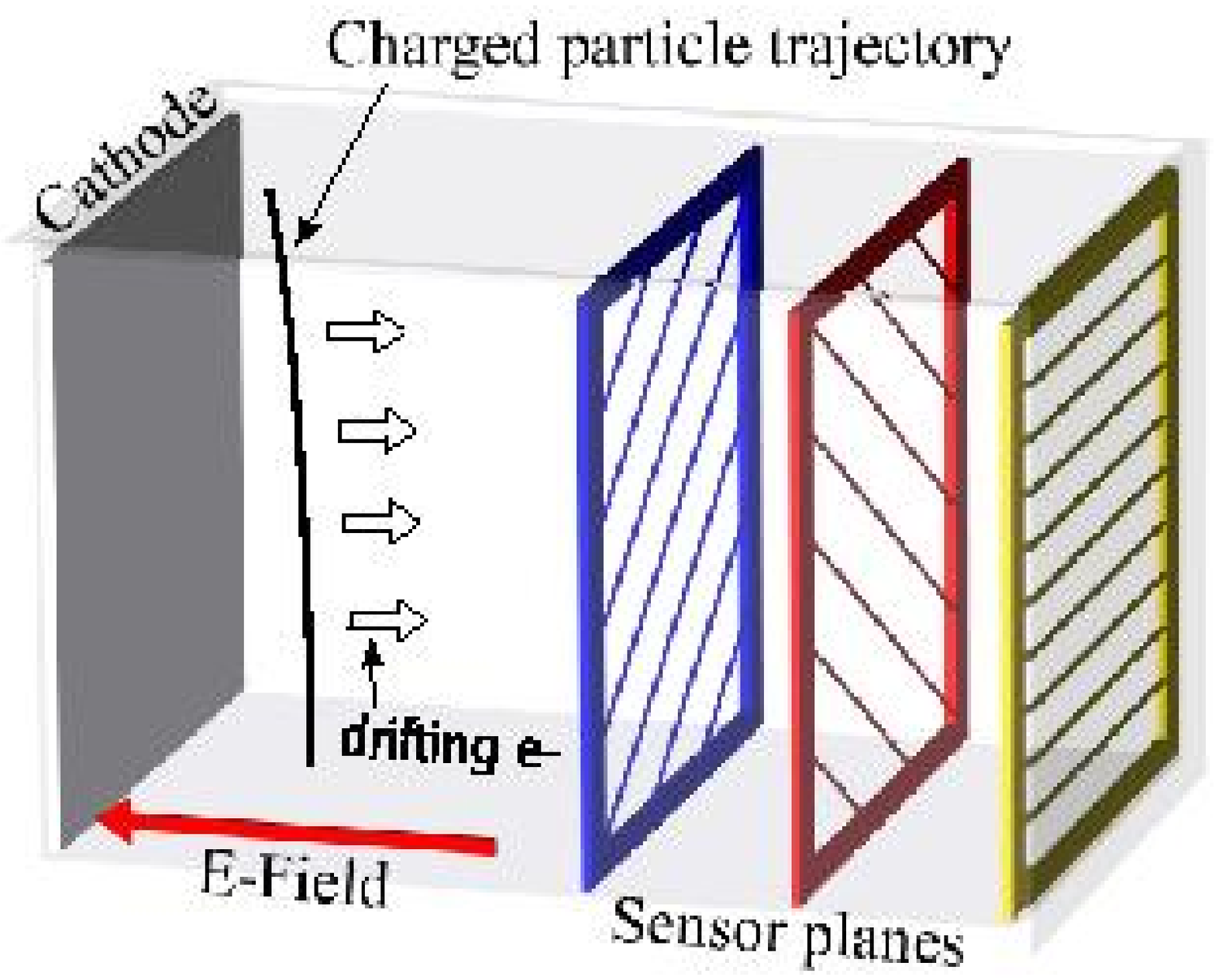}  \\[0.5cm]
   \includegraphics[width=12cm]{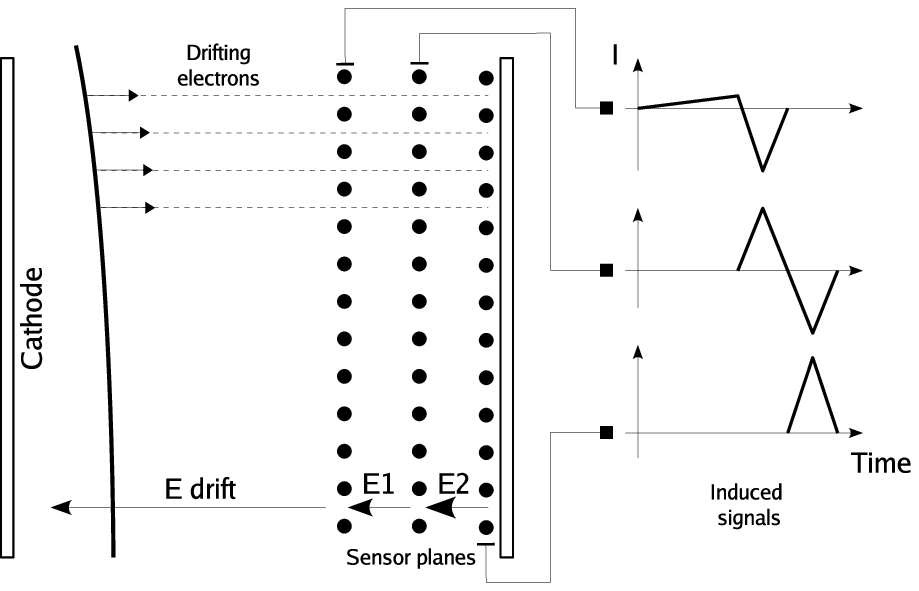}  \\
   \end{tabular}
\caption{\label{fig:SchemaTPCwires} Schematic view of a LAr TPC with three sensing planes.}
\end{center}
\end{figure}

The transparency of a wire plane or grid for electrons drifting along 
the electric field lines is a function of the ratio of the fields $E_1$ and $E_2$, 
in front and behind the grid respectively, and of the ratio $\rho = 2\pi r/p$ 
where $r$ is the wire radius and $p$ is the spacing between wires (\emph{pitch}). 
Full transparency is reached when the following condition is satisfied \cite{Bunemann}:

\begin{equation} 
\frac{E_2}{E_1}>\frac{1+\rho}{1-\rho}
\label{eq:bunemann}
\end{equation} 

This condition has to be balanced against the requirement that the grid has to
act as an electrostatic shielding between the space in front and behind it. If this
condition is satisfied, a drifting charge will be sensed by the electrodes behind the grid
only when the grid itself is crossed. Detailed calculations show that the shielding
power, $\sigma$, of a grid is approximately given by the following formula 
\cite{Bunemann}:
\begin{equation}
\sigma=\frac{p}{2\pi d}\ln \left(\frac{p}{2\pi r}\right)
\label{shieldingpower}
\end{equation}
where $d$ is the distance between the grid and the next electrode. Note, that shielding
factors deviating a few percent from 1 are acceptable.

The drifting electrons, following the electric field lines, can cross
a succession of several planes of parallel tensed wires oriented in
different directions (see Fig.~\ref{fig:SchemaTPCwires}), where the
condition~\eqref{eq:bunemann} is verified.
The electrostatic separation between the sensing wires allows a very sharp 
localization of the drifting charge, hence a good spatial resolution. 
Each of the wire planes provides a two-dimensional projection of the event image with a
common coordinate (the drift time), allowing the three-dimensional
reconstruction of the event. The basic three-dimensional pixel (or ``bubble'') size 
is determined by the wire pitch and the resolution in the drift direction.
The resolution cannot be improved by arbitrarily reducing the spacing of the wires.
In addition to the practical difficulties of precisely stringing wires at a pitch 
below $1\,mm$, there is a fundamental limitation: the electrostatic force between 
the wires is balanced by the mechanical tension, which cannot exceed a critical value. 
This gives the following approximate stability condition~\cite[Eq.~28.14]{PDG}:
\begin{equation}
\frac{p}{L}\geq 1.5 \times 10^{-3} V \sqrt{\frac{20\,g}{T}}
\label{wirestability}
\end{equation}
where $L$ is the wire length, $V[kV]$ is the voltage of the sense wire and $T[g]$ is the
tension of the wire in grams--weight equivalent.

%A minimum of two wire planes is required to fully determine the three dimensional 
%localization of the ``bubbles'', although additional planes can be used to resolve 
%ambiguities during the event reconstruction and improve the spatial resolution.

While approaching a plane, the electrons induce a current only on the
wires nearby; when moving away, a current of opposite sign is induced.
A detailed analysis of the induced signals for the geometrical
configuration displayed in Fig.~\ref{fig:SchemaTPCwires} has been
performed by Gatti et al.~\cite{GATTI}. Due to the shielding of the
adjacent grids, the signals have a limited duration.  Only the signals
on the wires of the first grid are ``prompt'', i.e. start as soon as
the ionizing particle crosses the chamber, and can be used to
determine the reference time for the electron drift ($t_0$). The time
$t_0$ can be also determined by the coincident detection of the
scintillation light produced in the argon, using a suitable 
system of photomultiplier tubes.

\subsubsection{Selection of the dielectric medium}
%%%
The liquid ionization medium has the advantage of a higher density and lower diffusion
compared to a gaseous medium. The higher density will permit to have an efficient target
for weakly interacting particles in a relatively small volume. The high stopping power of
the liquid leads to a short enough range of the (low energy) charged particles, to be
fully contained in the sensitive volume. For this type of events the LAr TPC works, in
addition to the imaging, as a calorimeter with fine granularity. LAr has a radiation
length of 14 cm and a nuclear interaction length of 83.6 cm giving good electromagnetic and
hadronic calorimetric properties. The high ionization density in the liquid gives enough
electron--ion pairs per unit length to detect directly the ionization charge without any
multiplication. Thus, the liquid has the advantage to give a high resolution spatial
tracking information, and, at the same time, acting as a continuous hadronic and
electromagnetic calorimeter. 
A summary of Physical and Chemical properties of argon can be consulted in 
Tab.~\ref{argonpropreties}.

A big difference for the signal measurement between gaseous and
liquid ionization media is given by the fact that there is no charge multiplication near
the sensing wires in a liquid. As a consequence of the short mean free path of drifting
electrons in liquid, they don't gain enough energy between the collisions to ionize other
atoms, and thus, cannot create an avalanche. The small amount of charge (about 20\,000
electrons for mips and 3 mm wire pitch) reaching the sensors must be read by a low noise
preamplifier with a high gain (see Sec.~\ref{sec:readout}).

 \begin{table*}[!ht]
 \begin{center}
 \begin{tabular}{|l|l|}
 \hline
 Atomic number & 18 \\ \hline
 Concentration in air & $0.934\%$\\ \hline
 \multirow{3}{*}{Naturally occurring isotopes} & ${}^{36}Ar=0.3365(30)\,\%$ stable\\ 
                                              & ${}^{38}Ar=0.0632(5)\,\%$ stable\\ 
                                              & ${}^{40}Ar=99.6006(30)\,\%$ stable\\ \hline
 Melting point (101325~Pa)& $83.8058\,K (-189.3$\textcelsius$)$\\ \hline
 Boiling point (101325~Pa)& $87.293\,K (-185.8$\textcelsius$)$\\ \hline
 \multirow{2}{*}{Density at boiling point (101325~Pa)}& $1.396\,kg/\ell$ liquid\\
                                                      & $5.79\,g/\ell$ gas\\ \hline
 Liquid heat capacity at boiling point (101325\,Pa)& $1.078\,kJ/kg/K$ liquid\\ \hline
 Latent energy of fusion at boiling point (101325\,Pa)& $161.0\,kJ/kg$ liquid\\ \hline
 $dE/dx_{min}$ for a mip & $2.12\,MeV/cm$ \\ \hline
 Critical energy (electrons) & $31.7\,MeV$ \\ \hline
 Mean excitation potential & $210\,eV$ \\ \hline
 Energy to produce an electron-ion pair & $23.6\,eV$\\ \hline
 Radiation length $X_0$ & $14.0\,cm$\\ \hline
 Moli\`{e}re radius & $9.28\,cm$\\ \hline
 Nuclear interaction length & $84.0\,cm$\\ \hline
 Maximal breakdown strength (depending on purity level) & $1.1-1.4\,MV/cm$\\ \hline
 $e^-$ Diffusion coefficient (89\,K) & $4.8\,cm^2/s$\\ \hline
 Recombination factor for mips ($\mu$) & $0.6$ at $0.5\,kV/cm$\\ \hline
 \end{tabular}
 \caption{\label{argonpropreties} Physical and chemical properties of argon.}
 \end{center}
 \end{table*}

\subsubsection{Calorimetry and particle identification}
%%%

Charged particles traversing the LAr sensitive volume produce
ionization electrons in a number proportional to the energy
transferred to the LAr. The ionization electrons drift along the
electric field lines to the wire planes pushed by the electric field
perpendicular to the wire planes, inducing a current on the wires near
which they are drifting while approaching the different wire
planes. Therefore, each wire of the read-out plane records the energy
deposited in a segment of the ionization track. Thus, the combined
spatial and calorimetric reconstructions, exploiting the fine
granularity and imaging capabilities of the detector, allows the
precise measurement of the energy loss per crossed distance ($dE/dx$)
and hence the identification of the ionizing particle.

%LAr is a non-compensating medium. However, in the described
%configuration, it appears as a completely homogeneous volume with very
%high read-out granularity. From the event visualization and from the
%local charge deposition density, it is possible to distinguish between
%electromagnetic and hadronic components of a shower and to
%approximately correct for the compensation factor.  

The complete calorimetric reconstruction of the events requires to
take into account the effects of the electron-ion recombination and
charge attenuation by electron attachment to electronegative
impurities. Both effects reduce the amount of collected charge with
respect to the one produced during the ionization. The recombination
of ion-electron pairs occurs immediately after the ionization, due to
their electrostatic attraction. The recombined electron-ion pairs
form excited atomic states that release the energy in the form of
(detectable) light. The magnitude of the recombination effect depends
on both the electric field and the ionization density
($dE/dx$). Charge attenuation is due to the attachment of the drift
electrons to the electronegative impurities, and depends also on the
electric field and on the concentration of electronegative impurities
present on the LAr volume. The precise knowledge of these two
phenomena allows their off-line unfolding from the calorimetric data.

\begin{figure}%[!ht]
\begin{center}
\begin{tabular}{c}
\includegraphics[width=10cm]{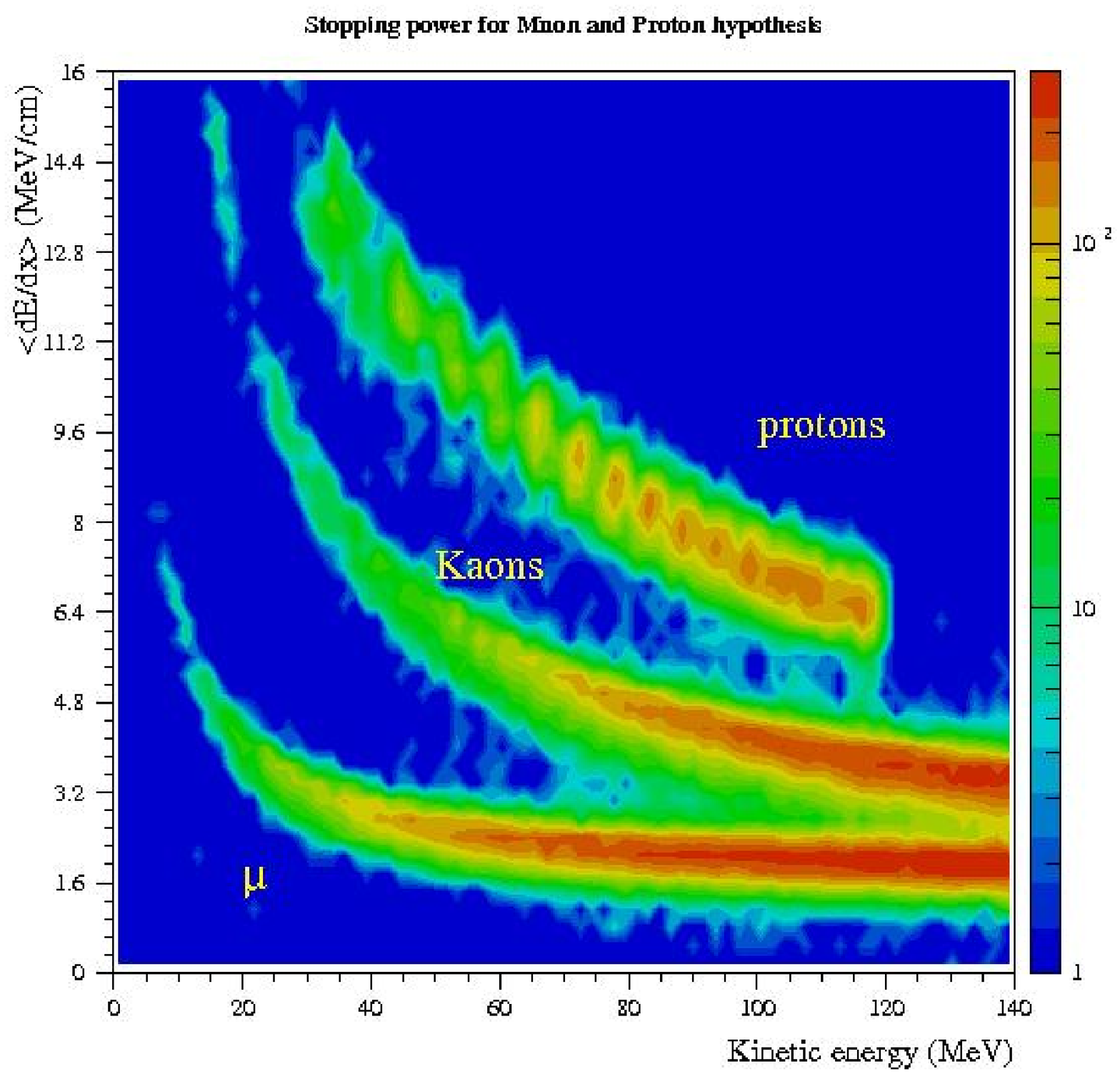} \\[0.5cm]
%\includegraphics[width=7.cm]{./Larintroplots/kaon} \\
%\multicolumn{2}{c}{\includegraphics[width=15cm]{./Larintroplots/PI2KASEP}}
\includegraphics[width=\textwidth]{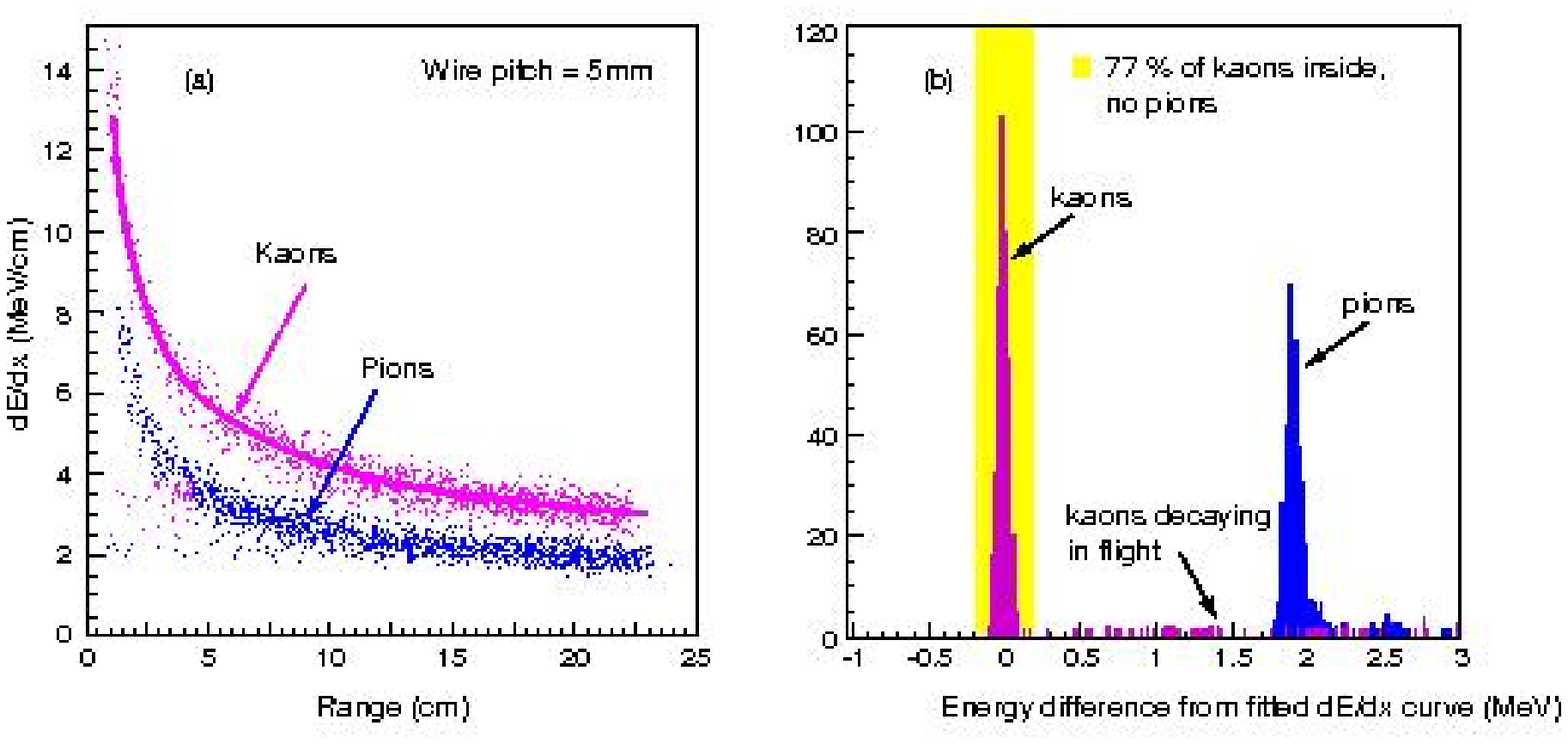}
\end{tabular}
\caption[Particle identification in ICARUS]
{(Top) Scatter plot showing the regions populated by fully simulated muons, 
kaons and protons on the $<dE/dx>$ vs. kinetic energy plane.
%(Top-right) Example of discrimination between single kaon and muon
%events (lines are drawn to guide the eye). 
(Bottom) Discrimination between kaon and pion events.}
\label{fig:kaon}
\end{center}
\end{figure}

If the granularity of the read-out electrodes is high enough, a
sampling of the energy loss per crossed distance ($dE/dx$) for several
points along the ionization tracks is possible. If, on the other hand,
the momentum of the particle is independently known, e.g.\ from its
curvature in a magnetic field, or from its range (for particles
stopping in the LAr sensitive volume only), or from multiple
scattering, we can determine the mass of the ionizing particle by
means of the Bethe-Bloch formula (see later section~\ref{sec:calrec}) 
which describes the energy loss per crossed distance ($dE/dx$),
and hence identify the ionizing particle. 
The particle identification capability~\cite{PID} is illustrated in 
Fig.~\ref{fig:kaon}, where the $dE/dx$ vs. kinetic energy behaviour 
of stopping muons, kaons and protons (top), and a sample of stopping 
pions and kaons (bottom), are compared.

\subsection{Technical considerations}
%%%%%%%%%%%%%%%%%%%%%%%%%%%%%%%%%%%%%

The particularities of the detection technique described in the two
previous sections, require special consideration of certain technical
aspects, namely:
\begin{itemize}
\item
A system to provide the time of the event occurrence ($t_0$) is needed
in order to trigger the data acquisition (DAQ) system and to determine
the absolute position of the event along the drift direction.
\item The electronic read-out system must be capable of detecting the signal
produced by a few thousand electrons.
\item
The level of purity of the LAr must be high enough so that the signal
is not strongly degraded along the drift path.
\end{itemize}
The present solutions provided by the ICARUS collaboration for each of
these aspects are summarized below.

\subsubsection{Trigger and $t_0$ determination}
%%%
The trigger system must be suited for the rare events to be detected
underground by ICARUS, and must provide the absolute time of the event
occurrence. In principle, the drift time difference between any two
points of the projected tracks allows to reconstruct the geometrical
topology of the events. However, the absolute position along the drift
direction must be known when correcting (off-line) the measured charge
for the effects of the charge attenuation. A precise knowledge of
$t_0$ is thus required to achieve a good energy resolution.

As we already mentioned in section~\ref{sec:imaging}, the signal on
the wire plane facing the drift volume starts as soon as the ionizing
particle crosses the detector, and the electron-ion pairs are
formed. This feature can be used to trigger the DAQ system (hence
referred as \emph{self-triggering} capability) and to provide the time
$t_0$. An alternative system can be provided by a set of
photo-multiplier tubes (PMT's) immersed in the LAr, which detect the
scintillation light produced in the LAr by the passage of an ionizing
particle. Scintillation photons are produced basically by two
different processes~\cite{IcaScintillation}:
\begin{itemize}
\item
by the direct excitation of an argon atom followed by an excited
molecule formation and de-excitation:
\begin{equation}
\mathrm{Ar}^* + \mathrm{Ar} \to \mathrm{Ar}_2^* \to 2\, \mathrm{Ar} + \gamma;
\end{equation}
\item
by the formation of a molecular state through recombination processes
between electrons and molecular ions:
\begin{equation}
\mathrm{Ar}^+ + \mathrm{Ar} \to \mathrm{Ar}_2^+ + e^- \to \mathrm{Ar}_2^* \to 2\, \mathrm{Ar} + \gamma
\end{equation}
\end{itemize}
In both cases the average photon wavelength is $\lambda =128$~nm, and
the emission occurs within a time window which depends on the excited
state. The de-excitation proceeds from two different states, with
decay constants of about 5~ns (fast component) and about 1~$\mu$s
(slow component), respectively. The detection of the scintillation
light requires the use of PMT's equipped with suitable windows (like
MgF$_2$, that transmits UV radiation down to 115~nm), or to provide
the PMT cathode with a proper wavelength shifting system. This trigger
system allows the detection of cosmic ray events with energies down to
a few hundred keV.

\subsubsection{Read-out system}
\label{sec:readout}
%%%

One of the main problems which this detection technique must cope with
is the relatively small amplitude of the signals. This is due to the
absence of charge amplification during the electron drift process or
near the anode.  Typically, 1~mm of a minimum ionizing track delivers
less than $10^4$ electrons in LAr. The signal is even smaller at low
electric fields due to the effects of the charge attenuation and the
electron-ion recombination. The imaging of ionizing events requires,
therefore, the use of low noise read-out electronics.

The present solution for the read-out system is the product of several
years of tests and experience gained by the ICARUS collaboration on
small scale prototypes~\cite{3tonfirst,3ton,3tonperf,3tonpuri}. 
The readout system is structured as a multichannel waveform recorder that
stores the charge information collected by each sense wire during the
drift of the electrons at a 2.5~MHz sampling rate. Each wire module
(16 wires) is equipped with a current integrating amplifier feeding a
10 bit flash ADC that samples the (multiplexed) signal with a
frequency of 40~MHz. 

\subsubsection{Argon purification}
%%%
\label{sec:purification}
In order to make long electron drift paths possible (of the order of
one meter or more) LAr must be ultra-pure. An electron in the liquid
undergoes about $10^{12}$ molecular collisions per second. Hence,
impurities with large attachment probability must be kept at very low
level, on the order or less than $0.1~$ppb (ppb=10$^{-9}$), to achieve
electron lifetimes in excess of a few milliseconds.  
The electro-negative impurities are mainly represented by oxygen, water,
carbon dioxide and, in minor concentrations, some chlorine and
fluorine compounds. 
If the impurity concentration is constant over the whole volume, the
charge decreases exponentially with the drift time:
\begin{equation}
Q(t)=Q(t_0) e^{-\frac{t}{\tau}}
\label{eq:puritydef}
\end{equation}
where $\tau$ is the mean lifetime of the electrons in argon. The lifetime is directly
connected to the impurity concentration $\rho$ by an inverse linear relationship
\cite{Buckley89}:
\begin{equation}
\tau[\mu s] \approx \frac{300}{\rho [ppb]}
\label{purityconvertion}
\end{equation}
Thus, the lifetime provides a direct measurement of the LAr impurity concentration 
($O_2$ equivalent).

Commercial LAr has a contamination level of the
order of a few ppm (ppm=10$^{-6}$) of oxygen equivalent, this corresponds to a mean
lifetime of only $0.3\,\mu s$, absolutely insufficient for a TPC.
A considerable amount of R\&D has been performed by the ICARUS collaboration 
in order to master the argon purification process \cite{bettini,600ton}.
LAr can also be contaminated inside the cryostat, by the degassing process of those
parts (walls, electrodes, cables, etc.) covered by the (hotter) argon
gas. Therefore, it is necessary not only to purify the LAr before the
filling of the cryostat, but also to ensure a continuous purification
inside the cryostat, forcing the LAr recirculation through a dedicated
purifying unit.

\section{The ICARUS detector}
\label{ICARUSproject}

The ICARUS collaboration is planning to build a large LAr TPC as a neutrino observatory
and a detector to search for nucleon decays \cite{ICARUS}. A possible detector would
contain 3000 tons of LAr, divided into five 600 ton (T600) modules; each T600 module
consists of two 300 ton half-modules with a common thermal insulation.
The first 300 ton module was successfully tested in 2001 at the surface of the Earth in
Pavia (Italy) \cite{600ton}; the first T600 module is being installed at the Gran Sasso
underground laboratory in Italy. The physics program contains the study of neutrino
oscillations with the long base line $\nu_{\mu}$-beam from CERN to the Gran Sasso~\cite{ICARUS}
with a search for $\nu_{\tau}$ appearance, measurements of solar, atmospheric and
supernova neutrinos and a sensitive search for nucleon decays.

\begin{figure}[!ht]
\vspace{0.5cm}
 \begin{center}
   \includegraphics[width=\textwidth]{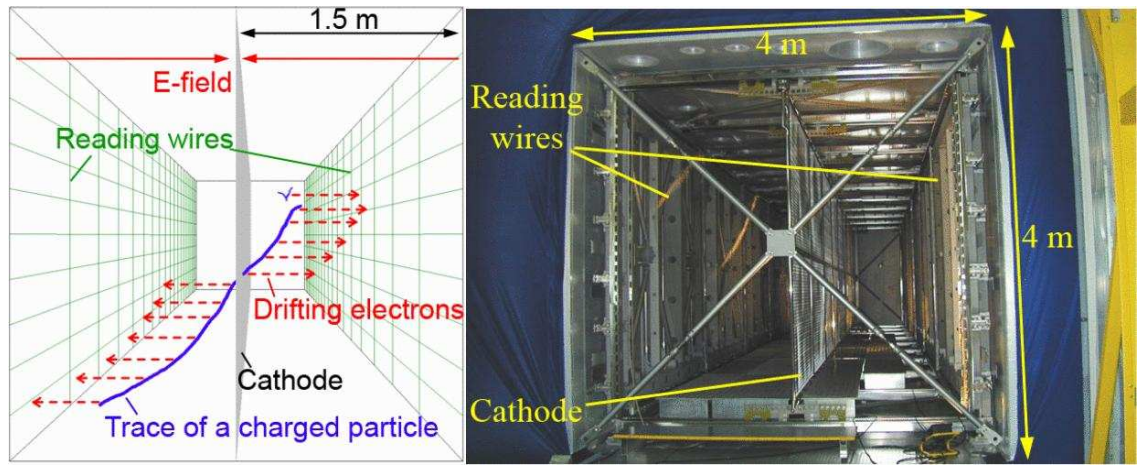}
 \caption[Picture of the open T300 ICARUS module during assembly.]
{\label{T600fig} Picture of the open T300 ICARUS module during assembly.}
 \end{center}
 \end{figure}

In an extensive R\&D program several small scale prototypes have been
constructed and operated, mainly at CERN, Pavia (Italy) and the Gran Sasso Laboratory
(Italy), in order to overcome the main technological problems in the establishment of the
ICARUS technology, and demonstrate many of the physics capabilities of the detector.

%\begin{figure}[!ht]
%\begin{center}
%\hspace{-0cm} 
%   \includegraphics[width=12cm]{./Larintroplots/5ktonICARUS}
% \includegraphics[width=14cm]{./Larintroplots/T3000}
%\caption{\label{5ktonfig} Artistic view of the ICARUS detector in the Gran Sasso laboratoy (Italy).}
%\end{center}
%\end{figure}

ICARUS T600 is a large cryostat divided into two identical, independent, adjacent
reservoirs (\textsl{half--modules}), containing more than 300\,tons LAr each (see Fig.~\ref{T600fig}). 
Both half--modules contain an internal detector (composed of two TPC's),
the field shaping system, monitors, probes and PMT's; they are externally surrounded by a
set of thermal insulation layers.

Outside the detector are located the read-out electronics (on top of the cryostat), and
the cryogenic plant composed of a liquid nitrogen (LN$_2$) cooling circuit and of a
system of purifiers needed to achieve the required LAr purity.

The inner detector structure of each half--module consists of two TPCs separated by a
common cathode. Each TPC is made of three parallel wire planes, 3 mm apart; the wires of
the first plane are horizontal and the orientation of the two other planes is $\pm$
60${}^{\circ}$ to the horizontal direction; the wire pitch is 3 mm.
%: The three wire
%planes of each TPC are held by a sustaining frame positioned onto the longest walls of
%the half--module. The number of planes in the detector has been chosen to be more than
%the minimal planes number (two for x-y reconstruction plus the time);
The third wire plane gives redundancy data for better pattern recognition. The total
number of wires in each half--module is 53\,248. The read-out of the signals induced on
the TPC wires by the electron drift allows the reconstruction of a full three-dimensional
(3D) image of the event; the 3D reconstruction of the event is obtained by correlating
the signals from the different planes at the same drift distance. Table \ref{ICARUS}
summarizes some technical data of the ICARUS T600  half--module (T300) detector.

 \begin{table}[!ht]
 \begin{center}
 \begin{tabular}{|p{4cm}|p{10cm}|}
 \hline
 Inner dimensions & $19.6\,m \times 3.6\,m \times 3.9\,m$ (length, width, height) \\ \hline
 Inner volume & $275.2\,m^3$ \\ \hline
 Active\footnotemark dimensions & $17.95\,m \times 3.0\,m \times 3.16\,m$ (length, width, height) \\ \hline
 Active volume & $170.2\,m^3$ \\ \hline
 Sensing planes & 1st Induction, 2nd Induction, Collection, 3~mm spacing \\ \hline
 Channel number & 53248 wires \\ \hline
 Electron drift length & $1.5\,m$ \\ \hline
 Maximal drift time & $1\,ms$ @ $500\,V/cm$ \\ \hline
 \end{tabular}
 \caption{\label{ICARUS} Some technical data of the ICARUS T600 half-module (T300).}
 \end{center}
 \end{table}

\addtocounter{footnote}{-0} \footnotetext{The active volumes are defined as the
parallelepipeds with the side surfaces equal to the wire chamber surface and with width
equal to twice the distance between the first wire plane (0${}^{\circ})$ and the
cathode.} \stepcounter{footnote}

An uniform electric field perpendicular to the wires is established in the LAr volume of
each half--module by means of a HV system, as required to allow and guide the drift of
the ionization electrons. The system is composed of a cathode plane, parallel to the wire
planes, placed at the centre of the LAr volume of each half-module at a distance of about
$1.5\,m$ from the wires of each side. The HV system is completed by field shaping
electrodes to guarantee the uniformity of the field along the drift direction, and by a
HV feedthrough to set the required voltage on the cathode. At the nominal voltage of
$75\,kV$, corresponding to an electric field of $500\,V/cm$, the maximum drift time in
LAr is about $1$~ms.

The top side of the cryostat hosts the exit flanges equipped with signal feedthroughs for
the electrical connection of the wires with the read-out electronics, and for all the
internal instrumentation (PMTs, LAr purity monitors, level and temperature probes, etc.).
The electronic chain is designed to allow for continuous read-out, digitization and
waveform recording of the signals from each wire of the TPC. It is composed of three
basic units serving 32 channels:
\begin{itemize}
\item The decoupling board receives analog signals from the TPC wires via vacuum-tight
feedthrough flanges and passes them to the analog board. It also provides biasing of the
wires and distribution of calibration signals.
\item The analog board houses the signal amplifiers, performs 16:1 multiplexing and the data
conversion (10 bit) at a 40 MHz rate. Thus every channel is sampled at $2.5$~MHz.
\item The digital board uses custom programmable chips (two per board) specially developed for
ICARUS, called DAEDALUS, that implement a hit finding algorithm. Each board receives the
multiplexed digital data via an external serial-link cable.
\end{itemize}

Ionization in LAr is accompanied by scintillation light emission. Detection of this light
can provide an effective method for absolute time measurement of the event as well as an
internal trigger signal. A system to detect this LAr scintillation light has been
implemented based on large surface (8\,in.) PMTs directly immersed in the LAr.

%The spatial reconstruction of ionizing tracks inside the LAr volume is performed by the
%simultaneous exploitation of the charge and of the light release following the energy
%loss processes of charged particles which cross the detector:
%\begin{itemize}
%\item electrons from ionization induce detectable signals on the TPC wires during their
%drift motion towards and across the wire planes (wire coordinate);
%\item UV photons from scintillation provide a prompt signal on the PMTs that allows the measurement
%of the absolute drift time and, hence, of the distance travelled by the drifting
%electrons (drift coordinate).
%\end{itemize}

%In this way, each of the planes of the TPC provides a two-dimensional projection of the
%event image, with one coordinate given by the wire position and the other by the drift
%distance. The various projections have a common coordinate (the drift distance).

The calorimetric measurement of the energy deposited by the ionizing particle in the LAr
volume is obtained by collecting information from the last of the three wire planes,
working in charge collection mode.

%\subsection{General overview of the detector.
% Results from a test in Pavia with cosmic rays. ICARUS physics
% program at LNGS: Solar and atmospheric neutrinos, $\nu$'s
% from supernovae, $\nu$'s from CNGS ($\nu_{\tau}$ appearance),
% measurement of $\theta_{13}$, search for nucleon decay}

% Purification (from paragraph 2.1): ... the use of a chemical
%filter\footnote{for ICARUS: Oxysorb (registered trademarks of Messer-Griesheim GmbH,
%Krefeld, Germany). In our experiment: fine metallic copper (copper monoxyde?) (see
%chapter \ref{secslowcontrol}} during the Liquid Argon filling and also sophisticated
%continuous purification, like in ICARUS. During the technical run with the ICARUS T600
%module (May--August 2001), the lifetime reached $1.8\,ms$, which corresponds to an
%impurity concentration of 0.17 ppb (part per billion) oxygen equivalent.

\chapter{The 50 liter Liquid Argon TPC}
\label{chap:50Ldetector}

The 50 liter Liquid Argon (LAr) Time Projection Chamber (TPC) is a detector 
built and successfully operated at CERN for R\&D purposes within the ICARUS programme. 
In the year 1997 it was 
exposed to the CERN neutrino beam for the entire SPS neutrino run period as proposed 
and approved at the SPSLC of January 1997~\cite{50Lrequest}. The detector, 
complemented with scintillators acting as veto, trigger counters and pre-shower 
counters, was installed in front of the NOMAD detector.
The collected data brought important information for a better understanding 
of the performance of Liquid Argon TPC's which should be useful for the entire 
ICARUS programme.

%The exposure of the prototype has given a substantial sample of quasi-elastic
%$\nu_\mu + n \rightarrow p + \mu^-$ events. In addition, further experience with 
%real neutrino events has been gained. 
%This experience has provided general information useful for
%the study of atmospheric neutrinos, proton decay and high energy neutrinos from CERN.
%For example the test has led to the optimization of the read-out chain in view of best
%extracting features of these events.

\section{The geometrical layout}
\label{sec:50Linfo}
The detector structure consists in a stainless steel cylindrical main vessel, 
70~cm diameter, 90~cm height, whose upper face is an UHV flange housing the 
feed-through's for vacuum, liquid Ar filling, high voltages 
and read-out electronics (Fig.~\ref{fig:50L_schema}). Inside the main vessel, 
an ICARUS type Liquid Argon TPC 
is mounted. The TPC has the shape of a parallelepiped whose opposite horizontal faces 
(32.5$\times$32.5~cm$^2$) act as cathode and anode, while the side faces, 47~cm long, 
support the field-shaping electrodes (Fig.~\ref{fig:G450L_TPC}). 
The mass of the Liquid Argon contained in the
active volume is 69~kg ($T=87$~K at 1~atm, $\rho=1.395$~g/cm$^3$).
Ionization electrons produced by the passage of charged particles drift vertically 
toward the anode by means of a constant electric field of 214~V/cm.

The read-out electrodes, forming the anode, are two parallel wire planes
spaced by 4~mm. Each plane is made up of 128 stainless steel wires, 100~$\mu$m diameter
and 2.54~mm pitch. The first plane (facing the drift volume) works in induction
mode while the second collects the drifting electrons (see Sec.~\ref{sec:imaging}). 
The wire direction on the induction plane runs orthogonally to that on the collection plane. 
The wire geometry is the simplest version of the ICARUS readout
technique~\cite{ICARUS} since both the screening grid and field wires in between 
sense wires have been eliminated. The wires are soldered on a vetronite frame which
supports also the high voltage distribution and the de-coupling capacitors.

\begin{figure}[t]
\begin{center}
  \begin {tabular}{cc}
    \includegraphics[width=7.cm]{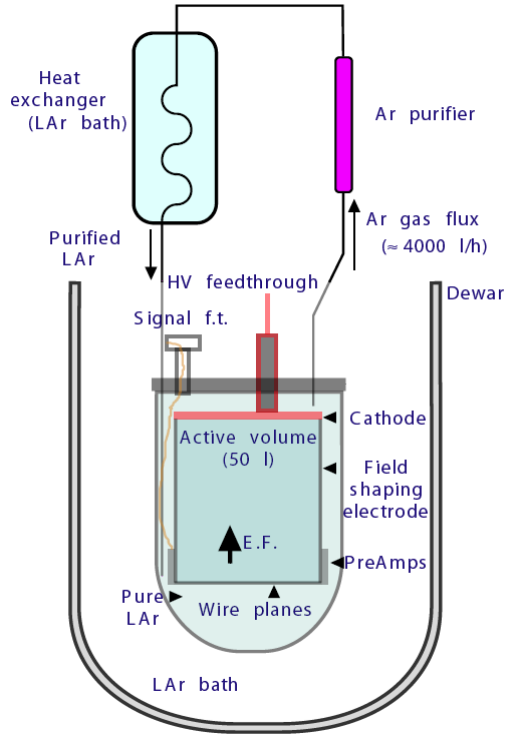} 
  \end{tabular}	
\caption{The 50 liter Liquid Argon Time Projection Chamber.}	
\label{fig:50L_schema}
\end{center}
\end{figure}

The cathode and the lateral field-shaping electrodes are copper
strips (5~mm thick and 1.27~cm wide) positioned on a vetronite support 
with printed board techniques. The support was glued on a honeycomb structure 
to ensure rigidity. The distance between two adjacent strips is 10~mm. 
The 10~kV drift voltage is distributed to the strips by means of 100~M$\Omega$ 
resistors. 
In Tab.~\ref{tab:50Ltech}, we have summarized the main technical data of the 
50L detector.

\begin{figure}[t]
\begin{center}
  \begin {tabular}{l r}
    \includegraphics[width=15cm]{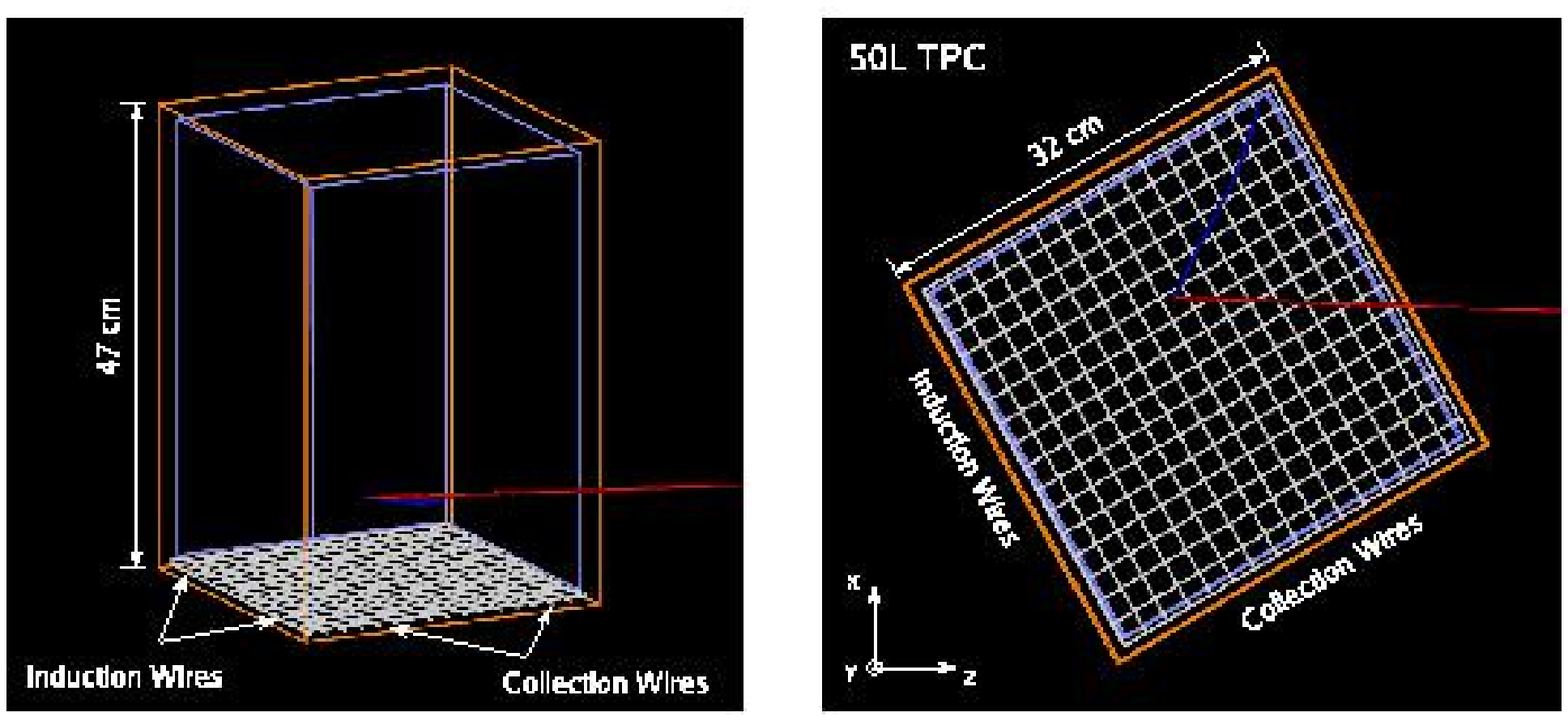}
  \end{tabular}	
\caption[Sketches of the 50 liter chamber]{Sketches of the 50 liter chamber.
Side (left) and top (right) views..}
\label{fig:G450L_TPC}
\end{center}
\end{figure}

\begin{table}[t]
\begin{center}
\begin{tabular}{|p{4cm}|p{10cm}|}
\hline
Inner dimensions      & $32.5\,cm \times 32.5\,cm \times 47.0\,cm$ (length, width, height) \\ \hline
Inner volume          & $49645\,cm^3\:\approx50\,l $                         \\ \hline
Active mass           & $67$~Kg                                              \\ \hline
Sensing planes        & Induction \& Collection (mutually orthogonal)        \\ \hline
Wire number           & 256 (128 per sensing plane)                          \\ \hline
Wire spacing (pitch)  & $2.54\,cm$                                           \\ \hline
Wire diameter         & $100~\mu m $                                         \\ \hline
Max. drift length     & $47\,cm$                                             \\ \hline
Electric field        & $214\,V/cm$                                          \\ \hline
Signal sampling rate  & @ $2.5\,MHz$ during $820\,\mu s$ (=2048 time samples)\\ \hline
\end{tabular}
\caption{\label{tab:50Ltech} Some technical data of the 50L detector.}
\end{center}
\end{table}

\section{The purification system}
\label{sec:electron_lifetime}

A crucial working parameter for a TPC operating with liquefied rare
gases is the lifetime of free electrons in the medium. Primary
electrons produced by the passage of charged particles drift toward
the anode crossing macroscopic distances. In the present case, for an
ionization electron created in the proximity of the cathode, the drift
path-length exceeds 46~cm.  The effectiveness of charge collection at
the readout plane is related to the purity of the Argon since
electron-ion recombination is mainly due to oxygen molecules present
in the LAr bulk~\cite{bettini}. The contamination of
electronegative molecules must be at the level of 0.1~ppb to allow
drifts over $\mathcal{O}(1)$ meters. 

This has been achieved using commercial gas purification systems to remove 
oxygen and polar molecules (H$_2$O, CO$_2$, fluorinated and chlorinated compounds) 
combined with ultra-high-vacuum techniques to avoid re-contamination of the 
liquid through leaks and with the use of low out-gassing materials for the 
detector components. Electron lifetimes higher than 3~ms (0.1~ppb) are easily and 
constantly reachable. Liquid phase purification allows fast filling of large detectors. 
Continuous recirculation of the liquid through the purifier during normal operation 
is used to keep the lifetime stable (against micro-leaks and out-gassing)
\cite{3ton,bettini,LArpuri}.

As mentioned in previous Sec.~\ref{sec:50Linfo}, the active part of the detector 
is located inside a stainless steel cylinder. The connection to the outside area is
obtained through a set of UHV flanges housing the signal, the high
voltage cables and the vacuum feed-through. The cylinder is positioned
into a 1~m diameter dewar partially filled with low purity Liquid
Argon acting as a thermal bath. 
The ceiling of the external dewar is in direct contact with air.  
The level of Argon in the dewar is so small that the pure Ar in the inner 
cylinder can evaporate as well. 

The detector is equipped with a standard ICARUS recirculation-purification 
system~\cite{3tonperf}. In this experiment the Argon purification is
carried out in the gaseous Argon phase through the recirculation system, 
while during the initial filling up stage is done in the Liquid Argon phase.
Once the filling up of the chamber finishes, the recirculation system takes
care of increasing the purity level first and then to keep it stable during
the data taking period.
\begin{figure}[t]
\begin{center}
\includegraphics[width=12cm]{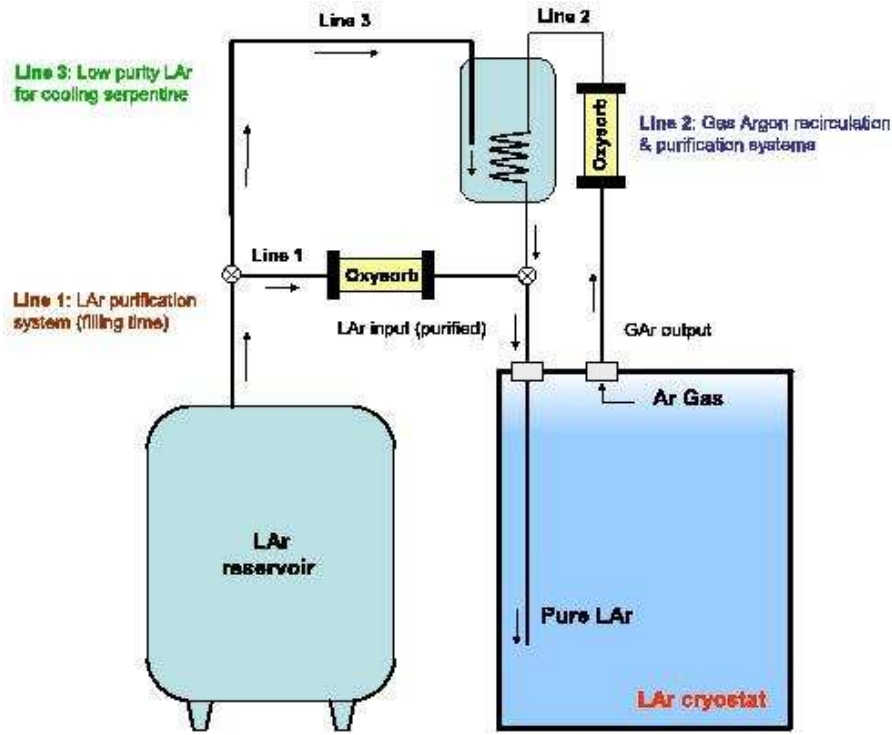} 
\caption
{\label{fig:50L_PurificationSystem} Scheme of the 50 liter 
Liquid Argon purification system.}
\end{center}
\end{figure}
This purification procedure in the recirculation phase is outlined 
in a simple way as follows (see Fig.~\ref{fig:50L_PurificationSystem}):
\begin{itemize}
\item The gaseous Ar in the active volume crosses the feed-through and 
reaches an Oxisorb filter for purification, thanks to pressure gradients. 
\item Now, the pure Argon goes through a cooling coil (inside a second buffer 
cooled by low-purity LAr) where it is liquefied again.
\item Finally, the pure LAr is collected back into the active volume. 
\end{itemize}

The Argon purity is characterized by the lifetime of the free
drifting electrons ($\tau$), defined by:
\begin{equation}
Q(t) \ = \ Q_0 e^{-\frac{t}{\tau}}
\label{eq:exp_law}
\end{equation}
$Q_0$ being the primary deposited ionization charge and $t$ the electron drift
time. The mean free path $\lambda$ of these drifting electrons is nothing but
the lifetime times the mean drift velocity, which have been measured in this
experiment (see Sec.~\ref{sec:driftvel}) to be 0.91~mm/$\mu$s.

\subsubsection{The purity monitor}
\label{sec:purmon}
The effectiveness of the purification system through direct
measurements of $\tau$ has been monitored during data taking by a
dedicated setup first developed by the ICARUS Collaboration in
1989~\cite{monitor_purity}. It consists of a small double-gridded
drift chamber located below the readout planes (see Fig.~\ref{fig:50L_PurityMonitor}). 
Electrons are generated near the cathode via photoelectric effect driven by a 20~ns
UV laser pulse. Each pulse produces a bunch of 10$^7$ electrons. They
drift toward the anode along the electric field lines and cross a 50~cm 
drift region between the two transparent grids. The ratio of the
induced signal in the proximity of the grids provides a real time
estimate of $\tau$ during data taking by means of the following relation:

\begin{equation}
R=\frac{Q_a}{Q_c}=\frac{T_c}{T_a}\frac{sinh(\frac{T_a}{2\tau})}{sinh(\frac{T_c}{2\tau})}\exp\biggl[-\frac{T_d+\frac{T_a+T_c}{2}}{\tau}\biggr]
\label{equ:lifetimerel}
\end{equation}
where $T_d$ is the time the drifting electrons spend from the first grid 
to the second, $T_c$ and $T_a$ are the time from the cathode to the first grid and
to the second grid to the anode respectively.

The relation~\eqref{equ:lifetimerel} can be understood in a very simple way:
When the electrons drift between cathode and window, the amplifier receives 
two equal and opposite currents that cancel each other. 
When the electrons drift between the window and the cathode grid a positive current 
flows in the amplifier; as a consequence its charge output increases linearly 
reaching a value $Q_c$, the charge leaving the window. 
When the electrons drift between the grids, 
no current flows in the amplifier (at least if the grids screen perfectly). 
Finally when the electrons drift between the anode grid and the anode, 
the current is negative. The output charge signal has a negative step, whose height 
is the charge $Q_a$ reaching the anode. 
If no charge is lost in the drift volume (zero impurity concentration), 
obviously $Q_a=Q_c$; if some charge is lost, corresponding to a lifetime $\tau$ 
the ratio of the charges is given by Eq.~\eqref{equ:lifetimerel}. In the usual case that
$\tau >> T_c,T_a$, we have a simpler expression:

\begin{equation}
R=\frac{Q_a}{Q_c}=\frac{T_c}{T_a}\exp\biggl[-\frac{T_d+\frac{T_a+T_c}{2}}{\tau}\biggr]
\label{equ:lifetimerels}
\end{equation}
An example of the charge output is shown in Fig.~\ref{fig:50L_PurityMonitor} (right). 
From a direct measurement of the purity monitor output signal 
(Fig.~\ref{fig:50L_PurityMonitor}), the collected charge $Q_a$, $Q_c$ 
and the time intervals $T_d$, $T_c$ and $T_a$ can be obtained and thus, the
electron lifetime $\tau$.
The initial lifetime during filling was about 100~$\mu$s. The recirculation/purification system,
circulating about 5 liters of LAr/hour allowed to increase this value to more
than 8~ms in three weeks (see Fig.~\ref{fig:50L_eLifetime}).
The total Argon consumption necessary to circulate the pure liquefied gas and to
compensate for the heat losses was $\sim$~200 liters per day.

\begin{figure}[!ht]
\begin{center}
  \begin{tabular}{ p{10cm} b{5cm} }
  \includegraphics[width=10cm]{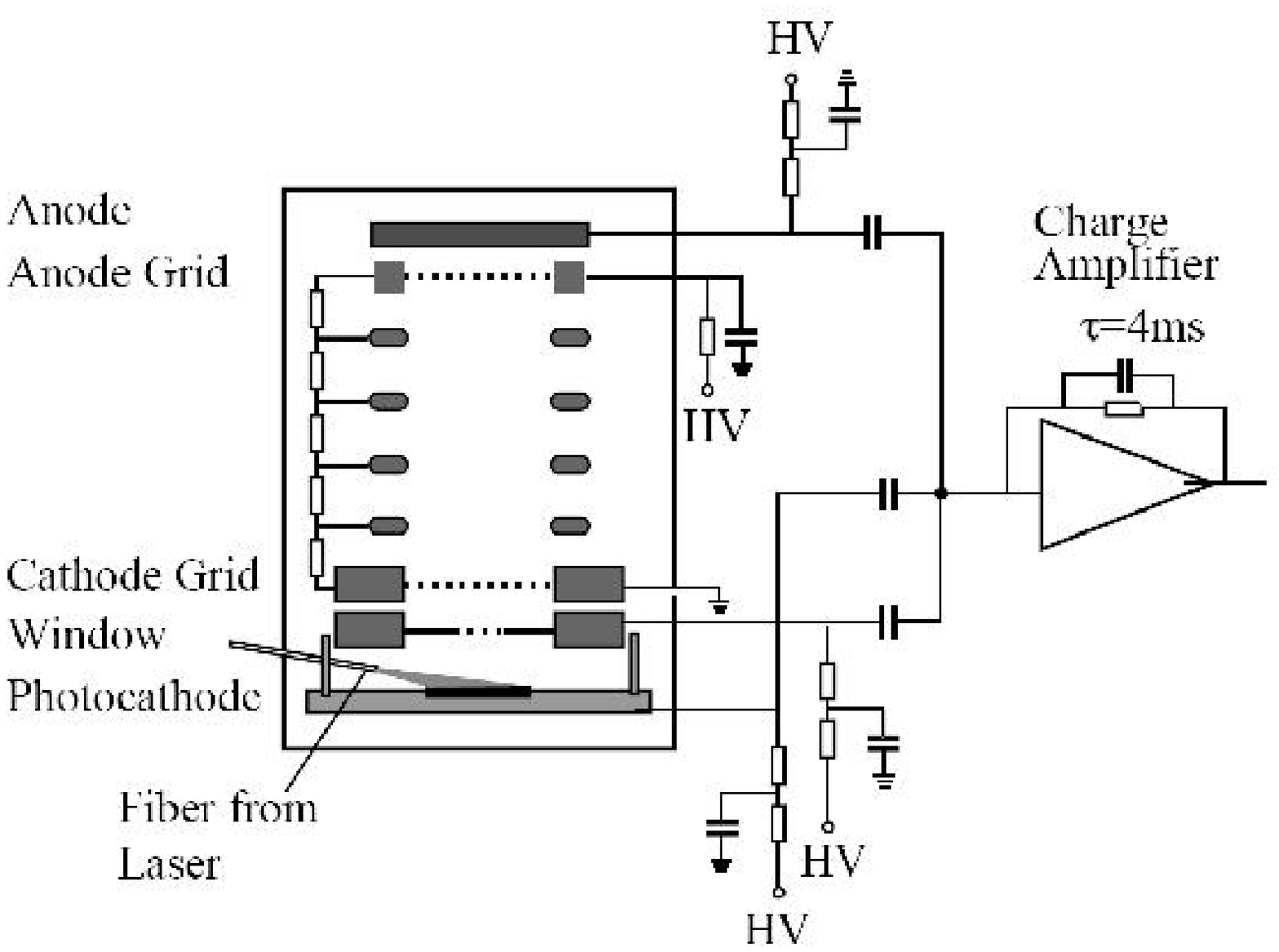} 
    &\includegraphics[width=4.5cm]{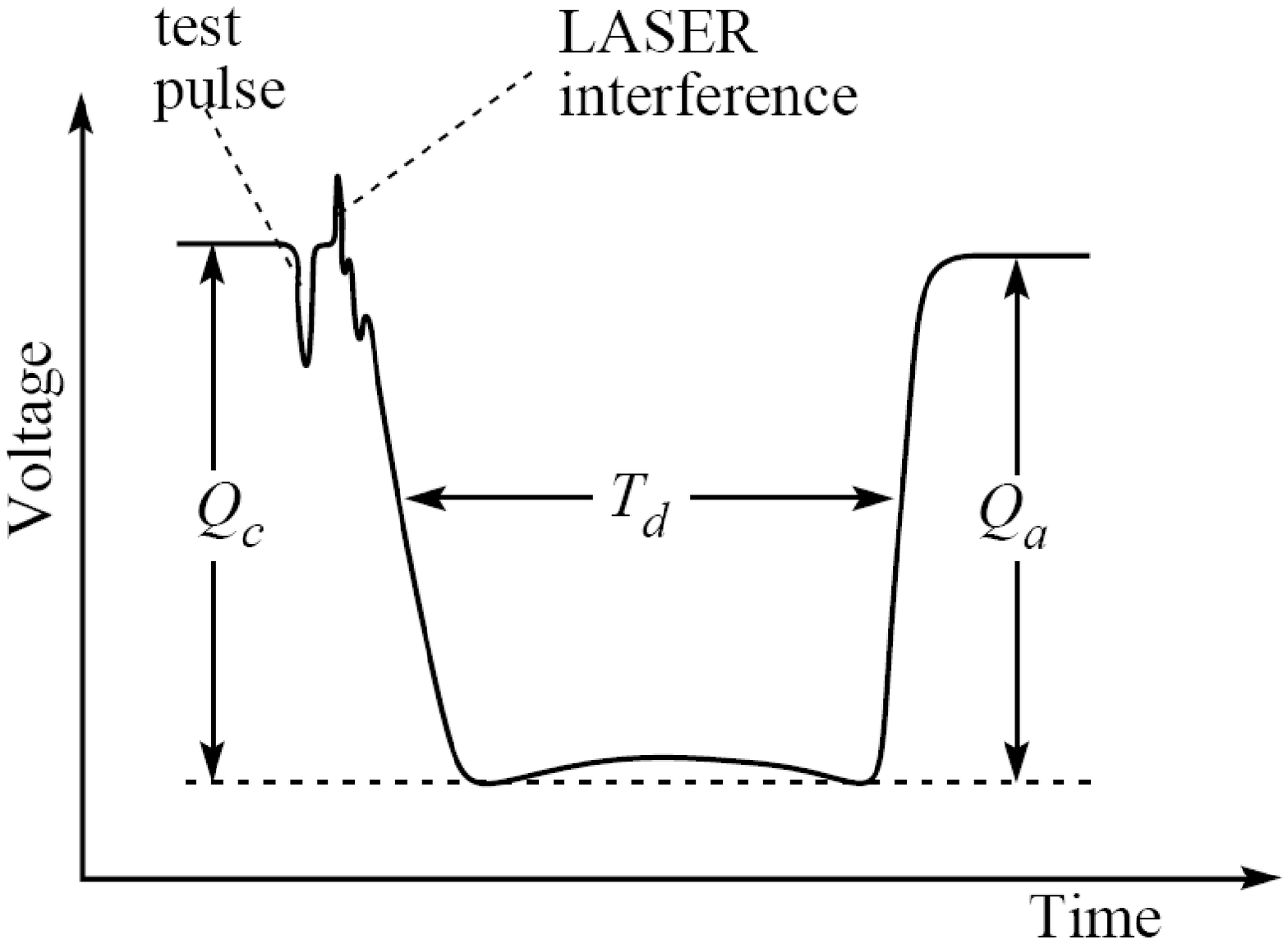} \\
  \end{tabular}
\caption[The purity monitor \& the charge output signal.]
{(Left) The purity monitor. 
(Right) The output charge signal.}
\label{fig:50L_PurityMonitor} 
\end{center}
\end{figure}

\begin{figure}[!ht]
\begin{center}
\includegraphics[width=12cm]{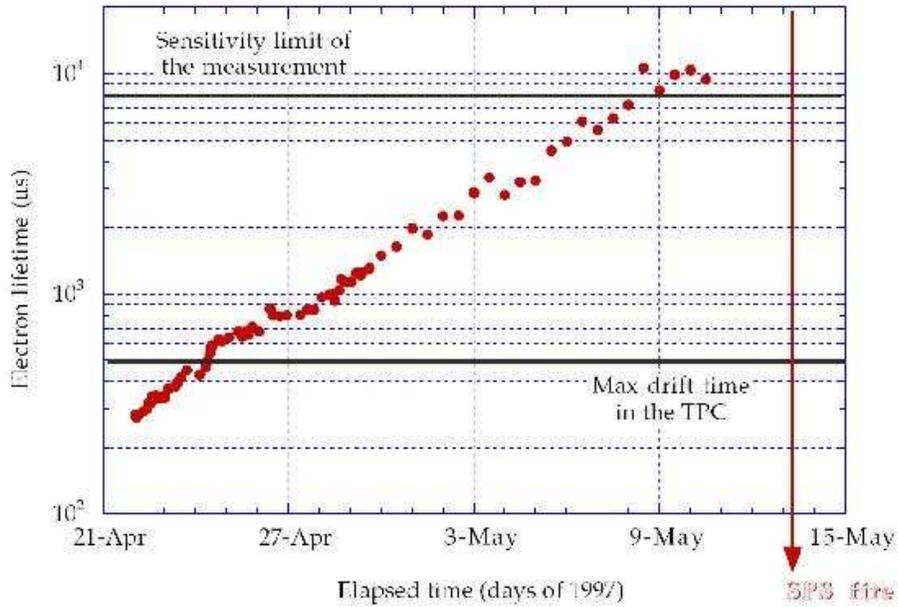} 
\caption[Lifetime of the drifting electrons in the 50 liter Liquid Argon TPC.]
{Lifetime of the drifting electrons in the 
50 liter Liquid Argon TPC. The filling of the chamber with LAr was performed on
the fourth of April 1997.}
\label{fig:50L_eLifetime} 
\end{center}
\end{figure}

An independent estimate of $\tau$ can be obtained monitoring the
charge attenuation for muons crossing the chamber at various heights 
(see~Sec.\ref{sec:lifetime2}). 
Both methods provided consistent results and the electron lifetime averaged
along the whole data taking period (9~months) turned out to be always higher than 10~ms.
This performance was excellent since the maximum drift time was about 500~$\mu$s
and therefore the attenuation of the ionization over the drift
distance can be considered negligible.

\section{The photosensitive dopant}
\label{sec:TMG}

As we discussed before, an ionizing particle in Liquid Argon produces, 
along the track, electron-ion pairs and excited atoms ($Ar^*$). 
The excited atoms decay, in few hundred nanoseconds, 
emitting ultraviolet photons with energy distribution peaked at 128~nm (9.7~eV). 
In the same time scale and depending on the ionization 
density and electric field, some of the pairs recombine also emitting UV photons. 
This produces saturation of the free charge as a function of the deposited energy 
and electric field. A possible solution to avoid this consists in dissolving in 
Liquid Argon a  photosensitive dopant able to convert back the scintillation light 
into free electrons, thus linearizing the dependence of the charge on the energy 
and the electric field. 

During this test, the Argon has been doped with Tetramethyl-Germanium
as photosensitive dopant because of the following advantages~\cite{TMG}:
\begin{itemize}
\item TMG is not absorbed by the Oxisorb filter in the recirculation system.
\item TMG can be easily purified to an electron lifetime level better than 10~$\mu$s.
\item TMG has a large photo-absorption cross section (62~Mbarn) and large quantum
efficiency (close to 100~\%); this implies that small quantities ($\approx$~ppm) 
of TMG are enough to convert all the de-excitation photons into electrons in the 
vicinity of the ionizing track.
\item TMG, at the ppm level, dissolves homogeneously in Liquid Argon.
\end{itemize}
The relation between the collected charge and the energy deposition was obtained
in the R\&D phase of the 3-ton ICARUS prototype (see~\cite{TMG})
from the analysis of the stopping muon and proton events at several electric fields and
TMG concentration of 1.3 ppm and 3.5 ppm. 
Fig.~\ref{fig:50L_TMG} illustrates such relation before and after the doping.

Argon excited atoms ($Ar^*$) produced by ionizing radiation lead to $Ar^*_2$ low excited
dimer formation through collision with $Ar$ atoms.
The rise-time corresponding to excimer formation and relaxation are in the 
sub-nanosecond range. For relatively low concentrations of the photodopant TMG
the process which drives the recovering of drifting electrons from the
scintillation light is a photoionization which follows the photoemission:
\begin{equation*}
\begin{split}
Ar^*_2 \ \rightarrow \ & Ar + Ar + \gamma  \\
\gamma + F \ \rightarrow \ & F^+ + e^-
\label{equ:sciTMG}
\end{split}
\end{equation*}

It is evident that the linearity of the detector response, after doping with TMG, 
is significantly improved especially at low fields where the recombination effect 
in pure Liquid Argon is higher. 
Continuous recirculation of the Argon through the purifier keeps the 
level of contamination stable against micro-leaks or out-gassing. 
Moreover no degradation of the space resolution has been measured, implying that 
the photo-conversion happens very close to the track.

\begin{figure}[t]
\begin{center}
\includegraphics[width=12cm]{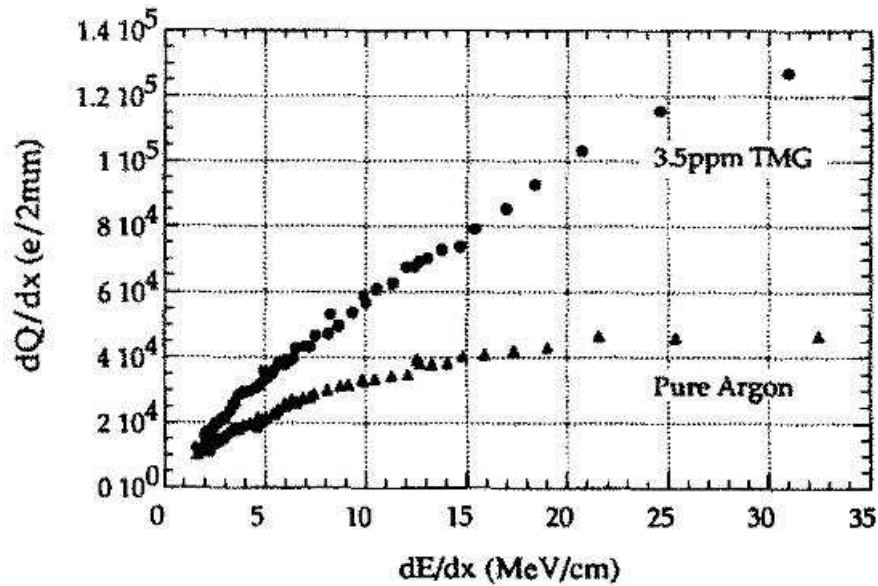} 
\caption[Correlation between charge and deposited energy with and without TMG]
{\label{fig:50L_TMG} Correlation between collected charge and deposited energy
at an electric field of 200 V/cm for pure Argon and with a TMG concentration 
of 3.5 ppm.}
\end{center}
\end{figure}

\section{Read-out set-up and data acquisition}
\label{sec:50Lreadout}
The mean charge per unit pitch (2.54~mm) for a mip crossing the chamber horizontally
corresponds to about $2.3 \times 10^4$ electrons ($\sim$4 fC) assuming
no charge loss along the drift volume. The smallness of the signal
and the absence of an amplification phase requires very low noise
charge pre-amplifiers. This has been reached by means of JFET technology.

\subsubsection{The pre-amplifiers}
\label{sec:preampli}
There is one preamplifier per read-out channel (256~wires), directly placed in the
vetronite frame which supports also the read-out wires.
Since the noise depends on temperature and input capacitance, a further improvement
is obtained placing the pre-amplifiers into the liquefied gas (at -186\textcelsius).
The polarization voltage (15~V) is distributed separately to each preamplifier 
to minimize the number of dead channels in case of failure of one of the components 
during data taking. 

Each one of these pre-amplifiers is compound by an integrator with a time constant of
$\tau_1 \sim 200~\mu s$ and a CR output impedance with time constant of $\tau_2 = 1~ms$:
The signal of the input current is integrated through the minimization of the input
impedance in conjunction with the appropriate frequency response of the preamplifier 
(this minimization also allows a reduction of possible interferences between the signals 
of the different channels).

\subsubsection{The amplifiers}
\label{sec:ampli}
The amplifiers gather the signal from the pre-amplifiers with typical
gain of 18~$mV/fC$ and an output impedance of 120~$\Omega$. They are of TL.026
type, with a differential stage at the input and the output followed of 
a {\it buffer} stage. There were located outside the chamber at room temperature.

During the whole data taking period the amplifiers have been operating in current
(or ``quasi-current'') mode, which allows a rather good visualization of the ionizing
events, with a signal to noise ratio of 11 for mip signals 
(see Fig.~\ref{fig:50L_WireSignal}) and no saturation of 
the electronic response even in the occurrence of highly ionizing events 
(i.e. electromagnetic showers).

\subsubsection{Data acquisition}
The amplified signal is brought up to a set of fast 8-bit ADC ({\it Analog to Digital Converter}). 
The signal is sampled with a 2.5~MHz frequency for a duration of 820~$\mu$s 
(2048 time samples), while the highest possible drift time (primary ionization 
at the cathode) amounts up to 500~$\mu$s (1250 time samples). 
Once a trigger is given, the data stream is recorded into a buffer. 
It consists on 256 read-out channels with 2048 time samples; 
the $t_0$ signal is also recorded as an squared pulse in a {\it virtual}
additional channel.
The arrival of a subsequent trigger causes the switch of the data stream to 
another buffer (up to 8 buffers are available). 
The multi-buffer writing procedure minimize the dead time for signal
recording to less than one sampling time (i.e. 500~$\mu$s). 
The events are written in raw-data format without any zero suppression;
they are stored locally on disk and automatically transferred to the main CERN tape
facility (CASTOR) using the network.

\subsubsection{The signal shape}
With this read-out configuration the visibility of the events appeared to be optimal
in both collection and induction views. In Fig.~\ref{fig:50L_WireSignal} the raw data
output visualization is shown for a multi-prong neutrino event. The optimization
of the analog electronic setup leads to rather clear and sharp raw images before
any additional filtering algorithms were applied. Besides, the induction signal has been
enhanced varying the ratio of the fields in the drift and gap regions.

The measured signals are induced by the moving charge of the drift electrons and 
are proportional to the drift velocity. 
Thus, strictly speaking the signal from the induction plane is bipolar: 
a small negative part from the slowly drifting electrons approaching the plane from 
the drift volume, and a much larger positive part induced when the electrons have 
passed through the plane and drift away much faster in the larger drift field 
among the induction and collection planes.

The careful optimization of the instrumental setup leads to a satisfactory level 
of performance.

\begin{figure}[!hb]
\begin{center}
  \begin{tabular}{ c }
  \includegraphics[width=14cm, height=16.5cm]{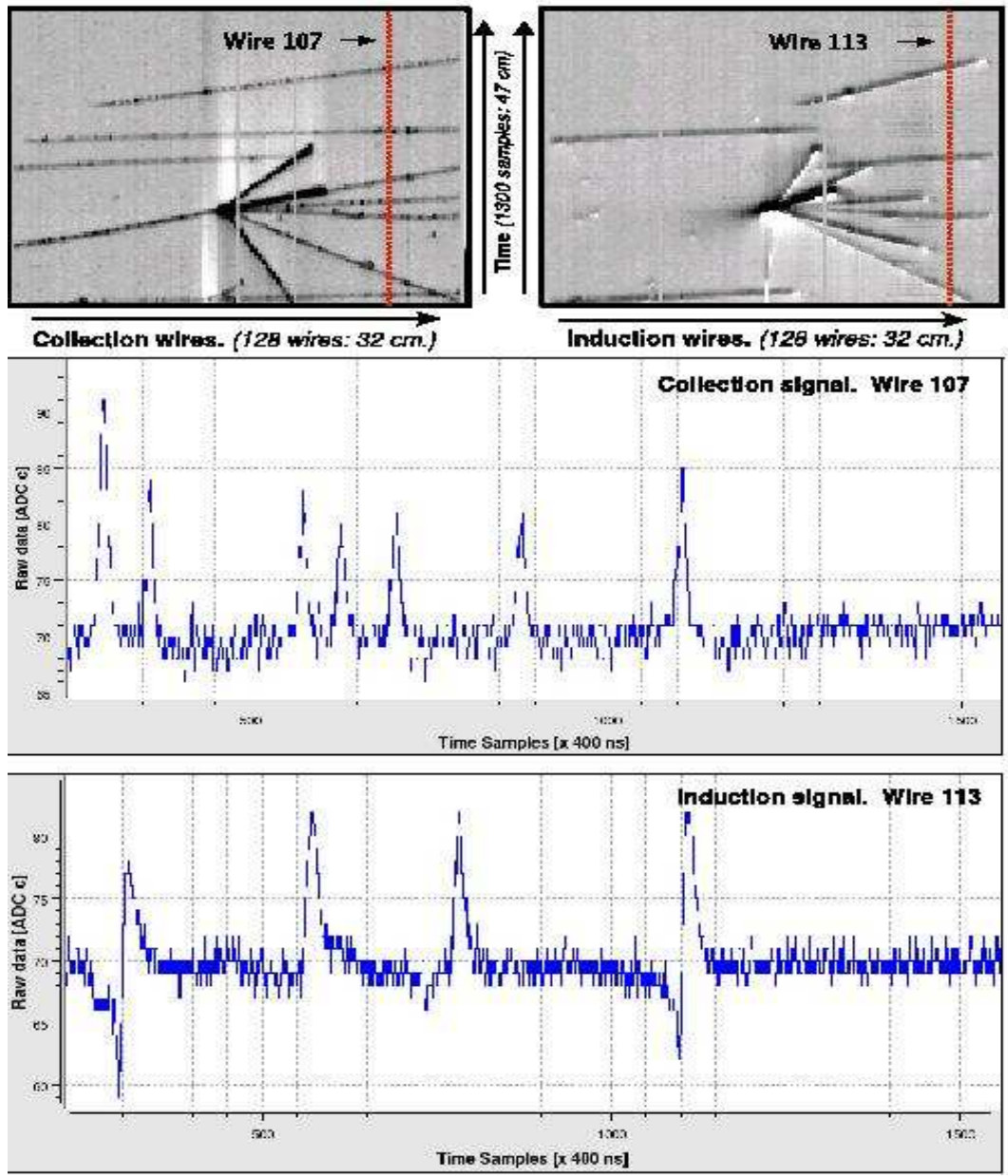}   \\ 
  \end{tabular}
\caption[Output visualization (Raw data) of a neutrino event.]
{\label{fig:50L_WireSignal} Output visualization of a multi-prong neutrino event in 
``quasi-current'' mode. [Top] The raw data from collection and induction wire planes.
Below two examples of a single channel (wire) electronic read-out in collection [middle]
and induction [bottom] modes.
}
\end{center}
\end{figure}

% LocalWords:  SPS SPSLC pre TPC's UHV through's atm de kV ppb

%%%%%%%%%%%%%%%%%%%%%%%%%%%%%%%%%%%%%%%%%%%%%
%%%%%%%%%%%%%%%%%%%%%%%%%%%%%%%%%%%%%%%%%%%%%
                                          %%%
\chapter{Off-line event reconstruction}   %%%
                                          %%%
%%%%%%%%%%%%%%%%%%%%%%%%%%%%%%%%%%%%%%%%%%%%%
%%%%%%%%%%%%%%%%%%%%%%%%%%%%%%%%%%%%%%%%%%%%%
\label{chap:recon}

%Charged particles traversing the LAr sensitive volume produce
%ionization electrons in a number proportional to the energy transferred
%from the particle to the LAr. The ionization electrons drift
%perpendicularly to the wire planes pushed by the electric field,
%inducing a current on the wires near which they are drifting while
%approaching the different wire planes. Therefore, each wire of the
%readout plane records the information of the energy deposited in a
%segment of the ionization track.

The goal of the reconstruction procedure is to extract the physical
information contained in the wire output signals, i.e. the energy
deposited by the different particles and the point where such a
deposition has occurred, to build a complete three dimensional spatial
and calorimetric picture of the event. The basic building block of a
track is called hit, defined as the segment of track whose energy is
read by a given wire of the readout wire planes. Therefore, the
spatial and calorimetric information of the track segment is contained
in the associated hit, and the sensitivity of the detector depends
entirely on the hit spatial and calorimetric resolutions.

In the present chapter we describe a method developed to perform the
hit spatial and calorimetric reconstruction. It is based on
the pioneer work done for the stopping muon sample
collected with the T600 detector~\cite{ricotesi}, however many 
new contributions and improvements have been developed during
this work in order to extend the reconstruction capabilities to
low multiplicity neutrino events collected with the 50L LAr
chamber (see Chapter~\ref{chap:50Lexperiment}).

The reconstruction algorithm proceeds through the following steps:
\begin{enumerate}
\item Hit identification (Sec.~\ref{sec:hit_id}): 
The hits are independently searched for
in every wire as output signal regions of a certain width above the
baseline output value.

\item Fine hit reconstruction (Sec.~\ref{sec:hitrecon}): 
The parameters defining the hit (position, height, area), 
which contain the physical information, are precisely
determined.

\item Cluster reconstruction (Sec.~\ref{sec:cluster}): 
Hits are grouped into common charge
deposition distributions based on their position in the wire/drift
coordinate plane.

\item Two-dimensional (2D) track reconstruction (Sec.~\ref{sec:2dtrack}): 
The 2D projections of the ionizing tracks are detected
taking care of interaction vertexes and splitting the clusters into chains of hits 
with smooth transitions between them.

\item Three-dimensional (3D) reconstruction (Sec.~\ref{sec:3drec}): 
The hit spatial coordinates are reconstructed using the association of hits from
different views to common 3D track segments. 

\item Calorimetric reconstruction (Sec.~\ref{sec:calrec}):
The spatial reconstruction of the tracks together with the information 
extracted from the fitting of the hits allows to precisely measure the energy 
deposited by a particle along its path.

\end{enumerate}

%The reconstruction method summarized in the previous steps can be attended by 
%means of a graphical user interface (GUI) which allows to the user to visualize
%and analyze the events, as well as to handle and assist the reconstruction 
%at any phase.

In the next sections, we describe in detail the different steps of the
hit reconstruction procedure.

%%%%%%%%%%%%%%%%%%%%%%%%%%%%%%
\section{Hit identification} %
%%%%%%%%%%%%%%%%%%%%%%%%%%%%%%
\label{sec:hit_id}

\subsection{Hit identification algorithm}
%%%%%%%%%%%%%%%%%%%%%%%%%%%%%%%%%%%%%%%%%

The hit identification aims at distinguishing signals produced by
ionization electrons from electronic noise. Hits are identified as
signal regions of a certain width with output values above the local
baseline (defined as the average output value in a given signal region). 
The hit identification algorithm typically acts over the
whole wire sample of a given readout plane. In the 50L configuration
there are 2 wire planes with 128 wires. Each wire is sampled 2048 times 
per event.
%Computing speed is a critical parameter for the hit search, above all
%for big LAr TPC where there are thousands of wires sampling long drift
%distances.
The hit identification is based on basic geometrical principles of
the wire signal and establishes rough preliminary bounds to the 
hit range along the drift coordinate, which will be of utmost 
importance in ulterior steps of the reconstruction procedure. 
No information from adjacent wires is used at this stage.

The algorithm loops over the selected group of wires (i.e.\ a whole
wire plane), skipping those with identified problems, such as large
noise conditions or disconnections, where no signal identification is
possible. Hits are searched for in the wire output signal after a low
frequency filter has been applied. The different steps are described
in the following paragraphs.

\subsubsection{Low \& High frequency noise filter}
%%%
Before the use of the hit identification algorithm is better to have
the best signal we can for this purpose. In order to remove the noise
contribution from the wire signals it is worth to perform a filter in frequencies.
Such a filter is based on the Fourier transform of the raw wire signal.
The Fourier spectrum reveals that the main contribution
of drifting electrons coming from a ionizing particle to the wire signal 
relies on quite low frequencies. The other contributions to the read-out 
signal are due to electronic noise, which can be partially removed performing
a suitable cut in frequencies. 
The Fourier transform of the wire signal in the 2048 output samples is
computed using the Fast Fourier Transform (FFT)
algorithm \cite{NUMERICAL_RECIPES}. 

The amplitude of the low frequencies is reduced by convoluting the 
result with the following function of the frequency $\nu$ (``soft'' step function):
\begin{equation}
S(f) = 1-\frac{1}{e^\frac{\nu-R}{\alpha}}
\label{eq:soft}
\end{equation}
where $R$ and $\alpha$ are the parameters determining, respectively,
the mean frequency radius and the transition region thickness. Setting
the parameters to $R=4.5$ kHz and $\alpha=0.9$ kHz effectively cuts
low frequencies up to about 6.5 kHz while reducing the diffraction
effect. This effect distorts the whole output range when a hit of
high amplitude and sharp edges is present in the wire. 
The FFT filter efficiently removes the 
low-frequencies modulation of the signal giving as a result more
uniform and constant noise baselines. 

In the same way, the FFT filter is used to deal with the annoying 
high-frequency noise removing the contribution of waves with a frequency
higher than 70~KHz. This make the search of hits much easier, above all
in the case of small signals which are little bigger than the electronic noise.

Fig.~\ref{fig:filter}~(top) shows the raw data before the application 
of the filter. Fig. \ref{fig:filter}~(bottom) shows the result of 
the application of the FFT filter with the cuts pointed above.

\begin{figure}[!ht]
\begin{center}
\includegraphics[width=14cm]{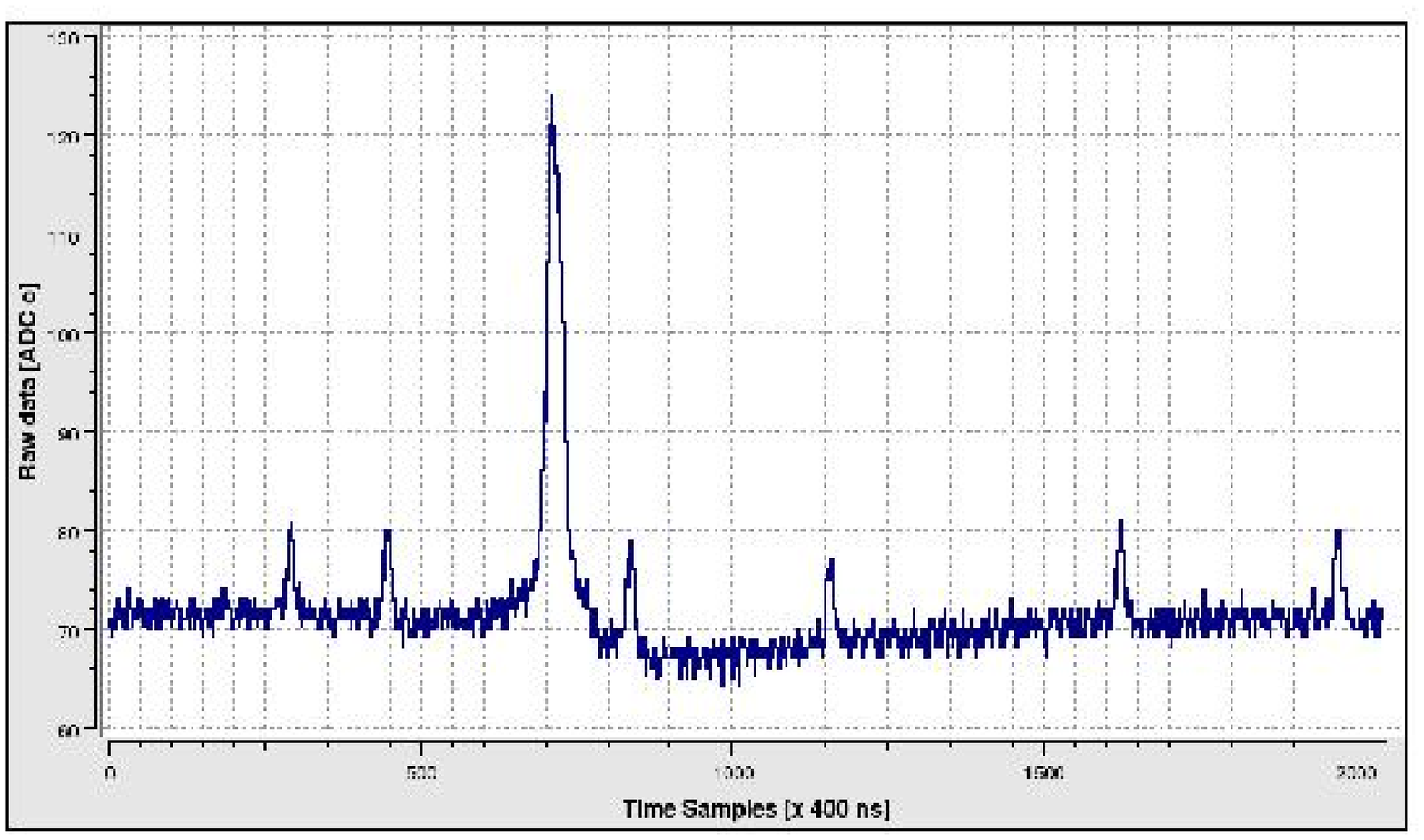}
\\[0.5cm]
\includegraphics[width=14cm]{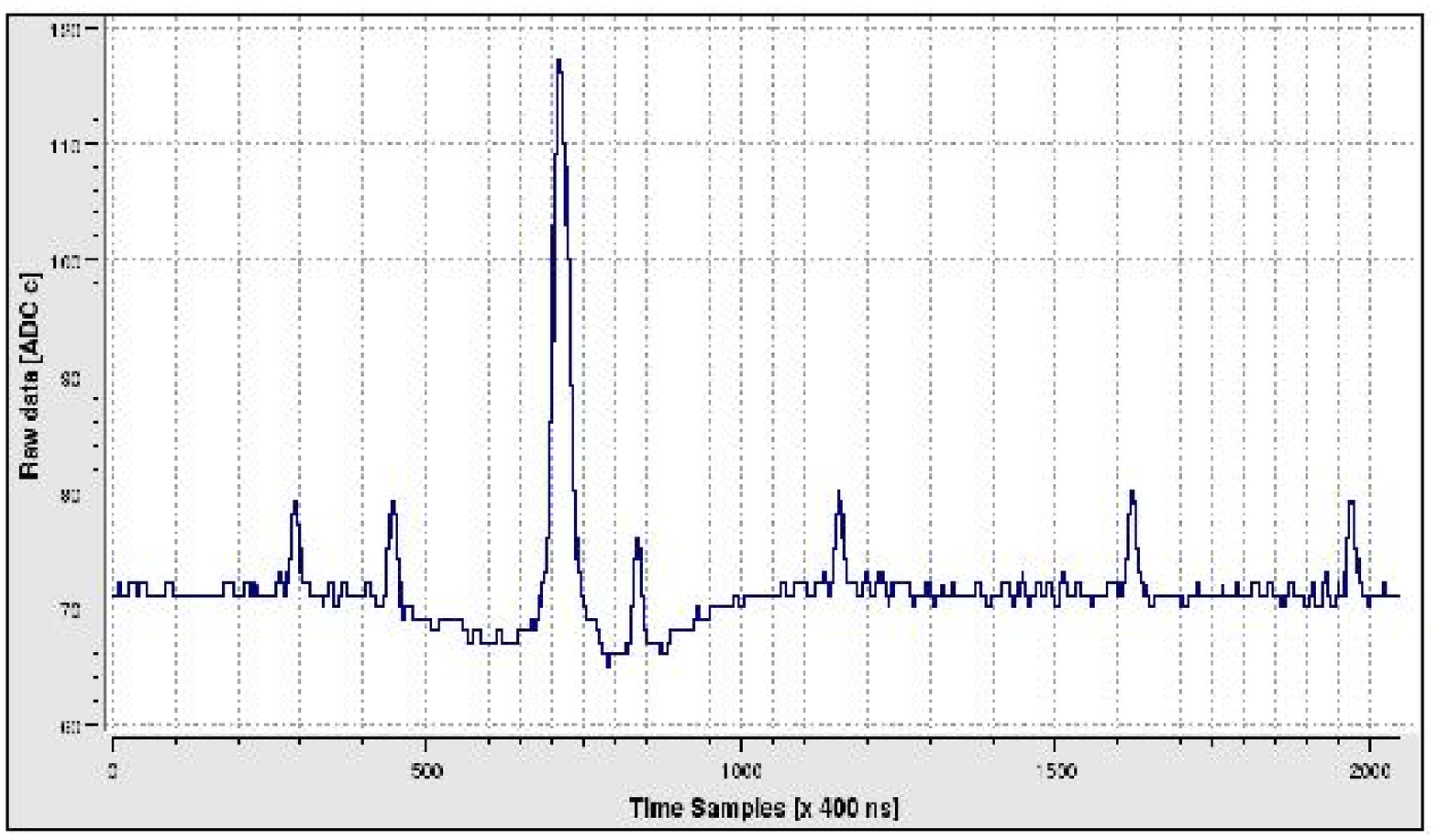}
\end{center}
\caption[Effect of the frequency filter on the wire signal]
{Effect of the frequency filter on the wire signal. 
(Top) Raw data with no filter applied. 
(Bottom) The filtered wire. The contribution of plane waves with frequencies 
below 6.5~KHz and above 70~KHz have been removed.}
\label{fig:filter}
\end{figure}

\subsubsection{Hit search}
%%%
The hit search proceeds on the filtered wire output signal. The
identification is based on the geometrical features of the wire output
in a given region, i.e.\ the presence of a relatively broad region of
output values above or below the baseline. The algorithm detects the pattern
of a hit in both Collection and Induction planes and determines the type
of the signal shape. This is very important for the Induction plane 
where different patterns of signals (unipolar or bipolar) appears depending on
the energy deposition. The behavior of Collection signals is always unipolar.
Correlations with the signal from neighboring wires are not
considered. 

For every wire, the hit search algorithm follows the following steps:
\begin{itemize}
\item Every output sample is compared with the baseline. 
A value above (below) the baseline by more (less) than $3-4$ ADC counts
triggers an unipolar (bipolar) hit candidate. 

\item The hit candidate is built with all the subsequent output samples
after a threshold is given. Depending on the detected type of hit,
the algorithm search for an unipolar or bipolar pattern in the wire signal.

\item The hit candidate is characterized by its width, i.e.\ the distance
(in drift samples) between the hit initial and final points. 
Rejection of fake (noise) candidates is achieved by imposing a minimal width
value.

\item Further rejection can be carried out depending on the expected shape of
the hit. For Collection wires, where the hits are expected to have an 
exponential falling slope, an extra requirement on the minimum distance 
from the peak position to the hit end is imposed. For Induction
wires, when a bipolar hit is detected we require in addition that the 
distance between the down and the up peak is below a minimum.
\end{itemize}

Fig.~\ref{fig:typicalhit} illustrates the two different types of hits
which we are dealing with and the parameters used to identify them.
Fig.~\ref{fig:goodpars} illustrates the performance of the 
hit searching algorithm in both wire planes. It is the result of 
an optimized choice of the parameters for the 50L readout.
As it was desired, the hit search procedure is able to detect the most
of the real hits, faking as less hits as possible.

\begin{figure}[!ht]
\begin{center}
\includegraphics[width=7.2cm]{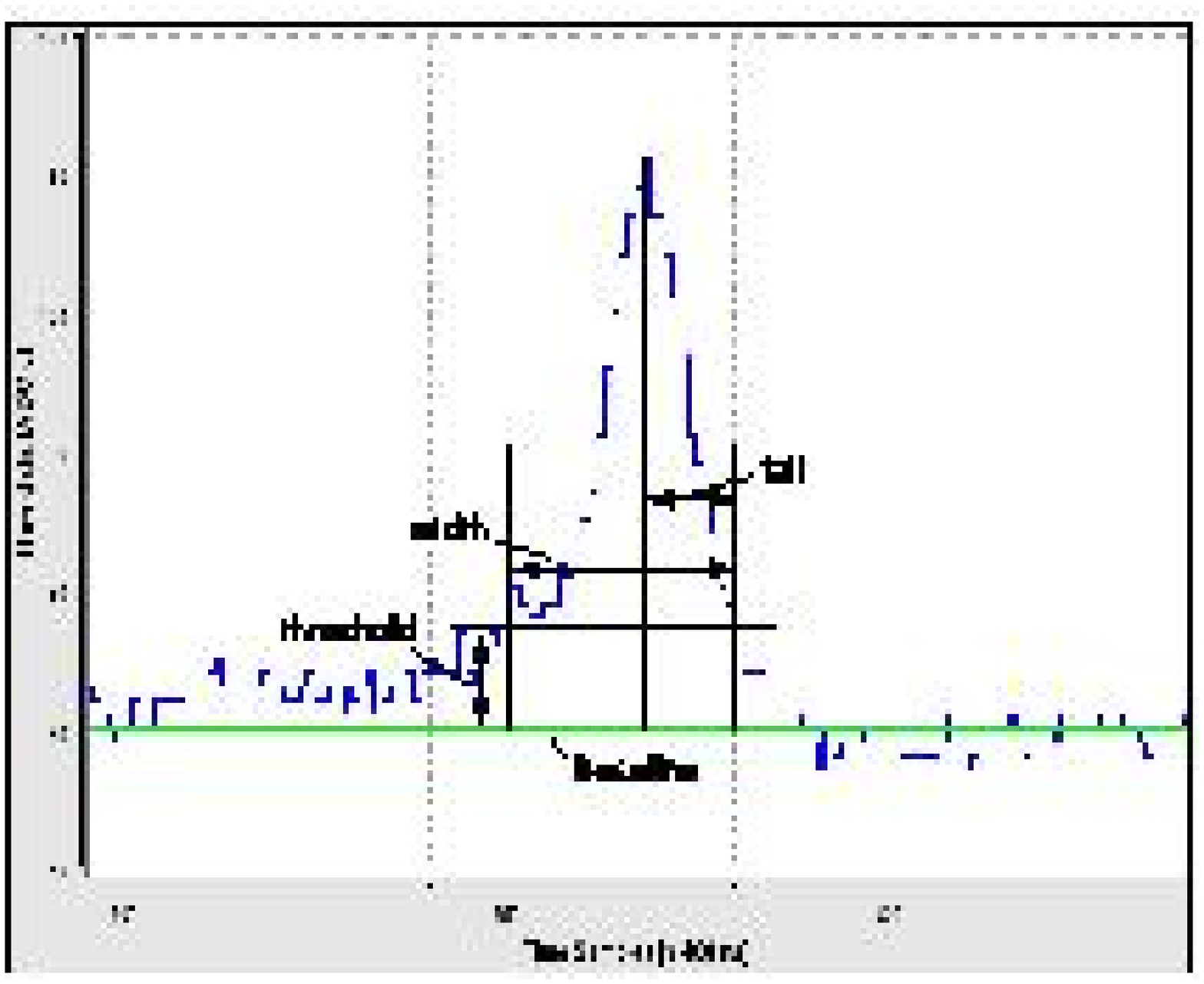}
\includegraphics[width=7.2cm]{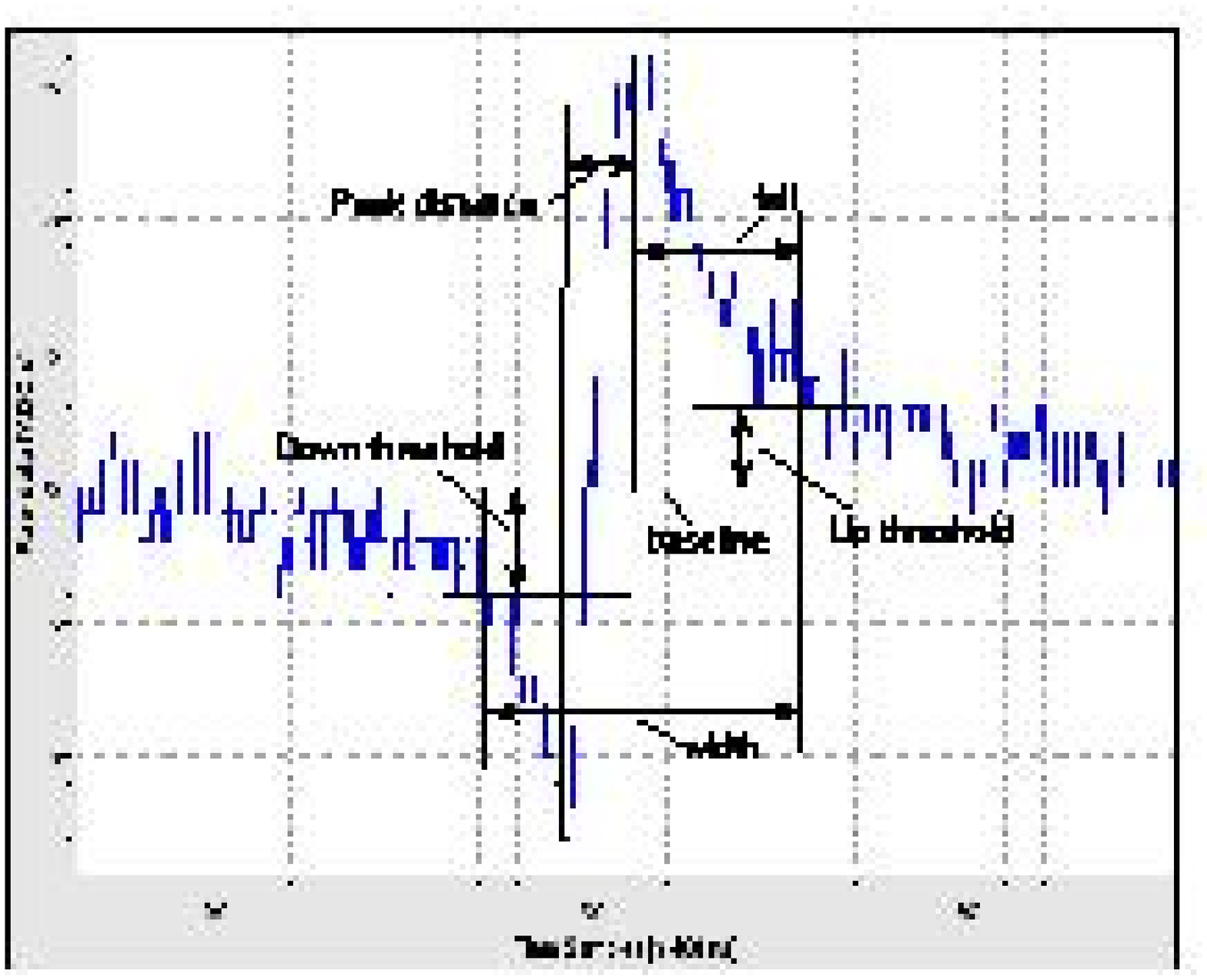}
\end{center}
\caption
[Typical hits in Collection and Induction wires]
{Example of a hit produced by an ionizing track on a Collection wire (left)
and on an Induction wire (right). 
Marked are the parameters used in the hit search.}
\label{fig:typicalhit}
\end{figure}

\begin{figure}[!ht]
\begin{center}
\includegraphics[width=\textwidth,height=10cm]{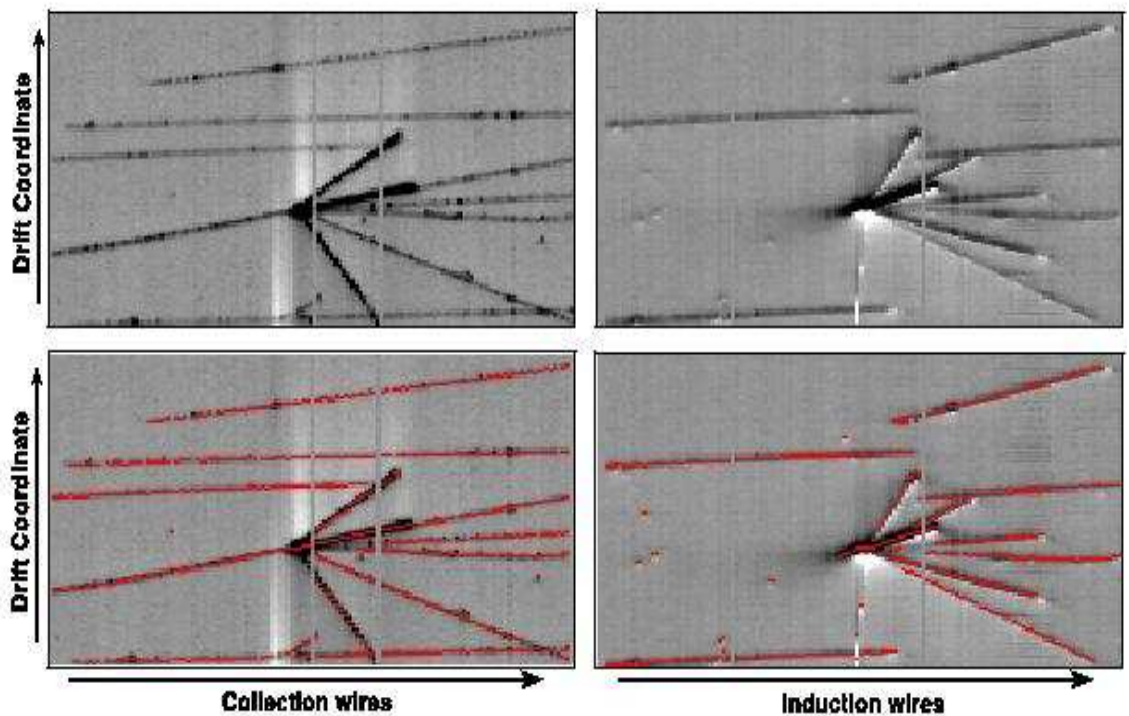}
\end{center}
\caption[Example of performance of the hit finding algorithm]
{Example of performance of the hit finding algorithm. 
(Top): gray scale mapping of the output signal in the 
Collection view (left) and the Induction view (right). 
(Bottom): in red, the hits found with an optimized choice of the hit 
searching parameters.}
\label{fig:goodpars}
\end{figure}

\subsubsection{Resolving close hits}
%%%
Close hits may appear when segments of tracks belonging to different
ionizing particles produce a signal in the same region of the
wire/drift plane (see Fig.~\ref{fig:closehits}~(left)), or when a single
ionizing particle travels with small angle with respect to the drift
direction (see Fig.~\ref{fig:closehits}~(right)). Identification of close
hits results in a better hit calorimetric and spatial reconstruction
of those tracks.

\begin{figure}[!ht]
\begin{center}
\includegraphics[width=7.2cm]{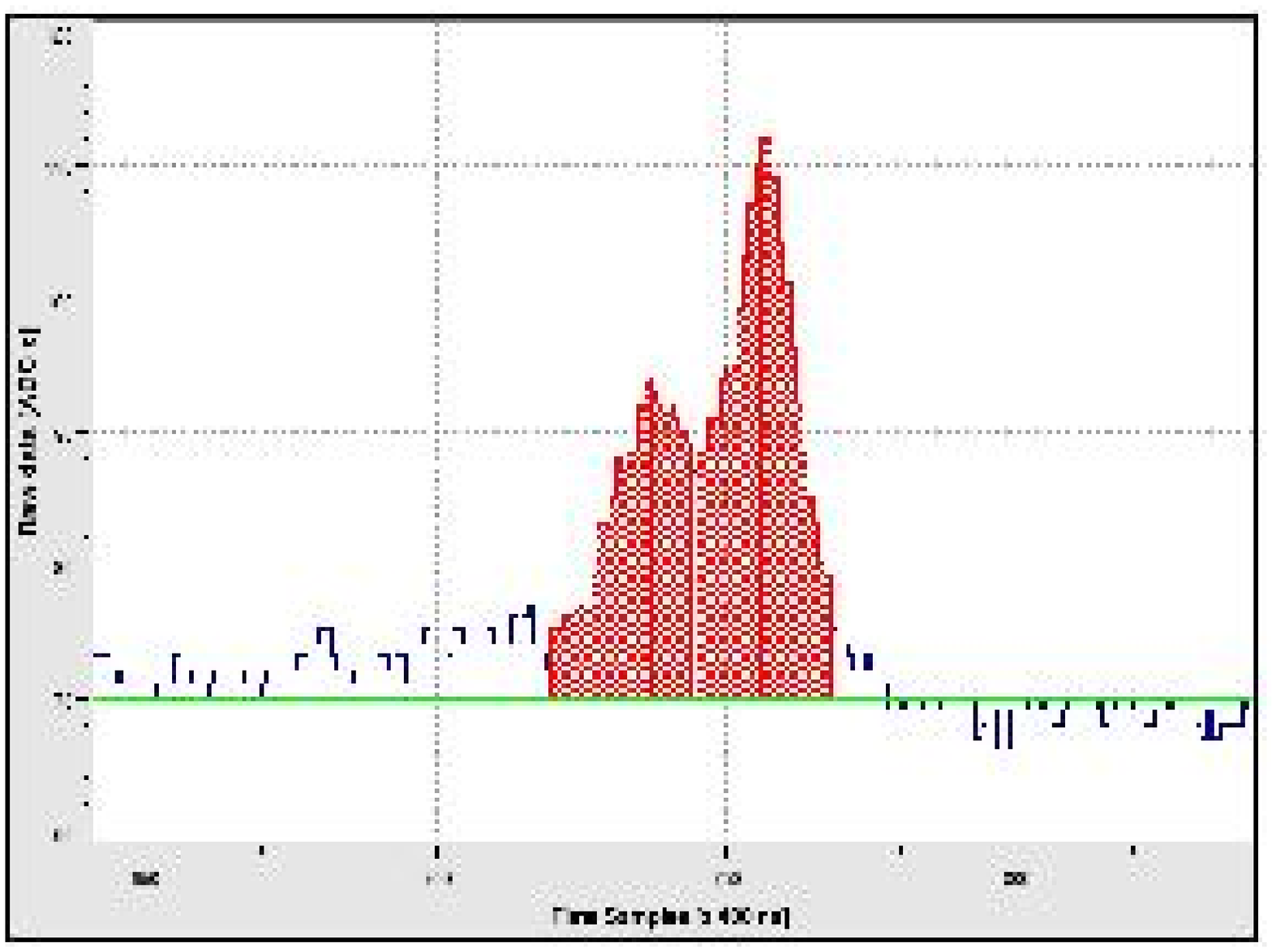}
\includegraphics[width=7.2cm]{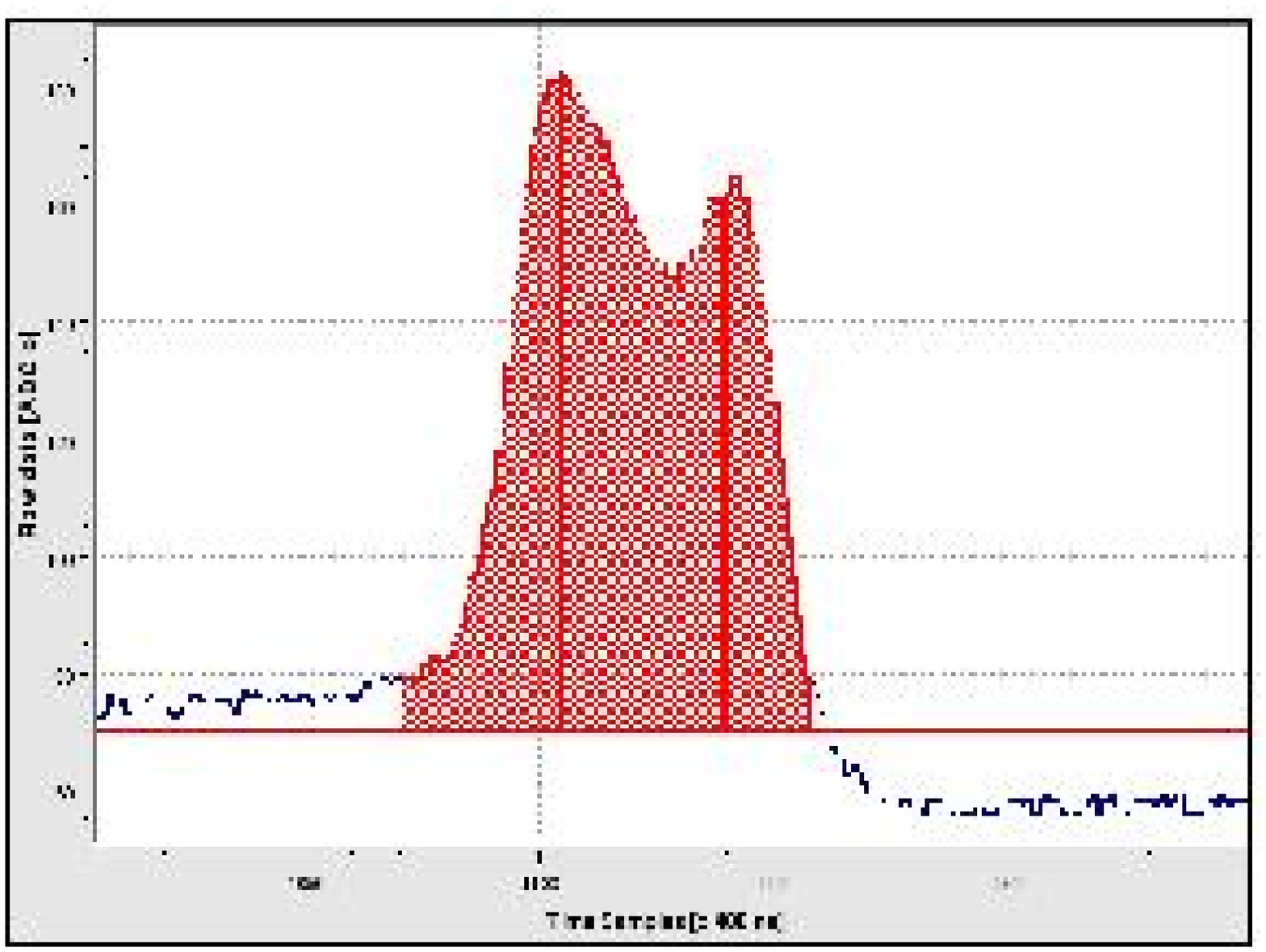}
\end{center}
\caption
[Example of close hits]{(Left) Example of close hits occurrences caused by
close tracks in the same region. (Right) Close hits from a single track in the
direction of the drift.}
\label{fig:closehits}
\end{figure}

The resolution of close hits is carried out within the hit finding
procedure. When a hit candidate is found, an embedded algorithm looks
for patterns which point out the presence of another close hit within 
the hit being analyze. 

In the case of Collection signals, when a falling and a rising slope 
are found consecutively within a hit candidate, the algorithm splits 
the candidate into two hits separated by the starting point of the 
rising slope. Although for bipolar signals in Induction wires this criteria is 
a little bit more complicated, the search for close hits is based on the
same strategy: the presence of a close hit is found when an unexpected 
change of slope is find out.
Once the close hits are found, the selection criteria are applied 
separately to both candidates.

\subsubsection{Hit parameterization after the hit finding algorithm}
%%%
After the search for hits in both wire planes they are stored together 
with their main geometrical features for further processing at the next 
stage of the reconstruction procedure.

\begin{figure}[!ht]
\begin{center}
   \includegraphics[width=7.2cm]{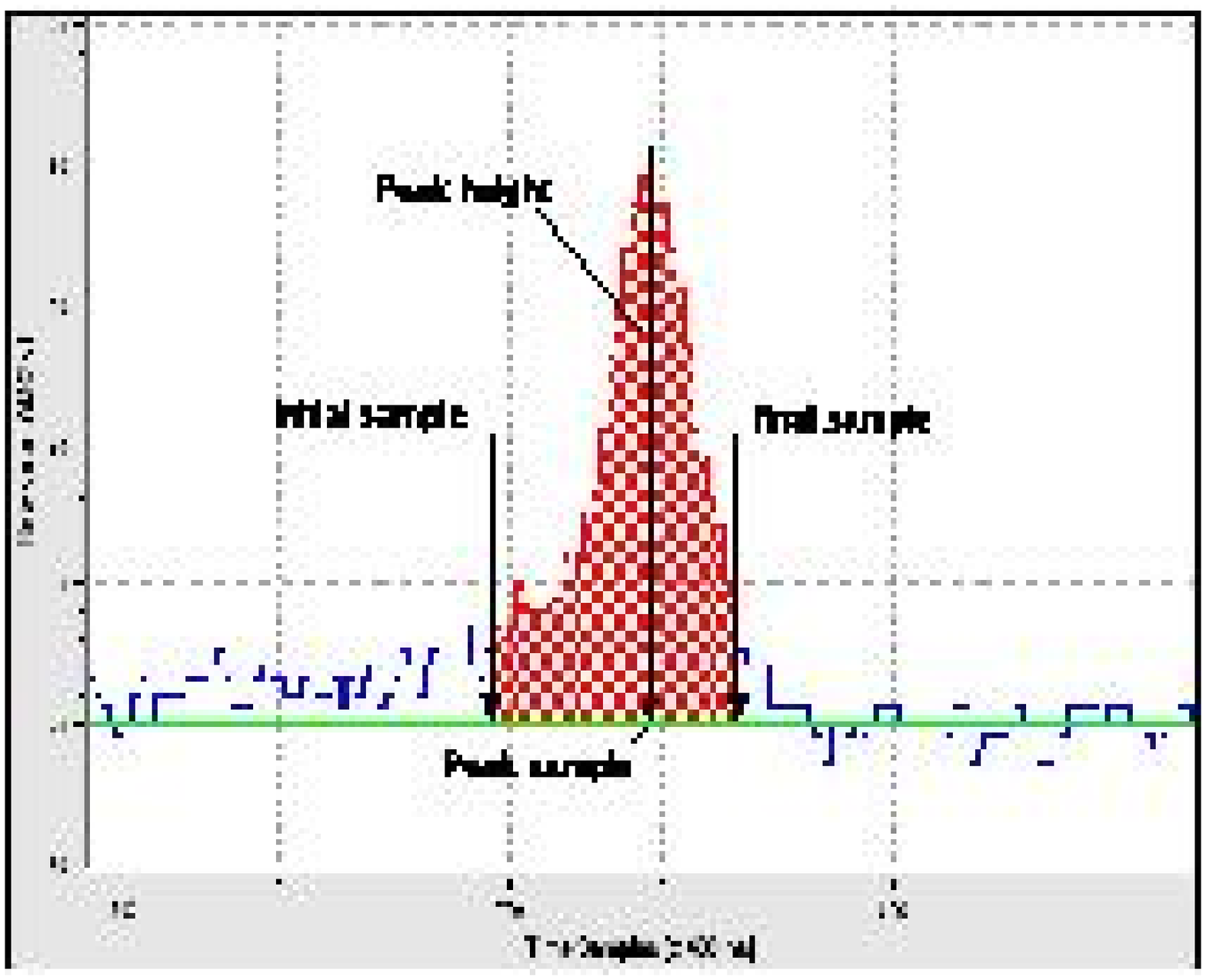}
   \includegraphics[width=7.2cm]{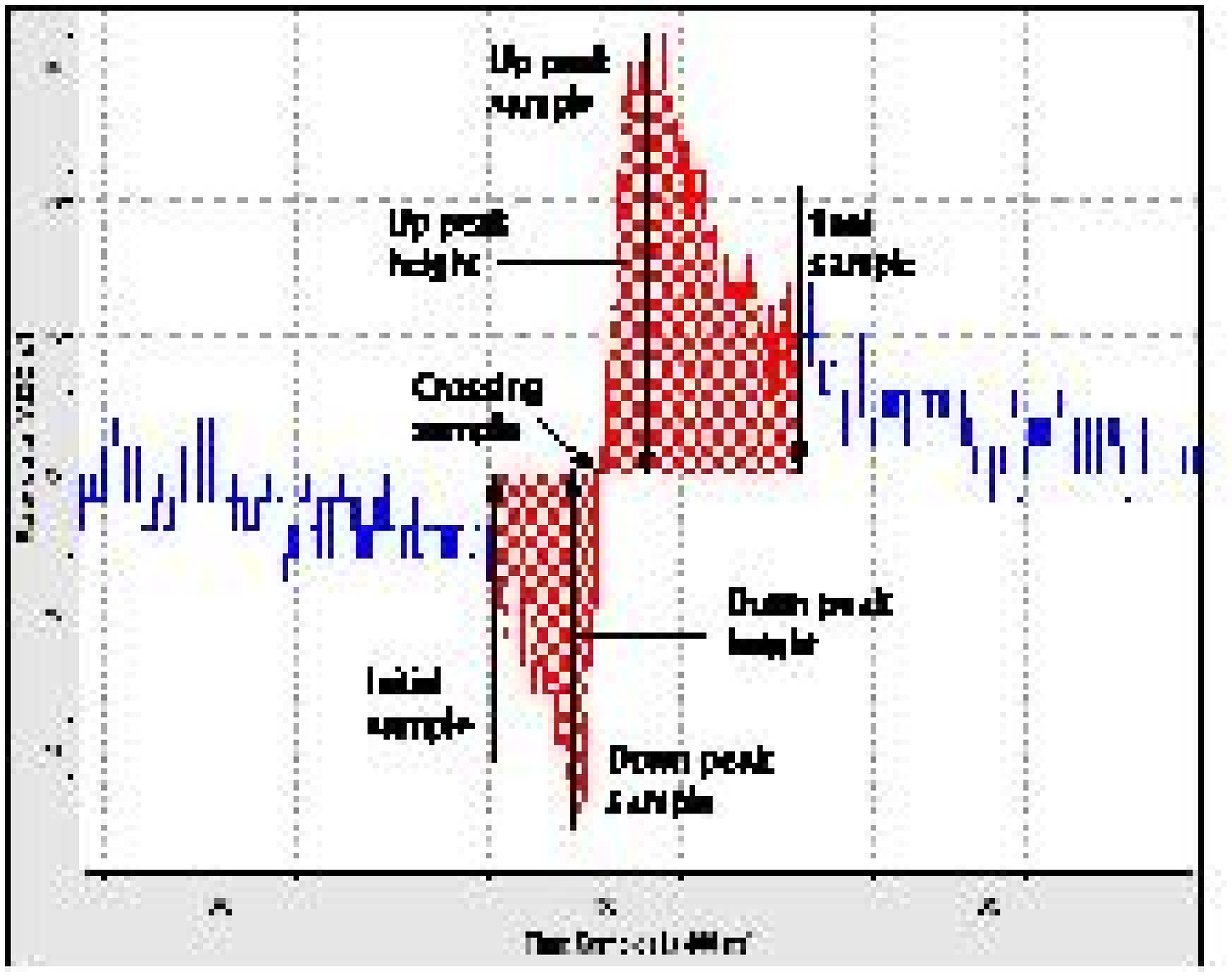}
\end{center}
\caption
[Parameters characterizing a hit after the hit finding algorithm]
{Parameters characterizing a hit after the hit finding algorithm for unipolar (left)
and bipolar signals (right).}
\label{fig:hitpars}
\end{figure}

The hit identification provides the following hit parameters for unipolar signals
(see Fig.~\ref{fig:hitpars}~(left)):
\begin{itemize}
  \item The view and wire indexes of the hit.
  \item The drift coordinate of the peak.
  \item The drift coordinate range of the hit, defined by the initial and
        final drift samples computed in the hit expansion.
  \item The height of the peak taken as the difference (in ADC counts) between
        the peak and the baseline. 
\end{itemize}

In the case of bipolar signals, in addition to the these parameters, the following ones
are also kept (see Fig.~\ref{fig:hitpars}~(right)):
\begin{itemize} 
  \item The drift coordinate of the down peak.
  \item The height of the down peak.
  \item The drift coordinate where the signal cross the baseline.
\end{itemize}

%%%%%%%%%%%%%%%%%%%%%%%%%%%%%%%%%%%
\section{Fine hit reconstruction} %
%%%%%%%%%%%%%%%%%%%%%%%%%%%%%%%%%%%

\label{sec:hitrecon}

Once the hit has been detected, a fine hit reconstruction is performed
aiming at extracting in an optimal way the parameters defining the hit
(position, height, area) which contain the physical information of
the original associated track segment. The hit spatial reconstruction
is entirely based on the determined hit peak position, whereas the hit
area in Collection wires is proportional to the energy deposited by 
the ionizing particle, and therefore constitutes the base of the 
calorimetric reconstruction.
In the present section we describe the fine hit reconstruction algorithm applied
to the typical unipolar signals from Collection plane. No fine hit reconstruction
is performed over bipolar Induction hits, and just the geometrical information from
the hit identification phase is enough for the next stages of the reconstruction method.

\subsection{Hit fit reconstruction method}
%%%%%%%%%%%%%%%%%%%%%%%%%%%%%%%%%%%%%%%%%%%%%%
\label{sec:finerec}
The adopted approach consists of performing a fit
of the output signal (the raw one) around the hit region with an analytical function 
that reproduces well the hit shape, and extracting the parameter
values from the fitted function. The aim is to extract from them,  
in the most accurate possible way, the hit peak position, height and area. 

The output signal within the hit window is fitted using the following analytical 
function of the drift time $t$:
\begin{equation}
f(t) = B+A\, \frac{e^\frac{-(t-t_0)}{\tau_1}}{1+e^\frac{-(t-t_0)}{\tau_2}}
\label{eq:fit}
\end{equation}
\noindent
where $B$ is the baseline, $A$ the amplitude, $t_0$ the point for
which the height of the function with respect to the baseline is equal to
$A$/2, and $\tau_1$ and $\tau_2$ are related to the falling and
rising characteristic times, respectively (see Fig.~\ref{fig:fit_fun}).

\begin{figure}[!ht]
\begin{center}
\includegraphics[width=8cm]{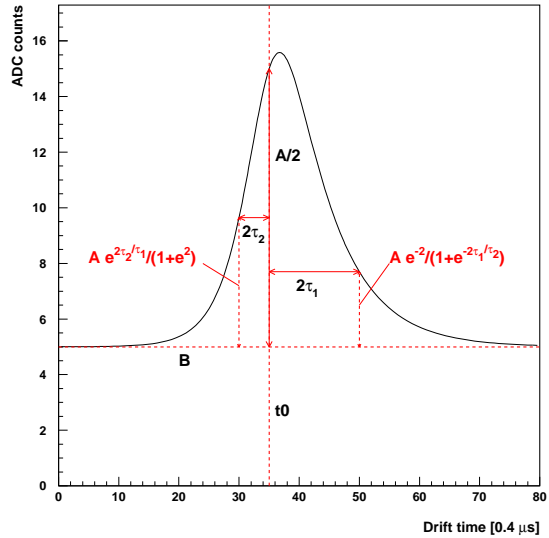}
\end{center}
\caption
[Function used in the hit fit]{Function used in the hit fit (see
Eq.~\eqref{eq:fit}) for the following parameter values:
$B$~=~5~ADC~cnts, $A$~=~20~ADC~cnts, $t_0$~=~14~$\mu$s,
$\tau_1$~=~3~$\mu$s, $\tau_2$~=~1~$\mu$s.}
\label{fig:fit_fun}
\end{figure}

For a given hit, the values of the fit parameters are those
obtained by minimizing the $\chi^2$ of the fit of $f(t)$ to the output
signal within the hit window. The minimization is performed using the
MINUIT package \cite{MINUIT}.

The hit area is obtained by numerical integration of the fit function
in the hit window, while the peak position ($t_{max}$) and height
($f(t_{max})$) are obtained analytically through the following analytical expressions:
\begin{equation}
\begin{split}
t_{max} & = t_0 + \Delta t                         \\
f(t_{max}) & = B + A\,\frac{e^\frac{-\Delta t}{\tau_1}}{1+e^\frac{-\Delta t}{\tau_2}}
\end{split}
\end{equation}with
\begin{equation}
\Delta t \equiv \tau_2 \ln \frac{\tau_1-\tau_2}{\tau_2} \nonumber
\end{equation}
Fig.~\ref{fig:hitfits} shows some examples of fitted hits
corresponding to different events occurring inside the LAr active volume. 

\begin{figure}[!ht]
\begin{center}
\begin{tabular}{cc}
\includegraphics[width=7.cm]{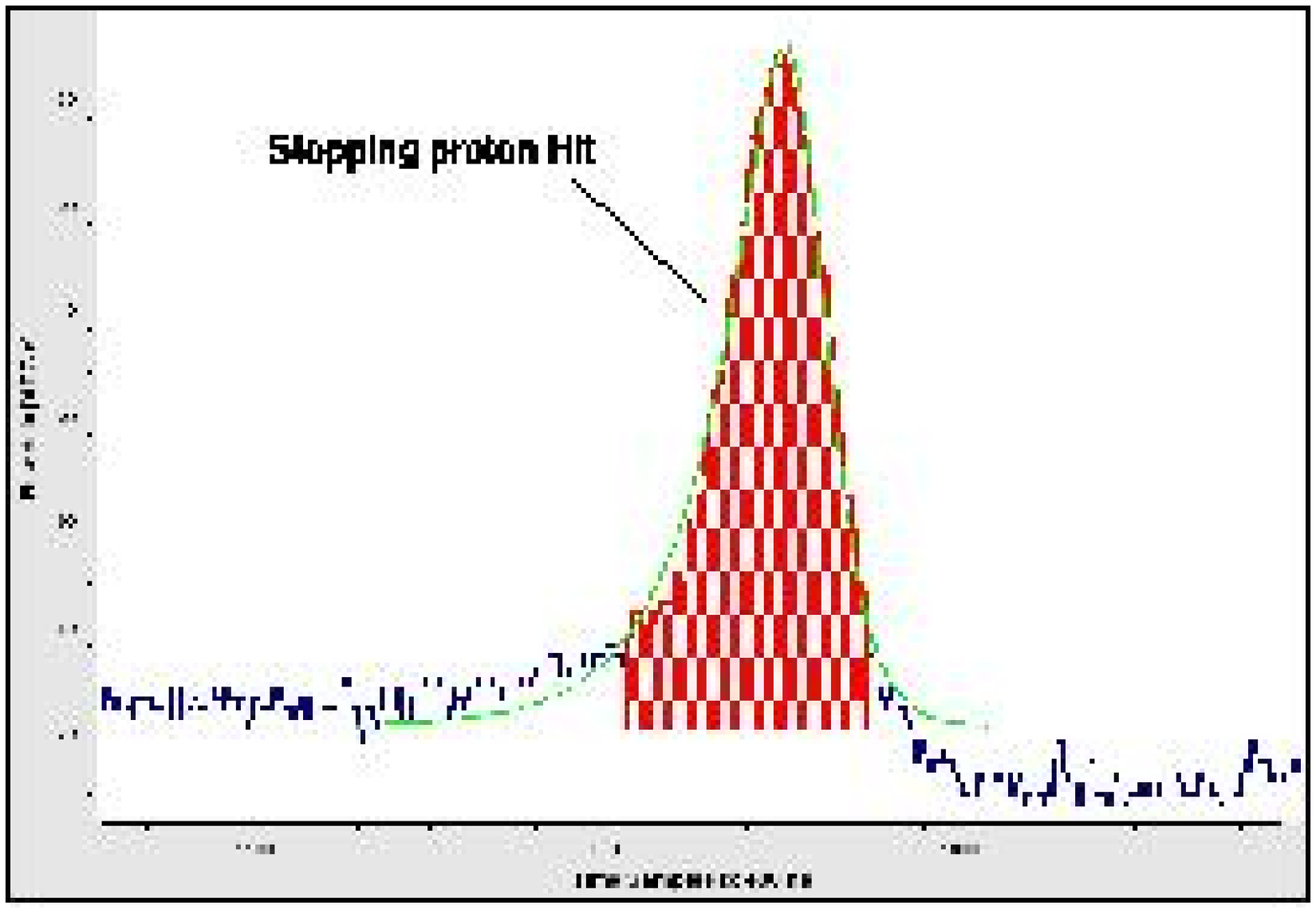} &
\includegraphics[width=7.cm]{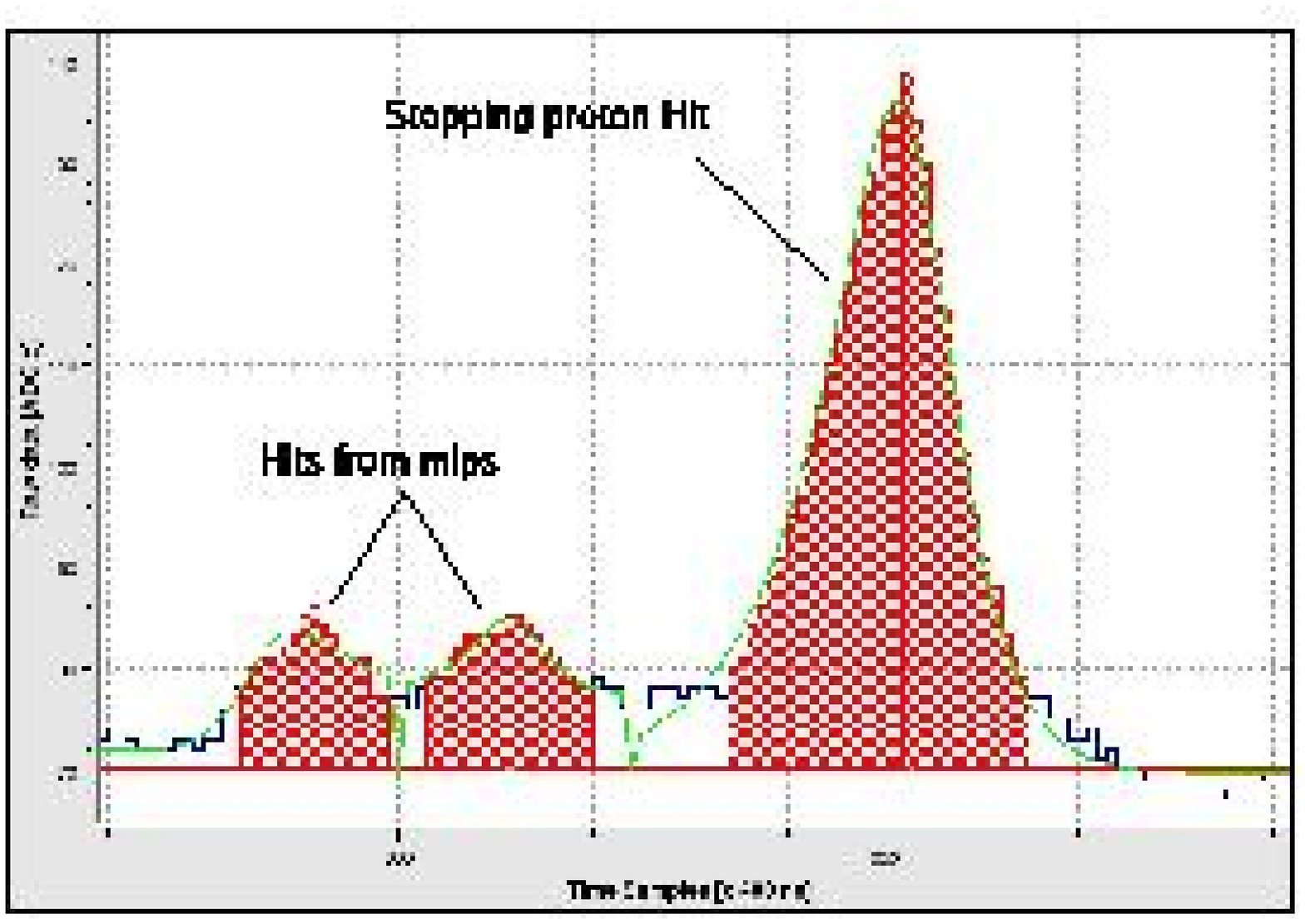} \\[0.5cm]
\multicolumn{2}{c}{\includegraphics[width=14.5cm]{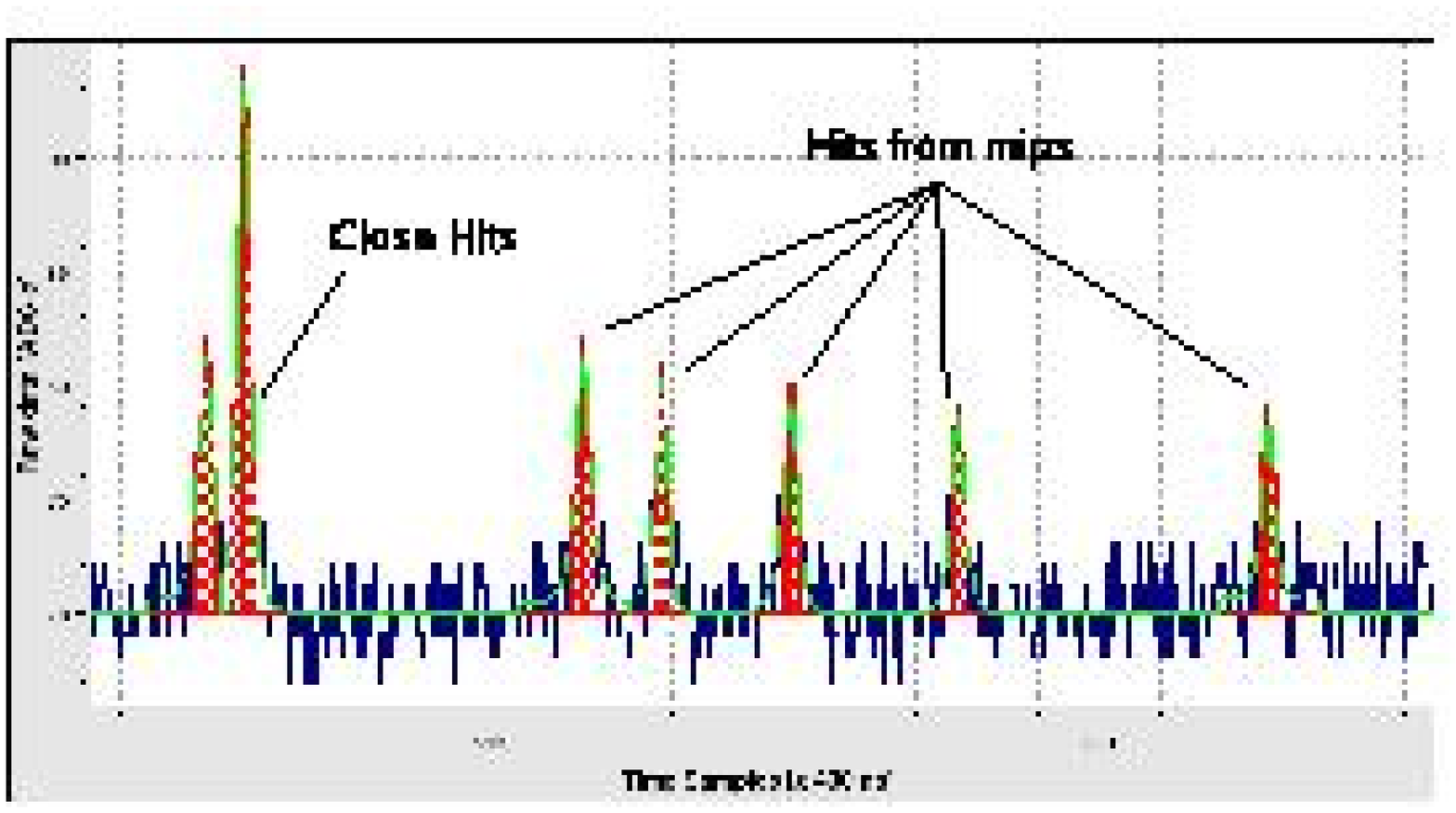}} \\
\end{tabular}
\end{center}
\caption[Examples of fitted hit signals]
{Fitted signal from a hit given by a proton near its stopping point~(top-left). 
The fitted hit of another stopping proton together with two hits from 
mips (top-right). An example of detected and fitted hits of a parallel 
shower of mips (bottom).}
\label{fig:hitfits}
\end{figure}

\subsubsection{Hit parameterization after the hit fine reconstruction algorithm}
%%%
The hit fine reconstruction adds the following quantities to the hits'
parameterization:
\begin{itemize}
\item The local mean value around the hit.
\item The bounds of the hit window, where the different reconstruction
procedures are carried out.
\item The peak position, the peak height and the hit area as computed by
the direct ADC counting method.
\item The peak position, the peak height and the hit area as computed by
the hit fit method.
\item The values of the fit parameters values.
\item The $\chi^2/n.d.f.$ of the fit.
\end{itemize}

%%%%%%%%%%%%%%%%%%%%%%%%%%%%%%%%%%%
\section{Cluster reconstruction} %%
%%%%%%%%%%%%%%%%%%%%%%%%%%%%%%%%%%%
\label{sec:cluster}

A cluster is defined as a group of adjacent hits within the wire/drift
coordinate plane. The goal of the cluster reconstruction is to
perform a first grouping of hits belonging to common charge deposition
distributions, such as tracks or showers. Clusters provide
identification criteria for the different patterns, and thus determine
which reconstruction procedure must be followed. 
Clusters also provide a criterion for discrimination between signal and noise hits 
as well as a method to recover undetected hits through an exhaustive search.

\subsection{Cluster reconstruction algorithm}
%%%%%%%%%%%%%%%%%%%%%%%%%%%%%%%%%%%%%%%%%%%%%
\label{sec:clusterrec}

\subsubsection{Linking the hits}
\label{sec:link}
Clusters are groupings of hits which are {\it close} to each other, and two 
hits are {\it close} when exists a \emph{link} between them. 
A \emph{link} is defined as a pair of hits coming from consecutive
wires and having overlapping drift coordinate ranges, where the drift
coordinate range of a hit is defined as the one computed by the hit
finding algorithm (see section~\ref{sec:hit_id}). Therefore, clusters
can be defined as groups of hits with links between them. 
The links allow to define the topological features of a hit or a hit grouping,
which will be of utmost importance during the 2D and 3D reconstruction
stages (Sec.~\ref{sec:2dtrack} and Sec.~\ref{sec:3drec}). 

To this purpose, a search for all the hits {\it close} to each other 
is performed, obtaining for any hit the set of those hits which are 
surrounding it. In such a way the links are built as logical connections
between a hit and its {\it close} hits. An example of how the links are
established is shown in Fig.~\ref{fig:linkscluster}.
In terms of these {\it links}, the bounds of a cluster are defined 
as those hits having one {\it link} at most.

\begin{figure}[!ht]
\begin{center}
   \includegraphics[width=7.2cm,height=6cm]{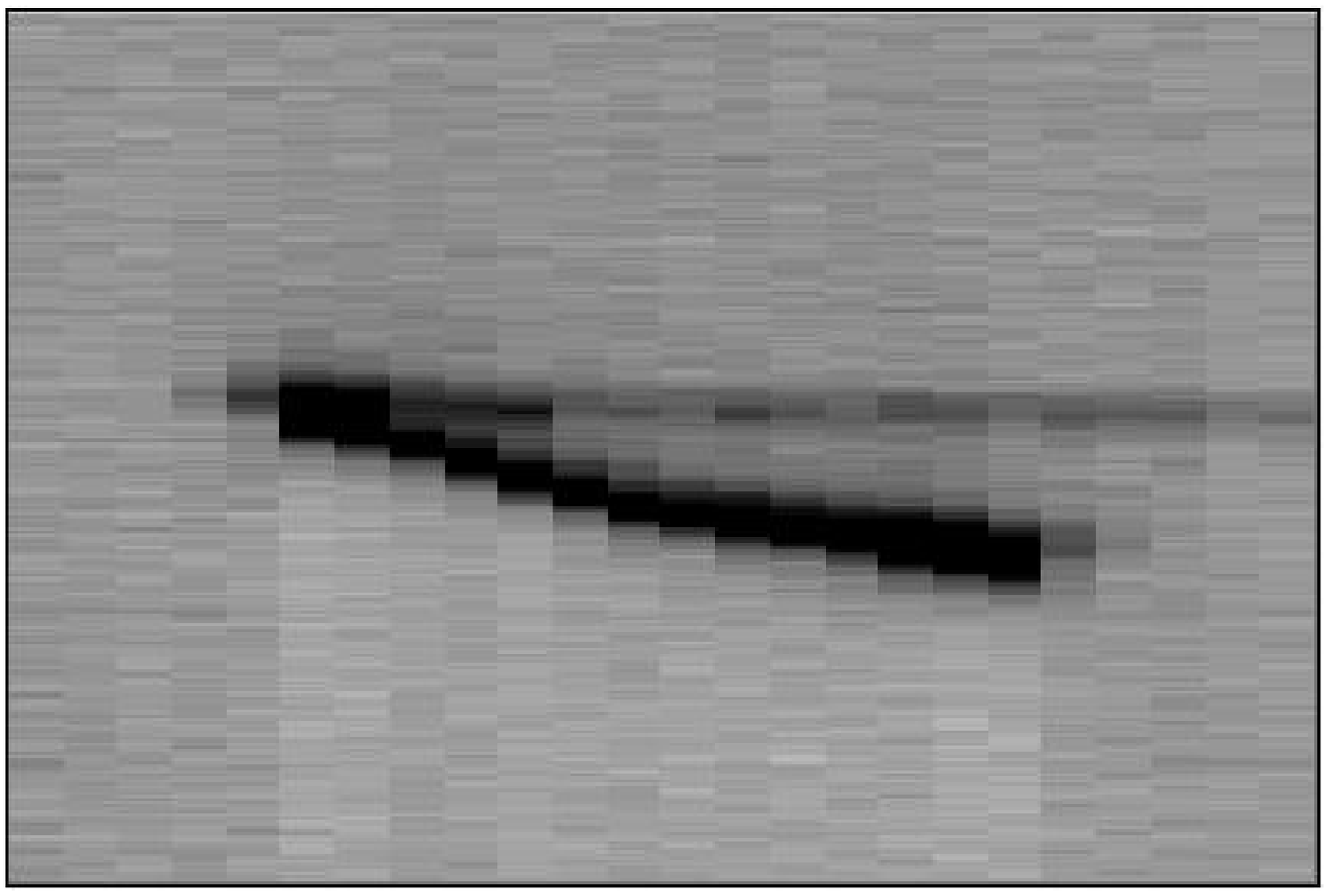}
   \includegraphics[width=7.2cm,height=6cm]{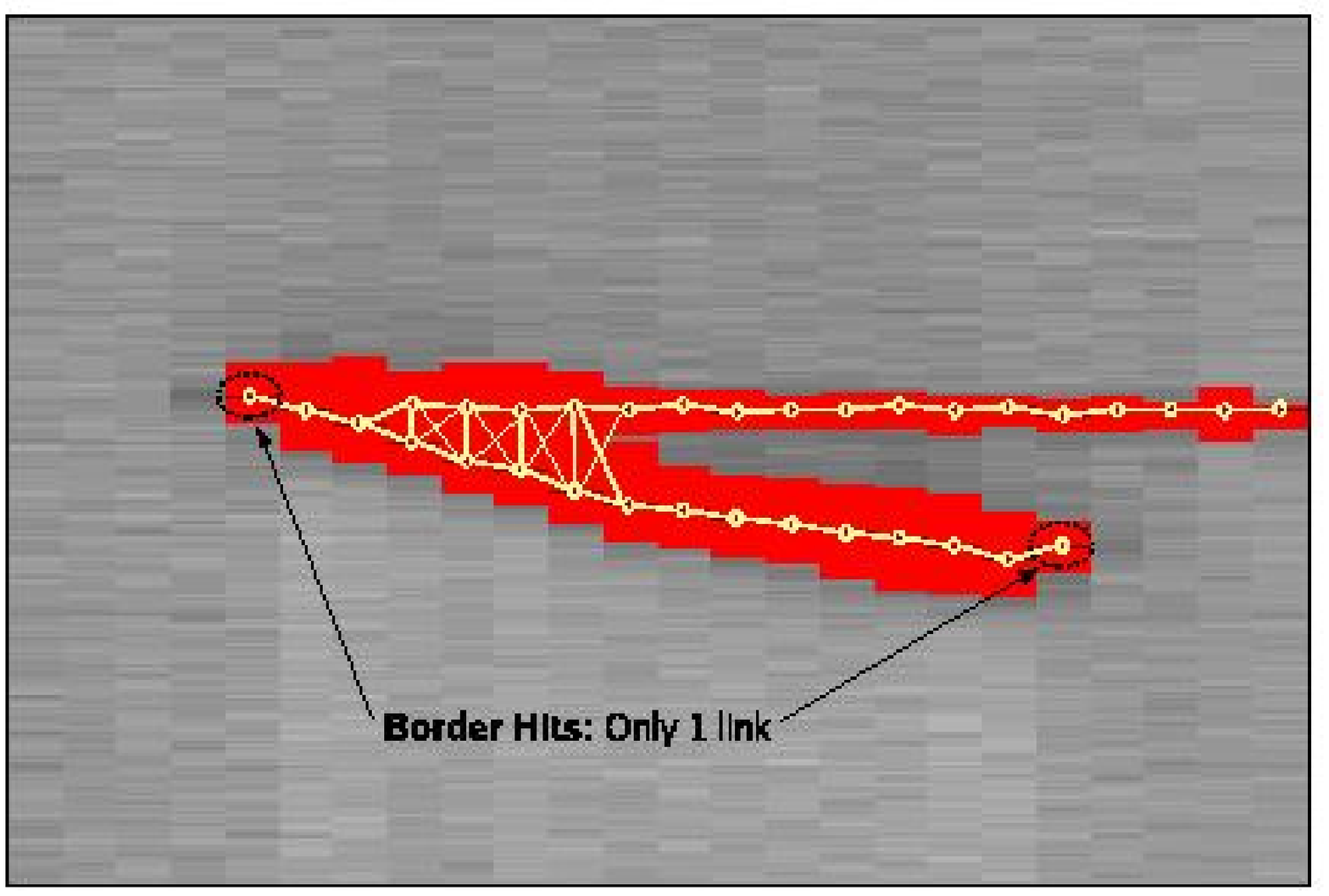}
 \end{center}
\caption
[Illustration of the hit linking procedure.]
{The pictures show the signal of a neutrino interaction in a region of
the Collection plane. The drift coordinate is in vertical and the wires 
in horizontal.
(Left) The raw data of a vertex from a neutrino interaction.
(Right) The found hits in red; the formed links between them are in yellow.
Because all the hits are connected by links to each other, they form a cluster.}
\label{fig:linkscluster}
\end{figure}

\subsubsection{Building the clusters}
Clusters are built in a two step approach. In the first step (pass~I),
preliminary clusters are built as groups of hits being connected by
links. The second step (pass~II), attempts to expand and merge
together the clusters found in pass~I, in order to reduce the number
of the resulting clusters. 
The reconstruction algorithm takes as input
a list of hits from a given wire plane together with their links,
and constructs all possible clusters out of them, 
using only the information obtained by the hit finding algorithm 
(see section~\ref{sec:hit_id}).

\begin{itemize}
\item[{\bf Pass I}] It is essentially established above: 
A preliminary grouping of the hits into clusters is performed:
Hits are grouped into a cluster if any of them is connected
by a {\it link} with at least another one belonging to the cluster.

\item[{\bf Pass II}] The goal of cluster reconstruction pass~II is to 
expand and join cluster fragments produced by pass~I. 
It proceeds as follows: the regions around the cluster borders are examined 
(where a cluster border is defined as a hit linked at most with one hit within the
cluster), searching for new hits which could be added to the cluster, 
hence reducing the cluster fragmentation caused by the misidentification of hits
during the hit search phase. 
Attempts for clusters merging/expansion are performed both along the wire 
and the drift time coordinate directions.
\end{itemize}

The algorithm for the cluster expansion proceeds in the following sequential steps:
\begin{itemize}
\item Cluster borders are identified as those hits with at most one link to
the cluster.
\item The direction of the search is determined as wire coordinate
increasing (decreasing) for borders linked with a hit
from the previous (next) wire. For single hit clusters, both
directions are consecutively inspected.
\item The slope of the search is determined as the difference between the
drift time coordinates of the cluster border hit and its linked hit. 
For single hit clusters, this slope is equal to zero.
\item New hits are searched for in the consecutive wire
along the determined direction using the same hit finding algorithm, but with
looser parameters. Those new hits which were {\it close} to a previously
existing one in the cluster are automatically joined to that cluster,
creating the corresponding new {\it link}. In this way a cluster is expanded
finding new misidentified hits in a clever and effective way: this exhaustive
search (with loose parameters) is performed just in the region where is expected 
to possibly have some new hits, but not in the overall data 
sample like in Sec.~\ref{sec:hit_id}.
\item The latter expansion can be continued until no more hits are found or when
it reaches to another existing cluster. In this case both clusters are merged.
\item The borders of the new resulting cluster are identified and the whole
process is repeated until no further expansion is possible.
\end{itemize}

In Fig.~\ref{fig:clusterrec} the reconstruction of cluster is exemplified. 
In Pass~I, the algorithm establishes links between
the hits and construct preliminary clusters; In Pass~II,
the algorithm starts the exhaustive search for hits in the directions pointed by
the yellow arrows (Fig.~\ref{fig:clusterrec} (top)). The final result is shown in
(Fig.~\ref{fig:clusterrec} (bottom)), where the bigger final cluster is marked 
in yellow (other clusters are not shown). It is worth to underline how misidentified 
hits are recovered in this second pass.

\begin{figure}[!ht]
\begin{center}
   \includegraphics[width=12.cm]{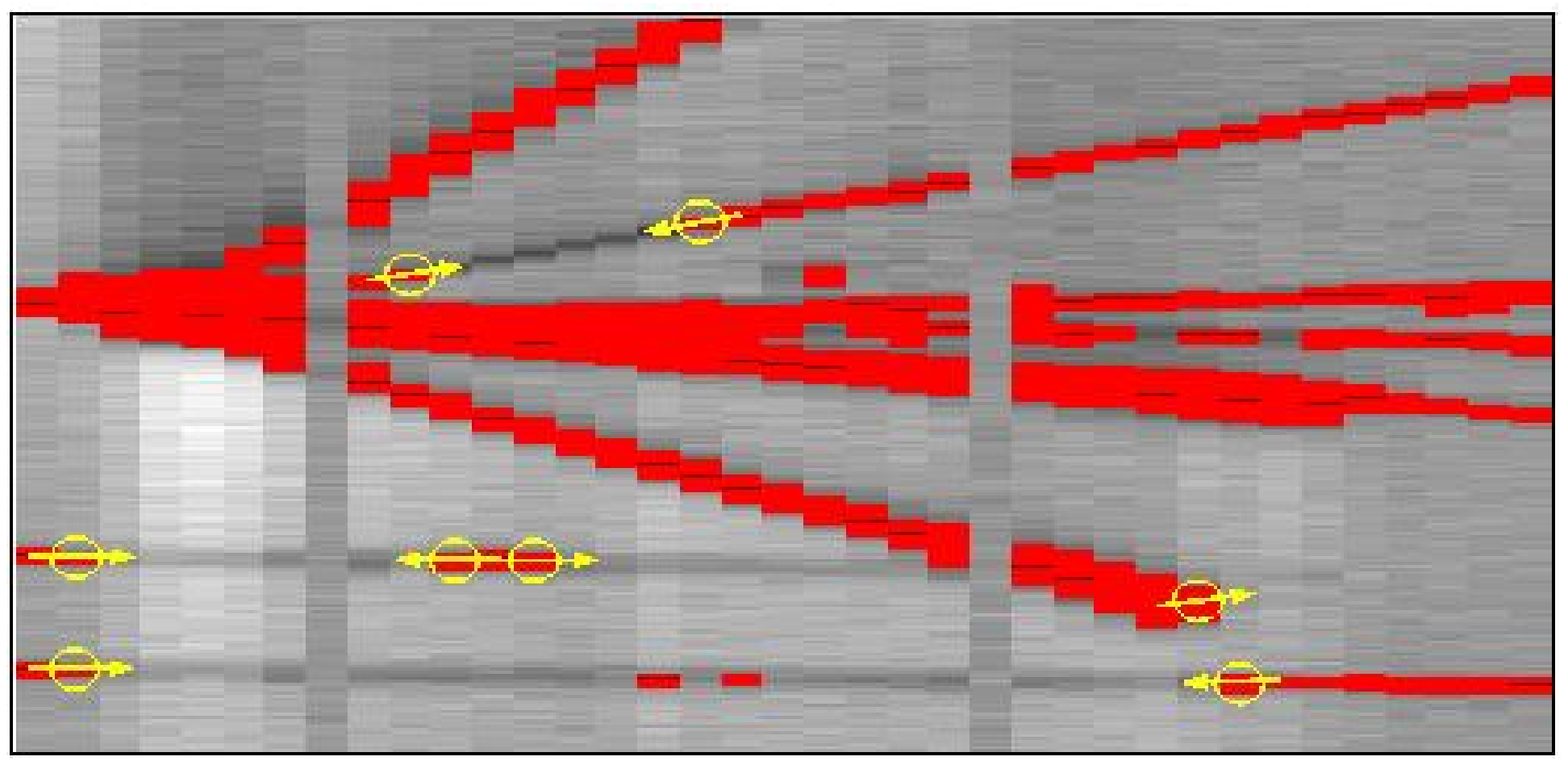} \\[0.5cm]
   \includegraphics[width=12.cm]{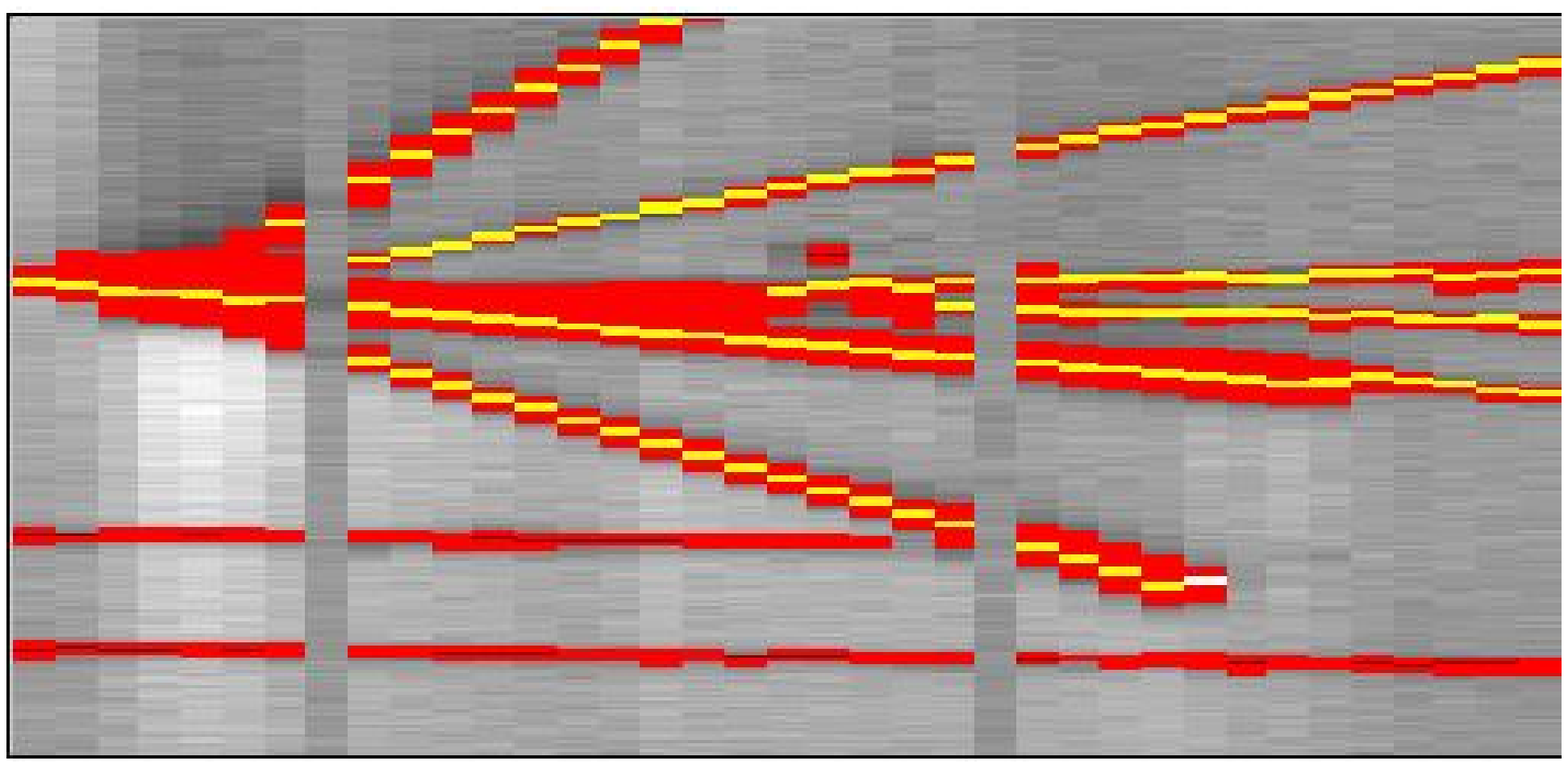}
\end{center}
\caption
[Illustration of the cluster reconstruction method]
{The pictures show the signal of a neutrino interaction in a region of
the Collection plane. The drift coordinate is in the vertical axis and the wires 
in the horizontal axis.
(Top): The raw data of some prong particles from a neutrino (DIS) interaction.
We represent in red the hits found before the cluster reconstruction procedure. 
Yellow circles and arrows show the border hits and the direction of the exhaustive 
hit search. (Bottom): The found hits (in red) after the cluster reconstruction method.
One of the built cluster is singled out with its hits in yellow.}
\label{fig:clusterrec}
\end{figure}

\subsubsection{Hit parameterization after cluster reconstruction}
Cluster reconstruction adds the following quantities to the hit parameters :
\begin{itemize}
\item The cluster identification number.
\item The number of links associated to the hit.
\item The list of hits to which the hit is linked.
\end{itemize}

These data are of utmost importance during the next step of the reconstruction
procedure (see Sec.~\ref{sec:2dtrack} and Sec.~\ref{sec:3drec}).

%%%%%%%%%%%%%%%%%%%%%%%%%%%%%%%%%%%
\section{2D track reconstruction} %%
%%%%%%%%%%%%%%%%%%%%%%%%%%%%%%%%%%%
\label{sec:2dtrack}

The 2D track reconstruction algorithms work separately in each wire plane in order 
to identify the 2D projections of the real ionizing tracks. 
The goal is to detect smooth chains of hits which can be identify as track segments. 
To this purpose two different approaches have been implemented: 
The {\it Tree Algorithm}~\cite{TreeFin} and the {\it Neural Network}
based approach~\cite{NNFin,NNCERN,NNFinFast}.
A proper recognition of 2D track projections in both (Collection and Induction) 
views will lead to vertex identification and to an efficient 3D track 
reconstruction procedure (Sec.~\ref{sec:3Dfrom2D}).

The description of the 2D track recognition methods and how they perform over the 
50L TPC data is presented in the following sections.

\subsection{Chains of links}
\label{sec:chains}
Both 2D finding methods operate into each one of the found 
clusters (see Sec.~\ref{sec:cluster}), retrieving the information of the 
hits and the links between them. Since a track segment can be thought of
as an unbroken chain of consecutive hits, all possible links between a hit 
and its adjacent are collected into chains\footnote{A chain is defined 
as a Collection of links consecutively connected}.
In principle there are many possible chains of links inside a cluster and only
some of them properly define a 2D track projection. 
The goal of the track finding methods is to efficiently selects the real
track projections from the whole sample of chains.
An example of some of the chains which could be formed from the links 
in a cluster is shown in Fig.~\ref{fig:ChainOfLinks}: The two chains at the bottom
would represent real 2D track projections.

\begin{figure}[!ht]
\begin{center}
  \begin{tabular}{ c c }
  \includegraphics[width=7cm,height=6cm]{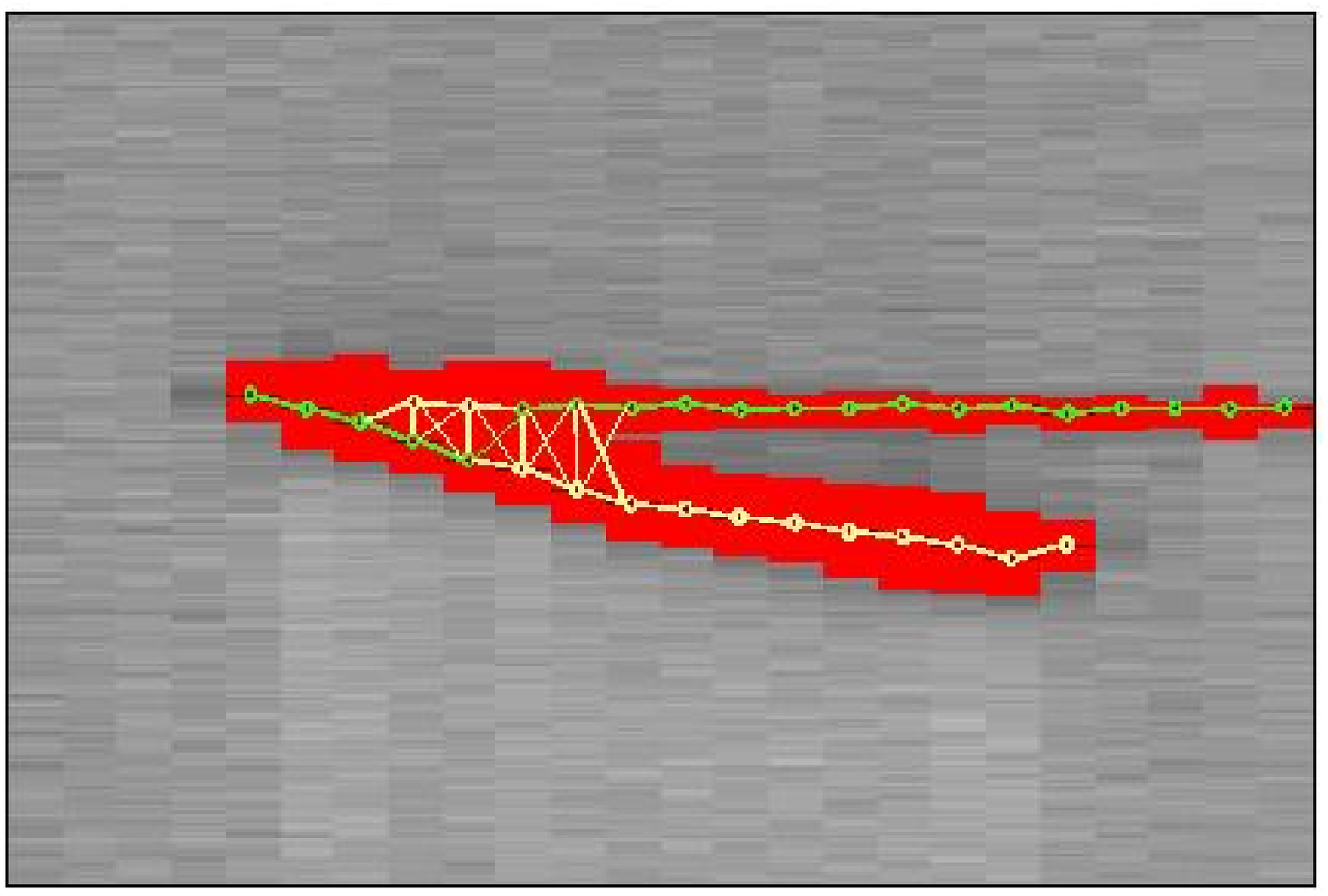} &
  \includegraphics[width=7cm,height=6cm]{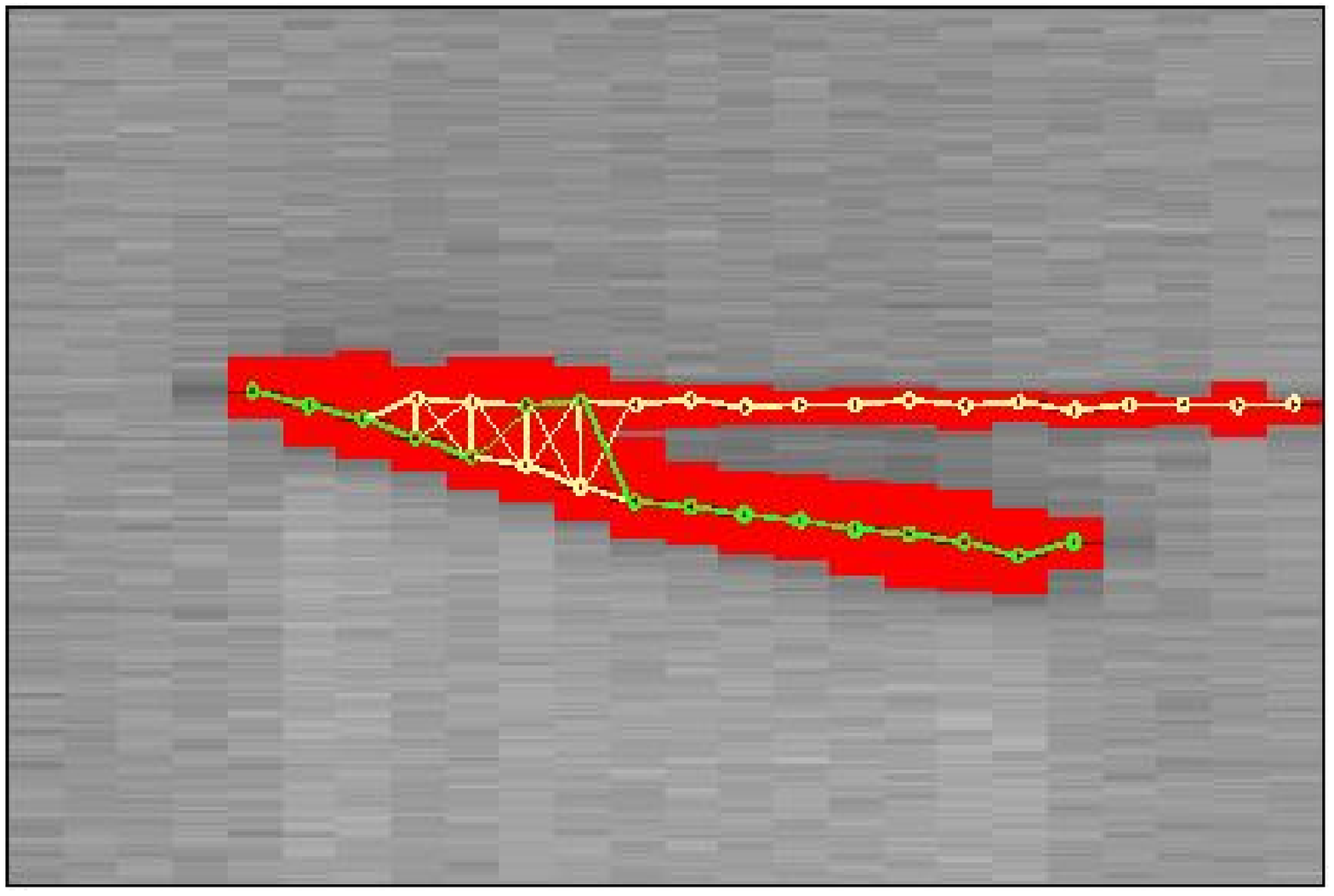} \\[0.3cm]
  \includegraphics[width=7cm,height=6cm]{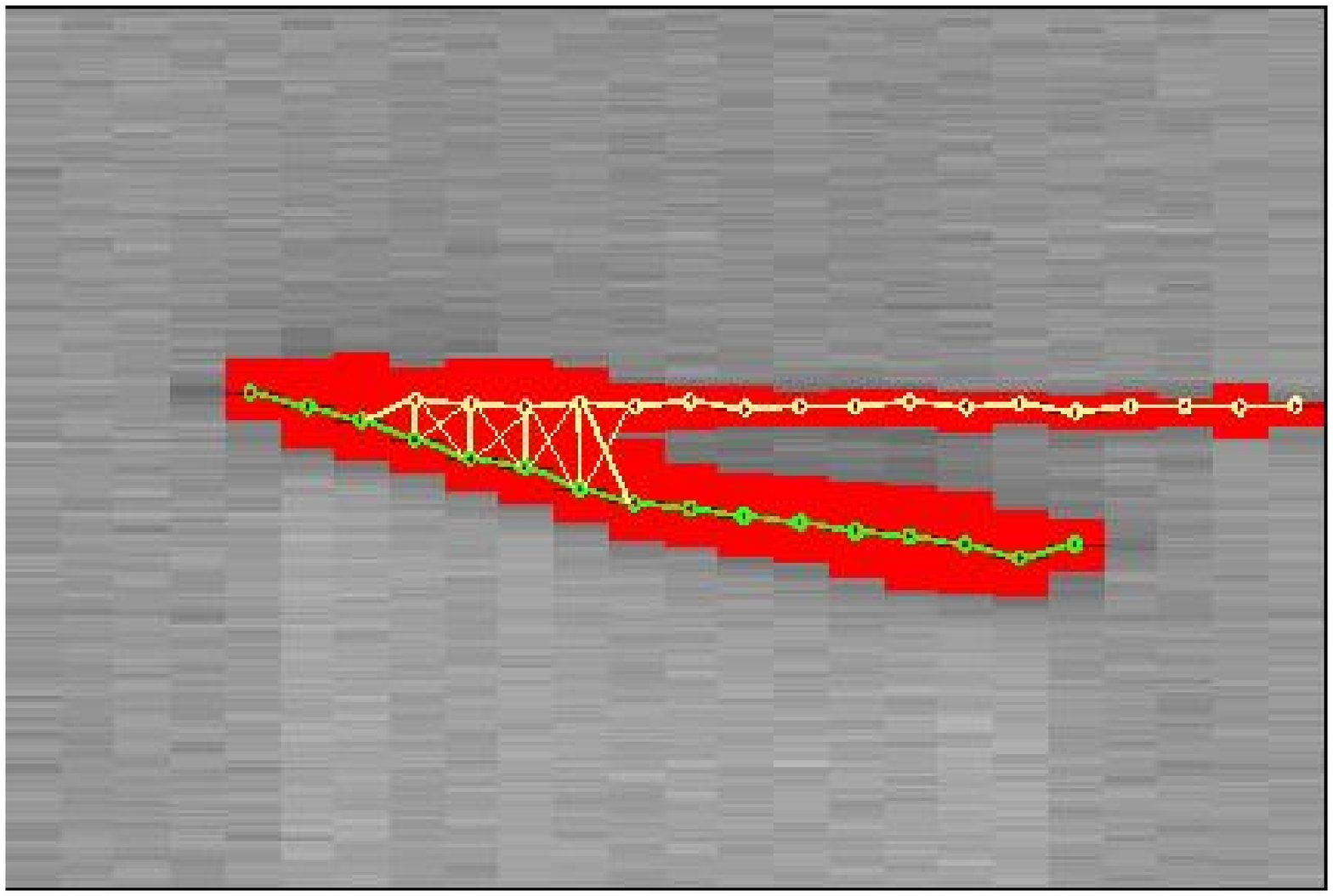} &
  \includegraphics[width=7cm,height=6cm]{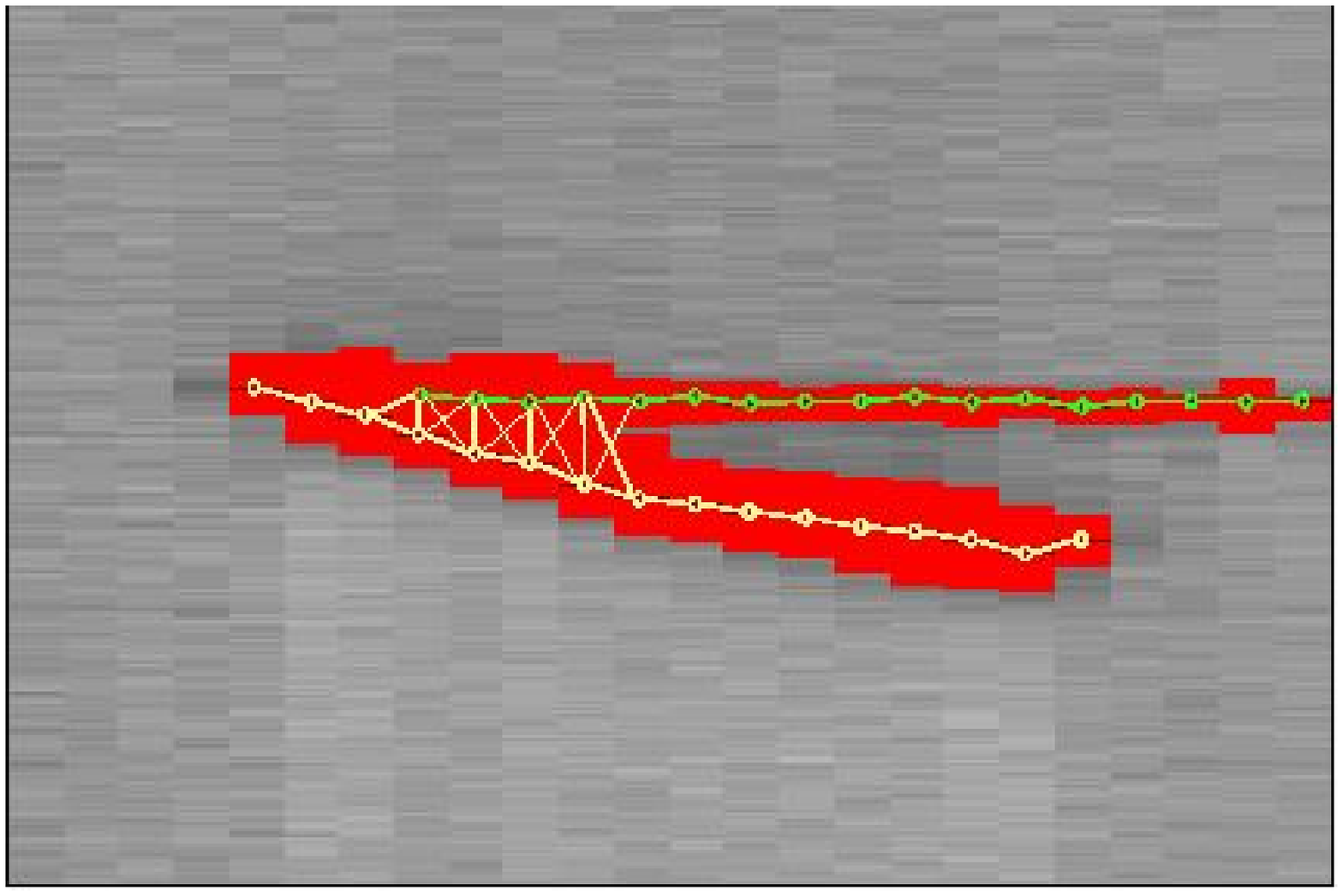} \\
  \end{tabular}	
\end{center}
\caption
[Examples of chains of links in a cluster.]
{Examples of some different chains (in green) which could be constructed from the 
links established (in yellow) in a cluster (see Fig.~\ref{fig:linkscluster}). 
The four chains shown are track candidates, but only the two ones
at the bottom would be real 2D track segments.}
\label{fig:ChainOfLinks}
\end{figure}

\subsection{The Tree Algorithm}
\label{sec:tree}
The Tree algorithm was successfully applied to the pattern recognition problem in
multi-layer track chambers (see~\cite{TreeFin}). Because of the hits from a ionizing 
particle in a LAr TPC also have a layered structure (the wires readout), the approach
to the track finding problem developed in \cite{TreeFin} is directly extensible 
to our purposes. 

The Tree algorithm works on the hits of a certain cluster using the
information provided by the links and constructs from them all the possible 
chains (candidates to track segments) in the following way:
\begin{itemize}
\item The algorithm starts from a link which contains a border hit (root link) 
and determines its elementary tree. 
The elementary tree of any given link is composed of this link, 
which is then called the trunk of the elementary tree (see Fig.~\ref{fig:Trees} (top)), 
and all links having the middle hit in common and with approximately the 
same values of track parameters 
(in this case we have only one track parameter: the slope of the link). 
These links are then called branches of the elementary tree.
Besides, all the possible elementary trees that can be consecutively connected with 
the starting one are also determined (see Fig.~\ref{fig:Trees} (right)).
\item The collected elementary trees are connected forming the full tree associated
to the root link. The links which define the limits of a full tree
are called leaves (see Fig.~\ref{fig:Trees} (bottom)). 
The algorithm recursively climbs the full tree from the root link, 
locating all chains from the root to any leaf but never finding 
the same chain twice. 
In this way tracks (or more precisely, reliable track candidates) can be recognized.
Because of the branches of any elementary tree must be 
compatible with the slope of the trunk, any chain built in this way also have 
smooth transitions between their links.
\item The real track segment is selected from the chains belonging to the full tree
by means of a selection criteria. In this case, a combination of two requirements
is imposed: The longest and smoothest chain is selected as
the track segment.
\item After the selection of the track, their links are stored, tagged
and they do not participate any more in the algorithm (under 
the assumption that a link is never shared by two different track segments).
Then, the algorithm continues processing in the same way the remaining links, 
starting from another root and detecting new track segments.
\end{itemize}

Examples of elementary trees are shown in Fig.~\ref{fig:Trees} (right)
The elementary trees can be combined to form a full tree as shown in 
Fig.~\ref{fig:Trees}(bottom). 
The correct track is immediately recognized as the smoothest chain of links 
in the full tree (chain in red in Fig.~\ref{fig:Trees} (bottom)).

\begin{figure}[!ht]
\begin{center}
  \includegraphics[width=\textwidth]{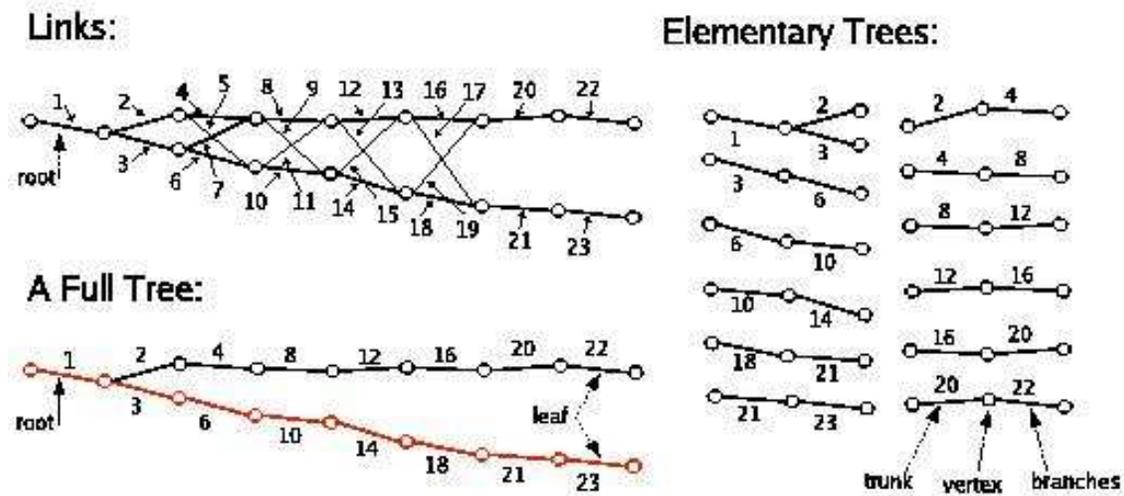}
\end{center}
\caption
[Illustration of the Tree algorithm for 2D track finding.]
{(Top) The starting point: a Collection of links in a cluster.
(Right) All the possible elementary trees.
(Bottom) The full tree is built chaining the elementary trees.
}\label{fig:Trees}
\end{figure}

\subsection{The Neural Network approach}
\label{sec:nntrack}
The Neural Network based computing techniques are powerful tools for pattern 
recognition problems\footnote{See \cite{Bishop} for a general introduction on neural
networks foundations},
which have been successfully applied to many problems in experimental particle
physics\footnote{See \cite{NNCERN} for an specific discussion on particle physics
problems}. 
The track finding problem is a task of this type: given a set of hits (signals) 
in space, the aim is to reconstruct particle trajectories subject to 
smoothness constraints. 
The Neural Network approach for the track finding in the LAr TPC here presented 
is based on the work developed in \cite{NNFin} and \cite{NNFinFast}, where
an algorithm based on the Hopfield-style recurrent neural network~\cite{NNHopfield}
is discussed.

A neural network consists of processing units, which are called neurons, and 
connections between them. The state (activation) of a neuron is a function of 
the input it receives from other neurons. The output of a neuron is usually 
equal to its activation. Therefore, the state of each neuron depends 
on the state of the neurons from which it receives input and on the strength (weight)
of these connections. Neural nets can learn representations by changing the 
weights of the connections between the neurons. 
In the following, the method specifically developed for the LAr TPC is described.

\subsubsection{The neural network algorithm}
The links between the hits in a cluster play the role of the neurons,
which can be in two possible states $S_{ij}=1$ if the link $i \rightarrow j$
is part of the track segment and $S_{ij}=0$ if this is not the case.
The connection between the neurons (links) is established through a term 
called synaptic strength $T_{ijkl}$, which involves the $(ij)$ link with the
$(kl)$ one. The dynamics of such a Hopfield type neural system is governed
by means of the minimization of the following energy function:
\begin{equation}
E = -\frac{1}{2}\sum_{ijkl}T_{ijkl}S_{ij}S_{kl}
\label{eq:NNenergy}
\end{equation}
The following step is to find a suitable form of the synaptic strengths $T_{ijkl}$
which performs the required task. 

Because of a track with $N$ hits can be considered as a set of $N-1$ links with a 
smooth shape and without bifurcation, the synaptic strenghts $T_{ijkl}$ 
have been chosen to take care of this requirement:
\begin{equation}
T_{ijkl} = T_{ijkl}^{(cost)} + T_{ijkl}^{(constraint)}
\label{eq:tterms}
\end{equation}
where $T_{ijkl}^{(cost)}$ is such that short adjacent links with small relative angles
are favored:
\begin{equation}
T_{ijkl}^{(cost)} = \delta_{jk}\frac{cos^{m}\theta_{ijl}}{r_{ij}^{n}+r_{kl}^{n}}
\label{eq:tcost}
\end{equation}
where $m$ is an odd exponent, $\theta_{ijl}$ is the relative angle between the
link $(ij)$ and the $(jl)$, $r_{ij}$ is the length of the $(ij)$ link and $n$ an 
integer (see Fig.~\ref{fig:Neuron}).
With this choice the energy of \eqref{eq:NNenergy} will blow up
for long links and large angles.
\begin{figure}[!ht]
\begin{center}
  \includegraphics[width=8cm]{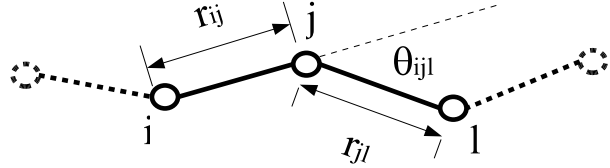}
\end{center}
\caption
[Definition of the neural network parameters associated to links.]
{Definition of link lengths $r_{ij}$ and angles $\theta_{ijl}$ between links.}
\label{fig:Neuron}
\end{figure}

The constraint term $T_{ijkl}^{(constraint)}$ consists of two parts:
\begin{equation}
T_{ijkl}^{(constraint)} = T_{ijkl}^{(1)}+T_{ijkl}^{(2)}
\label{eq:tconst}
\end{equation}
The first term in \eqref{eq:tconst} takes care of the requirement that there should be
no bifurcating tracks: two segments with one identical index should not be ``on''
at the same time. This is done with the following choice for $T_{ijkl}^{(1)}$:
\begin{equation}
T_{ijkl}^{(1)} = -\frac{1}{2}\alpha[\delta_{ik}(1-\delta_{jl})+\delta_{jl}(1-\delta_{ik})]
\label{eq:tconst1}
\end{equation}
The other term, $T_{ijkl}^{(2)}$, ensures that the number of neurons that are ``on'' is
roughly equal to the number of hits $N$. It has the generic form:
\begin{equation}
T_{ijkl}^{(2)} = -\frac{1}{2}\beta\biggl[\sum S_{kl} - N\biggr]^2
\label{eq:tconst2}
\end{equation}
This configuration of the synaptic strenghts obviously is symmetric ($T_{ijkl}=T_{klij}$)
and with vanishing diagonal elements ($T_{ijij}=0$).

The final expression for the energy function is:
\begin{equation}
E = -\frac{1}{2}\sum\delta_{jk}\frac{cos^{m}\theta_{ijl}}{r_{ij}^{n}+r_{kl}^{n}}S_{ij}S_{kl}
    +\frac{1}{2}\alpha\sum_{l \neq j}S_{ij}S_{il}
    +\frac{1}{2}\alpha\sum_{k \neq i}S_{ij}S_{kj}
    +\frac{1}{2}\beta\biggl[\sum S_{kl} - N\biggr]^2
\label{eq:NNenergy2}
\end{equation}
The $\alpha$ and $\beta$ factors are Lagrange multipliers entering in the definition of
the energy function.
At a given time the state of the system is given by the vector $\mathbf{S}=(...,S_{ij},...)$,
while the dynamics of the system is determined by the $\mathbf{T}$~matrix of synaptic 
connections. 
By construction, the solution of the problem of track finding is the state of the 
neurons $\mathbf{S}$ which minimizes the energy function \eqref{eq:NNenergy2}.
The local updating rule which brings the system to a local minimum through the 
gradient descent method of the energy function is \cite{NNHopfield}:
\begin{equation}
S_{i}=sign\biggl(\sum_{j}T_{ij}S_{j}\biggr)
\label{eq:NNlocalmin}
\end{equation}
However, in general is necessary to reach the global minimum of the system for a better
performance. One way to get this is to expose the system to a noisy environment, in which
the the state vector $\mathbf{S}$ obeys the Boltzmann distribution:
\begin{equation}
P(\mathbf{S})=\frac{1}{Z}\exp^{-E(\mathbf{S})/T}
\label{eq:NNboldist}
\end{equation}
where the partition function $Z$
\begin{equation}
Z=\sum_{\mathbf{S}}\exp^{-E(\mathbf{S})/T}
\label{eq:NNpartfun}
\end{equation}
is the sum over all possible states and $T$ is the temperature of the system.

For an efficient and fast perform of the system relaxation to the minimum, is usual to 
make use of the mean-field theory (MFT) approximation \cite{NNmeanfield}
and get for the thermal average of $S_{ij}$:
\begin{equation}
V_{ij}\equiv\langle S_{ij}\rangle_{T}=\tanh\biggl(-\frac{1}{T}\frac{dE}{dV_{ij}}\biggr)
      =\tanh\biggl(\frac{1}{T}\sum_{kl}T_{ijkl}V_{kl}\biggr)
\label{eq:NNvi}
\end{equation}
Note that the local update rule in \eqref{eq:NNlocalmin} can be 
obtained from \eqref{eq:NNvi} taking the $T\rightarrow 0$ limit.
$V_{ij}$ can take any value between -1 and 1, and is more convenient to limit
the range to the interval $[0,1]$ in other to get no contribution from deactivated
neurons. To this purpose, the following linear transformation is taken
\begin{equation}
V_{ij}=\frac{1}{2}\biggl[1+\tanh\biggl(-\frac{1}{T}\frac{dE}{dV_{ij}}\biggr)\biggr]
\label{eq:NNvi2}
\end{equation}
The final update rule for the neurons is obtained combining Eqs.~\eqref{eq:NNvi2} 
and \eqref{eq:NNenergy2}:
\begin{equation}
V_{ij}=\frac{1}{2}\biggl[1+\tanh\biggl(
     \frac{1}{T}\sum_{k}\frac{cos^{m}\theta_{ijk}}{r_{ij}^{n}+r_{jk}^{n}}V_{jk}
    -\frac{\alpha}{T}\biggl(\sum_{k \neq j}V_{ik}+\sum_{k \neq i}V_{kj}\biggr)
    -\frac{\beta}{T}\biggl(\sum_{kl}V_{kl} - N\biggr)
    \biggr)\biggr]
\label{eq:NNupdaterule}
\end{equation}
Given an initial state of the neurons, the dynamics of the system is fully 
determined by \eqref{eq:NNupdaterule}. 

\subsubsection{Algorithm operation}
The neural network described above can be applied to the 2D track recognition problem 
in a certain cluster. It takes as input the links between the hits (neurons) as they 
were introduced in Sec.~\ref{sec:link} (each hit is just linked with its {\it close} 
hits) and, after the iteration process described by the update rule in
\eqref{eq:NNupdaterule}, the set of ``active'' links will define the track projections.
The algorithm operates as follows:
\begin{itemize}
\item All the neurons (links) are first initialized to an active state, $S_{ij}=1$
(\ref{fig:NNevol} (top)).
This configuration performs better than a random initialization in the particular
case of the interactions we have in the LAr TPC. 
This is because of such state is closer (in general) to the optimum 
state, which minimizes \eqref{eq:NNenergy2}.
\item The update of the neurons is performed asynchronously through Eq.~\eqref{eq:NNupdaterule}. 
This means that a neuron is updated after the other: 
the current updating neuron ``sees'' the other neurons in their actual state.
\item The process is iterated until a certain level of convergence is reached.
The adopted criteria for convergence is based on the changes of the activation
state of all neurons after each iteration:
\begin{equation}
\frac{1}{N}\sum_{ij}\lvert V_{ij}^{(I}-V_{ij}^{(I-1)}\rvert \leq 1\times10^{-5}
\end{equation}
\item After convergence, the state of a neuron $S_{ij}$ is activated if its 
mean value $V_{ij}$ is bigger or equal than $0.5$ and deactivated if lower 
(\ref{fig:NNevol} (middle)).
\item Finally, the 2D tracks projection are identified as chains of consecutive 
activated links (Fig.~\ref{fig:NNevol} (bottom)).
\end{itemize}

There are some free parameters in the definition of the update rule in 
\eqref{eq:NNupdaterule}: The $\alpha$ and $\beta$ factors can be modified in order 
to give more or less relative importance to the different
terms in \eqref{eq:NNupdaterule}. The exponents in the $T_{ijkl}^{(cost)}$ term of
\eqref{eq:tcost} rule the smoothness requirement for the track segments.
These parameters have to be defined in such a way that short and almost straight
links are favored. 
The configuration adopted in this work is the following: $m=5$, $n=2$, 
$\alpha=0.4$, $\beta=0.2$, which gives good quality of the results and fast convergence.
However, the behavior of the network does not depend strongly on the parameters and
other configurations are equally good.
On the other hand, the temperature $T$ determines the speed of the convergence. 
High temperatures means slower convergence but gives the system more chances 
to scape from local minima than in the case of low temperatures.

\begin{figure}[!ht]
\begin{center}
  \includegraphics[width=10cm]{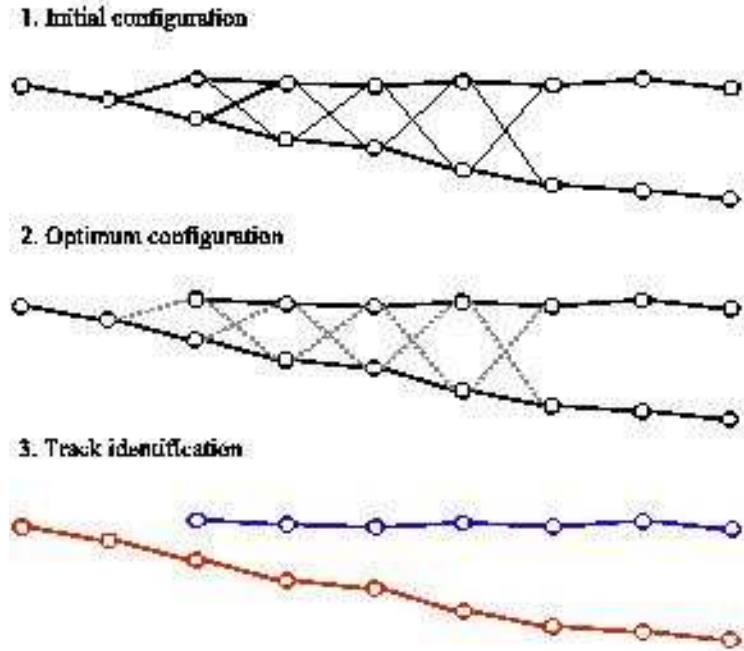}
\end{center}
\caption
[Illustration of the neural network algorithm for 2D track finding.]
{(Top) The starting point: All neurons enabled.
(Middle) The optimum configuration by means of the updating 
rule in \eqref{eq:NNupdaterule}.
(Bottom) 2D track segments are found as chains of consecutive enabled links.}
\label{fig:NNevol}
\end{figure}

\subsubsection{Gathered information after 2D track finding}
\label{sec:trackinfo}
Both methods introduced above efficiently detect 2D track segments inside 
a cluster of hits:
tracks are recognized as chains of hits with smooth transitions (links) 
between them. This association of hits into 2D track structures
provides a method to detect interaction vertexes in each wire plane, which
will be of utmost importance at 3D reconstruction stage. 
Although track projections are not necessarily straight, they are fitted to a straight line
which gives logical ``support'' and allows to extrapolate the track segments.
In addition, the 2D track detection allows to assign to each hit a track 
identification number as well as topological information related with their
surrounding hits and the track to which they belong. In such a way, 
a track bound can be easily defined in the same way as was done for cluster
bounds (see Sec.~\ref{sec:link}): The bounds of a track are those hits
having one link at most with another hit belonging to the same track 
(it can have however links with another hits of other tracks).
Besides, a track is an unidimensional object which can be parameterized like 
a curve: every hit occupies a given position in the chain.
In Fig.~\ref{fig:Tracks} (top-left) two tracks segments have been found in the cluster:
the hits colored in green (blue) belong to the same track, in light green (light blue) 
the bounds of the tracks have been marked as well. 

\subsubsection{Hit parameterization after 2D track reconstruction}
%%%%%%%
The information gathered by the 2D track finding procedure is stored as 
the following parameters in the hit structure:
\begin{itemize}
\item The track identification number.
\item The number of links with other hits within the same track segment.
\item The list of hits to which the hit is linked within the same track segment.
\item The position of the hit in the track chain.
\end{itemize}

\subsubsection{Track expanding and vertex finding}
This information gathered after the 2D track finding procedure can be used
to achieve two important tasks: Track expanding and vertex finding.
A vertex, in a certain wire projection, can be defined as the point where two or more
tracks meet. Because it is difficult to resolve the different hits in the 
vicinity of a vertex, in most of the cases, the track segments do not reach to 
the intersection of the tracks (see Fig.~\ref{fig:Tracks} (top-left)). 
This technical issue can be tackled in the following way:
\begin{itemize}
\item The fitted track is extrapolated from the bound hits trying to find
another track bound in its extrapolated path (Fig.~\ref{fig:Tracks} (top-right)).
\item If such a bound is found, the track is expanded by means of {\it virtual}
hits until that bound hit (Fig.~\ref{fig:Tracks} (bottom-left)). 
{\it Virtual} hits (represented in pink in 
Figs.~\ref{fig:Tracks} bottom-right and bottom-left) are only logical 
and do not provide any calorimetric information, they are created to complete 
the geometrical reconstruction of the tracks.
\end{itemize}

\begin{figure}[!ht]
\begin{center}
  \begin{tabular}{ c c }
  \includegraphics[width=7cm,height=6cm]{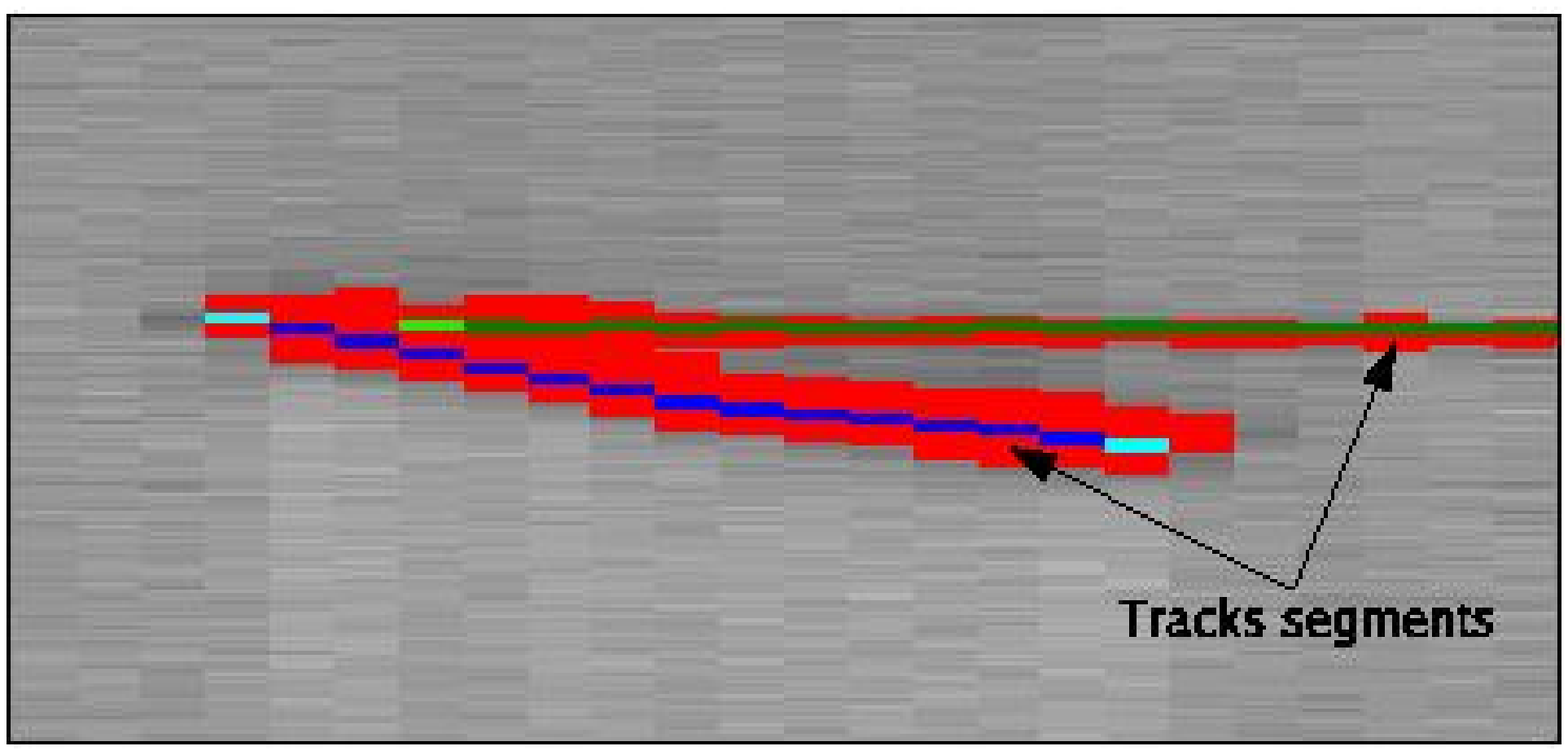} &
  \includegraphics[width=7cm,height=6cm]{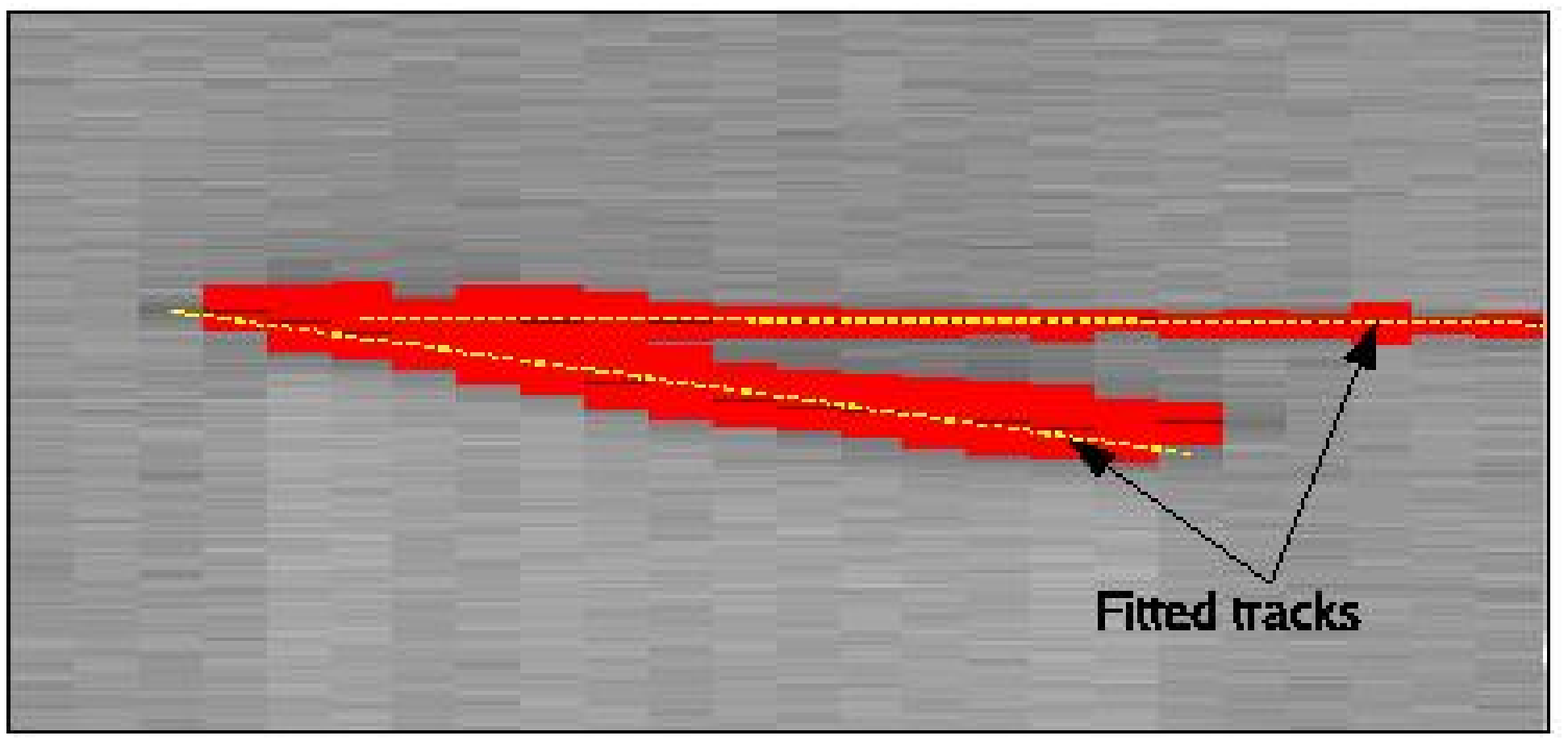} \\[0.3cm]
  \includegraphics[width=7cm,height=6cm]{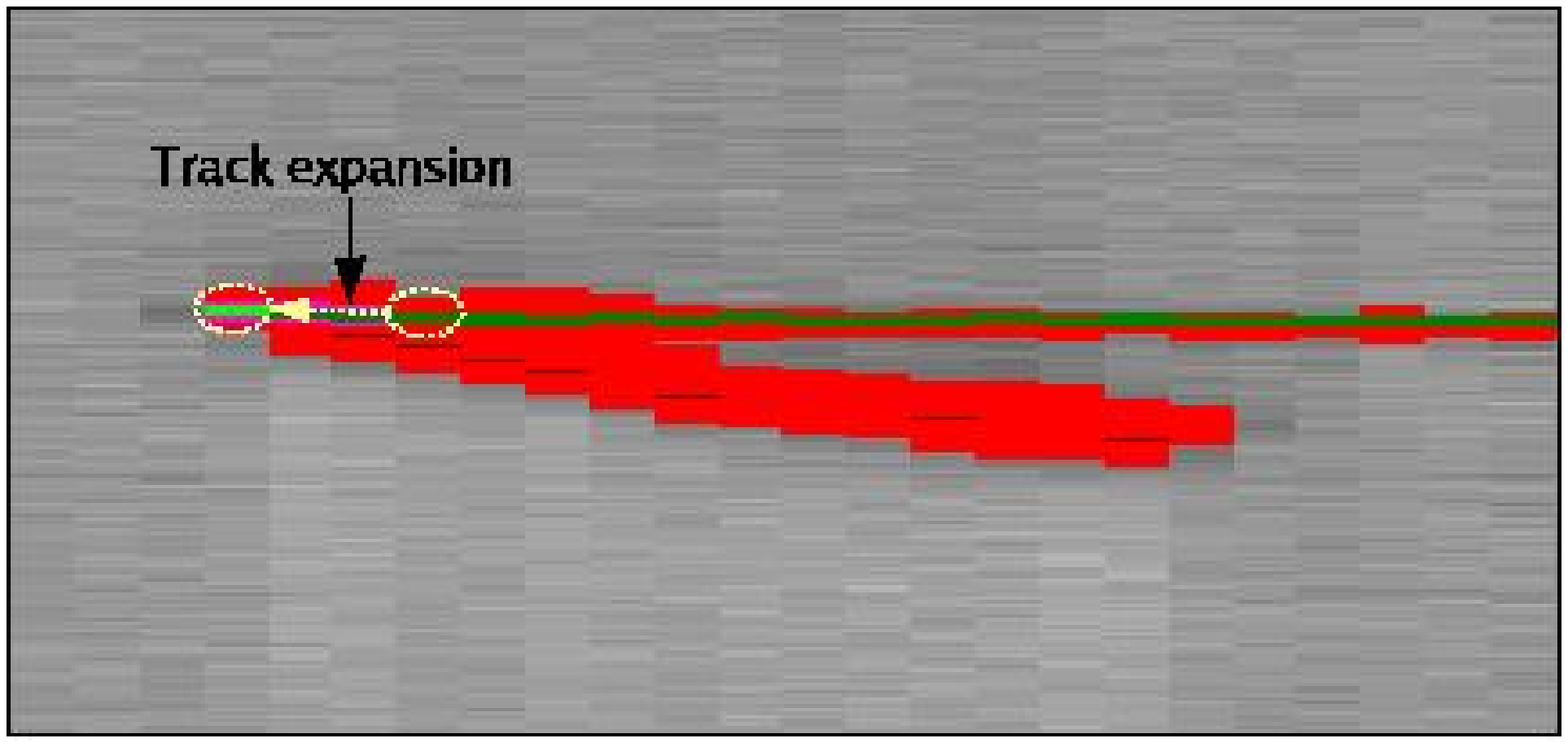} &
  \includegraphics[width=7cm,height=6cm]{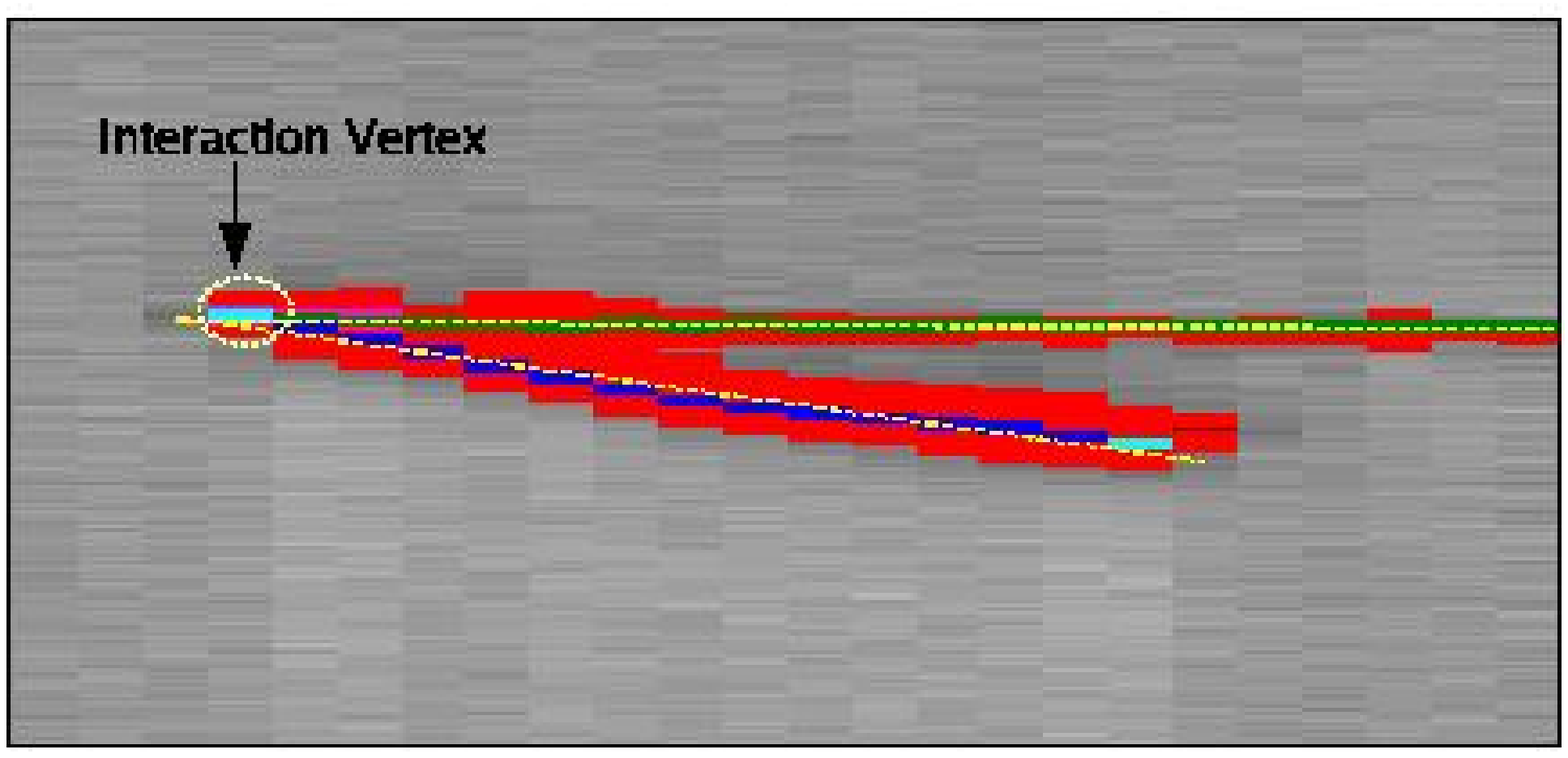} \\
  \end{tabular}	
\end{center}
\caption
[Example of found tracks in a cluster]
{The track finding procedure finds two tracks in the cluster of 
Fig.~\ref{fig:linkscluster}: 
(Top-left) The two found tracks are highlighted in green and blue respectively. 
(Top-right) The tracks are fitted to a straight line.
(Top-left) If the extrapolation of a track connects with the bound hit of another
one, the track is expanded until that hit.
(Bottom-right) Final step, the track segments are properly reconstructed and the 
interaction vertex is recognized.}
\label{fig:Tracks}
\end{figure}

The latter procedure completes the 2D track finding approach in a proper way,
providing the right coordinates for the track segments.
Finally, vertexes are easily identified as those border hits in the track which
have a neighbor which is also a border hit of another track.
In Fig.~\ref{fig:TracksDIS}, the track finding method performance is shown:
The eight ionizing tracks of a deep-inelastic (DIS) neutrino event have been
detected, expanded and the interaction vertex have been recognized.

\begin{figure}[!ht]
\begin{center}
  \includegraphics[width=12cm,height=8cm]{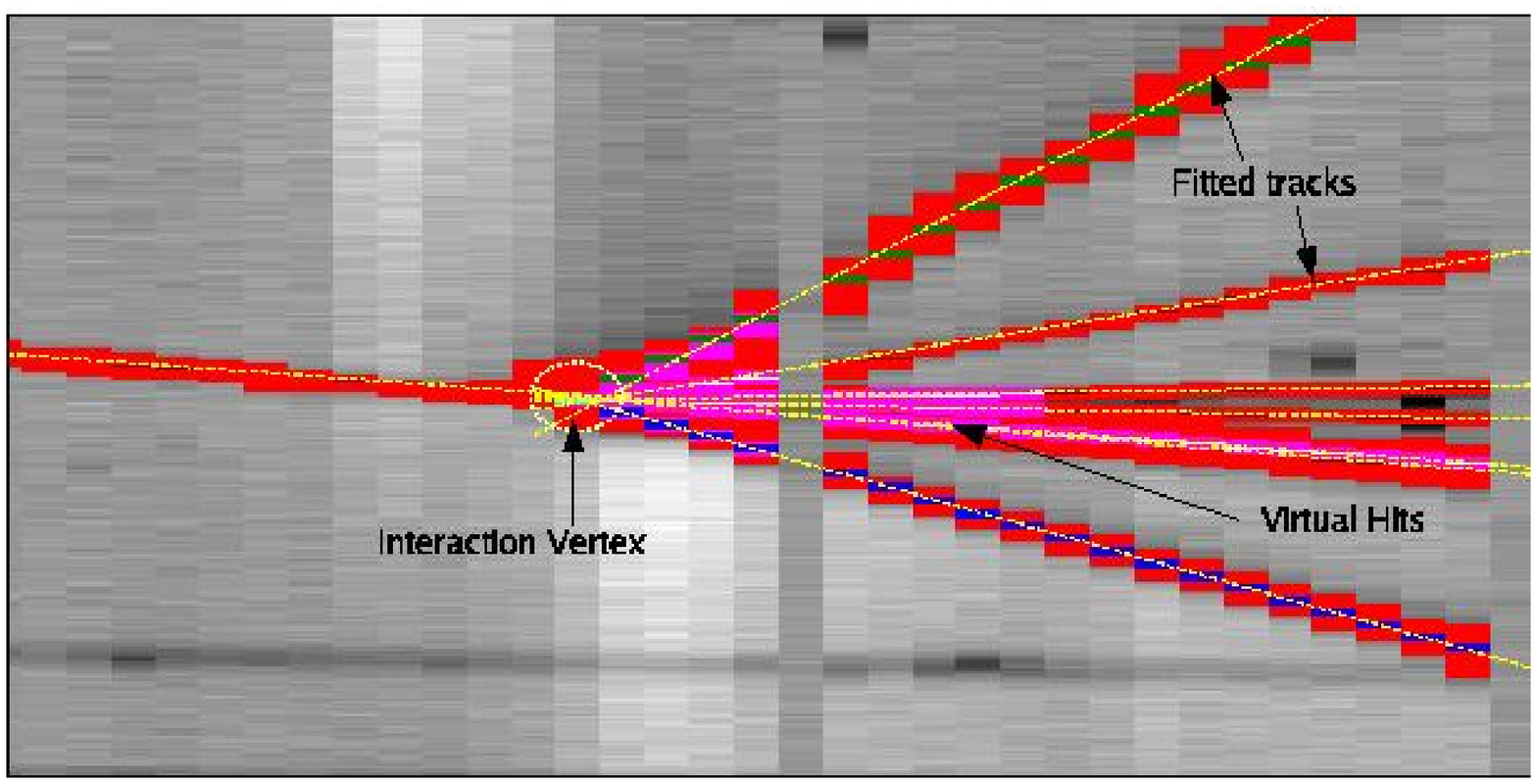}
\end{center}
\caption
[Example of the track finding performance]
{Example of the track finding performance over a DIS neutrino interaction 
cluster with eight prong particles. Two of the tracks are highlighted in green and 
blue respectively. The fitted lines are also shown, as well as the virtual hits 
(in pink) used to expand the tracks till the interaction vertex.}
\label{fig:TracksDIS}
\end{figure}

%%%%%%%%%%%%%%%%%%%%%%%%%%%%%%%%%%%
\section{3D reconstruction}      %%
%%%%%%%%%%%%%%%%%%%%%%%%%%%%%%%%%%%
\label{sec:3drec}

The goal of the three dimensional (3D) reconstruction is the
determination of the space coordinates of the different energy
deposition segments produced by the ionizing tracks traversing the
LAr sensitive volume. Each wire plane constrains two spatial degrees
of freedom of the hits, one common to all the wire planes (the drift
coordinate) and one specific for each plane (the wire coordinate). The
redundancy on the drift coordinate allows the association of hits
from different planes to a common energy deposition, and together with
the wire coordinates from at least two planes, allows the hit spatial
reconstruction of the hit.

The 3D reconstruction algorithm developed for the LAr TPC has to different 
modes of operation:
\begin{itemize}
\item The first one collects the hits within a cluster and tries to match them
with the hits of complementary views.
\item The second one does the same, but its search of complementary hits 
is based on the detected 2D track segments (Sec.~\ref{sec:2dtrack})
of different views: it tries to match first the 2D tracks projections and thereafter
match the hits belonging to them.
\end{itemize}

In the following sections we describe both approaches to the 3D reconstruction. 
For latter usage, we start summarizing the
expressions to transform from the wire/drift coordinates
system into the Cartesian system for the 50L TPC.

\subsection{Transforming wire/drift coordinates into 3D Cartesian coordinates}
%%%%%%%%%%%%%%%%%%%%%%%%%%%%%%%%%%%%%%%%%%%%%%%%%%%%%%%%%%%%%%%%%%%%%%%%%%%%%
\label{sec:transform}

We define the local Cartesian reference frame (see Fig.~\ref{fig:refsys})
centered at the bottom left corner of the electronic readout of the 
LAr sensitive volume. Because both the wire direction on
the Induction and Collection planes run orthogonally each other,
the transformation to the Cartesian reference frame is very simple.
Thus, coordinates $x$, $y$ and $z$ are associated directly to the Induction wires
(in decreasing number direction), to drift time and to the Collection wires 
(increasing number direction) respectively.
We explicitly quote below these transformations:
\begin{equation}
\begin{split}
x & = (N_{wires}-1-(w_{ind}+0.5))~p \\
y & = s~(v_{d}/f)             \\
z & = (w_{col}+0.5)~p               
\end{split}\label{eq:3dcoord}
\end{equation}
\noindent
where the constants entering the formulae are the following: 
$N_{wires}$~($= 128$~wires), the number of read wires in Induction plane;
$p$~($= 2.54$~mm), the wire pitch at LAr temperature;
$f$~($= 2.5$~MHz), the readout sampling frequency;
and $v_{d}$~($=0 .91$~mm/$\mu$s), is the electron drift velocity
(see Sec.~\ref{sec:driftvel}).
Finally, $w_{ind}$, $w_{col}$ and $s$ are related to the indexes of the wires 
recording the hit in Induction and Collection wire planes, and the drift time 
coordinate (in number of samples), respectively. 

\begin{figure}[!ht]
\begin{center}
\includegraphics[width=9cm,height=10cm]{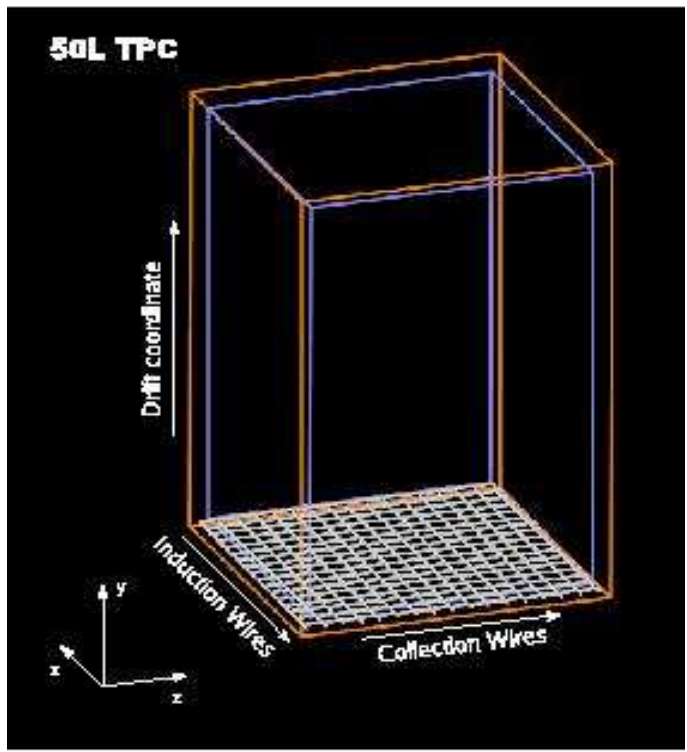}
\end{center}
\caption[Sketch of the 50L TPC and the Cartesian local reference frame]{
Sketch of the 50L TPC and the Cartesian local reference frame.}
\label{fig:refsys}
\end{figure}

\subsection{3D reconstruction algorithm hit by hit}
%%%%%%%%%%%%%%%%%%%%%%%%%%%%%%%%%%%%%%%%%%%%%%%%%%
\label{sec:3dalgo}
The hit 3D reconstruction algorithm takes as input the sample of hits
belonging to any reconstructed cluster from an arbitrarily selected
view (called \emph{main view}), and searches for hits from
complementary secondary views corresponding to the same ionizing track
segment. Such a hit is called \emph{associated} hit. The combined
information from a hit and its associated hit from one or more
complementary views allows the determination of the hit spatial
coordinates.

The algorithm proceeds in three steps (passes). In pass~I, all
possible associations between hits from the main and the complementary
views are carried out. The associations are established based on the
redundancy of the drift coordinate and the relations between hits
within the clusters (links). Pass~II searches for deficient
associated hits, redefining their position based on the neighboring
hits. Finally, in pass~III, the optimal spatial coordinates
for unassociated hits are computed.

\subsubsection{Pass~I}
\label{sec:3dalgopI}
The goal of pass~I is to identify the maximum possible number of
associations for hits from the main view, combining single-view (links)
and multi-view (associations) relations between hits.

The first and most important step is the search for associations of the
cluster bound hits (i.e. hits with at most one link to the cluster). 
This provides a starting reference point for the association
of the remaining hits of the cluster. Cluster bound hits are required
to be associated to cluster bound hits from the complementary view.
The remaining hits are looped over following the cluster links.  
In first approximation, a given hit $h_1$ linked to a hit $h_2$ is
associated to a hit $h^c_1$ from the complementary view only if
$h^c_1$ is linked to a hit $h^c_2$ associated to $h_2$. If the
previous condition cannot be fulfilled, $h_1$ is associated to the hit
association candidate $h_1^c$ closest to $h^2_c$.

The hit association candidates are those hits from a complementary
view with equal drift coordinates (within a tolerance of
$5$ drift samples) and compatible wire coordinates. 

\subsubsection{Pass~II}
%%%%%%%%%
\label{sec:3dalgopII}

The inefficiencies of the hit finding and cluster reconstruction
algorithms (see sections~\ref{sec:hit_id} and \ref{sec:cluster})
together with the presence of ambiguities in the determination of the
hit association result in a non negligible level of bad hit
associations and left-over hits after pass~I. 

In order to increase the hit association rate and quality, pass~II
searches and repairs failures on hit associations.  For this, we
loop over the hits starting from cluster bounds and following the
links within the cluster. The following algorithm is applied:

\begin{itemize}
\item For every two consecutive (linked) hits, $h_a$ and $h_b$, having both
an associated hit in a complementary view, the position in space of
the next hits are compared to the extrapolation defined by the
straight line joining $h_a$ and $h_b$.
\item All hits ($h_1, ..., h_n$) found until the first hit $h_{n+1}$
whose distance to the extrapolated line is smaller than $1$~cm are flagged for an 
ulterior 3D coordinates (re)assignment. 
Flagged hits are freed when no $h_{n+1}$ hit fulfilling the previous condition is 
found before a cluster bound is reached.

\item Spatial coordinates of the flagged hits are (re)computed in the
following way: coordinate $y$ is computed by using the corresponding
expression from Eq.~\eqref{eq:3dcoord}; coordinates $x$ and $z$ are
interpolated using the straight line parameters defined by the projection of $h_b$
and $h_{n+1}$ in the $(x,z)$ plane.
\end{itemize}

This algorithm repairs in an efficient way the position of those hits
coming from straight line or small angle tracks, yielding no effect on
hits coming from large angle scattering tracks. A third pass is needed
in order to reduce to zero the rate of left-over hits.

\subsubsection{Pass~III}
%%%%%%%%%
\label{sec:3dalgopIII}

Hits remaining  unassociated after pass~II are reconstructed in space during
pass~III, in order to reach the required 100$\%$ efficiency on the hit
3D reconstruction. Pass~III algorithm proceeds as follows:
\begin{itemize}
\item
Starting from every unassociated hit $h_0$, the hits in the cluster
are looped over, following the link path in both increasing and
decreasing wire coordinate directions. For each direction, the loop
continues until the first associated hit (reference hit) is found.

\item
If reference hits at both sides of $h_0$ ($h_a$ and $h_b$) are found
before the cluster bounds are reached, the spatial coordinates of all
unassociated hits between them are computed as in pass~II.

\item
If one of the reference hits is not found, the spatial coordinates of
the unassociated hits are computed using a straight line extrapolation.
The straight line parameters are defined by the found reference ($h_a$ or $h_b$) 
and the closest associated hit met after following the links in the same direction 
we used to find the reference hit.

\end{itemize}

\subsection{3D reconstruction by 2D track matching}
\label{sec:3Dfrom2D}
This approach to the 3D reconstruction of hits is completely based on the information 
provided by the 2D track projection finding method. The effort spent in finding 
the 2D tracks leads to a very clear, simple and efficient 3D reconstruction algorithm.
It works as follows:

\begin{itemize}
\item Starting from every track segment $t_i$ in the \emph{main view}, the algorithm
tries to match the track bounds with the bounds of any another track segment in 
the complementary view~\footnote{Obviously, the different projections of the same
ionizing track must have equal drift coordinates of their bounds 
(up to a difference of $5$ drift samples).}.

\item If a complementary track $t^c_i$ is found, both projections are 
associated to each other and the algorithm begins the matching of the hits 
within the tracks.

\item  Taking advantage of the track segments structure, the matching of the 
hits is performed exploiting the position of the hits within the track
\footnote{Every track has a structure defined by the chain of hits and every 
hit occupies a certain position in that chain (see Sec.~\ref{sec:trackinfo}).}: 
The association of hits from complementary tracks is based
on the drift coordinate like in Sec.~\ref{sec:3dalgopII} and on the relative 
hit position in the track. 
Given a hit $h_i$ belonging to a certain track $t_i$
in the main view, its associated hit $h^c_i$ is searched from the hits in 
the associated track $t^c_i$ in the following way:
All the hits in $t^c_i$ whose drift time is compatible 
with the drift time of $h_i$ are collected; if there are more than a single candidate,
the hit $h^c_i$ whose relative position in $t^c_i$ is closer to the relative position
of $h_i$ in $t_i$ is selected.
\end{itemize}

In Fig.~\ref{fig:2dtrackmatch} we show an example of how this algorithm works on a 
low multiplicity neutrino event: Tracks projections are correctly determined in both
views and then matched by time coincidences of their borders.

\begin{figure}[!ht]
\begin{center}
\includegraphics[width=14cm]{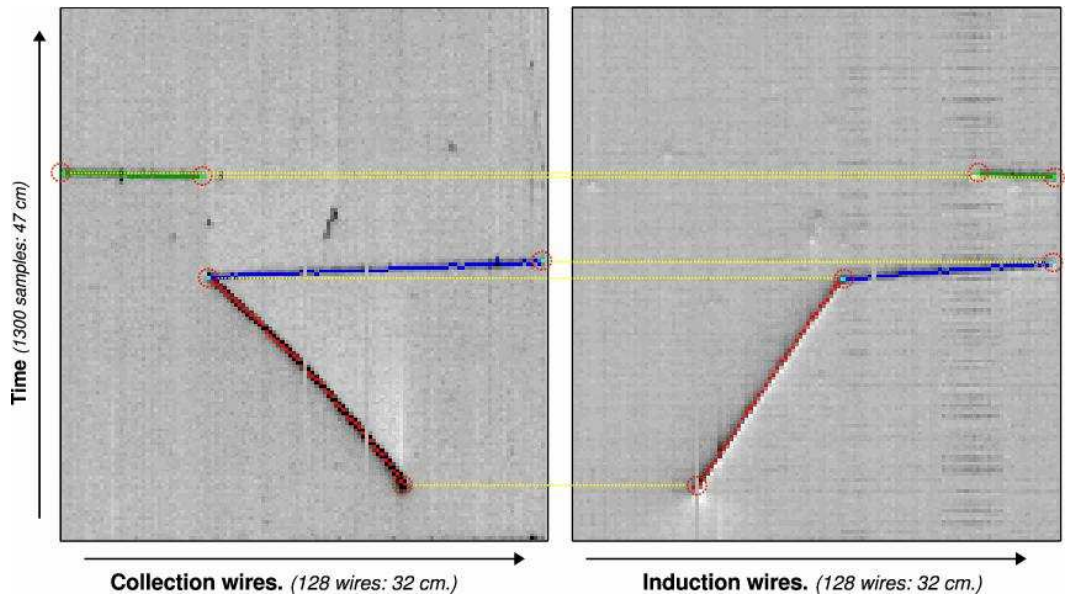}
\end{center}
\caption[Example of the 2D track matching procedure]
{Example of the 2D track matching procedure over a low-multiplicity neutrino 
interaction. The tracks projection are correctly found and matched.}
\label{fig:2dtrackmatch}
\end{figure}

This simple algorithm has an excellent performance when the track projection 
segments are correctly associated. For the rest of the cases where any track
segment can not be properly matched, the 3D reconstruction is performed using the 
hit by hit method (see Sec.~\ref{sec:3dalgo}). 

\subsubsection{Hit parameterization after 3D reconstruction}
%%%%%%%
The picture of the hit is completed by adding the following
information to the hit parameterization:

\begin{itemize}
\item
The associated hit(s) from the complementary view(s) and its complementary track
segment (if any).
\item 
The space coordinates, $x$, $y$ and $z$ in the Cartesian reference frame.
\end{itemize}

\subsection{Reconstruction performance}
%%%%%%%%%%%%%%%%%%%%%%%%%%%%%%%%%%%%%%%%%%%%%%%%%%%%%%%%

The 3D reconstruction of hits performs efficiently on events of low
multiplicity composed of several well-defined tracks (not necessarily straight). 
Since it is purely based on geometrical criteria, the
reconstruction method is not meant to be carried out on events with
more complex energy deposition patterns, such as electromagnetic or
hadronic showers. For these kind of events, a method combining spatial
and calorimetric relations among hits needs to be developed.

This section is essentially dedicated to present some examples of the performance 
of the spatial reconstruction algorithm operating on some neutrino events collected 
during the running of the 50L TPC exposed to the CERN neutrino beam 
(Chapter~\ref{chap:50Lexperiment}). The 3D reconstruction of a $\nu_\mu$~CC
event with four charged particles in final state is shown in Fig.~\ref{fig:DISrec1}.
The tools developed in this work have been successfully applied to more complex 
topologies, as the case of the $\nu_\mu$~CC event shown in Fig.~\ref{fig:DISrec2},
where a deep-inelastic neutrino event with eight charged particles in final state
is fully reconstructed.

\begin{figure}%[!]
\begin{center}
\includegraphics[width=15cm]{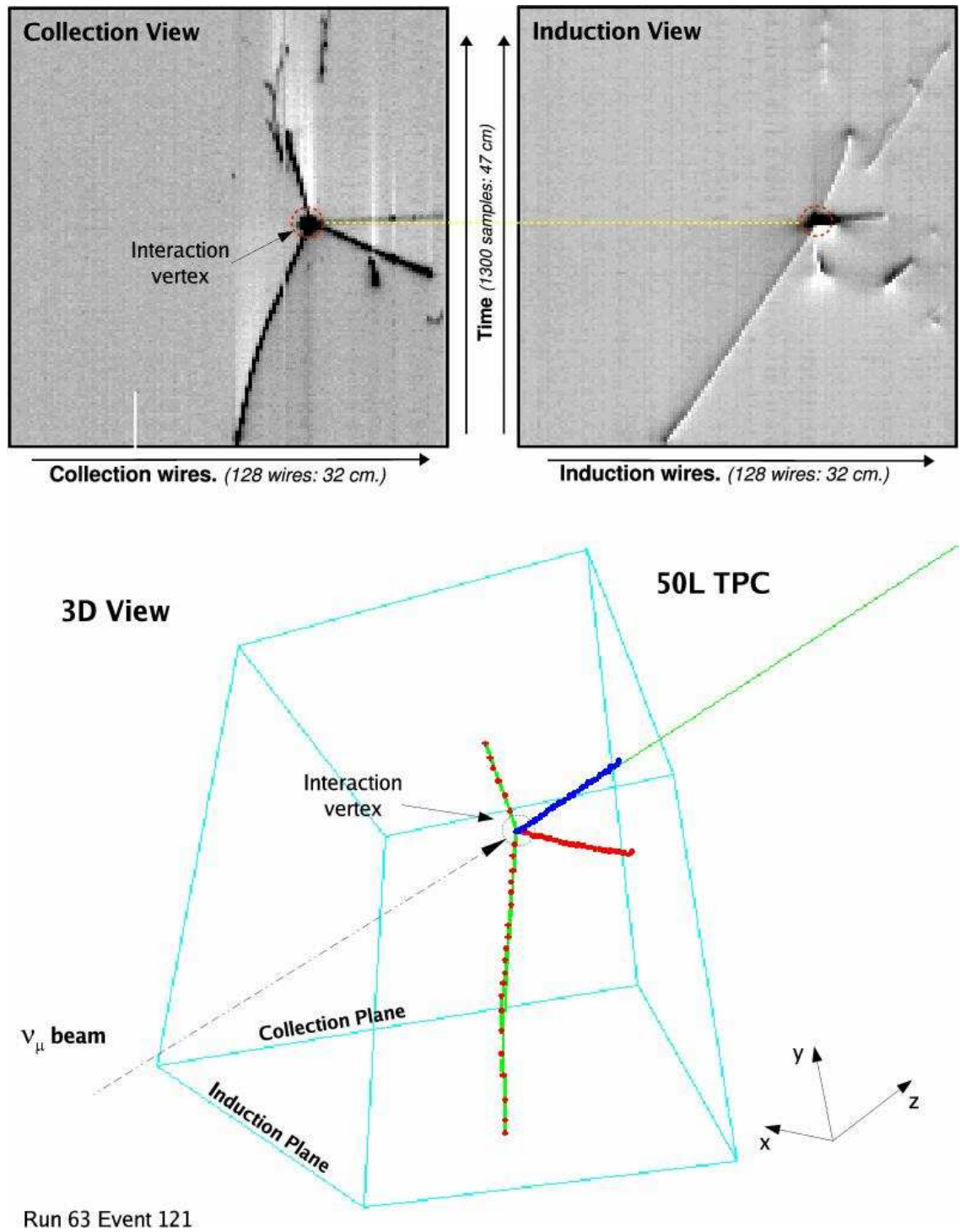}
\end{center}
\caption[Example of the 3D reconstruction of a 4-prong $\nu_\mu$ CC event]
{Example of the 3D reconstruction of a 4-prong $\nu_\mu$ CC event. (Top) 
The two raw data views of a $\nu_\mu$~CC interaction which breaks a nucleon into 
four visible charged particles. (Bottom) Three dimensional view of the 
reconstructed event.}
\label{fig:DISrec1}
\end{figure}

\begin{figure}%[!]
\begin{center}
\includegraphics[width=15cm]{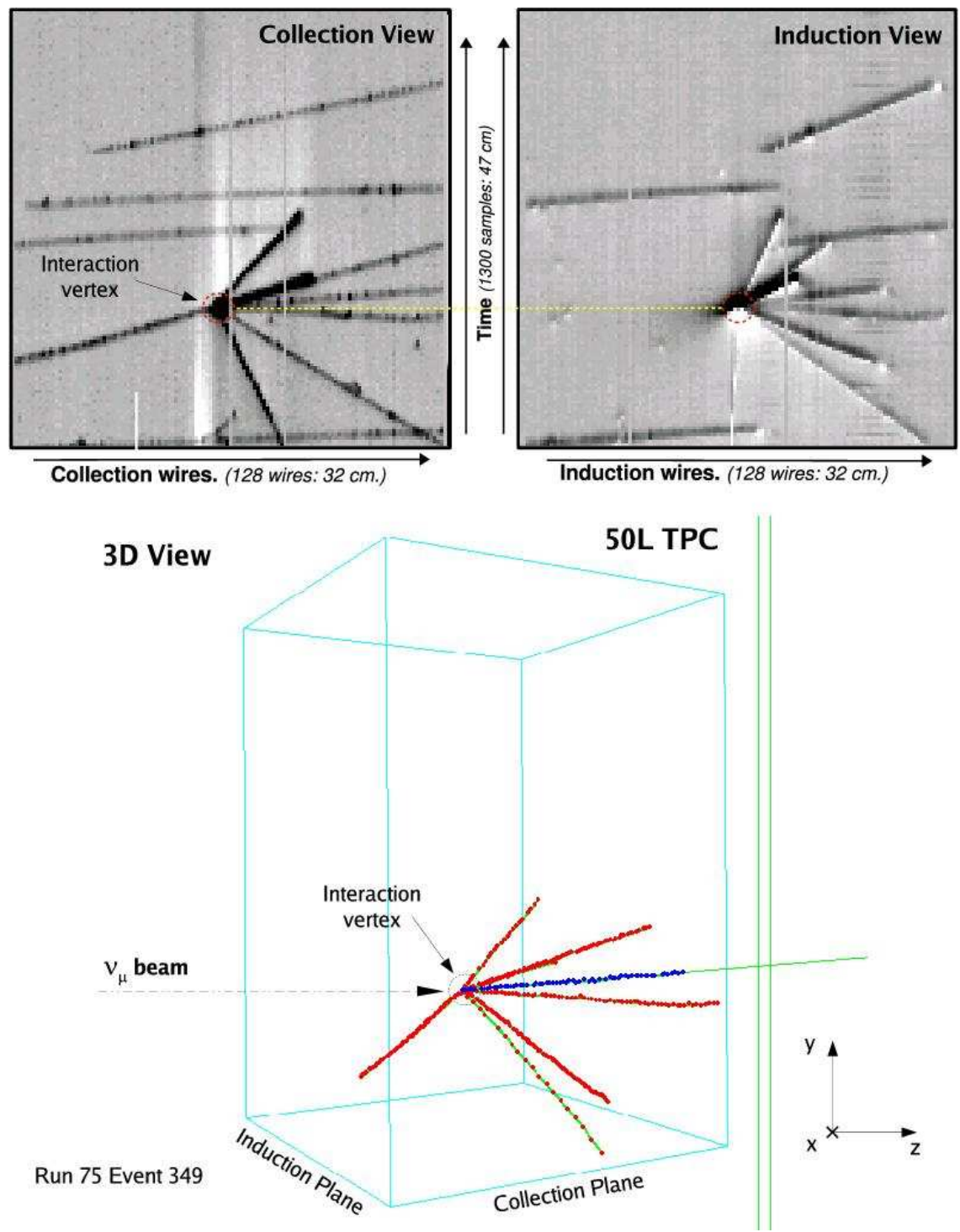}
\end{center}
\caption[Example of the 3D reconstruction of a multiprong deep-inelastic $\nu_\mu$~CC 
event]
{Example of the 3D reconstruction of a multiprong deep-inelastic $\nu_\mu$ CC event.
(Top) The raw data from Collection and Induction views. (Bottom) The three dimensional
trajectories of the eight primary particles are successfully recovered.}
\label{fig:DISrec2}
\end{figure}

\newpage
\section{Calorimetric reconstruction}
\label{sec:calrec}
A proper spatial reconstruction of the tracks together with the information 
extracted from the fitting of the hits allows to precisely measure the energy 
deposited by a particle along its path. This capability of LAr detectors is of
utmost importance since allows the energetic reconstruction of every stopping 
particle, as well as its identification based on the analysis of its energy loss
pattern.

\subsection{The average energy loss ($\bar{\Delta}$)}
\label{sec:average}

The average particle energy loss $-dE/dx$ is usually given in units of 
$MeV/(g/cm^2)$. The length unit $dx$ (in g/cm$^2$) is commonly used,
because the energy loss per surface mass density $dx = \rho \cdot ds$
with $\rho$ (in g/cm$^3$) being the density of absorbing material and $ds$ the
length (in cm) is largely independent of the properties of the material.
This length unit $dx$ consequently gives the surface mass density of the 
material.

According to the Bethe-Bloch formula, the average energy loss is given by the
following Eq.~\cite{Leo,Grupen}:

\vspace{0.1cm}
\begin{equation}
\begin{split}
- \frac{dE}{dx} & =
2 \pi N_A r^2_e m_e c^2 \; \frac{Z}{A} \; \frac{z^2}{\beta^2} \; \rho \; \cdot
\Bigl[ \ln \Bigl( \frac{2 m_e c^2 \beta^2 \gamma^2 W_{max}}{I^2} \Bigr)
              - 2\beta^2 -\delta  \Bigr] \\
                & = \frac{C_1}{\beta^2} \cdot
\Bigl[ \ln \Bigl( C_2 \beta^2 \gamma^2 W_{max} \Bigr)
              - 2\beta^2 -\delta  \Bigr]
\end{split}\label{eq:bethe}
\end{equation}
\noindent
with
\begin{itemize}
\item[-] $2 \pi N_A r^2_e m_e c^2 = 0.1535$~MeV cm$^2$/g;
\item[-] $z$: charge of the incident particle in units of $e$ ($z$=1 for muons);
\item[-] $Z,A$: atomic number and atomic weight of the absorber 
 ($Z$=18, $A$=39.948 for argon);
\item[-] $m_e, r_e$: electron mass (0.511~MeV) and classical electron radius
($2.817 \cdot 10^{-13}$~cm); 
\item[-] $N_A$: Avogadro number ($6.022 \cdot 10^{23}$ mol$^{-1}$);
\item[-] $I$: Ionization constant, characteristic of the absorber material
(188~eV for argon);
\item[-] $\beta = v/c$ of incident particle ($\gamma = 1/\sqrt{1-\beta^2}$);
\item[-] $\rho$: density of absorbing material (1.4~g/cm$^3$ for liquid argon);
\item[-] $W_{max}$: maximum energy transfer in a single collision (see below);
\item[-] $\delta$: density correction (see below);
\end{itemize}

For all practical purposes, Eq.~\eqref{eq:bethe} in a given material is a
function only of $\beta$ and only a minor dependence on $m_0$ (the mass of the
ionizing particle) at the highest energies is introduced through $W_{max} \;$. 
The expression as computed for muons in liquid argon 
is shown by the top curve in Fig.~\ref{fig:depbethe} exhibiting a minimum 
of 2.1~MeV/cm around $p_{\mu} =$~360~MeV.

The maximum transferable kinetic energy to an electron depends on the mass
and the momentum of the incident particle. It can be probed that for
relativistic particles ($E_{kin} \sim E$ and $p \sim E$, with $c=1$):

\begin{equation}
W_{max}(\gamma) = \frac{2m_e\beta^2\gamma^2}{1+2\gamma m_e/m_0 +(m_e/m_0)^2} 
                \approx \frac{E^2}{E + m_0^2 /2m_e}
\label{eq:wmax1}
\end{equation}
\noindent
Values for the density correction $\delta$ can be approximated by~\cite{Leo}:

\begin{tabular}{c|ll}
           & 0                                & $      X < X_0 $ \\
$\delta =$ & $4.6052 \, X + C + a(X_1 - X)^m$ & $X_0 < X < X_1 $ \\
           & $4.6052 \, X + C$                & $      X > X_1 $ \\
\end{tabular}

\vspace{0.2cm}
\noindent
where $X = log_{10}(\beta\gamma)$. The quantities $X_0, X_1, m, a$ and $C$
depend on the absorbing material. For liquid argon we assume the following 
values~\cite{yellowreport}: 
$X_0$=0.201, $X_1$=3, $m$=3, $a$=0.196 and $-C$=5.212.

It is also of interest to consider the mean energy loss excluding
energy transfers greater than some cutoff $T_{cut}$ (\emph{truncated}
mean energy loss)~\cite{PDG}:
\begin{equation}
-\left. \frac{dE}{dx}\right|_{T<T_{cut}} = 
2\, \pi\, N_a\, r_e^2\, m_e c^2\, \rho\,
\frac{Z}{A}\, \frac{1}{\beta^2} \left[\ln \frac{2\, m_e c^2\, \gamma^2\,
\beta^2\, T_{upper}}{I^2} - \beta^2 \left(1+\frac{T_{upper}}{W_{max}} \right) - \delta\right]
\label{eq:truncbethe}
\end{equation}
where $T_{upper}$ is the minimum between $T_{cut}$ and $W_{max}$, so
that the equation reproduces the Bethe-Bloch equation for $T_{cut} >
W_{max}$. In practice, the cut-off $T_{cut}$ is given by the
energy threshold for $\delta$-ray detection. 

Finally, in practical situations, the mean or average value is given by:
\begin{equation}
\bar{\Delta} = \int_0^{\Delta x} -\frac{dE}{dx}\, dx
\label{eq:deltadedx}
\end{equation}
where $dE/dx$ comes from the Bethe-Bloch formula (Eq.~\eqref{eq:truncbethe}), and
$\Delta x$ is the distance within we are measuring the energy loss.

\subsection{The most probable energy loss ($\Delta_{mp}$) }
\label{sec:mp}
The energy loss distribution for thin absorbers is strongly asymmetric, 
due to the intrinsic statistical nature of 
the ionization process. The fluctuation of the energy loss around the
mean value are well described by the {\it Landau} distribution~\cite{Landau}.
 According to the Landau's theory, the probability $f_L( x , \Delta )$ 
that a singly charged particle loses an energy $\Delta$ when traversing an 
absorber of thickness $x$ is expressed as:

\begin{equation}
f_L( x , \Delta ) = \frac{1}{\xi} \cdot \phi (\lambda)
\label{eq:landau}
\end{equation}
where
\begin{equation}
\phi (\lambda) = \frac{1}{\pi} 
\int^{\infty}_0 e^{ - u \ln u - \lambda u} \; \sin \pi u \; du
\label{eq:phi}
\end{equation}
\noindent
$\lambda$ is the deviation from the most probable energy loss $\Delta_{mp}$ :
\begin{equation}
\lambda = \frac{\Delta - \Delta_{mp}}{\xi}
\label{eq:lambda}
\end{equation}
where $\Delta$ is the actual energy loss in a layer of thickness $x$ and

\begin{equation}
\xi = 
0.1536 \cdot \frac{Z}{A} \; 
\frac{1}{\beta^2} \cdot \rho \; x \hspace*{0.5cm} [MeV]
\label{eq:xi}
\end{equation}
($\rho$ in g/cm$^3$ and $x$ in cm)

 For the particular case of LAr TPC's, $x \; (\equiv \Delta z)$ is the effective
portion of the track exposed to the wire, also called {\it track pitch
length}.

The {\it most probable energy loss} is calculated to be~\cite{Landau}:
\begin{equation}
\Delta_{mp} = \xi \Bigl[ \ln \frac{\xi \; 2 m_e c^2 \beta^2 \gamma^2 }{I^2}
              - \beta^2 + 1 - \gamma_E \Bigr]
            = \xi \Bigl[ \ln \frac{\xi}{\epsilon} + 0.198 - \delta \Bigr]
\label{eq:dmp}
\end{equation}
where $\gamma_E = 0.577$ is Euler's constant, $\delta$ is the density effect
and:

\begin{equation}
\ln \epsilon = \ln \frac{ (1 - \beta^2) I^2}{2 m_e c^2 \beta^2} + \beta^2
\label{eq:epsilon}
\end{equation}
Therefore, for a given material the most probable value energy loss only 
depends on the muon momentum and on the track pitch length. 
Eq.~\eqref{eq:dmp} represents a very good approximation for the most
probable energy loss of charged particles in thin absorbers on the vicinity
($\beta\gamma \sim 4$) and beyond the ``minimum-ionizing region''.
In practice, $\Delta_{mp}$ is extracted
by fitting the energy loss distribution with a Landau-Gaussian
convoluted function, where the Gaussian accounts for the resolution in
the charge measurement, mainly due to the electronic noise. The
Landau-Gaussian convolution is defined by the following expression:
\begin{equation}
f(\epsilon) = \frac{K}{\sqrt{2\pi}\sigma}\, \int_{-\infty}^{+\infty}
f_L(\epsilon')\, e^{-\frac{(\epsilon-\epsilon')^2}{2\sigma^2}}
d\epsilon'
\label{eq:landau-gaus}
\end{equation}
where $K$ is an overall normalization factor depending on the
statistics, $\sigma$ is the width of the convoluted Gaussian
distribution, $\epsilon$ the energy and $f_L$ is the Landau
distribution, given by Eq.~\eqref{eq:landau}.

% ------------------------------------------------------------------------
\begin{center}
\begin{figure}[!ht]
\begin{center}
\includegraphics[width=10cm]{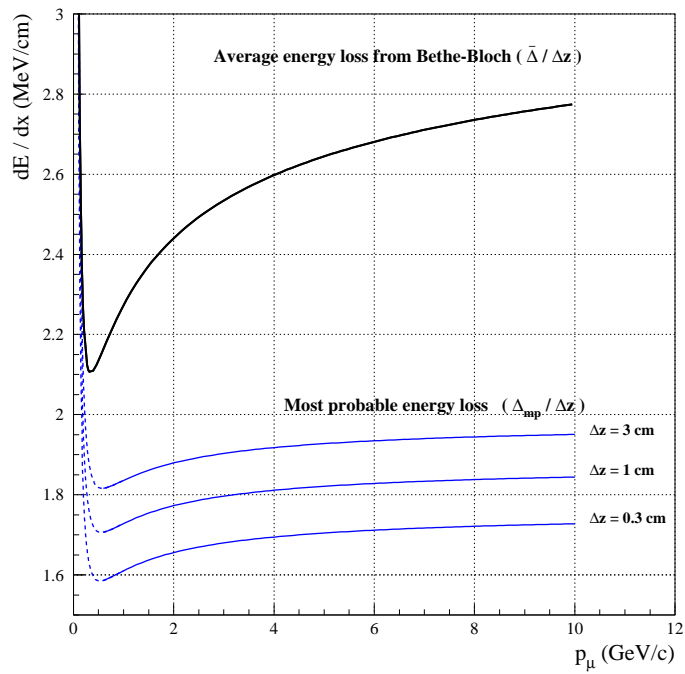}
\caption[Average energy loss $-dE/dx$ from Bethe-Bloch formula]
{Average energy loss $-dE/dx$ from Bethe-Bloch formula (top line)
compared to the {\it most probable} energy loss for three values of the 
track pitch length $\Delta z$ (bottom lines).}
\label{fig:depbethe}
\end{center}
\end{figure}
\end{center}
% ------------------------------------------------------------------------

\subsection{Average vs. most probable energy loss}
\label{sec:avvsmp}
In many practical situations, is necessary to parameterize the measured charge 
of a certain ionizing particle. In principle, we could do that by means 
of either \dmean~or \dmp, however, the use of the mean value presents two
technical problems, one intrinsic to the definition of this quantity,
the other related to the limited available statistics. On one
hand, the fluctuations of the energy loss are determined by a few
high-energy $\delta$-rays, and thus \dmean~depends on the $\delta$-ray
detection threshold ($T_{cut}$) entering
expression~\eqref{eq:truncbethe}. This quantity is not a priori known
since it has a strong dependence on the performance of the spatial
reconstruction algorithm. On the other hand, since
\dmean~strongly depends on the Landau tail, large statistics are
needed in order to significantly populate the tail and hence obtain a
reliable estimation of \dmean. None of the previous considerations
affects the determination of \dmp.

On the other hand, the use of \dmp~as energy loss estimator presents an additional
problem: while \dmean~is linear with the crossed distance (\Dx), as
shown in Eqs.~\eqref{eq:bethe} and \eqref{eq:deltadedx}, 
\dmp~behaves as $\Delta x \ln\Delta x$, as seen from Eqs.~\eqref{eq:dmp}
and~\eqref{eq:xi}. This means that the energy loss per crossed distance
measured at hits from track segments of equal kinetic energy but
different associated crossed distances yields Landau distributions with
equal \dmean~but different \dmp~values. 
The effect is shown in Fig.~\ref{fig:depbethe},
where the mean and most probable energy losses per crossed distance are
plotted as a function of the muon momentum for various $\Delta z$ values.

\subsection{Particle identification}
\label{sec:pid}
The function~\eqref{eq:bethe} as computed for protons, kaons, pions and muons
in liquid Argon, is shown by the curves in Fig.~\ref{fig:dEdxVsT}.
The lines follow the expected analytical ``average'' behaviour only.
The experimental points spread around
the corresponding curve for each type of particle because of the
fluctuations of the energy loss by catastrophic energy loss processes,
i.e. by interactions with high energy transfers, and because of the
detector resolution on the kinetic energy and $dx$ measurement.

 At non-relativistic energies, $dE/dx$ is dominated by the overall
$1/\beta^2$ factor and decreases with increasing velocity until
about $\beta \gamma \approx 4$  (or $\beta \approx 0.96$),
where a broad minimum of ionization is reached (a particle at
this point is known as a {\it mip}).

 From the figure, it is clear that for {\it energies below the minimum ionizing value},
each particle exhibits a $dE/dx$ curve which, in most cases, is distinct
from the other particle types because of the mass difference.
This is precisely the characteristic that has been exploited in this work
as a means for identifying particles in this energy range.

 The minimum value of $dE/dx$ is almost the same for all particles
of the same charge and, as the energy increases beyond this point,
the term $1/\beta^2$ becomes almost constant and all curves seem to
converge together to very similar $dE/dx$ values.
As soon as the detector resolution on the measured
kinetic energy and on the track pitch are taken into account,
the fact that all particles behave so close above the m.i.p.
region independently on the mass, make them indistinguishable
by means of the $dE/dx$ technique.

\begin{figure}[!ht]
\begin{center}
 \includegraphics[width=10.cm]{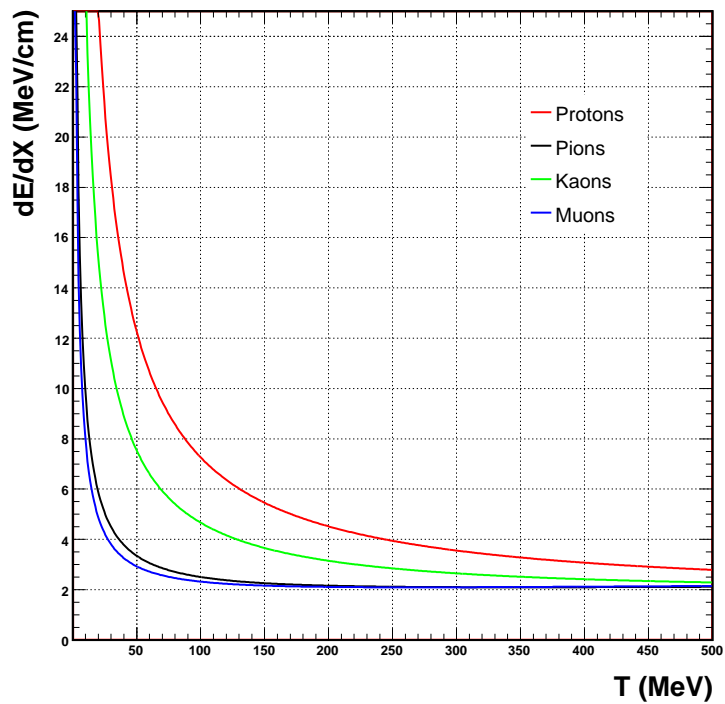}
 \caption[The stopping power $dE/dx$ as function of Kinetic Energy for
  different particles in LAr.]
 {The stopping power $dE/dx$ as function of Kinetic Energy for
  different particles in LAr.}
 \label{fig:dEdxVsT}
 \end{center}
\end{figure}

\subsection{Energy calibration}
\label{sec:ecal}

The energy calibration consists of transforming the charge associated to a hit
into energy units. The ionization charge is precisely measured in the 
Collection wire plane (see Sec.~\ref{sec:hitrecon}). 
The energy $E_i$ associated to a given hit is:
\begin{equation}
E_i \ = \ \frac{CW}{\mathcal{R}} \ Q_i(t)
\label{eq:QtoE}
\end{equation}%
where $C$ is the calibration factor; $W$ is the average energy needed for the 
creation of an electron-ion pair (i.e. 23.6 eV); 
$\mathcal{R}$ is the electron-ion recombination factor
and $Q_i(t)$ is the corrected charge by the electron lifetime from Eq.~\eqref{eq:exp_law}
in Sec.~\ref{sec:electron_lifetime}.
The calibration factor $C$ converts the detector measuring units (ADC counts) 
into charge units (fC), while the recombination factor $\mathcal{R}$
depends on the absorber medium, on the applied electric field and on the density
of released charge, i.e. on $dE/dx$. 

Thus, the energy calibration consist on determining $C$ and finding
a suitable model which describes the recombination factor $\mathcal{R}$ for a 
given detector configuration.

\subsubsection{The electron recombination factor}
\label{sec:recfactor}
Ionization electrons produced by the passage of a charged particle
through the LAr volume have a non negligible probability to recombine
with one of the positively ionized atoms. In such a case, the energy
released in the ionization appears in the form of photons and we say
that the drift electrons have been quenched.  The recombination
probability depends on the absorber, on the applied electric field,
and on the density of released charge, i.e. on $dE/dx$. This
dependence may be intuitively understood as follows: the higher the
field the faster the electrons and ions are separated, and the less
time is available for the recombination; on the other hand, the larger
the $dE/dx$, the higher the number of neighbouring positive ions which
are available for an electron to recombine.

The dependence of $\mathcal{R}$ with $dE/dx$ can be modeled by Birk's law
\cite{Birks} which has been successfully applied
to LAr detectors in \cite{Rec_factor}:
\begin{equation}
\frac{dQ}{dx} \ = \ \frac{a\, \frac{dE}{dx}}{ 1+k_B\, \frac{dE}{dx} }
\label{eq:birks}
\end{equation}%
$dE/dx$ being the pure Bethe-Bloch ionization loss,
$a$ and $k_B$ are constants depending on the applied
electric field and $k_B$ is the Birks coefficient accounting for quenching 
due to recombination. The recombination factor is the
ratio between the measured and the produced (theoretical) charge:
\begin{equation}
\mathcal{R}=\frac{\Delta Q^{meas}}{\Delta Q^{th}}
\end{equation}
Therefore, Eq.~\eqref{eq:birks} can be rewritten in terms of the
recombination factor by noting that $a$ has units of charge per
energy, so that it can be expressed as the inverse of the factor $CW$
(see Eq.~\eqref{eq:QtoE}) times a dimensionless multiplicative
constant ($b\equiv a\,CW$) ranging between 0 and 1. 
Multiplying both sides of the equation by $CW$ and dividing by $dE/dx$ we obtain:
\begin{equation}
\mathcal{R} = \frac{b}{1+k_B \frac{dE}{dx}}
\label{eq:recomb}
\end{equation}
A precise knowledge of the recombination factor is of the utmost
importance since it determines the absolute energy calibration of the
detector.

In Sec.~\ref{sec:50Lcal}, we present the analyses for the energy calibration 
for the 50L LAr TPC.

\subsection{Measurement method}
\label{sec:meascal}

From the previous discussions, it is clear that the determination
of the energy loss per crossed distance can be obtained by means of the measurement
of the deposited energy for every hit together with the calculation of its associated
track pitch length. In the following we discuss some aspects of the procedure 
to achieve this task in the framework or LAr TPC detectors.

\subsubsection{The track pitch length}
The length of the track portion associated to a hit $h_i$ (hereafter called
the \emph{track pitch length}: $\Delta z_i$) depends on the wire pitch and the orientation
of the track with respect to the wire plane. It can range from the
wire pitch length (for track pieces parallel to the wire pitch direction) 
to distances of the order of the chamber's side (for tracks perpendicular to the pitch direction). 
Thus, the track pitch length can be written in terms of the track direction in 
a given hit:
\begin{equation}
\Delta z_i  =  \frac{p}{\cos{\theta_i}}
\label{eq:dx}
\end{equation}
being $p$ the wire pitch and $\theta_i$ the angle between the track at the hit $h_i$
position respect to the direction of the wires (see Fig.~\ref{fig:trkseg}).

\subsubsection{Track segmentation}
In order to make an estimation of the track pitch length at any given hit $h_i$
belonging to a track, a partition of the track into segments of a certain number 
of hits is performed. Since a track is not straight in general,
the aim of this procedure is to obtain a parameterization of the trajectory of 
the track in terms of straight line segments, 
which locally approximate the real track direction.
Given a certain track segment with $n$ hits belonging to it, its direction is
determined by fitting a three dimensional straight line to these $n$ points. 
Besides, an additional constraint is imposed to this fit: 
any track segment uses the result to the fit of the previous calculated one
in order to have smooth transitions between the different parameterized lines, i.e.
the fitted lines belonging to consecutive segments must have one point in common.
As a result, we obtain a partition of the track trajectory into straight segments, 
which are consecutively connected defining a trajectory (see Fig.~\ref{fig:trkseg}).

\begin{figure}[!ht]
\begin{center}
  \includegraphics[width=14cm]{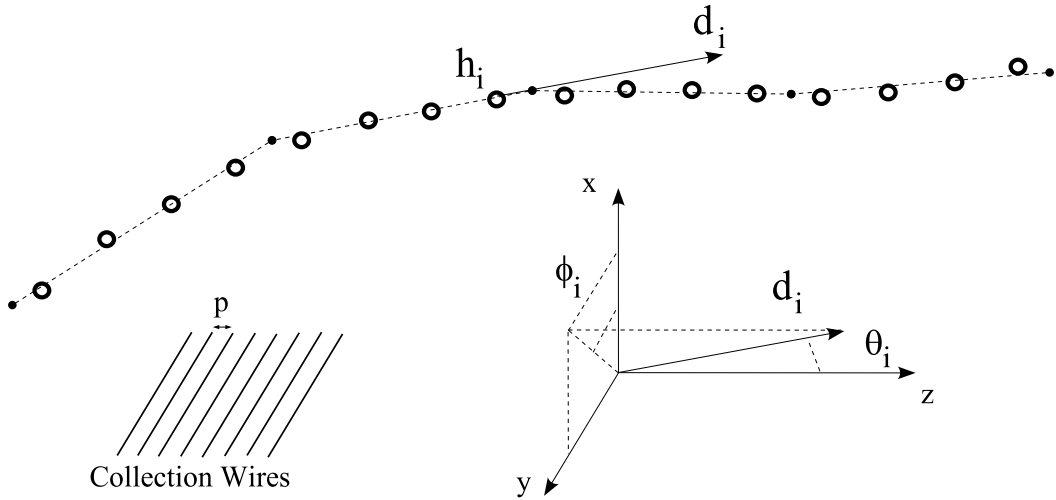}
\end{center}
\caption
[Illustration of the track segmentation procedure]
{Illustration of the track segmentation procedure and the track pitch determination.}
\label{fig:trkseg}
\end{figure}

The direction associated to any track segment is determined by the parameters entering
in the linear fitting, given in terms of a unit vector $\mathbf{d}_i$, or in terms
of the angles $\theta$ and $\phi$
\begin{eqnarray}
\theta_i & = & \arccos\,(d_i^z) \nonumber \\
\phi_i & = & \arctan\,\left(\frac{d_i^y}{d_i^x}\right)
\end{eqnarray}
where the index $i$ may refer to a hit or a segment, and $\mathbf{d}_i
= \{d_i^x,d_i^y,d_i^z\}$ is the corresponding associated direction
vector.

Hence, the energy deposition of a given hit is calculated from the fitted area in 
Sec.~\ref{sec:finerec}, while its associated pitch length is obtained from the
track segmentation procedure through Eq.~\eqref{eq:dx}.

Finally, the mean energy loss per crossed distance (Eq.~\eqref{eq:bethe})
measured for every hit is approximated by:
\begin{equation}
\left(-\frac{dE}{dx}\right)_i = \frac{E_i}{\Delta z_i}
\label{eq:dedxexp} 
\end{equation}
with $E_i$ and $\Delta z_i$ the energy (Eq.~\eqref{eq:bethe})
and track pitch (Eq.~\eqref{eq:dx}) associated to hit $h_i$.

\subsubsection{Range determination}
The segmentation procedure also provides an optimal method to compute 
the track \emph{range}\footnote{The distance a particle can penetrate before it loses 
all its energy.}. From a theoretical point of view, the mean range for a 
stopping particle might be calculated through the integration of the Bethe-Bloch 
formula \eqref{eq:bethe} along the particle's path:
\begin{equation}
 R = \int_{0}^{E_0} \; \frac{dE}{dE/dx}  
\label{eq:range}
\end{equation}
where integrations limits go from the initial point with kinetic energy $E_0$ until
the stopping point. If one assumes that the energy loss is continuous, the distance a
particle can penetrate before it loses all its energy is a well
defined number, the same for all identical particles with the same
initial energy. Fig.~\ref{fig:kinvsrange} shows the energy--range
curves for different particles calculated by a numerical integration
of the Bethe-Bloch formula, which yields the approximate path length
traveled. As in the $dE/dx$ vs. kinetic energy case (see Fig.~\ref{fig:dEdxVsT}),
the curves give the expected ``average'' behavior.

\begin{figure}[!ht]
\begin{center}
\includegraphics[width=10cm]{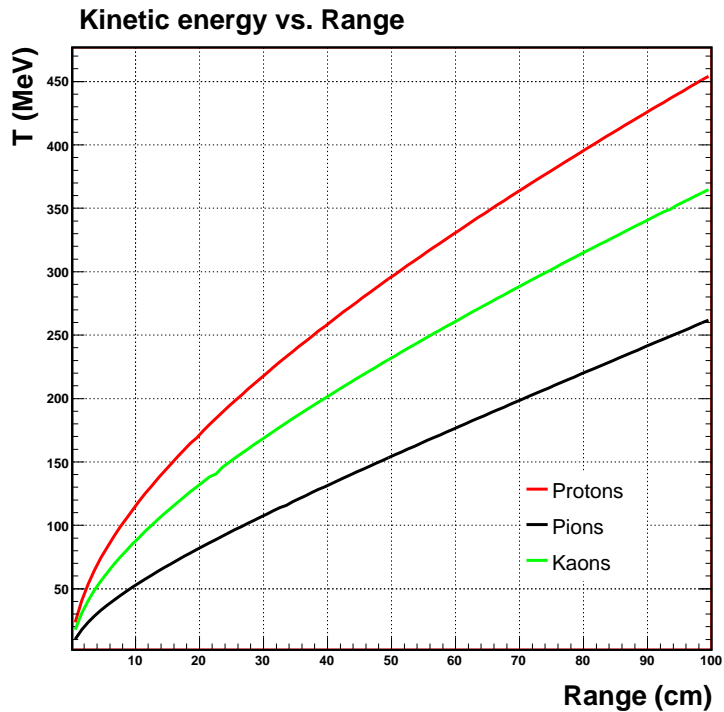}
\caption[Kinetic energy vs. range from Bethe-Block]
{Kinetic energy vs. range from Bethe-Block.}
\label{fig:kinvsrange}
\end{center}
\end{figure}

The definition of a range is complicated because
of the multiple Coulomb scattering in the material and catastrophic
energy loss processes (both ignored in Eq.~\eqref{eq:range}), which
cause the particle to follow a zigzag path through the absorber
and lead to substantial range straggling.
The range, defined as the straight-line length, will generally
be smaller than the total zigzag path length. 
The track segmentation method leads to a better estimation of the ``real'' range:
\begin{equation}
R = \sum_{i=1}^{N_s} r_i
\label{eq:exprange}
\end{equation}
being $r_i$ the straight-line length of each one of the segments composing the track,
and $N_s$ the total number of such segments. 

% LocalWords:  cnts mips ij ijkl strenghts jk ijl jl NNenergy tconst ik klij il
% LocalWords:  ijij kj MFT dE dV NNlocalmin NNvi ijk NNupdaterule tcost intra
% LocalWords:  unidimensional unassociations maxdistance unassociation dx MeV
% LocalWords:  ds mol eV truncbethe du TPC's priori fC bethe

%%%%%%%%%%%%%%%%%%%%%%%%%%%%%%%%%%
%%%%%%%%%%%%%%%%%%%%%%%%%%%%%%%%%%

\chapter{The 50L TPC exposed to The CERN Neutrino Beam}   
\label{chap:50Lexperiment}
%%%%%%%%%%%%%%%%%%%%%%%%%%%%%%%%%%
%%%%%%%%%%%%%%%%%%%%%%%%%%%%%%%%%%

Within the broad physics programme of the ICARUS project, an important
issue is the study of oscillations through the detection of
accelerator neutrino interactions. The possibility to observe
$\nu_\mu \to \nu_e$ and $\nu_\mu \to \nu_\tau$ by means of kinematic
criteria is known to be limited, among others, by the knowledge we have of 
nuclear effects (Fermi motion, nuclear re-scattering and absorption, etc.). 
Therefore it is very important to acquire experimental data in order 
to tune the existing Monte-Carlo models. In addition, the exposure of a LAr TPC
to a neutrino beam is mandatory to demonstrate the high recognition 
capability of the technique and gain experience with real neutrino events. 

In 1997, the ICARUS collaboration together with a group from 
INFN and Milano University~\cite{Stefano} proposed to expose a 50 liter LAr TPC 
to the multi-GeV wide band neutrino beam of the CERN West Area Neutrino Facility
(WANF)~\cite{wanf}, during the NOMAD~\cite{NOMAD} and CHORUS~\cite{CHORUS} data taking.
The test was part of an  R\&D program for a medium baseline
$\nu_\tau$ appearance experiment~\cite{icarus-cern-mi}. The idea
was to collect a substantial sample of quasi-elastic (QE) interactions 
($\nu_\mu$ + n $\to$ p + $\mu^-$) to study the following physics items: 
\begin{itemize} 
\item Measurement of the acoplanarity and missing transverse momentum
in events with the $\mu$-p topology in the final state, in order to
assess Fermi motion and proton re-scattering inside the nucleus. 
\item Appearance of nuclear fragments (short tracks and {\it blobs}
around the primary interaction vertex) in quasi-elastic events. 
\item A preliminary evaluation of e/$\gamma$ and e/$\pi^0$
discrimination capability by means of the specific ionization measured
on the wires at the beginning of the candidate track. This measurement
is limited by the size of the chamber.
\end{itemize}

The data collected in 1997 offer the unique opportunity to study 
nuclear effects in the Argon nucleus and to assess the identification and
reconstruction capability of a LAr TPC for low-multiplicity neutrino events. 
The work presented in this thesis shows, for the first time, a comprehensive set of results 
from the 1997 test (see \cite{50Lold,tesi_alessandro,tesi,curioni} 
for preliminary results). The structure of the chapter is as follows: the
experimental setup and the calibration measurements of the 50 liter TPC
are discussed in Sec.~\ref{sec:Setup} and Sec.~\ref{sec:calibration},
respectively. At the beginning of Sec.~\ref{sec:Analysis_QE} the event
reconstruction and particle identification are detailed, 
thereafter the analysis of a \goldens of quasi-elastic 
$\nu_\mu$ charge current (CC) interactions is presented together with a 
comparison with theoretical expectations. 
Finally, in Sec.~\ref{sec:xsection} we measure the quasi-elastic
cross section for $\nu_\mu$ CC interactions.

\section{The experimental setup}
\label{sec:Setup}

The LAr TPC was placed on a platform 4.5 meters above ground, right in between 
the CHORUS and NOMAD detectors (Fig.~\ref{fig:setup}). 
Full details of the geometrical and technical features of the 50L TPC 
have been treated in Sec.~\ref{sec:50Linfo}.
As mentioned there, the active part of the detector is located inside
a stainless steel cylinder. The connection to the outside area is
obtained through a set of UHV flanges housing the signal, the high
voltage cables and the vacuum feed-through. The cylinder is positioned
into a 1~m diameter dewar partially filled with low purity Liquid
Argon acting as a thermal bath and it is rotated 30$^\circ$ along
the vertical axis with respect to the nominal beam direction to reduce
the number of particles crossing just one readout wire.

The modest size of the LAr TPC fiducial volume ($\sim$50 liter), coupled with 
the high energy of the WANF $\nu$ beam (Sec.~\ref{sec:beam}), made necessary 
a muon spectrometer downstream the TPC. The NOMAD detector was the perfect choice
to achieve this task: A coincidence with the NOMAD DAQ 
was set up to use the detectors located into the NOMAD magnetic dipole as 
a magnetic spectrometer (Sec.~\ref{sec:NOMAD}). 
The experimental setup was completed with additional 
counters for the trigger and veto systems (Sec.~\ref{sec:trigger}).

\begin{figure}[!ht]
\centering
\includegraphics[width=\textwidth]{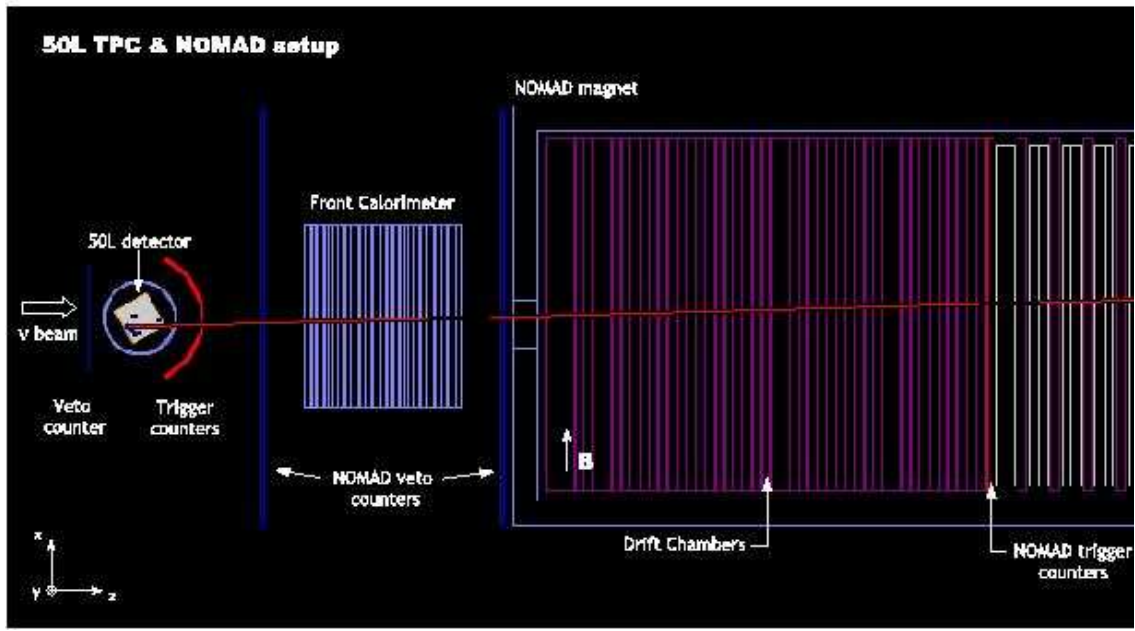}
\caption{A sketch of the experimental setup (top view).}
%Relative sizes are not to scale.}
\label{fig:setup}
\end{figure}

\subsection{The Neutrino Beam}
\label{sec:beam}
The chamber has been exposed to the $\nu$ beam produced at the CERN
West Area Neutrino Facility (WANF~\cite{wanf,flux_nomad}).
The geometry of the WANF is shown in Fig.~\ref{figure:beam_wanf}.
The Super Proton Synchrotron (SPS) at CERN supplied a beam of 450~GeV protons 
which was directed onto a beryllium target, thereby producing a large number of 
secondary hadrons. In every accelerator cycle (14.4~s) the protons used to 
produce neutrinos were extracted from the SPS in two spills, which were separated by 
2.7~s. Each spill had a full width at half maximum of 3~ms and 
contained about $1.8\times 10^{13}$ protons. 
The number of protons incident on the target was monitored by a beam current 
transformer (BCT). The number of produced neutrinos was often expressed in units 
of protons on target (pot). 

\begin{figure}[!ht]
\begin{center}
\includegraphics[width=\textwidth]{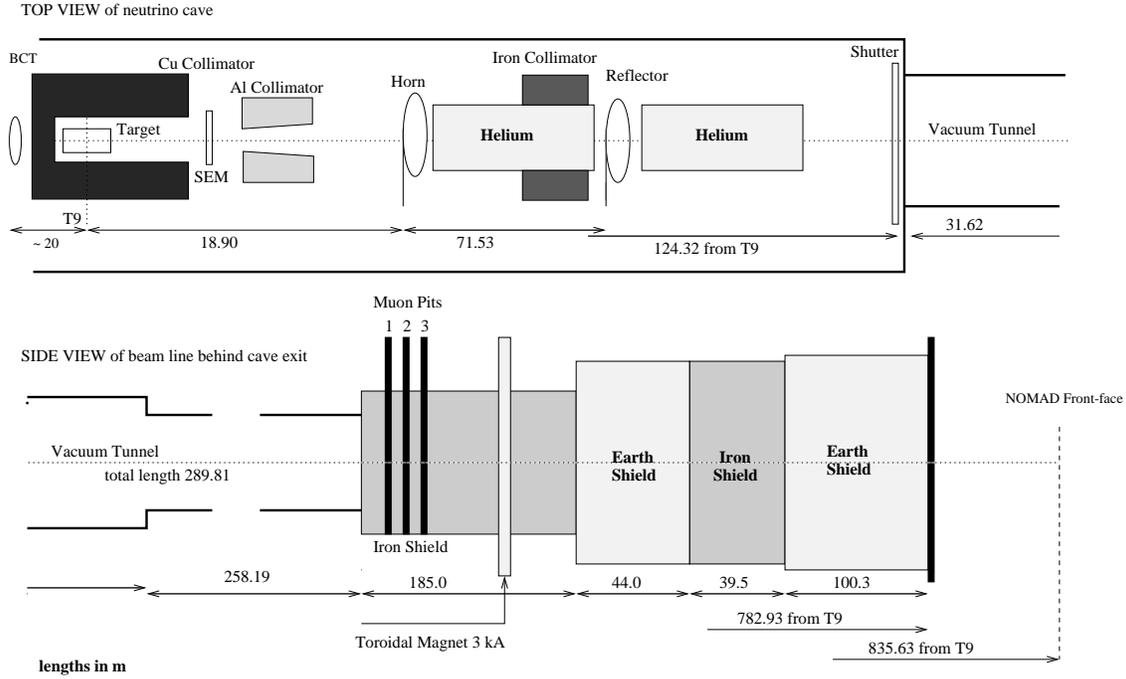}
\caption[Schematic layout of the WANF beam line]{Schematic layout of the WANF beam line 
(all lengths in meters).}
\label{figure:beam_wanf}
\end{center}
\end{figure}

The multiplicity and charge asymmetry of the secondary particles were measured by 
secondary emission monitors (SEMs) placed before and after the target. 
The beam focusing elements, downstream of the target, were large magnets, 
designed to collimate the positively charged secondary particles, while deflecting 
the negatively charged particles away from the beam axis. Most of the focusing was 
done by the first element, called the horn. The second element, the reflector, 
focused the smaller angle particles which had been missed by the horn. Two helium 
bags were installed in order to increase the neutrino flux by reducing multiple 
scattering and secondary interactions along the beam. Downstream of the focusing 
elements was a vacuum decay tunnel of 290~m length in which the positively 
charged particles could decay and, thereby, produce the neutrinos. The major 
contributions to the $\nu_{\mu}$ flux came from the reactions:
\begin{eqnarray*}
\pi^+ &\rightarrow & \mu^+ + \nu_{\mu} \\
K^+ &\rightarrow & \mu^+ + \nu_{\mu}.
\end{eqnarray*}
The charge conjugate reactions also occurred, giving rise to a $\overline{\nu}_{\mu}$ 
component in the beam. 
Earth and iron shielding directly after the decay tunnel filtered out all but 
neutrinos and some muons. Silicon detectors in several pits within the shielding region
measured the muon multiplicity and could be used to monitor the beam and determine the 
absolute flux. An additional toroidal magnet, which was placed after the muon pits, 
deflected most of the remaining muons.
The resulting neutrino beam reached the NOMAD detector 835~m downstream of 
the proton target with the composition and properties shown in 
Fig.~\ref{fig:nuFlux} and Tab.~\ref{table:nuFlux} \cite{flux_nomad}. 
The predictions of the neutrino fluxes are based on Monte Carlo simulations. 
For further reference on the WANF layout and the alignment procedures, 
see \cite{WANFallig}.
\begin{figure}[!ht]
\begin{center}
\includegraphics[width = 13cm]{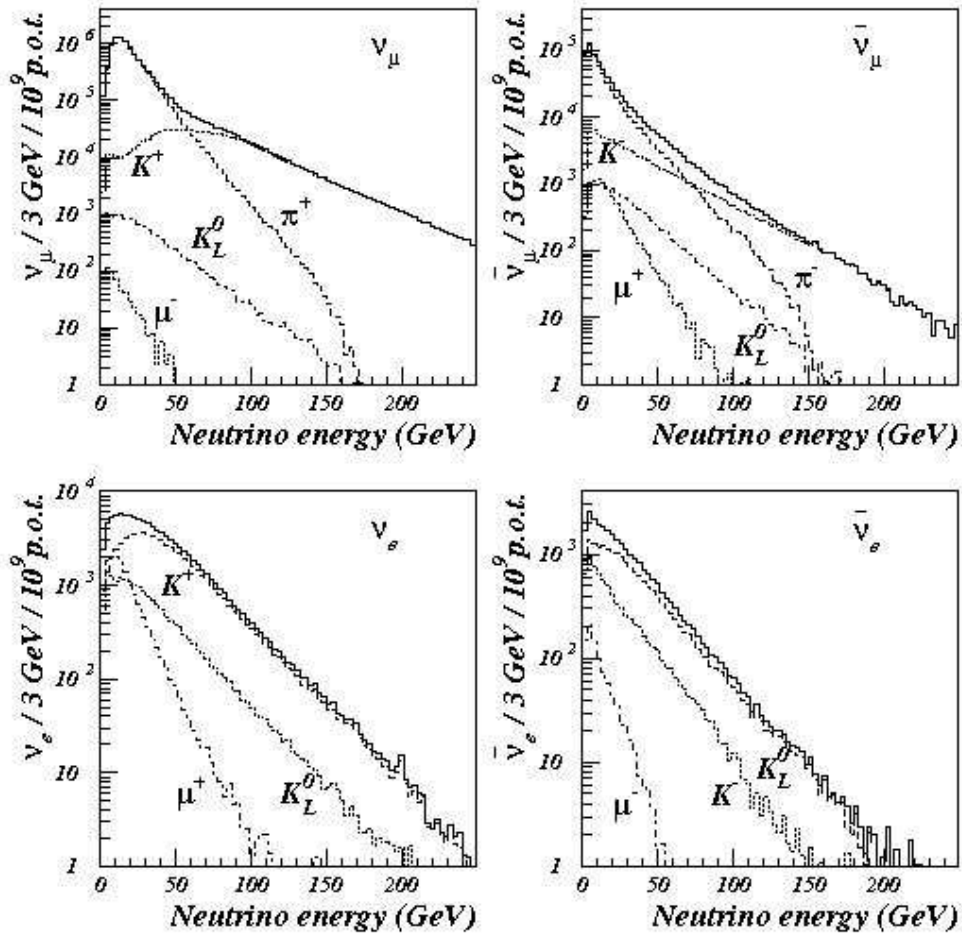}
\caption[Composition and energy spectrum of the CERN SPS neutrino beam]
{Composition of the neutrino beam and the $\nu_{\mu}$, $\overline{\nu}_{\mu}$, 
$\nu_{e}$ and $\overline{\nu}_{e}$ energy spectra at NOMAD, within the transverse 
fiducial area of $260\times260~cm^2$. The Figures are taken from \cite{flux_nomad}.}
\label{fig:nuFlux}
\end{center}
\end{figure}
%%%%

\begin{table}[!ht]
\begin{center}
%\begin{tabular}{|>{$}c<{$}|>{$}c<{$}>{$}c<{$}|>{$}c<{$}>{$}c<{$}|>{$}c<{$}>{$}c<{$}|>{$}c<{$}>{$}c<{$}|>{$}c<{$}>{$}c<{$}|}\hline
\begin{tabular}{|c|c|c|c|c|c|c|c|c|c|c|}\hline
\text{Neutrino} & \multicolumn{2}{c|}{Neutrino} &\multicolumn{8}{c|}{Source} \\ 
\text{Species} & \multicolumn{2}{c|}{Flux}& \multicolumn{2}{c|}{$\pi^+$ or $\pi^-$} & \multicolumn{2}{c|}{$K^+$ or $K^-$} &  \multicolumn{2}{c|}{$K^0_L$} & \multicolumn{2}{c|}{$\mu^+$ or $\mu^-$} \\  \hline
& \text{{Rel.\ Abund.}} & $\langle E_{\nu} \rangle$ & $\%$ & $\langle E_{\nu} \rangle$& $\%$ & $\langle E_{\nu} \rangle$ & $\%$ & $\langle E_{\nu} \rangle$ & $\%$ & $\langle E_{\nu} \rangle$ \\ \hline  
$\nu_{\mu}$ & 1.0 & 24.3 & 90.4 & 19.1 & 9.5 & 73.0 & 0.1 & 26.8 & $<$ 0.1 & 11.4 \\ \hline
$\overline{\nu}_{\mu}$ & 0.0678 & 17.2 & 84.0 & 13.8 & 12.8 & 38.1 & 1.9 & 26.9 & 1.2 & 17.0 \\ \hline
$\nu_{e}$ & 0.0102 & 36.4 & - & - & 68.0 & 41.8 & 17.8 & 30.3 & 13.6 & 16.8 \\ \hline
$\overline{\nu}_{e}$ & 0.0027 & 27.6 & - & - & 25.1 & 22.8 & 68.2 & 30.4 & 3.5 & 11.1 \\  \hline
\end{tabular}
\caption[Composition of the neutrino beam]{Composition of the neutrino beam 
and the sources of the different components \cite{flux_nomad}.}
\label{table:nuFlux}
\end{center}
\end{table}

The mean energy of the $\nu_\mu$ reaching the detectors is 24.3~GeV when integrating 
over the active NOMAD area ($2.6\times 2.6$ $m^2$), while contamination from other flavors 
are below 7\% for $\bar{\nu}_\mu$ and $\sim 1$\% for $\nu_e$~\cite{flux_nomad}.
Being the TPC centered on the beam axis and covering a smaller surface, 
a harder neutrino flux is expected, with a mean energy of 29.5~GeV 
(see Fig.~\ref{fig:nuFlux50L}).

\begin{figure}[!ht]
\begin{center}
\includegraphics[width=9cm]{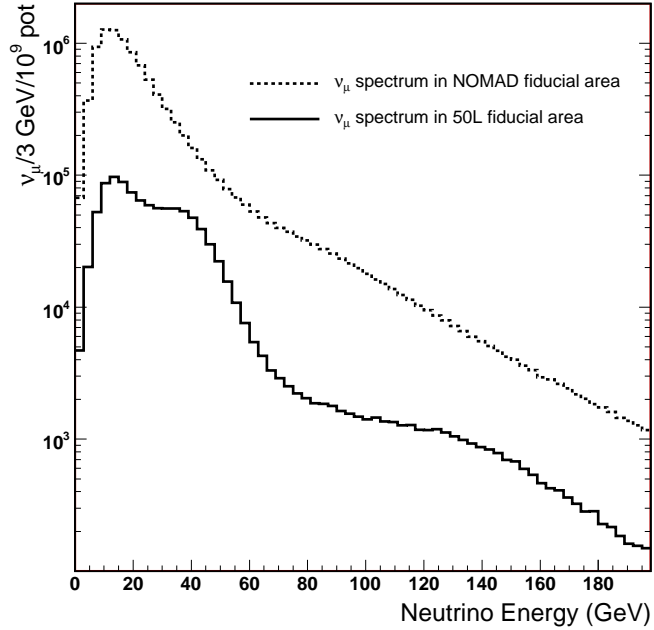}
\caption[$\nu_\mu$ energy spectra on 50L (NOMAD) fiducial area]
{$\nu_{\mu}$ energy spectra on 50L (NOMAD) transverse fiducial area.}
\label{fig:nuFlux50L}
\end{center}
\end{figure}

\subsection{The NOMAD detector}
\label{sec:NOMAD}
The NOMAD detector \cite{NOMAD} was composed of several independently 
working sub-detectors. A schematic view of the detector is shown in 
Figure \ref{figure:NomadDetector} along with the coordinate system. 
The sub-detectors were placed along the beam-line with positional sensitivity 
transverse to the beam direction. 
The \emph{veto} system upstream of the detector allowed a fast identification of muons 
entering the detector. Following the muon veto, there was the 
\emph{forward calorimeter} (FCAL) which was used as a target for studies requiring very 
high statistics while sacrificing some resolution in the vertex finding. 
The \emph{drift chambers} (DCH) in the center of the detector served both as an active 
target and a very precise tracking device; hence the most interesting of the detected 
interactions occurred there. 
Further downstream, the \emph{trigger planes}, the \emph{transition radiation detector} (TRD), 
the \emph{preshower} (PRS) and the \emph{electromagnetic calorimeter} (ECAL) were situated. 
These inner detectors (DCH to ECAL) were all surrounded by the old UA1 dipole 
magnet which  provided a constant magnetic field of $0.4$~Tesla. 
The \emph{hadronic calorimeter} (HCAL) followed 
downstream of the inner detectors and at the very end there were the \emph{muon chambers}.
The $x$ and $y$ axes of the co-ordinate system were centered in the middle of the 
first drift chamber. The neutrino beam entered with an angle of 42~mrad with respect to 
the $z$ axis. 

In what follows, we describe some of the NOMAD subdetectors that were key for the 
experiment and the analyses described in this work. 

\begin{figure}[!ht]
\begin{center}
\includegraphics[width = 15cm]{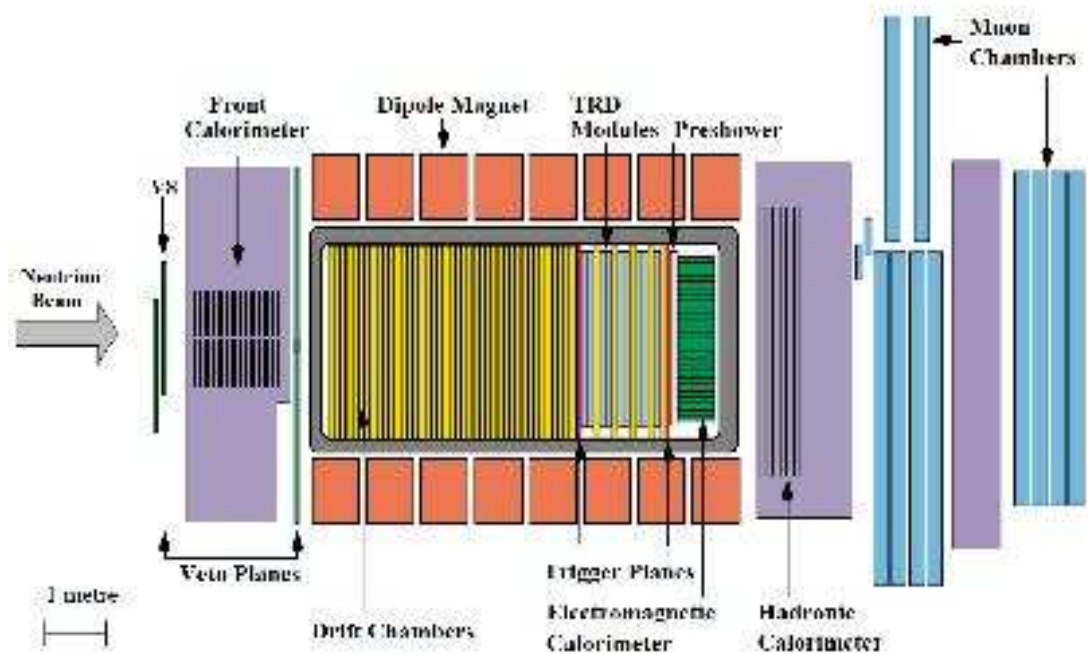}
\caption[Side view of the NOMAD detector]{Side view of the NOMAD detector.}
\label{figure:NomadDetector}
\end{center}
\end{figure}

%\subsubsection{The Magnet}
%The NOMAD detector reused the dipole magnet from the former UA1 experiment which 
%is described in \cite{MAGNET}. The magnet coil was made from aluminum 
%and provided a near constant field of 0.4~Tesla which required a current of 5713~kA. 
%The field, which was perpendicular to the $z$ axis throughout the DCH and the TRD 
%region, allowed an accurate momentum reconstruction for charged particles traversing 
%the central detector. 
%The two iron supports, which acted as yokes, had been instrumented as calorimeters. 
%The front pillar was used to form the forward calorimeter, the back pillar formed 
%the hadronic calorimeter.

\subsubsection{The Veto Counters}
\label{sec:Veto}
Being the most upstream detector, the veto consisted of 59 scintillator counters 
and covered an area of $5.4 \times 5~m^2$. It was used to detect muons traveling 
with the neutrino beam and charged particles from upstream interactions, 
and prevented them from causing valid triggers. An additional veto plane was 
positioned between the FCAL and the DCH. Charged particles from interactions 
downstream of the veto, e.\ g.\ in the magnet coil or the support structures of 
the drift chambers, were removed during analysis by constraints on the vertex position.

The scintillators were read out by photo-multipliers and a veto signal was 
constructed from a logical OR of all the individual counters.  The efficiency of the
veto was constantly monitored and found to be stable at about $96~\%-97~\%$. 
The contribution of the veto to the overall dead time of the experiment during 
the two neutrino spills amounted to about 4~\%. 
A detailed description of the veto system can be found in \cite{trigger_NOMAD}. 

\subsubsection{The Forward Calorimeter (FCAL)}
\label{sec:FC}
The forward calorimeter, being the instrumented front pillar of the magnet, 
was used as a massive active target and allowed neutrino physics studies including 
di-muon production and neutral heavy lepton searches. The FCAL consisted of 23 iron 
plates, which were 4.9~cm thick and separated by 1.8~cm air gaps. The first 20 gaps 
were instrumented with scintillator counters. The dimensions  of the scintillators 
were $175 \times 18.5 \times 0.6~cm^3$. Five consecutive counters along the beam 
direction were bunched together via light guides to form a module which was 
read out at both ends by photo-multipliers (see Figure \ref{figure:FCAL}). 
Ten such modules were arranged vertically to form a stack. There were four stacks 
aligned along the beam axis. The active region of the FCAL had a mass of 17~tons, 
was about 5 nuclear interaction lengths in depth and had an active area transverse 
to the beam of $175\times 190~cm^2$. Further details can be found in \cite{NOMAD}.

\begin{figure}[!ht]
\begin{center}
\includegraphics[width = 100mm,clip]{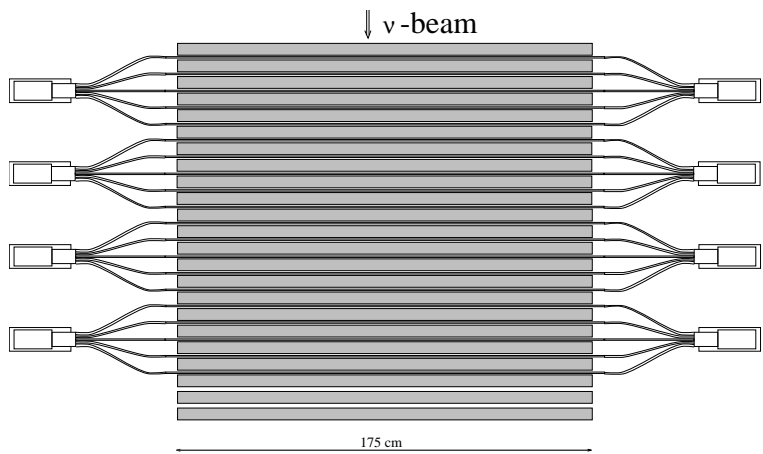}
\caption[Top view of the FCAL]{Top view of the FCAL.}
\label{figure:FCAL}
\end{center}
\end{figure}

%%%%%%%%%%%%%%%%%%%%%%%%%%%%%%%%%%%%%%%
\subsubsection{The Drift Chambers (DCH)}
\label{sec:DCH}
%%%%%%%%%%%%%%%%%%%%%%%%%%%%%%%%%%%%%%%
A crucial part of the detector were the drift chambers which provided at the same 
time the target material and the track and vertex reconstruction of the charged 
particles. 
%This double function implied conflicting requirements for the walls 
%of the drift chambers, for they had to be as massive as possible to guarantee 
%a large number of neutrino interactions and as light as possible to minimize 
%multiple scattering of particles, secondary particle interactions and photon 
%conversions.
%In order to reduce the radiation length for a given target mass, the chambers 
%were made of low density and low atomic number materials which resulted in less 
%than 1~\% of a radiation length in the inactive area between two consecutive 
%measurements.   
The full drift chamber region consisted of 11 modules, each made up of four 
individual drift chambers (composed of low density, low atomic number materials). 
In addition to these 44 drift chambers, there 
were another 5 chambers in the TRD region, providing additional tracking. 
%The chambers were built on panels made of aramid fibers in a honeycomb structure 
%and sandwiched between two kevlar-epoxy resin skins. These skins ensured the 
%mechanical rigidity and flatness over the area of $3\times 3~m^2$. Each drift 
%chamber comprised four panels with three gaps of 8~mm in between. 
The chambers were filled with a mixture of argon (40~\%) and ethane (60~\%), 
and equipped with sense wires.
%Since the chambers were not completely gas tight, the gas was 
%circulated permanently in a closed circuit with a purifier section that 
%removed oxygen and water vapour.
The volume with maximal acceptance and reconstruction efficiencies was referred 
to as fiducial volume and had the dimensions of $2.6\times 2.6 \times 4~m^3$ 
with a mass of 2700~kg. The overall density of $100~kg/m^3$ was about equal 
to the density of liquid helium. 
%Table \ref{table:DCHcomposition} shows the atomic composition of the 
%fiducial volume in detail.

%\begin{table}[!ht]
%\begin{center}
%\begin{tabular}{l|c|c}\hline
%             & \text{Proportion}            & \\
%Atom         & \text{of Weight (\%)} & \text{Atomic Weight} \\ \hline \hline
%carbon       & 64.3                         & 12.011     \\
%oxygen       & 22.13                        & 15.999    \\
%nitrogen     & 5.92                         & 14.007   \\
%hydrogen     & 5.14                         &  1.008    \\
%aluminum     & 1.71                         & 26.982  \\
%chlorine     & 0.31                         & 35.453    \\
%silicon      & 0.27                         & 28.086    \\
%argon        & 0.19                         & 39.948     \\
%copper       & 0.03                         & 63.546     \\ \hline
%Total        & 100.00                       &            \\ \hline \hline
%protons      & 52.43                        &            \\
%neutrons     & 47.57                        &            \\ \hline
%\end{tabular}
%\caption[Composition of the drift chamber fiducial volume]{Composition of the 
%drift chamber fiducial volume \cite{MBisch}.}
%\label{table:DCHcomposition}
%\end{center}
%\end{table}

The sense wires in the three gaps of a drift chamber were oriented with -5, 0 
and +5~degrees with respect to the magnetic field direction, corresponding to 
the NOMAD $x$ axis. The three ionization signals produced by a particle traversing 
a drift chamber allowed a hit positioning accurate to about 1.5~mm in the $x$ 
direction, due to the angles between the wires. Transverse to the wires ($y$ direction)
a position accuracy of $150~{\mu}m$ was reached by the separation of the wires 
and a constant gas drift velocity. The position along the beam line ($z$ direction) 
was determined by the position of the wire planes which was constantly monitored. 
The efficiency of each wire to record a hit was typically 97~\%. The three 
dimensional information from each drift chamber could be used to reconstruct 
the helical path of charged particles through the magnetic field. 
The curvature allowed the measurement of the momentum of the particles through 
the relation
\begin{equation} 
p \cos{\theta_{\lambda}} = q B r
\end{equation}
where $p$ is the momentum and $q$ the charge of the particle, $\theta_{\lambda}$ 
the pitch angle of the helix, $r$ the radius of curvature and $B$ the magnetic field. 
The momentum resolution was a function of the particle momentum and track length. 
For muons and charged hadrons with normal incidence to the measuring planes the 
momentum resolution could be parametrized by
\begin{equation} 
{\left(\frac{\sigma_p}{p}\right)}^2 =  {\left(\frac{a}{\sqrt{L}}\right)}^2 + {\left(\frac{bp}{\sqrt{L^5}}\right)}^2
\label{equ:resolution}
\end{equation}
with
\begin{equation} 
\begin{array}{rcl}
a & = & 0.05~m^{1/2} \\
b & = & 0.008~m^{5/2}(GeV/c)^{-1}
\end{array}
\end{equation}
and $L$ denoting the length of the track. The first term in Eq.~\eqref{equ:resolution} 
corresponds to the error due to multiple-scattering, the second term arises from the 
single hit resolution. A detailed description of the drift chamber is 
given in \cite{NOMADDC}.

\subsection{The Trigger and Veto system for the 50 liter set-up}
\label{sec:trigger}
The trigger is provided by a set of scintillators located between the
chamber and the NOMAD apparatus (Fig.~\ref{fig:setup}). 
Each of the 3 trigger scintillator counters has a dimension of
110$\times$27~cm$^2$. They were positioned in a half-circle 60~cm
beyond the center of the chamber.  Incoming charged particles are
vetoed by 5 scintillators mounted in front of the chamber, 50~cm
before its center, and by the last scintillator plane of CHORUS.  
The latter vetoes particles deflected by the CHORUS magnetic field
entering the chamber at large angles with respect to the nominal beam
direction.

The trigger requires the coincidence of the SPS beam spill and at
least one of the trigger scintillators, vetoed by the scintillators
put in front of the chamber and the CHORUS plane. This trigger is put
in coincidence with the two trigger scintillator planes of NOMAD 
(T1 and T2 in Fig.~\ref{fig:setup})~\cite{trigger_NOMAD}. 
Moreover, a trigger is rejected if the NOMAD acquisition system is in BUSY mode or
if the delay with respect to the previous trigger is lower than
500~$\mu$s (``drift protection'')\footnote{This condition inhibits the
occurrence of event overlaps in the multibuffer readout system.}. 

The request of a charged particle triggering the chamber locally and 
reaching NOMAD up to T1 and T2 inhibits the acquisitions of neutral-current and $\nu_e$
charged-current events, limiting the sample of the present test to
$\nu_\mu$ charged-current interactions.  The trigger efficiency has
been monitored during data taking through a dedicated sample of
through-going muons. It turned out to be 97\% averaged over the whole
data taking period.

\section{Calibration of the TPC}
\label{sec:calibration}

\subsection{Measurement of the electron drift velocity}
\label{sec:driftvel}
A fundamental parameter for the reconstruction of ionizing events occurring
inside the fiducial volume of the 50L TPC is the velocity of the drifting electrons.

An absolute determination of the drift velocity and its uniformity
along the fiducial volume of the chamber has been obtained exploiting
the external scintillators and the additional information from NOMAD.
A dedicated trigger selecting through-going muons (``mip sample'')
has been put into operation adding two additional scintillators
(30$\times$30 cm$^2$) before and after the external dewar. The
relative position of the chamber with respect to the NOMAD reference
frame has been obtained by residual minimization of the track
parameters recorded both by the TPC and the NOMAD data acquisition
system. The drift velocity is obtained by fitting the
absolute vertical position of the muon as reconstructed by NOMAD
trackers versus the drift time measured locally by the TPC 
(Fig.~\ref{fig:DriftVelocity}).
The corresponding drift velocity is 
\begin{equation}
v_d = 0.905 \pm 0.005~mm/\mu s 
\end{equation}
The measurement of the drift velocity is essential because it defines the 
``pitch'' in the drift coordinate: Since the signal is sampled with a 2.5~MHz 
frequency for a duration of 820~$\mu$s (2048 time samples), 
the drift pitch amounts up to 0.36~mm. 
The maximum drift distance of 47~cm is covered by a drifting electron in 520~$\mu$s.

\begin{figure}[t]
\begin{center}
\includegraphics[width=10cm,height=9cm]{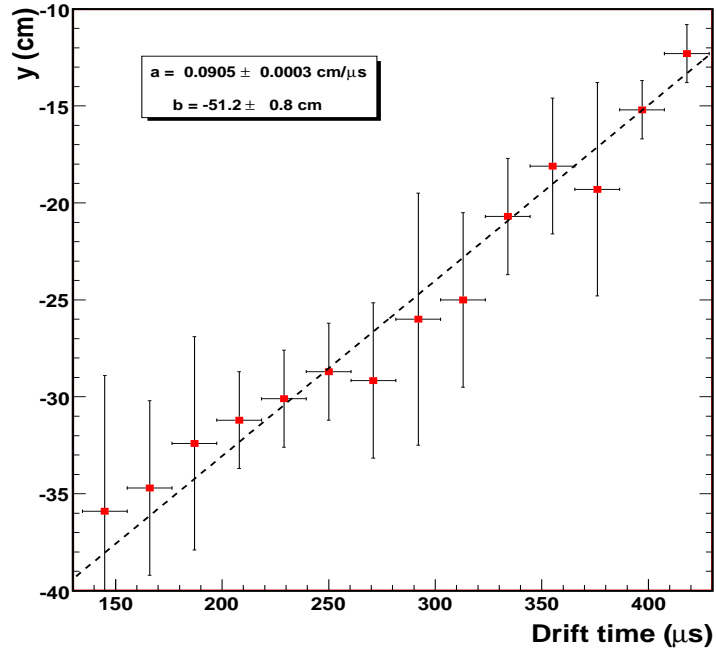}
\caption[Measurement of the electron drift velocity]
{Measurement of the electron drift velocity.}
\label{fig:DriftVelocity}
\end{center}
\end{figure}

\subsection{Measurement of the electron drift lifetime}
\label{sec:lifetime2}
The attenuation of the drifting electric charge towards the anode is a crucial
working parameter for the TPC. It essentially depends on the purity levels in the
Liquid Argon and on the applied electric field. Thanks to a purity monitor placed within
the pure LAr bath (see Sec.~\ref{sec:purmon}), it was possible to 
obtain direct measurements of the drifting electron lifetime $\tau$.
During the data taking period the impurity levels of oxygen kept themselves stable, 
getting values of $\tau$ higher than 10~ms.

In this section, we perform an independent measurement of $\tau$ based on the spatial
and calorimetric reconstruction of a sample of muons crossing the chamber at 
different heights (see Fig.~\ref{fig:SMuons}).
This sample of muons was gathered from the data collected during the exposition to the
WANF beam, by means of an automatic scanning and reconstruction algorithm.
Over the whole data, the algorithm was able to collect a big enough number of 
correctly reconstructed through-going muons. They are essentially muons coming 
from $\nu_\mu$~CC interactions occurring before the fiducial volume of the TPC, 
uniformly distributed in height and crossing the chamber with an approximately 
constant value of their drift coordinates. 
%In Fig.~\ref{fig:SMuons} we show some 
%xexamples of the through going muons used for this analysis.

\begin{figure}[t]
\begin{center}
\includegraphics[width=15cm]{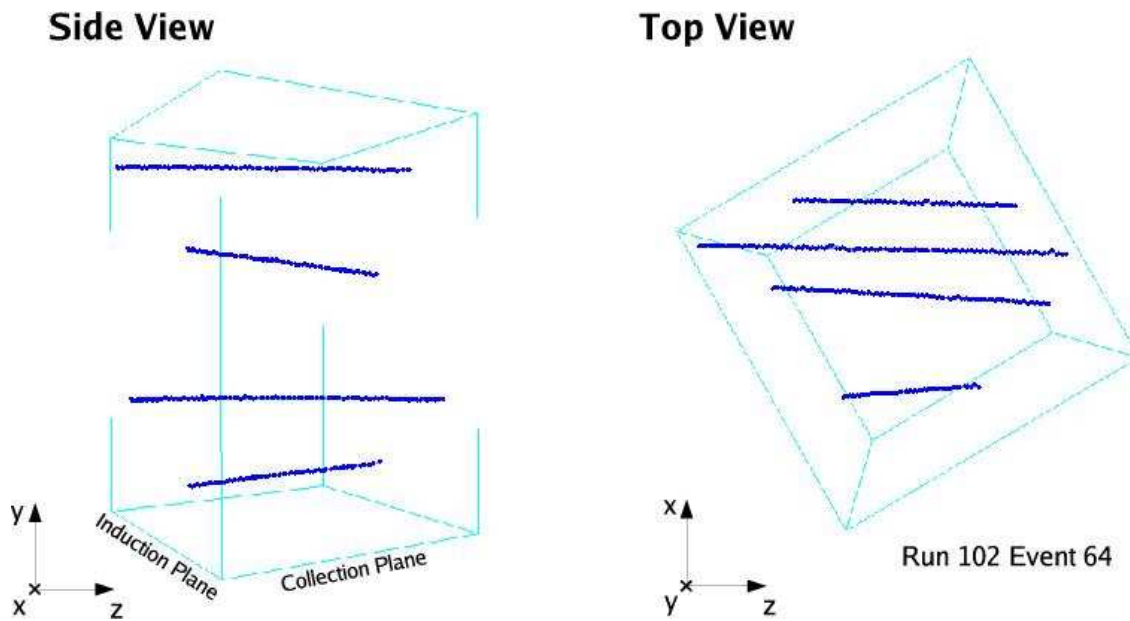}
\caption[Example of the 3D reconstruction of parallel through-going mips]
{Example of the 3D reconstruction of parallel through-going mips.}
\label{fig:SMuons}
\end{center}
\end{figure}

The spatial and calorimetric reconstruction tools allows to measure the energy loss 
per crossed distance ($dE/dx$) hit by hit for every muon (see Sec.~\ref{sec:calrec} and
Eq.~\eqref{eq:dedxexp}). 
The most probable value (\dmp) is used as a suitable estimator of the energy loss for
each one of those muons (see Sec.~\ref{sec:mp}). As it was already discussed in 
Sec.~\ref{sec:avvsmp}, \dmp~depends on the crossed distance associated to the sampling
hits of the track, but shows a slight dependence on the kinetic energy. 
Being the sample of mips used in this analysis horizontal and straight muons 
coming in the beam direction, 
their track pitch length hardly varies from the mean value: $\Delta z = 0.29\pm0.01~cm$.
The latter ensures that the expected \dmp~is essentially the same for all the muons
in the sample (see Fig.~\ref{fig:dEdxVsKin}).

\begin{figure}[t]
\begin{center}
\includegraphics[width=10cm]{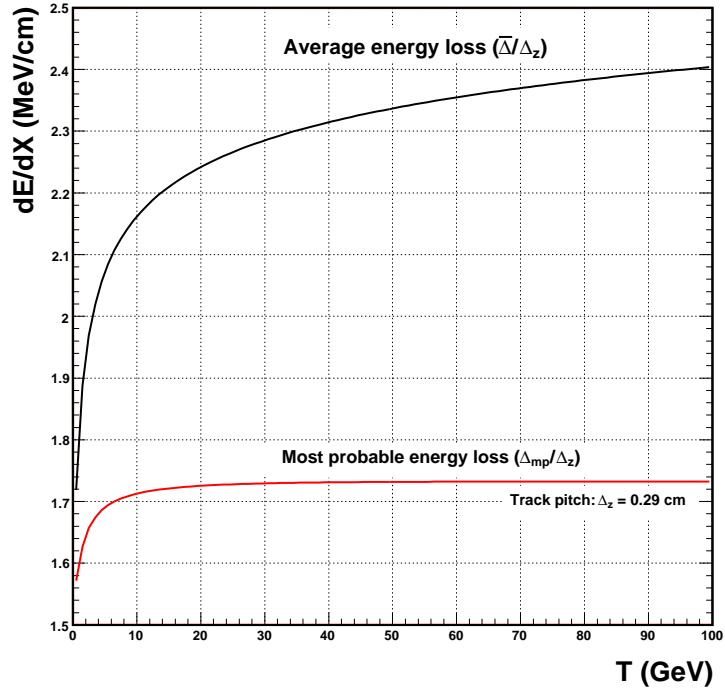}
\caption[Average and most probable energy loss as a function of kinetic energy.]
{Average and most probable energy loss as a function of kinetic energy.}
\label{fig:dEdxVsKin}
\end{center}
\end{figure}

In order to measure the charge attenuation effect due to the LAr impurities, we analyze
the dependence of \dmp~as a function of the drift coordinate
The procedure is as follows:
\begin{itemize}
\item A $dE/dx$ histogram of every muon is obtained.
\item Every histogram is fitted to a convoluted Landau-Gaussian distribution, 
in order to obtain the most probable value of the distribution (\dmp). 
On the other hand, the average drift coordinate ($T_{mean}$) of the track is computed 
as well.
\item A two dimensional distribution of the \dmp~versus the mean drift time is built.
\item Finally, the latter distribution is converted into the profile form  of
Fig.~\ref{fig:tau} and then, fitted to a straight line.
\end{itemize}

From the slope of the curve in Fig.~\ref{fig:tau} is possible to obtain an accurate
estimation of the electron drift lifetime during the data taking stage. 
When the drift time $t$ is much smaller than $\tau$, the exponential attenuation 
of the drifting charge (Eq.~\eqref{eq:exp_law}) can be approximated by an straight 
line, being the slope the inverse of the lifetime. Thus, from the fitted data 
in Fig.~\ref{fig:tau} we have a value for the slope of $p1=-0.015\pm0.007 MeV\mu s/cm$, 
which leads to a lifetime of $\tau = 120~\mu s$ with a relative error of $50$\%. 
Such a big error comes from the flatness of $p1$. This means that we are over the 
sensitivity limit  of the measurement and that the attenuation over the drift 
distance is negligible.

\begin{figure}[t]
\begin{center}
\includegraphics[width=10cm,height=9cm]{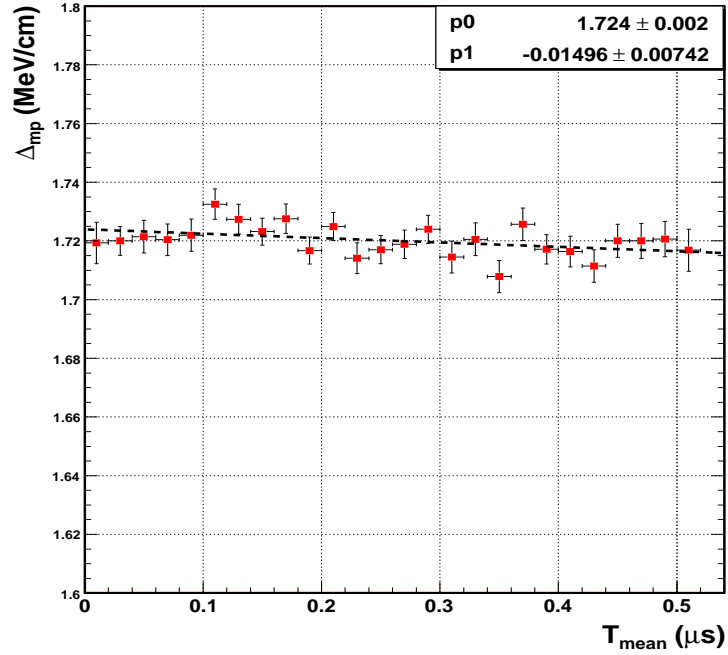}
\caption[Most probable value of $dE/dx$ vs drift time]
{Most probable value of $dE/dx$ vs drift time.}
\label{fig:tau}
\end{center}
\end{figure}

\subsection{Beam alignment}

In order to obtain a correct reconstruction of the 
events is mandatory to determine precisely the position of the chamber with respect
to the beam.
The alignment has been done moving the local reference system (RS) of the TPC
(see Sec.~\ref{sec:transform}) to the NOMAD reference system and then, to the 
beam. Since NOMAD alignment with respect to the beam was very precisely
measured, this error source is less important than the one we will have for 
the alignment of the TPC with respect to NOMAD.

We call $\theta$ the rotation in the $z-x$ plane and $\phi$ the rotation
in the $y-z$ plane. In terms of these angles, the alignment of NOMAD with 
respect to the beam is written as~\cite{NOMAD}
\begin{equation}
\begin{split}
\theta_{N}& = 0~mrad \\
\phi_{N}& = 42~mrad = 2.41^\circ 
\end{split}\label{eq:NomadAngles}
\end{equation}

\subsubsection{Alignment of the TPC respect to NOMAD}
In order to determine the alignment parameters of the 50L TPC respect to the
NOMAD RS, we made use of the same mip sample of Sec.~\ref{sec:driftvel} used 
for the electron drift velocity determination. 
There are 268 through-going muons correctly measured and 
reconstructed both by NOMAD and the LAr TPC. 
The relative angles $\theta$ and $\phi$ are then calculated through the minimization
of the difference between the direction angles measured by the 50L and NOMAD 
respectively.
Final distributions after the minimization are shown in Fig.~\ref{fig:50L-NOMAD_align}.
The value of the alignment parameters are determined as the ones which make these 
distributions to be centered at zero:
\begin{equation}
\begin{split}
\theta & = 30.08 \pm 0.06^\circ \\
\phi   & = - 0.34 \pm 0.03^\circ \\
\end{split}\label{eq:50LAngles}
\end{equation}

\begin{figure}[t]
\begin{center}
\begin{tabular}{c c}
  \includegraphics[width=7.cm,height=6.8cm]{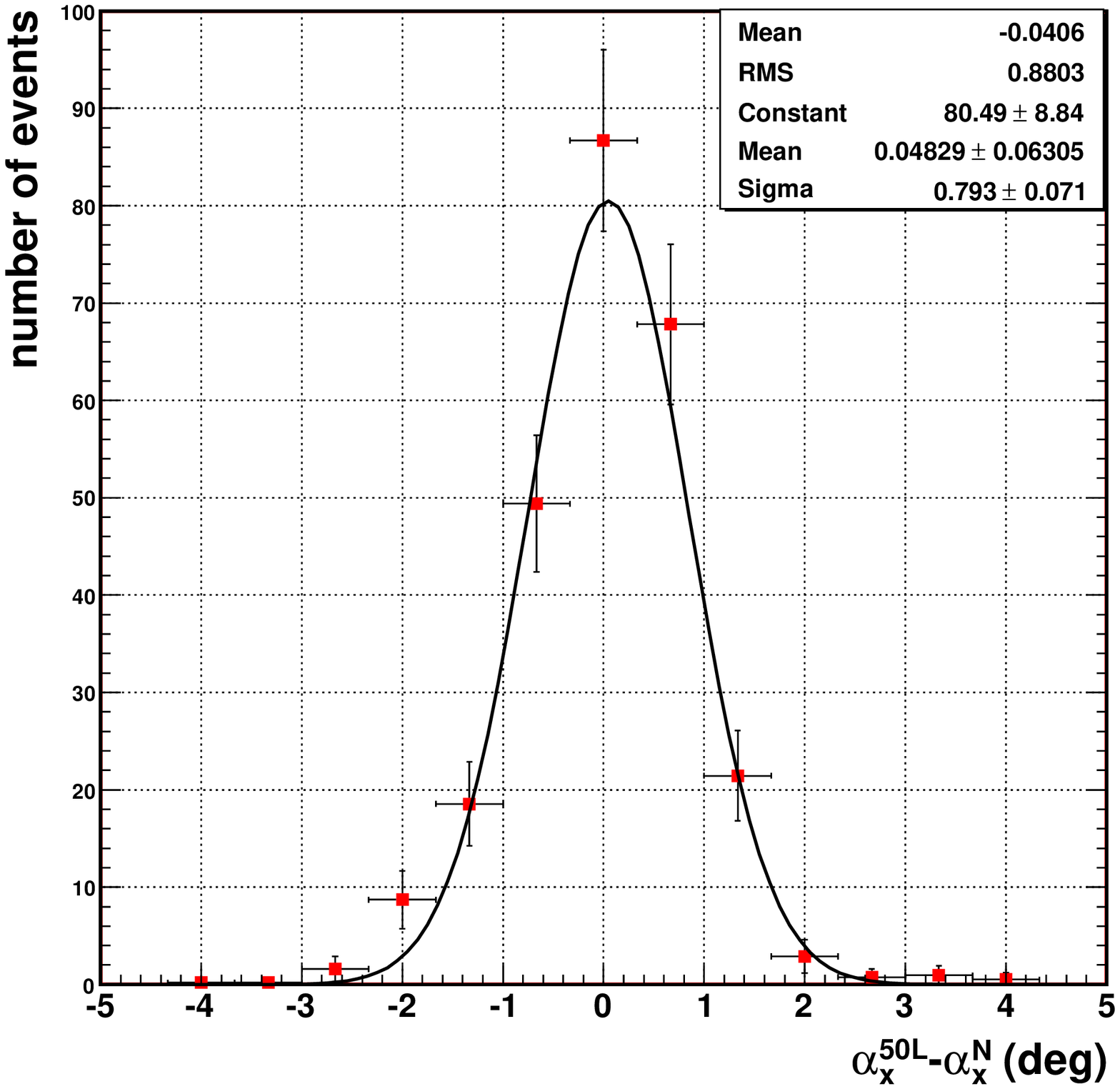} &
  \includegraphics[width=7.cm,height=6.8cm]{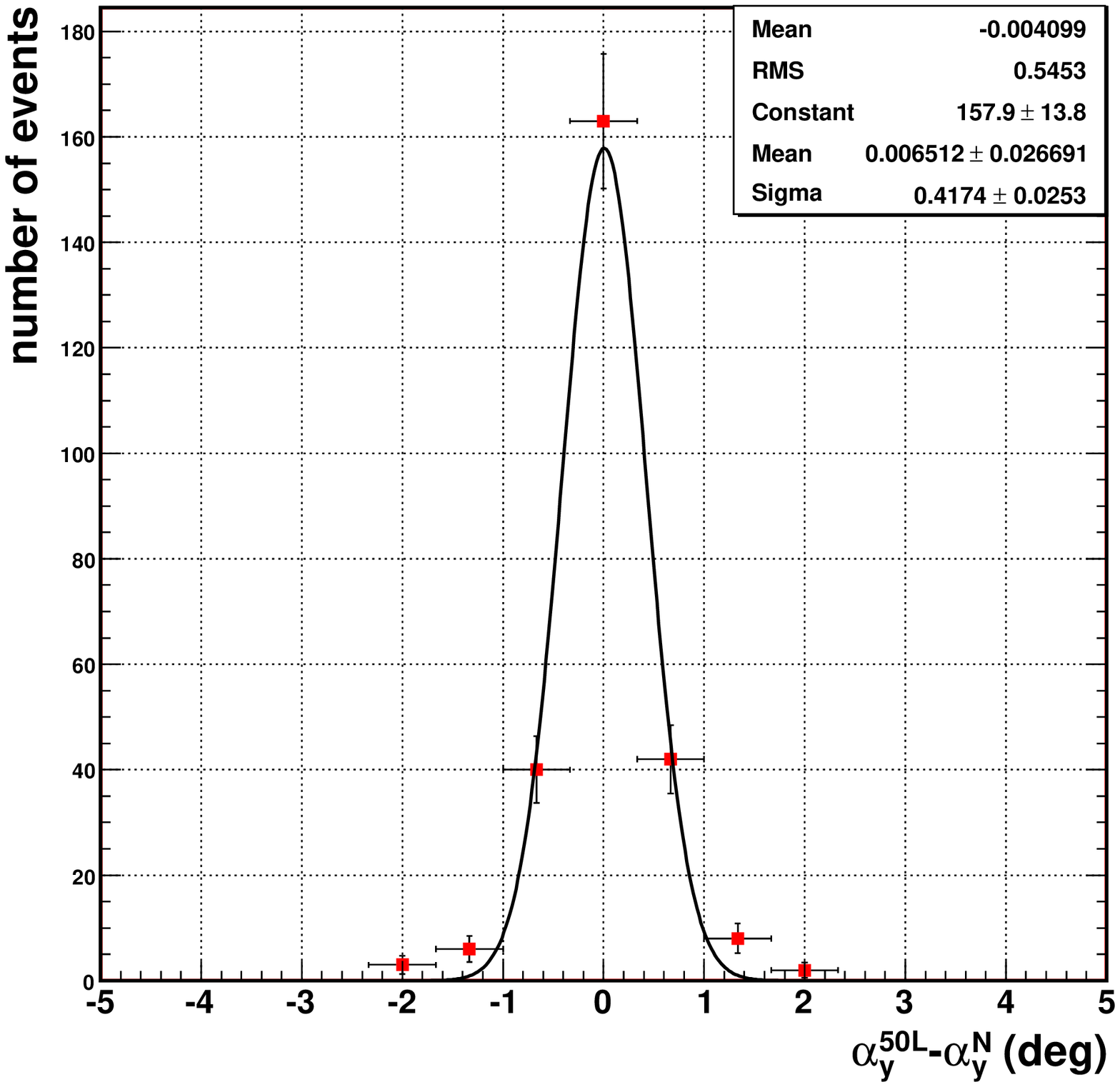}
\end{tabular}
\caption[Difference of the measured 50L and NOMAD angles along $x$ and $y$ axes]
{Difference of the measured 50L and NOMAD angles along $x$ and $y$ axes}
\label{fig:50L-NOMAD_align}
\end{center}
\end{figure}

\subsection{Energy calibration}
\label{sec:50Lcal}
As it was discussed in Sec.~\ref{sec:ecal}, the energy calibration consists 
in expressing the charge associated to a hit in energy units. The measured 
charge is related to the real deposited energy by means of a unit conversion constant
($CW$) and the recombination factor ($\mathcal{R}$ in Eq.~\eqref{eq:QtoE}).
In this section, we measure the energy calibration factors for the 50L TPC under
the assumption that the recombination factor is well described by Birk's law
in Eq.~\eqref{eq:birks}.

\subsubsection{The \emph{overall} calibration factor}
The calibration factor $C$ converts the detector measuring units (ADC counts) 
into charge units (fC), and is usually obtained from a set of dedicated test pulses, 
simulating signals ranging from single to several tens of minimum ionizing particles
~\cite{ELIFETIME,ricotesi}. It was determined for the collection wires during the set-up
phase before the data taking, when read-out was running in ``charge'' mode.
As it was commented in Sec.~\ref{sec:50Lreadout}, the final (and optimal) configuration 
adopted for the read-out worked in ``current'' mode which, unfortunately, 
did not have a calibration test pulse to measure $C$.

Due to this lack of information, what we measure instead is the so called \emph{overall} 
calibration factor $\alpha \equiv CW/R$ in Eq.~\eqref{eq:QtoE}.
To this purpose, we have collected a sample of around 3000 through-going muons 
coming from $\nu_\mu$ CC interactions. The spatial and calorimetric reconstruction
of those muons allows to measure precisely the deposited charge per unit of length 
($dQ/dx$) as described in Sec.~\ref{sec:calrec}.
We show in Fig.~\ref{fig:dqdx_muons} the measured distribution of $dQ/dx$ for 
the 3000 mouns in the sample:
The data is fitted to a convoluted Landau-Gaussian function to obtain the most probable
energy loss for mips in terms of detector charge units: 
$\bar{\Delta}_{mp} = 216.7 \pm 0.1$~ADC/cm.\footnote{The bar in $\bar{\Delta}_{mp}$ 
means the value is given in detector charge units.}
The most probable energy loss slightly depends on the energy for the range of the 
muons being considered (see Fig.~\ref{fig:dEdxVsKin}).
From this measurement, $\alpha$ can be obtained assuming that according to
the theoretical prediction for muons in the range of energies of the beam, 
the most probable energy loss is $\Delta_{mp} = 1.736 \pm 0.002$~MeV/cm, therefore:
\begin{equation}
\alpha_{mip} \ \equiv \ \frac{CW}{R_{mip}} \ = \ 
\frac{\Delta_{mp}}{\bar{\Delta}_{mp}}\Biggr\rvert_{mip}
\ = \ 8.01 \pm 0.01 \times 10^{-3} \ MeV/ADC
\label{eq:cal_factor}
\end{equation}
\begin{figure}[t]
\centering
\includegraphics[width=10cm]{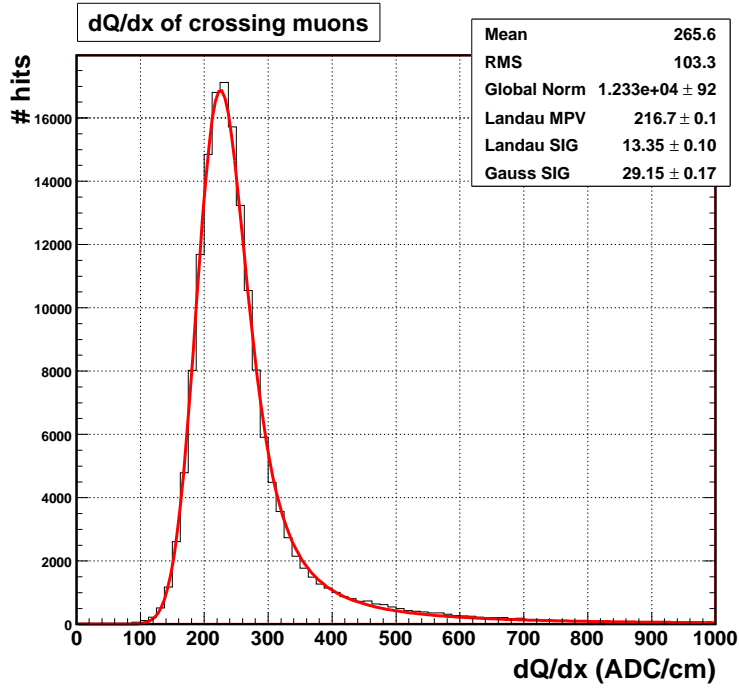}
\caption[$dQ/dx$ of crossing muons.]
{$dQ/dx$ of crossing muons. The distribution is fitted to a convoluted 
Landau-Gaussian distribution.}
\label{fig:dqdx_muons}
\end{figure}

\subsubsection{The recombination factor}
As we discussed in Sec.~\ref{sec:recfactor}, the recombination factor $\mathcal{R}$ 
of drifting electrons strongly depends on the ionization density and this dependence 
of $\mathcal{R}$ with $dE/dx$ can be modeled by Birk's law (Eq.~\eqref{eq:birks}):
\begin{equation*}
\frac{dQ}{dx} \ = \ \frac{a\, \frac{dE}{dx}}{ 1+k_B\, \frac{dE}{dx} }
\end{equation*}
The size of the range of the fully contained protons collected in the \goldens
considered in Sec.~\ref{sec:Analysis_QE},
offers the opportunity to precisely test the energy loss pattern in the active
medium. 
Fig.~\ref{fig:dqdxvsrange} shows the observed charge per unit length
($dQ/dx$) versus the residual range for protons\footnote{The
residual range is the actual range minus the range already covered at
the time of the deposition of a given $dQ/dx$.}. 
Dots are direct measurements
from the reconstructed hits of the proton tracks. From this distribution 
we estimate the most probable value of $dQ/dx$ for each bin of the residual range. 
Finally, from this data we fit $k_B~= 0.035 \pm 0.001$~cm/MeV
(\eqref{eq:birks}) for the proton hypothesis (continuous line), constraining
the value of $a$ to make $\mathcal{R}$ compatible with Eq.~\eqref{eq:cal_factor} 
at mip region: 
\begin{equation}
a = (1+k_B\Delta_{mp}\rvert_{mip}) / \alpha_{mip}
\end{equation}

Based on the knowledge of the Birk's law parameters, a precise
determination of the deposited energy can be obtained through direct charge 
measurements by means of Eq.~\eqref{eq:birks}. Moreover, pure Bethe-Bloch ionization loss
can be corrected by the detector response in the region of high $dQ/dx$, where quenching
effects are more sizable. This correction is important for particle identification
in LAr, which exploits the different energy loss of particles near the stopping point.

In summary, we have precisely measured all the calibration parameters needed for a 
proper kinematical reconstruction of the $\nu$ events occurring in the TPC. 
The electron drift velocity allows to translate time samples into the vertical 
coordinate $y$, and the electron lifetime is needed to correct the attenuation of the 
measured charge. The alignment angles are very important to properly describe 
the kinematics on the transverse plane respect to the beam direction. Finally, 
the recombination factor precisely describes the calorimetric picture of the events.
In Tab.~\ref{tab:50Lcalib}, we summarize the measured values of the calibration 
parameters for the 50L detector.

\begin{figure}[t]
\begin{center}
\includegraphics[width=10cm]{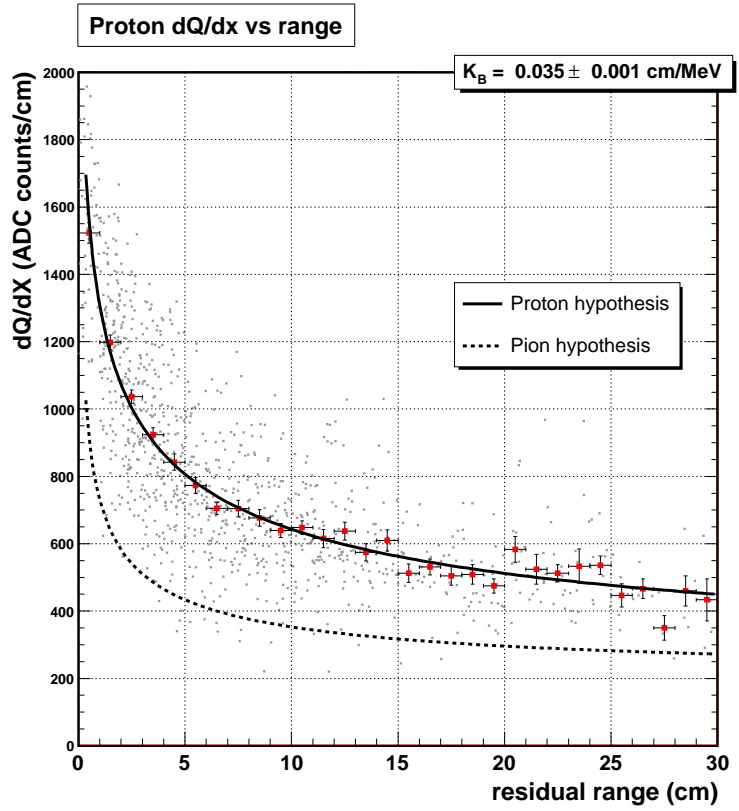}
\end{center}
\caption[$dQ/dx$ as a function of the residual range for protons.]
{$dQ/dx$ as a function of the residual range for protons.
Dots are direct measurements from the reconstructed hits of the proton tracks, while
the points with errors bars are the estimated most probable of $dQ/dx$ for each bin 
of the residual range. Data is fitted with a Bethe-Bloch corrected by the detector 
response function (Birk's law) in the hypothesis of pion (dashed line) and proton 
(full line)}
\label{fig:dqdxvsrange}
\end{figure}

\begin{table}[t]
\begin{center}
\begin{tabular}{|l|l|}
\hline
Drift velocity        & $0.905 \pm 0.005$~mm/$\mu$s                           \\ \hline
Electron lifetime     & $> 10$~ms \emph{(over the sensitivity limit)}             \\ \hline
Alignment angles      & $\theta=30.08\pm0.06^\circ\quad\phi=-0.34\pm 0.03^\circ$ \\ \hline
$\alpha_{mip}$        & $8.01\pm0.01\times 10^{-3}$~MeV/ADC                  \\ \hline
$k_B$                 & $0.035\pm0.001$~cm/MeV                               \\ \hline
\end{tabular}
\caption{\label{tab:50Lcalib} Calibration parameters of the 50L detector.}
\end{center}
\end{table}

\section{Analysis of quasi-elastic $\nu_\mu$ charge current interactions} 
\label{sec:Analysis_QE}

\subsection{Data taking and event selection}
\label{sec:Sel_QE}

The data set recorded with the 50L chamber amounts to
$1.2\times 10^{19}$ protons on target. 
As mentioned in Sec.~\ref{sec:Setup}, the trigger efficiency was monitored
during data taking and its integrated value is 97\%. Additional losses
of statistics are due to the TPC (3\%) and NOMAD (15\%) dead time and
to detector faults. These contributions add up to give an effective
lifetime of 75\%. Over the whole data taking period around 70000 triggers 
were collected in the TPC from which 20000 have at least 
a reconstructed muon possibly transversing the fiducial volume 
(see Sec.~\ref{sec:muon_rec}). From all these a priory $\nu_\mu$ CC candidates,
a visual scanning reveals that around half of them show a vertex in the 
fiducial volume, the rest are $\nu$ interactions in the surrounding LAr bath 
(which also give trigger) plus a contamination of crossing muons due to veto 
inefficiencies. Therefore, we have a set of around 10000 $\nu_\mu$ CC events.

\subsubsection{The {\bf golden sample}}
From the whole sample of $\nu_\mu$ CC events, we have selected a set of 86 
events called the \goldens. They are events having a clear two prong topology 
with a mip leaving the chamber in the beam direction and an identified contained 
proton. The main features of the golden sample are the following:
\begin{itemize}
\item An identified proton of kinetic energy larger than 40~MeV, fully contained 
in the TPC and one muon whose direction extrapolated from NOMAD matches the outgoing 
track in the TPC (see Sec.\ref{sec:muon_rec}).  
\item The distance of the primary interaction vertex to any of the
TPC walls has to be greater than 1~cm. 
\item The muon candidate track projected onto the wire plane must be longer 
than 12 wire pitches. 
\item The presence of other stopping particles is allowed as
far as their visible range does not exceed the range of a proton of 40
MeV kinetic energy. 
\item No other tracks than the identified muon can leave the TPC.
\item Events with at least one converted photon with an energy deposition in the 
fiducial volume greater than 10~MeV are also rejected. 
\end{itemize}
The tightness of these selections makes the \goldens a very
clean topology for visual scanning (Fig.~\ref{fig:Golden_rec}~(Top)).

\begin{figure}%[!ht]
\centering
\includegraphics[width=15cm]{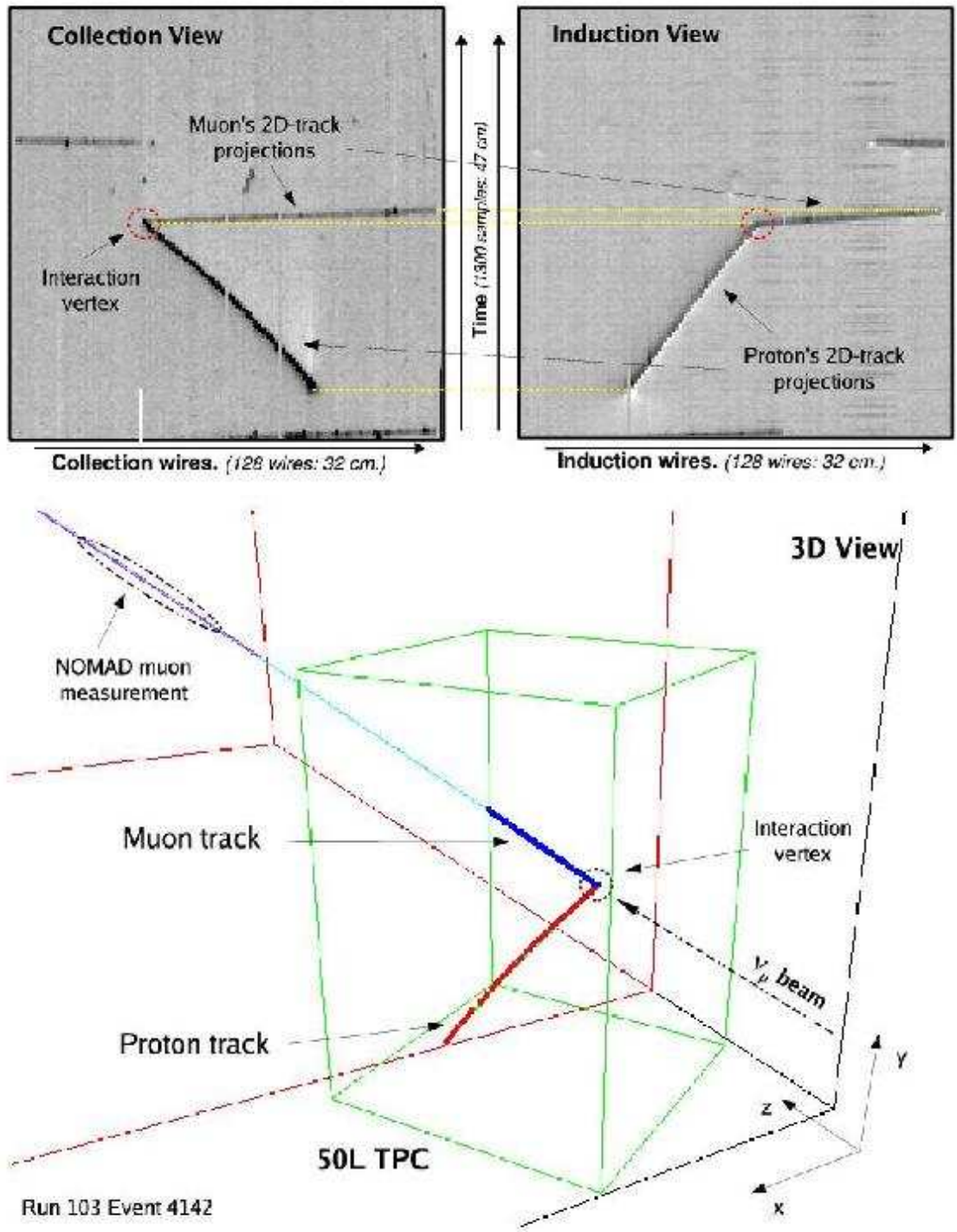}
\caption[Example of the 3D reconstruction of a low-multiplicity (\emph{golden}-like) 
$\nu_{\mu}$ CC event]
{Example of the 3D reconstruction of a low-multiplicity (\emph{golden}-like) $\nu_{\mu}$ CC event.
(Top) The raw image from collection and induction wire planes:
Hits and 2D track projections have been identified. 
(Bottom) Three dimensional view of the reconstructed event  
embedded in a 3D recreation of the experimental setup described in 
Sec.~\ref{sec:Setup}. }
\label{fig:Golden_rec}
\end{figure}

The request of containment is very severe since the chamber volume is small.
On the other hand, for a kinetic energy ($T_p$) below 40~MeV the proton range is 
comparable with the wire pitch and neither the proton momentum nor the interaction
vertex can be reconstructed with due precision. However, for
$T_p>40$ MeV the $\pi^{\pm}$/p misidentification probability is
negligible (see Sec.~\ref{sec:prec}). 

\subsection{Proton reconstruction}
\label{sec:prec}
Proton identification and momentum measurement were
performed using only the information provided by the TPC, following the 
reconstruction method explained in detail in Chap.~\ref{chap:recon}.

\subsubsection{Proton identification}
In Sec.~\ref{sec:pid}, we discussed that the average energy loss
is a function of the mass and the kinetic energy of the particle.
For the case of particles stopping in the fiducial volume of the chamber,
a direct relation between their kinetic energy and their covered range 
(i.e. the distance a particle can penetrate before it loses all its energy)
is obtained by means of the integration of Eq.~\eqref{eq:bethe} along the particle's
path (see Sec.~\ref{sec:meascal}). 

The discrimination between protons and charged pions is performed exploiting their
different energy loss behavior as a function of the range.
For the case of candidates stopping in the fiducial volume of the chamber,
as the ones of the \goldens, p/$\pi^\pm$ separation is
unique (see Fig.~\ref{fig:proton_dqdxvsR}): $dQ/dx$ is measured along 
the proton candidate track and is compared with the different particle
hypothesis. 
Monte-Carlo studies reveal that almost 100\% of protons are 
identified as such, on the basis of their $dQ/dx$ shape on the vicinity of the
stopping point; in addition, the fraction of pions and kaons misidentified
as protons is negligible, showing thus the superb particle identification capabilities
of the Liquid Argon TPC technique.

\begin{figure}[!ht]
\begin{center}
 \includegraphics[width=10cm]{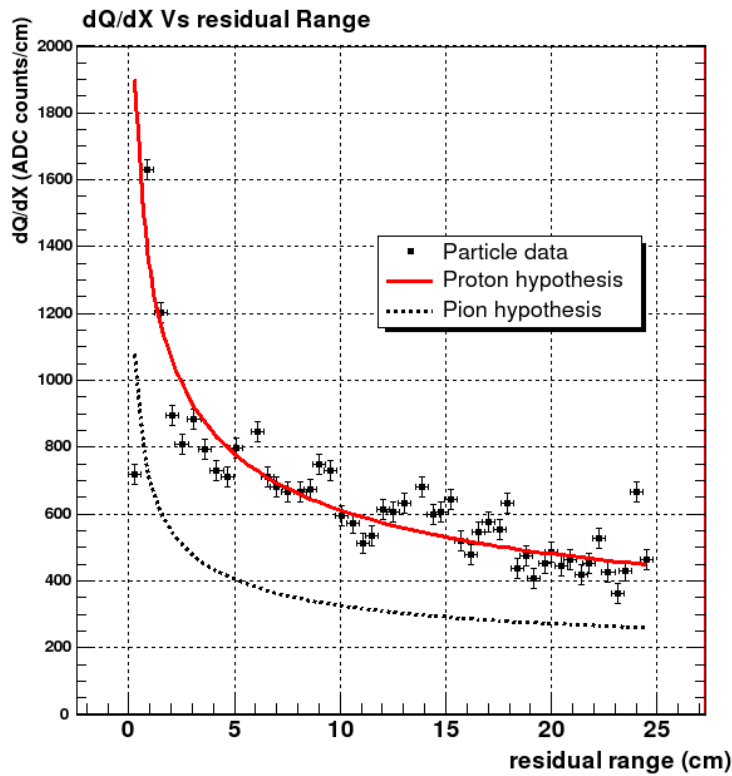} 
\caption[$dQ/dx$ as a function of the residual range for a reconstructed 
stopping proton]
{$dQ/dx$ as a function of the residual range for the
reconstructed proton in the event shown in Fig.~\ref{fig:Golden_rec}: 
continuous (dashed) line represents the expected behavior of protons (pions) 
in the detector.}
\label{fig:proton_dqdxvsR}	
\end{center}
\end{figure}

\subsubsection{Momentum measurement}
Once we identify a contained proton, its kinetic energy is calculated
from range, which only depends on the spatial reconstruction of the tracks
(see Sec.~\ref{sec:meascal}).
In this case, the momentum uncertainty is dominated by 
the finite pitch of the wires (2.54 mm)\footnote{The equivalent pitch in the
vertical direction (drift direction) is much smaller due to the high
sampling rate of the fast ADC ($360\ \mu$m).} and the performance of the
reconstruction tools. 
For an uncertainty of $\pm 1$ wire in the determination of the starting and 
ending points of the projection of the track in 
both wire planes, and neglecting the error in the drift coordinate, a
rough estimation of the error in the total length of the track is given by:

\begin{equation}
\frac{L\sqrt{2}\times 2.54~mm}{L_{tot}}
\end{equation}\label{eq:rerror}
being $L$ the projection of the track in the collection/induction plane and
$L_{tot}$ the total track length. The error in the range determination 
is automatically translated into an error in the energy measurement.
This has been tested using a Monte-Carlo simulation of the detector:
The resolution on the kinetic energy ($T_p$) varies from
3.3\% for protons with $T_p=50$ MeV up to 1\% for $T_p$ larger than
200 MeV. 

The angular resolution in the collection or induction view
depends on the number $N$ of wires hit by the particle along its path; it is
\begin{equation}
\sigma \simeq 0.36\sqrt{12}/(2.54\ N^{3/2})
\label{eq:angerror}
\end{equation}
which for a proton covering 10 wires ($N=10$) means an error of 15 mrad.

\subsection{Muon reconstruction}
\label{sec:muon_rec}
As mentioned in Sec.~\ref{sec:Setup}, the modest size of the 50L TPC
made necessary the use of NOMAD as an external muon spectrometer
in order to be able to measure the prong muons produced in 
$\nu_\mu$~CC interactions.
The kinematic reconstruction of the outgoing muons exploits the
tracking capability of the NOMAD drift chambers (see Sec.~\ref{sec:DCH}). 
An event triggering the chamber will have at least one penetrating track 
reaching the T1 and T2 trigger
scintillators bracketing the TRD of NOMAD (Sec.~\ref{sec:trigger}). 
The corresponding track, nearly horizontal at the entrance of the NOMAD 
drift chamber volume, is reconstructed with an average momentum precision 
given by Eq.~\eqref{equ:resolution} in Sec.~\ref{sec:DCH}. 
A 10~GeV horizontal muon crossing all the chambers
($L\sim 5$~m) is reconstructed with a precision of 2.2\%.  

\subsubsection{Muon matching}
The reconstructed particle is traced back to the TPC accounting for the
magnetic field and the presence of the forward NOMAD calorimeter (Sec.~\ref{sec:FC}).  
The latter introduces an additional - in fact, dominant - source of
uncertainty due to multiple scattering (MS) in the iron plates of the
forward calorimeter (see Sec.~\ref{sec:FC}). 
For small scattering angles ($\theta \ll 1$~rad), 
the MS uncertainty on the transverse momentum is 
independent of $p$ and it turns out to be $\sim 140$~MeV. 
The correctness of the back-tracing
procedure has been cross checked comparing the direction angles of the
particles belonging to the mip sample, as measured by the TPC, with
the corresponding quantity from NOMAD\footnote{Due to the high
sampling rate of the TPC, the angular resolution in the $y-z$ plane,
i.e. in the vertical plane along the nominal beam direction
($z$-axis), is dominated by the NOMAD uncertainty.}. 

\begin{figure}[!ht]
\begin{center}
\includegraphics[width=14cm]{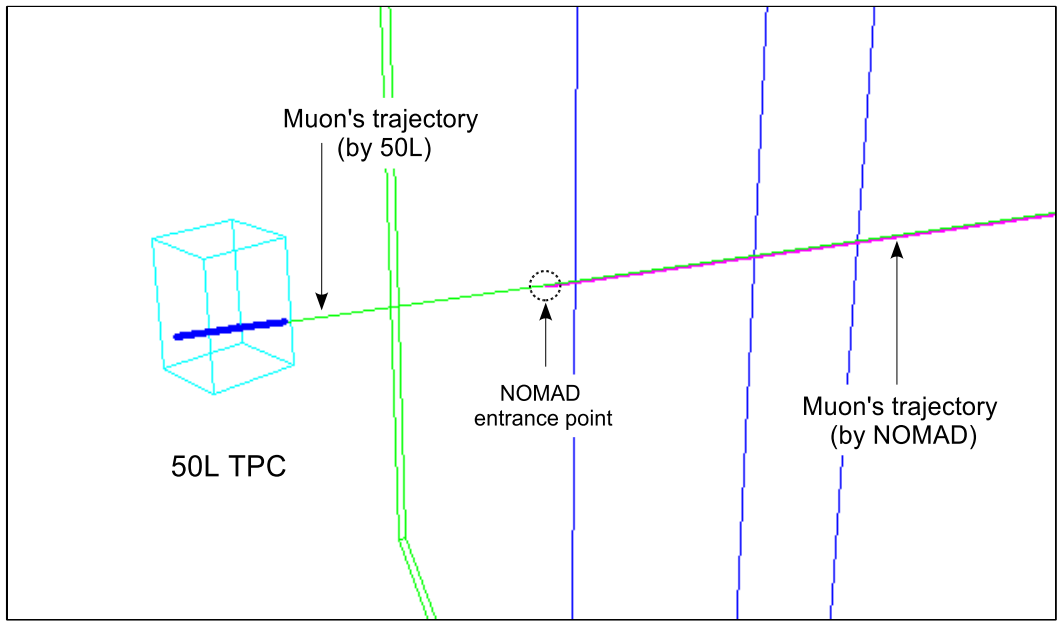}
\caption[Example of the 50L-NOMAD matching for penetrating muons]
{Example of the 50L-NOMAD matching for penetrating muons.}
\label{fig:50L-NOMAD_match}
\end{center}
\end{figure}

In Fig.~\ref{fig:50L-NOMAD_match} we show an example of the measurement of
a single muon: the traced-back muon predicted from the NOMAD measurement is shown
in pink, while the trajectory of the same muon extrapolated from the TPC data 
(blue points) is drawn in green.
Both measurements are compared to each other at the same $z$ position located at 
the entrance of the NOMAD detector, in order to assure
that the NOMAD muon is the same that traversed the TPC: 
Taking into account the errors in both measurements, we require that 
the difference between the traced-back muon and 
the original one is lower than 10~cm at NOMAD entrance point and their difference 
in angle below 100~mrad. 
As we comment in Sec.~\ref{sec:montecarlo}, the 
Monte-Carlo simulation reproduces both 50L and NOMAD measurements and thus, 
it allows to test the efficiency of the latter cuts, showing that more than
$97~\%$ of the muons fullfil those conditions (Fig.~\ref{fig:AngdiffVsP}), 
being the rest low-energy muons strongly affected by multiple scattering 
in the front calorimeter.

\begin{figure}[!ht]
\begin{center}
\includegraphics[width=8cm]{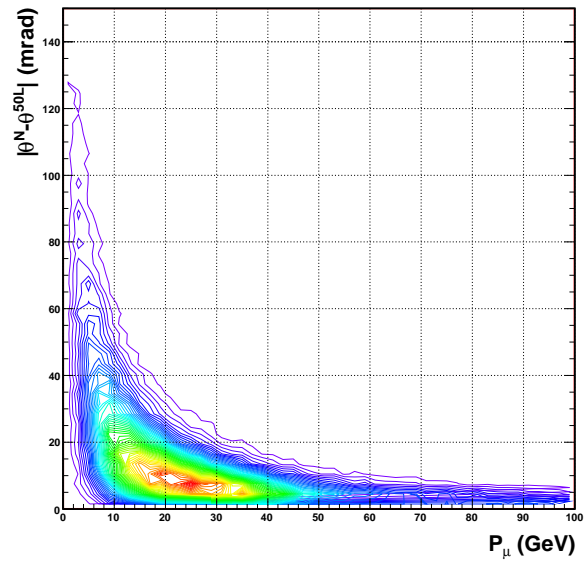}
\caption[Difference between the 50L and NOMAD measurements of muon direction as a function
of the momentum.]
{Difference between the 50L and NOMAD simulated measurements of muon direction as a 
function of the momentum. In real events, we reject events with a difference higher 
than 100~mrad.}
\label{fig:AngdiffVsP}
\end{center}
\end{figure}

\begin{figure}[!ht]
\begin{center}
\includegraphics[width=8cm]{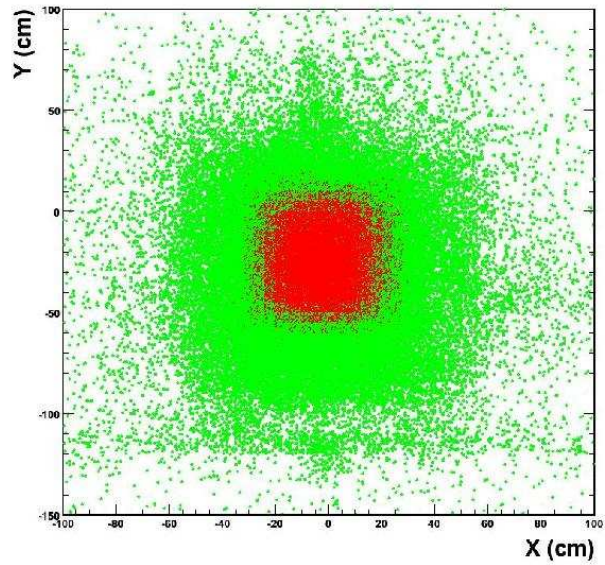}
\caption[NOMAD entrance point of triggering muons]
{The entrance point to NOMAD of all triggering muons (in green) and just the 
ones coming from the fiducial volume (in red).}
\label{fig:Npoint}
\end{center}
\end{figure}

The kinematic reconstruction of the outgoing muon from NOMAD 
allowed us to identify the events produced outside the fiducial volume 
that trigger the chamber:
The vast majority are $\nu$ interactions in the dewar or in the thermal 
bath plus a contamination of crossing muons due to veto inefficiencies. 
In Fig.~\ref{fig:Npoint} we show the distribution of the entrance point
for all the triggering muons measured in NOMAD (in green) and just the ones 
which possibly transversed the active volume of the TPC (in red), which amount
up to $30~\%$ of the whole sample of triggering muons. As we commented at the
beginning of Sec.~\ref{sec:Sel_QE}, around half of them were effectively 
originated in $\nu$ interactions inside the chamber.

\subsection{The Monte-Carlo data sample}
\label{sec:montecarlo}
Geometrical acceptance, background subtraction and QE selection efficiencies 
have been evaluated using a Monte-Carlo simulation. 
Neutrino-nucleus interactions are generated using NUX~\cite{nux}.
We use FLUKA~\cite{FLUKA} to simulate nuclear effects. 
NUX-FLUKA has successfully reproduced NOMAD data and therefore it is a
reliable tool for the study of neutrino-nucleus
interactions~\cite{FLUKA2}.
100000 $\nu_\mu$ CC QE and non-QE events have been generated simulating 
the WANF beam for the 50L chamber geometrical acceptance (see Fig.~\ref{fig:nuFlux50L})
and final prompt particles emerging from neutrino-nucleus
interactions are tracked through the 50 liter TPC and the NOMAD
detectors by means of a GEANT4 simulation~\cite{geant}. 
In Fig.~\ref{fig:setup} we show an output of a quasi-elastic $\nu_\mu$ CC event: 
The leading proton is fully contained within the TPC, while the prong muon goes
out the TPC, passing through the front calorimeter, entering in the magnet with
the drift chambers and finally giving a trigger in the scintillators planes. 
In order to efficiently reproduce the geometrical acceptance, the geometry and materials
of the different sub-detectors have been simulated with care, using the detailed 
technical information presented in Sec.~\ref{sec:Setup}.
The energy deposition of all final particles is digitized emulating the TPC 
wire read-out, in order to reproduce detector resolutions and offline reconstruction 
efficiencies.

Thanks to the simulation, full information about the final state particles and their
energy deposition in LAr is provided. This information can be accessed both at
particle level and at hit level, and besides, detailed information of secondary interactions
is also available. Outside the TPC the simulation just take care of the trajectory 
of the muon, keeping the incoming position and momentum before the front calorimeter
and immediately after, when the first of the drift chamber is hit. Finally,
the impact position (if any) of the muon in the scintillator planes is recorded 
in order to decide if it gives trigger. The momentum measurement by NOMAD
is also simulated and tracked-back to the LAr chamber taking into account multiple 
scattering effects in the front calorimeter (see Sec.~\ref{sec:muon_rec}). 

To exemplify the results of the simulation and reconstruction tools, we show in 
Fig.~\ref{fig:MCres} the visible energy (left) and the transverse missed momentum
(right) of the \goldens compared to the Monte-Carlo expectation. 
We see that the resolution on the energy measurement is rather accurate given the 
good precision measurement that both the TPC and NOMAD can achieve for the proton 
and the muon energy respectively. About the kinematical reconstruction in 
transverse plane, the missed momentum is utterly dominated by the muon reconstruction, 
since it is affected by the presence of the front calorimeter (Sec.~\ref{sec:muon_rec}).
On the other hand, the momentum direction of fully contained protons are very 
accurately measured thanks to the excellent 3D reconstruction capabilities of the 
LAr TPC.

\begin{figure}[!ht]
\begin{center}
  \begin{tabular}{c c}
    \includegraphics[width=7.cm,height=6.8cm]{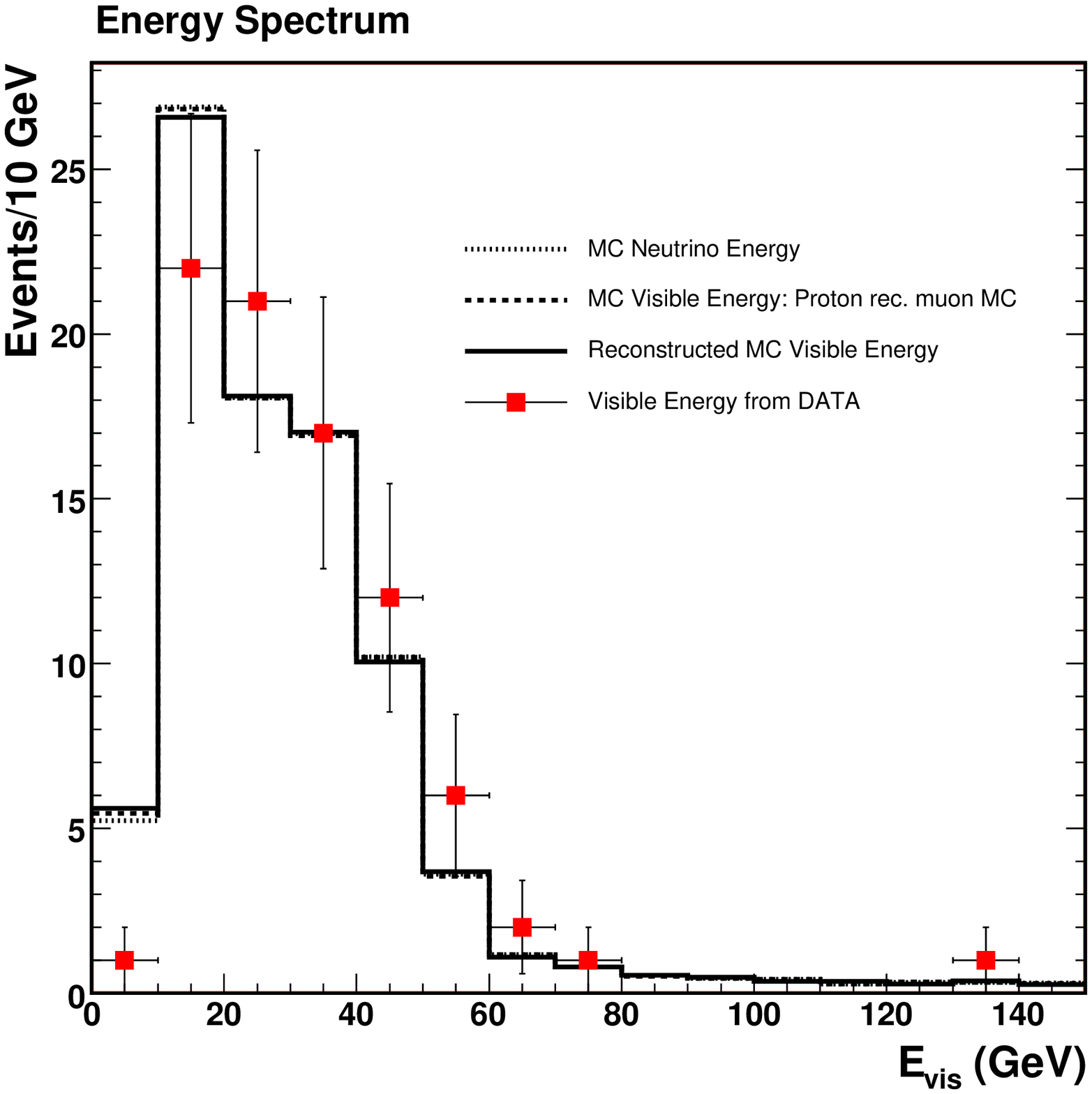} &    
    \includegraphics[width=7.cm,height=6.8cm]{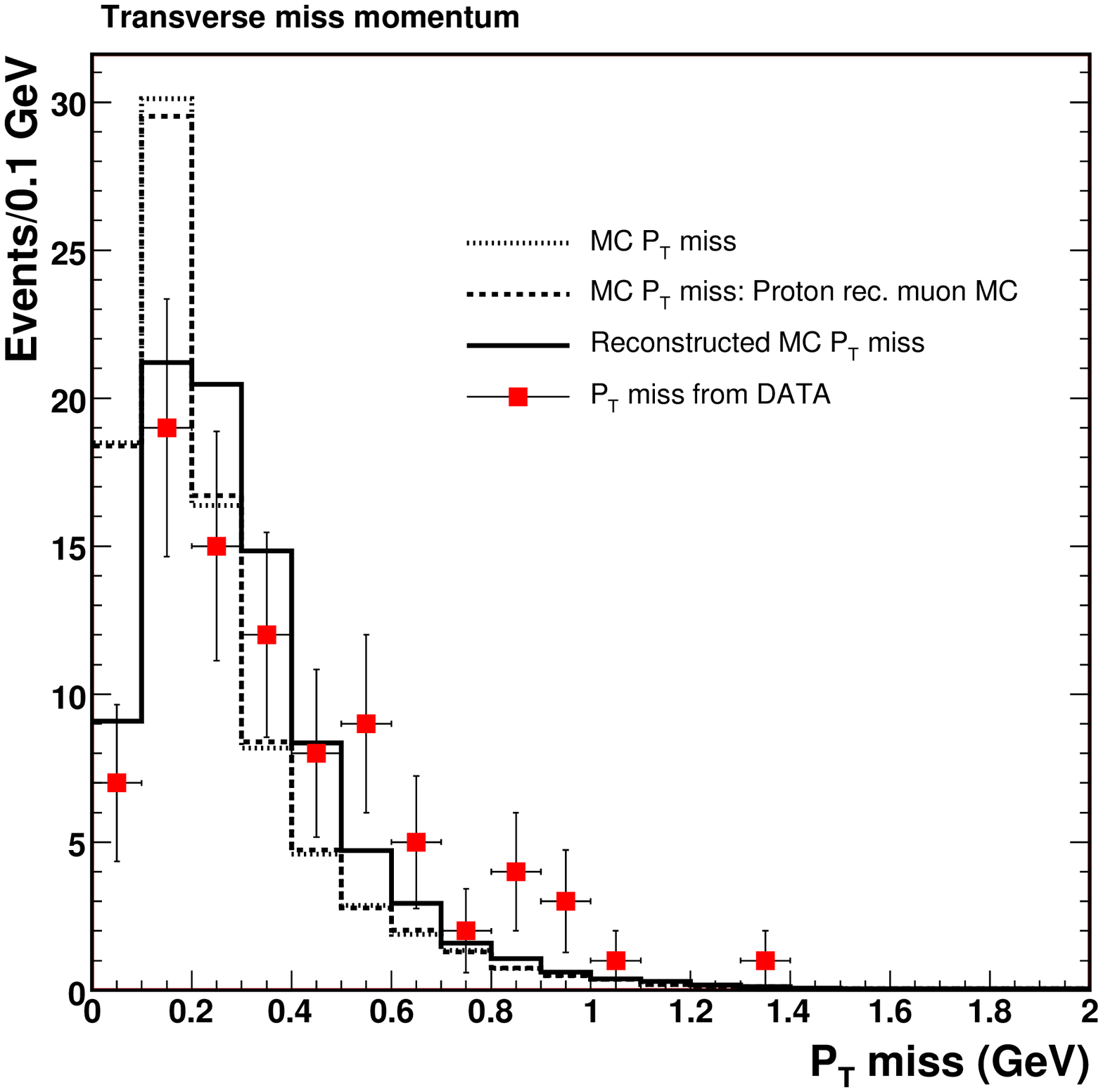}      
  \end{tabular}	
\end{center}
\caption[Visible energy and transverse miss momentum reconstruction.]
{\label{fig:MCres}
Visible energy (left) and transverse miss momentum (right) for QE {\it golden} events. 
The squared markers correspond to data. The continuous (dashed) line represents 
the reconstructed (generated) events for Monte-Carlo neutrino QE interactions 
normalized to the \goldens data. 
The dotted line corresponds to the same quantities computed by adding the 
reconstructed proton momentum to the generated muon momentum. }
\end{figure}
%---------------------------------------------------------------

\subsection{Geometrical acceptance}
\label{sec:acc}
Due to the large amount of material between the TPC and the NOMAD trackers 
(more than 11 interaction lengths),
%and the small distance between the
%chamber and the calorimeter, the $\pi/\mu$ misidentification
%probability is negligible for the present study, even accounting for
%possible $\pi$ decays in flight. 
there is an important fraction of discarded events: The prong muons leaving
the chamber after a $\nu_\mu$ CC interaction can be stopped or deflected
by the front calorimeter giving as a result a non-triggered or 
a bad-matched event. Obviously, the latter affects essentially to low energy 
and large angle muons. Neither the energy loss nor the multiple scattering in the front 
calorimeter are enough to disturb more energetic muons. The latter effect is shown
in Fig.~\ref{fig:acc} (right) where we plot the geometrical acceptance as a function
of the muon momentum for QE and DIS events separately.

In pure QE interactions, the fraction of the incoming neutrino momentum carried
by the prong muon is nearly equal to $1$ due to the low energy transferred to the 
nucleon. Thus, the muon spectra slightly differs from the spectra of interacting
neutrinos, which peaks around $20$~GeV (Fig.~\ref{fig:acc} (left)).
On the contrary, non-QE (DIS~$+$~RES) interactions are characterized by a high 
momentum transfer with the nucleus, which gives as a result a muon spectra shifted 
to low momentum values (Fig.~\ref{fig:acc} (left)).
The convolution of the muon momentum distributions with their 
respective geometrical acceptance leads to a very significant difference between the 
total acceptance for QE events (96\%) and non-QE events (72\%).

\begin{figure}[!ht]
\begin{center}
  \begin{tabular}{c c}
    \includegraphics[width=7.cm,height=6.8cm]{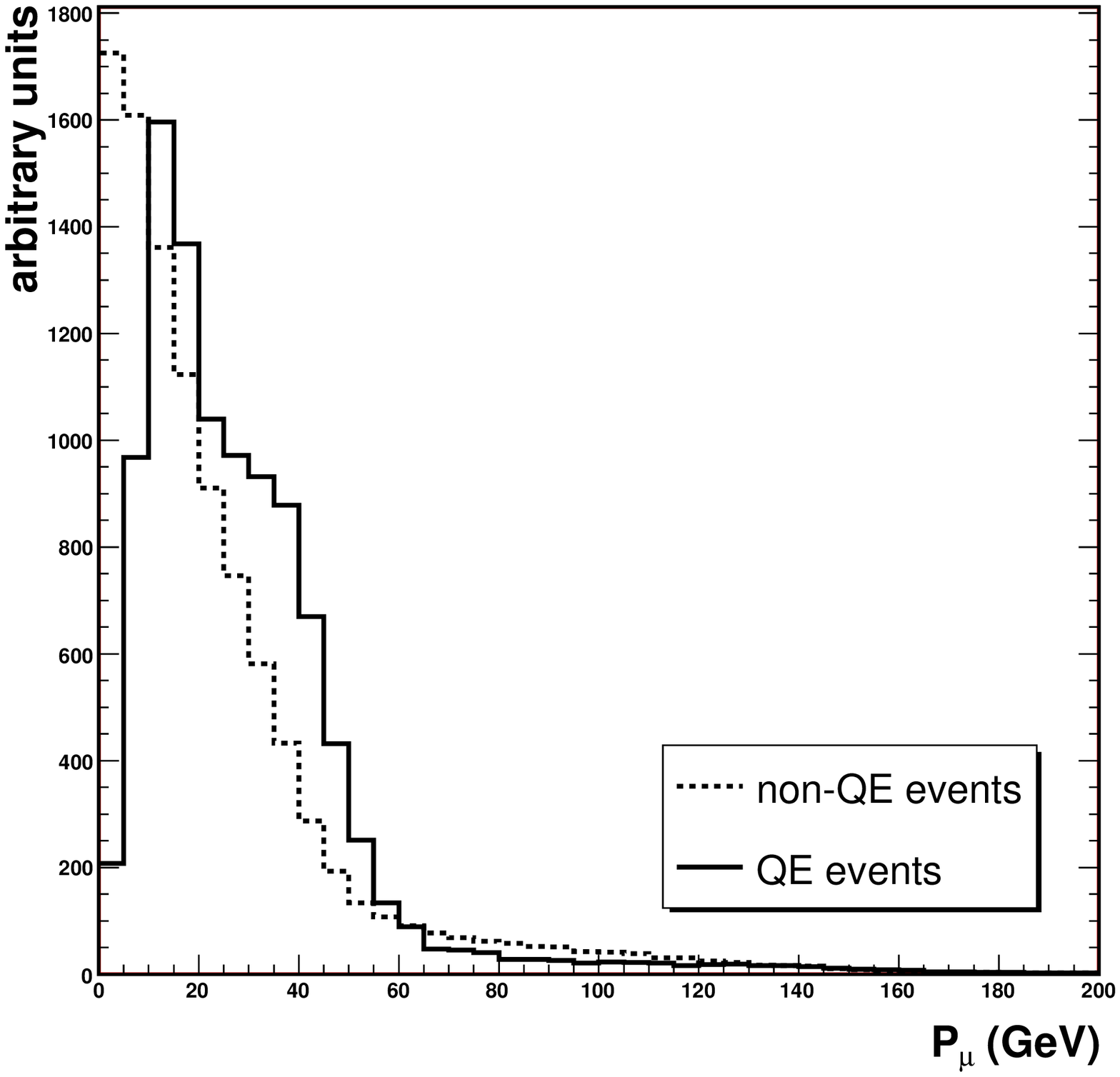} &
    \includegraphics[width=7.cm,height=6.8cm]{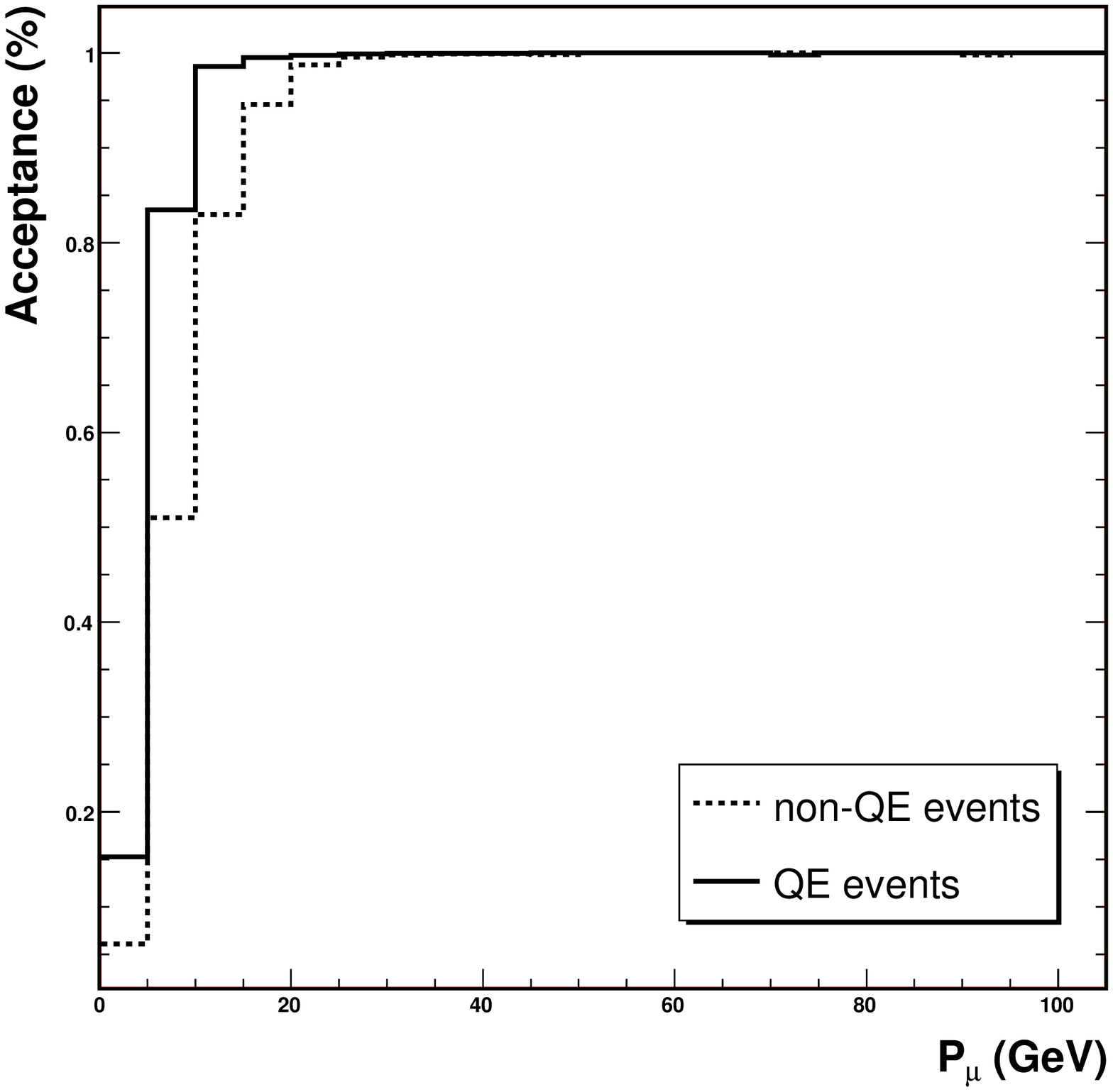} 
  \end{tabular}		
\caption[Muon momentum distribution and geometrical acceptance]
{(Left) Simulated distributions in muon momentum for QE (solid line) and non-QE
events (dotted line). These distributions are normalized to the same (and arbitrary) 
number of events. (Right) Geometrical acceptance as a function of the muon momentum for 
QE (solid line) and non-QE events (dotted line)}
\label{fig:acc}
\end{center}
\end{figure}

\subsection{Background estimation}
\label{sec:background}

The \goldens contains pure 
QE interactions ($\nu_\mu \ n \ \rightarrow \mu^- \ p$) plus 
an intrinsic background dominated by resonant production followed by pion 
absorption in the nucleus ($\nu_\mu \ p \ \rightarrow \Delta^{++} \mu^- 
\rightarrow \mu^- \ p \ \pi^+ \,, \nu_\mu \ n \ \rightarrow \Delta^{+} \mu^- 
\rightarrow \mu^- \ p \ \pi^0$). There is also an instrumental background due to 
resonance (RES) and deep-inelastic (DIS) events where final
state neutral particles ($\pi^0$'s, $n$'s and $\gamma$'s) scape undetected.
On the other hand, the 
$\nu_\mu \ n \ \rightarrow \Delta^{+} \mu^- \rightarrow \mu^- \ n \ \pi^+$
contamination is negligible in this tightly selected sample due to the superb
$\pi^{\pm}$/p identification capabilities of the LAr TPC.

Thanks to the Monte-Carlo information we can properly simulate the selection cuts 
imposed to the \emph{golden} events (see Sec.~\ref{sec:Sel_QE}) over the full 
sample of generated events. Imposing such selection criteria, an algorithm 
efficiently selects the \emph{golden} events from the whole sample of 
simulated QE interactions. We found that 16.6~\% of them are \emph{golden}, 
the rest are events with more particles in the final state
due to nuclear re-interactions or having a not contained proton.\footnote{The
automatic selection efficiency was cross-checked by means of visual scanning in a
smaller sample of simulated data.}

On the other hand, we have analyzed the sample of $\nu_\mu$ CC deep-inelastic 
(DIS) and resonant (RES) events in order to evaluate inefficiencies of the vetoing 
selections for $\nu_\mu~n~\rightarrow \Delta^{+} \mu^- \rightarrow \mu^-~p~\pi^0$ 
(i.e. the probability to miss both the decay photons of the
$\pi^0 \rightarrow \gamma \gamma$ or the $e^+e^- \gamma$ system in
case of $\pi^0$ Dalitz decay) and for events with several neutral particles
(neutrons and energetic photons from nuclear interactions) which scape from the detector
(see Fig.~\ref{fig:resonance}).
The search for this irreducible background in the DIS plus RES samples reveals 
that 0.14\% are \emph{golden}-like events. 

\begin{figure}[!ht]
\centering
\includegraphics[width=14cm]{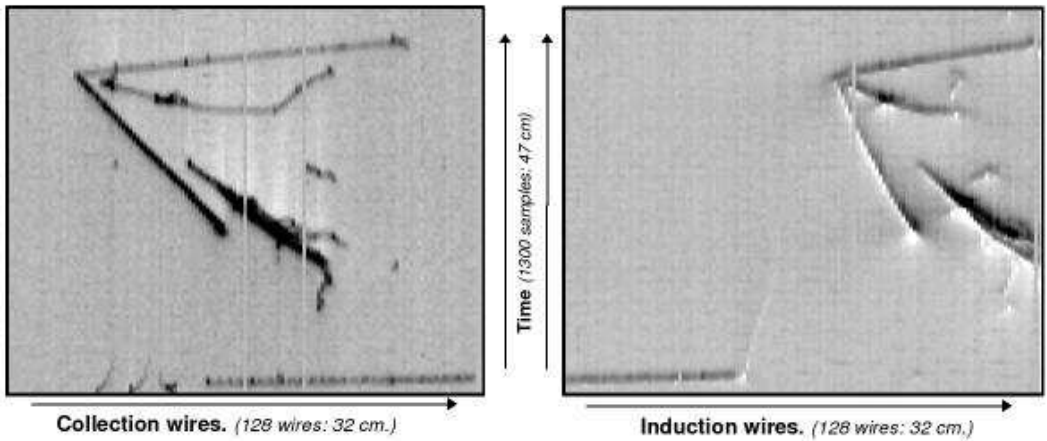}
\caption[The raw image of a $\nu_\mu \ n \rightarrow \mu^- \Delta^+
\rightarrow \mu^- \ p \ \pi^0$ event]
{The raw image of a low multiplicity real event in the
collection (left) and induction plane (right). The event
is reconstructed as ($\nu_\mu \ n \rightarrow \mu^- \Delta^+
\rightarrow \mu^- \ p \ \pi^0$) with a mip leaving the chamber, an
identified stopping proton and a pair of converted photons from the
$\pi^0$ decay. When these photons escape from the chamber, the event is 
tagged as a \emph{golden} event.}
\label{fig:resonance}
\end{figure}

Taking into account the relative weight between QE and
non-QE events\footnote{Respect to the total number of $\nu_\mu~$CC events, 
in our simulations, 2.3\% are QE, 91.5\% are DIS and the remaining 6.2\%
corresponds to RES events.
The simulation of the latest is not done treating each resonance individually,
an average of them is taken into account instead.},
we find that the total expected contamination of the \goldens
is 28\% (around 40\% of them correspond to events with an escaping
$\pi^0$). In Table~\ref{table:relvalues}, we summarize the parameters used
in background estimation.

The Monte-Carlo prediction for escaping $\pi^0$ 
can be checked using test samples that consist of \emph{golden} events with one 
($N_1$) or two ($N_2$) converted photons pointing to the interaction
vertex. Assuming the gamma identification probability to be
uncorrelated for the two photons, we have
$N_1=2N(1-\epsilon_\gamma)\epsilon_\gamma$ and
$N_2=N\epsilon_\gamma^2$, $N$ being the (unknown) overall rate of $p \
\mu^- \ \pi^0$ final states and $\epsilon_\gamma$ the photon
identification efficiency.  After the scanning of the test samples,
$\epsilon_\gamma = (N_1/2N_2 +1)^{-1}$ turned out to be (43$\pm$ 9) \%.
Hence, the probability of missing both gammas is (32$\pm$10) \% to
be compared with the MC calculation of 20.4\%. 

The estimated background is of utmost importance in the analyses of QE interactions
being presented.
It is key for a proper estimation the QE cross section calculation 
(Sec.~\ref{sec:xsection}).
Furthermore, as we will discuss in Sec.~\ref{sec:Kin_QE}, adding this sizable
background leads to a much better description of the event kinematics.

\begin{table*}[!ht]
 \begin{center}
  \begin{tabular}{|l|c|c|}
   \cline{2-3}\cline{2-3}
   \multicolumn{1}{c|}{}  &  QE       & non-QE     \\\hline
   Relative weight        & $2.3$~\%  & $97.7$~\%  \\
   Geom. acceptance       & $96$~\%   & $72$~\%    \\
   \emph{Golden} fraction  & $16.6$~\% & $0.14$~\%   \\
   \hline
  \end{tabular}
  \caption[Summary of parameters for background estimation.]
{Summary of parameters for background estimation.}
\label{table:relvalues}
\end{center}
\end{table*}

\subsection{Event rates}
\label{sec:Ev_rates}

We use Monte-Carlo to estimate the expected number of
events in the detector. The simulation of the beam predicts a flux of 
$2.37 \times 10^{-7}\,\nu_\mu$/$cm^2$/p.o.t. over the TPC exposing area. 
This flux convoluted with the neutrino cross sections 
and scaled to the fiducial mass of the detector (65.3~Kg of LAr) gives an event rate of 
$2.05 \times 10^{-15}\,\nu_\mu$CC/p.o.t.
Now assuming a total exposure of $1.2\times 10^{19}$ p.o.t. and an effective lifetime 
of 75\% (see Sec.~\ref{sec:Sel_QE}), the total number of $\nu_\mu$ CC is equal 
to 18450. Out of these 18450 events, only muons above a certain momentum threshold and 
within angular acceptance will trigger the detector.
The latter parameters for the experimental configuration are summarized in 
Table \ref{table:ExpPar}.
Taking the real data muons that triggered the system, we saw that most of 
them are above 8 GeV in momentum and below 300 mrad in angle. 
If we apply those cuts to the Monte-Carlo samples, we end up
with 18450 $\times$ 0.023 (QE fraction) $\times$ 0.96 (efficiency 
of the acceptance cuts for QE), equal to 400 quasi-elastic events.
For DIS+RES events: 18450 $\times$ 0.977 (fraction of DIS+RES) $\times$ 0.72 
(efficiency of acceptance cuts) gives 12600.
In total we expect 13000 events to be compared with the 10000 $\nu_\mu$ CC we have 
after visual scanning of {\it good triggers} (see Sec.~\ref{sec:Sel_QE}). 

The expected number of \emph{golden} events is obtained taking the 16.6\% 
(golden fraction) of the total QE, and adding the corresponding 28\% due to background 
contribution (see Sec.~\ref{sec:background}), which finally gives 
$85\pm9(stat)\pm13(sys)$
\emph{golden} events to be compared with the 86 we observe. The systematic uncertainty
is dominated by the fraction of QE events (2.3\%) and the beam simulation (8\%)
\footnote{A detailed discussion about the different sources of systematic errors can
be found in Sec.~\ref{sec:xsection}, where the $\nu_\mu$~CC QE cross section is 
calculated.}.

\begin{table*}[!ht]
 \begin{center}
  \begin{tabular}{|l|r l|}
   \hline
   Neutrino Flux             &$2.37 \times 10^{-7}$  & $\nu_\mu$/$cm^2$/p.o.t.\\
   Interaction rate          &$2.05 \times 10^{-15}$ & $\nu_\mu$~CC/p.o.t.      \\
   Total exposure            &$1.2\times 10^{19}$    & p.o.t.                   \\
   Effective detector lifetime &        $75$           & \%                       \\
   \hline
   \end{tabular}
  \caption[Summary of the experimental configuration parameters for event counting]
{Summary of the experimental configuration parameters for event counting.}
\label{table:ExpPar}
\end{center}
\end{table*}

\subsection{Analysis of the kinematics of quasi-elastic interactions}
\label{sec:Kin_QE}
In spite of the limited statistics (86 events), the \goldens provides
information on the basic mechanisms that modify the kinematics of
neutrino-nucleus interactions with respect to the corresponding
neutrino-nucleon process. Nuclear matter perturbs the initial state of
the interaction through Fermi motion; it also affects the formation of
the asymptotic states through nuclear evaporation, hadronic
re-scattering or hadronic re-absorption. Several kinematic variables 
are only marginally affected by nuclear effects; in this case, the
corresponding distributions can be reproduced once the $\nu$-nucleon
interaction is corrected for Fermi motion and Pauli blocking.
%e.g. a Saxon-Woods potential for the nucleus\footnote{In this
%model, nucleons are treated as a Fermi degenerate gas located in a
%potential well of the form $E(r) = E_0/(1+e^{(r-R)/a}) $ where $a$ is
%0.6~fm; $E_0=46$~MeV is the maximum well depth and $R$ is the nuclear
%radius ($R=3.6$~fm for the Argon).}. 
Clearly, purely leptonic variables belong to this category. 
There are also a few hadronic variables whose distribution is strongly influenced by
the selection cuts but show limited sensitivity to nuclear effects. In
particular, the proton kinetic energy is bounded by the $T_p>40$~MeV
cut and the requirement of full containment in the fiducial volume.
This distribution is shown in Fig.~\ref{fig:proton_var} together with
the transverse momentum of the proton for the \emph{golden sample}.
Similarly, in Fig.~\ref{fig:muon_var} we show the kinetic energy
and the transverse momentum distributions of the muon. 
Figs.~\ref{fig:proton_var} and \ref{fig:muon_var} can
be used as a consistency check. They demonstrate that Monte-Carlo reproduces
the kinematic selection performed during the scanning and analysis of
the \emph{golden} channel. They also show that our selected sample contains 
a non-negligible contamination from non-QE events. As already indicated, we 
estimate this contamination to be 28~\%.

\begin{figure}[!ht]
\begin{center}
  \begin{tabular}{c c}
    \includegraphics[width=7.2cm,height=7.5cm]{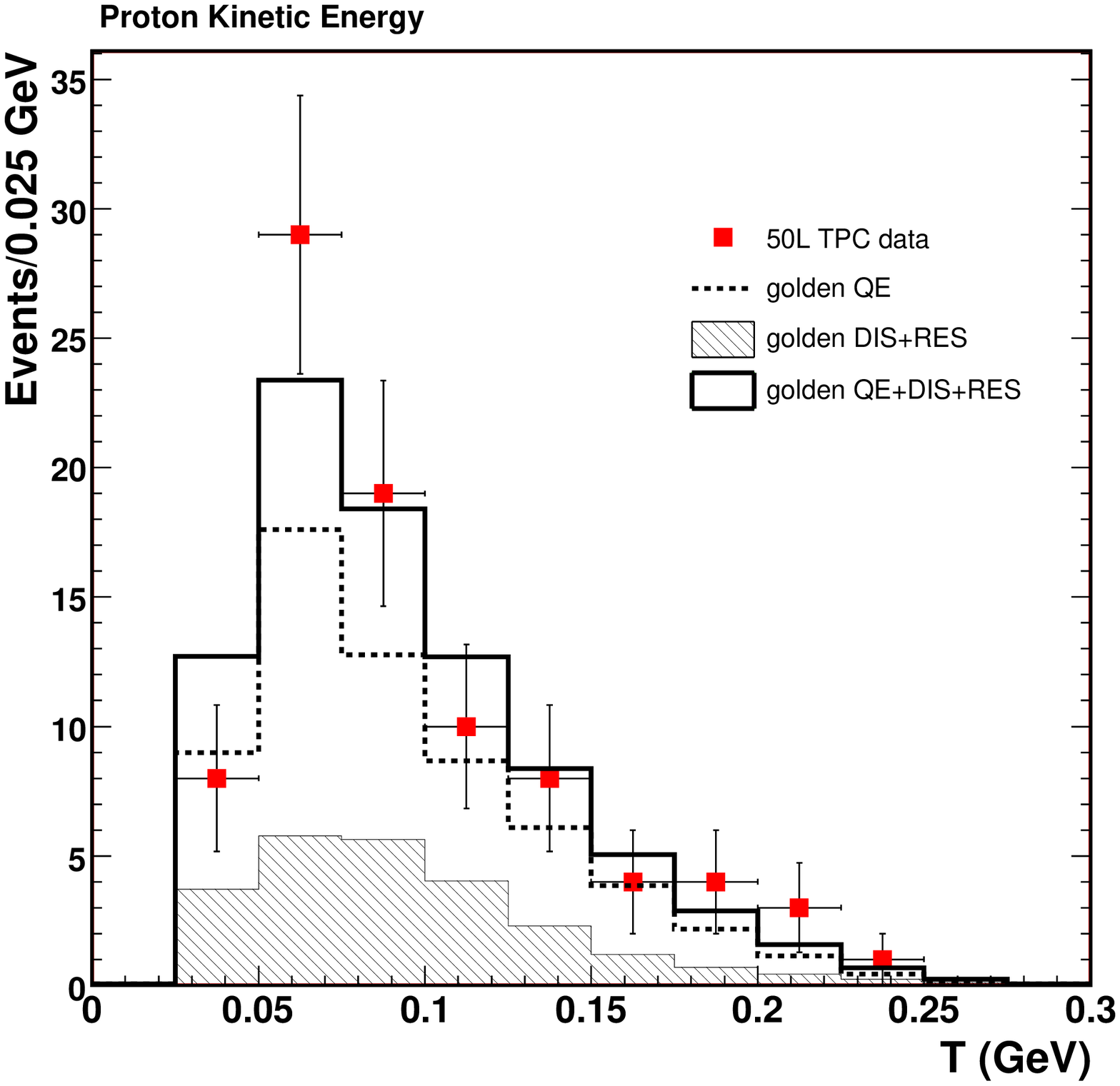} &
    \includegraphics[width=7.2cm,height=7.5cm]{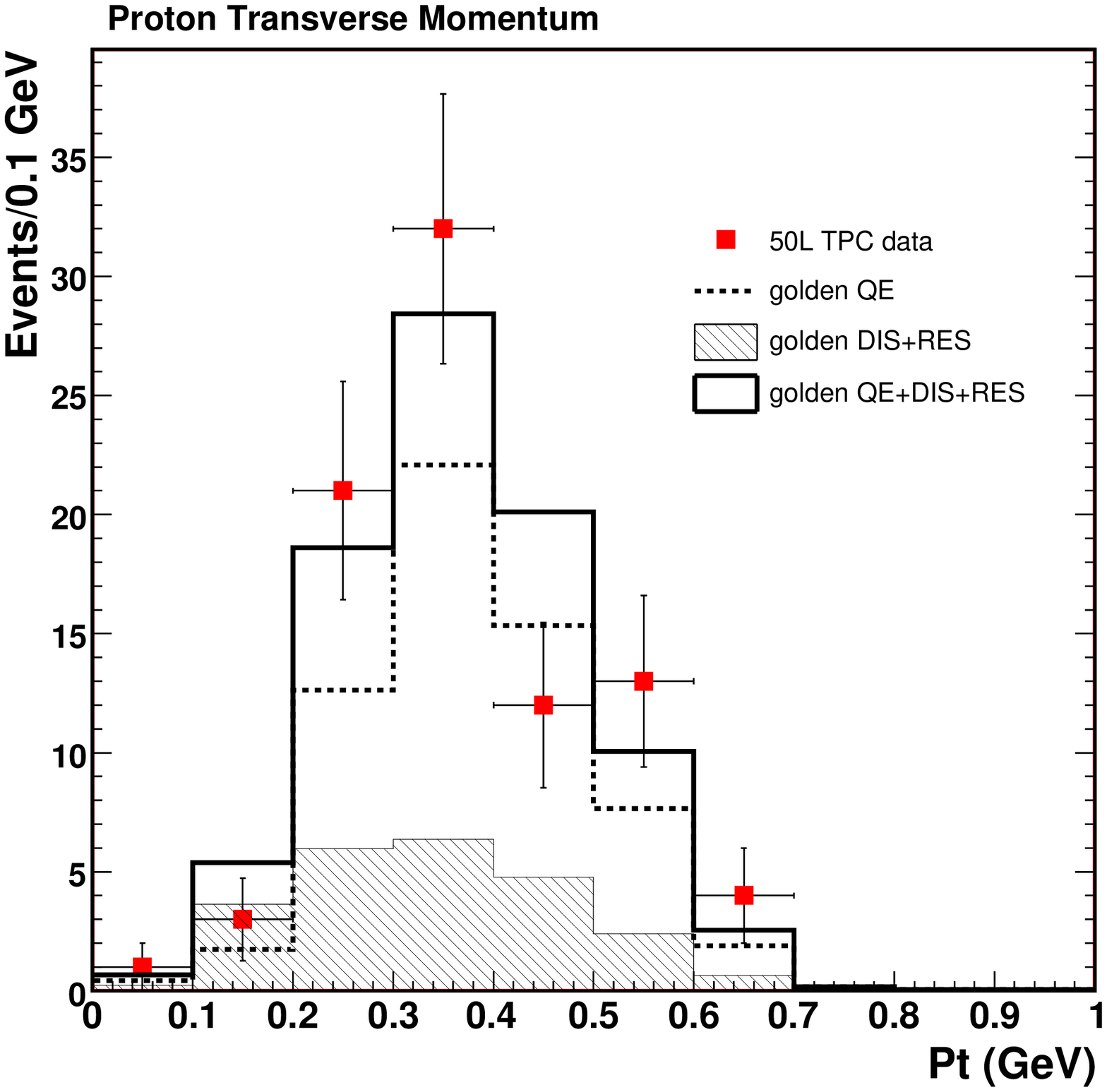} 
  \end{tabular}		
\caption[Distribution of the proton kinetic energy and transverse momentum 
 for the \goldens]
{Distribution of the proton kinetic energy and transverse momentum for the 
\goldens (squared marks). 
Dotted histogram represents the expectation from simulated Monte-Carlo QE events, 
while the hatched one represents the simulated background from non-QE events. 
Both contributions are summed in the filled histogram and normalized to the 
\goldens data. }
\label{fig:proton_var}
\end{center}
\end{figure}

\begin{figure}[!ht]
\begin{center}
  \begin {tabular}{cc}
    \includegraphics[width=7.2cm,height=7.5cm]{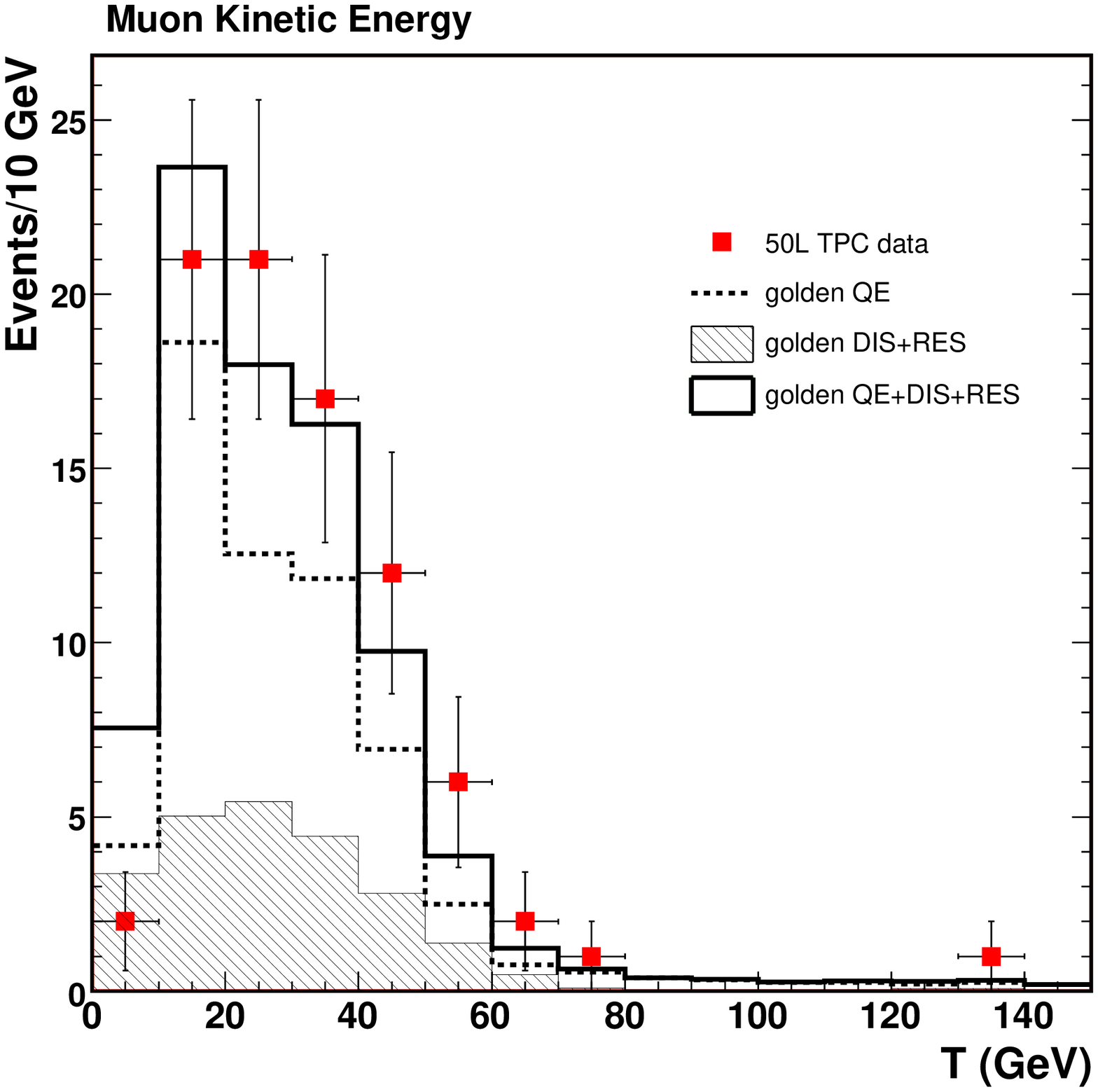} &
    \includegraphics[width=7.2cm,height=7.5cm]{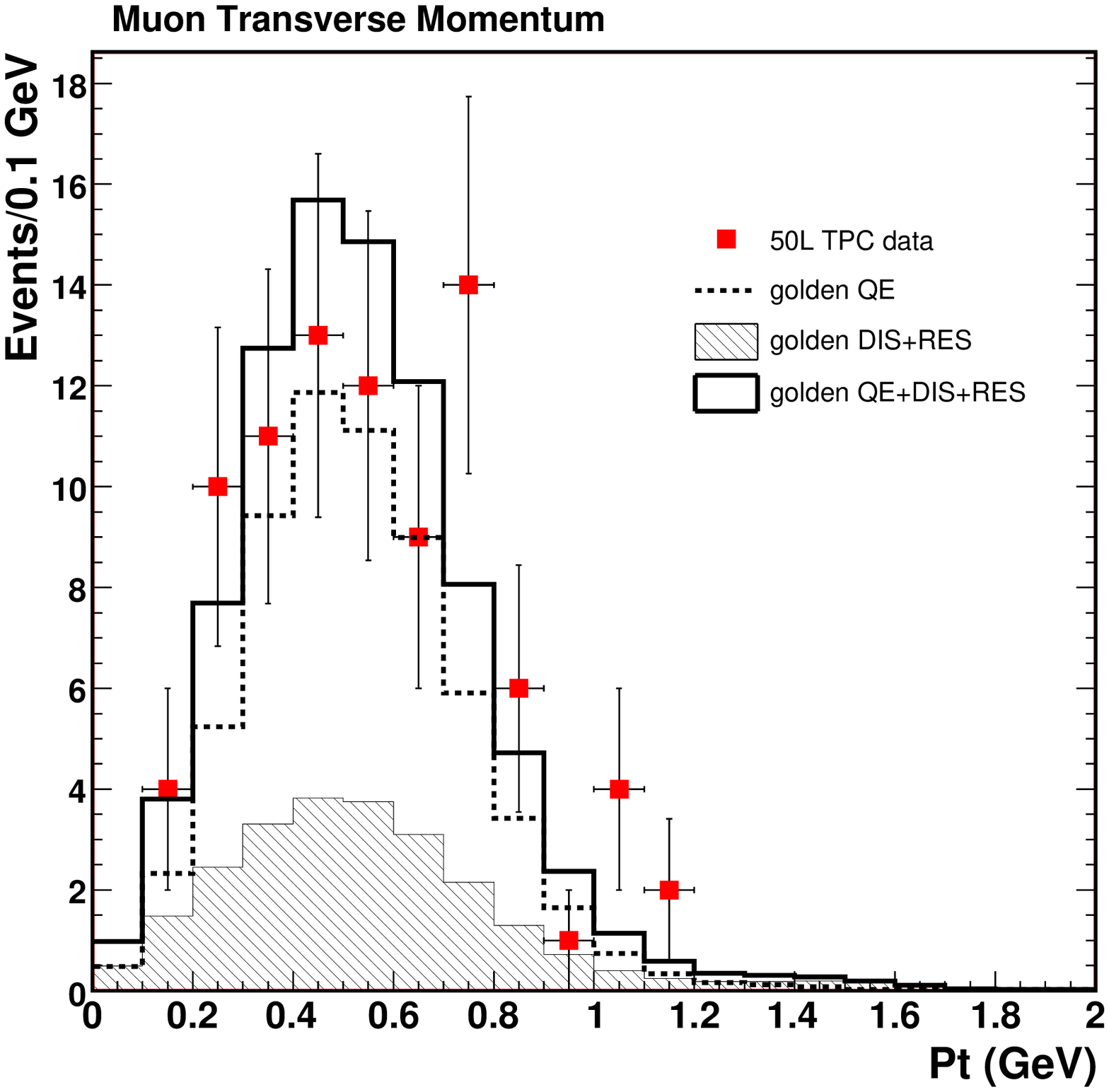} 
  \end{tabular}	
\caption[Distribution of the muon kinetic energy and transverse momentum 
for the \goldens.]
{Distribution of the muon kinetic energy and the muon transverse momentum 
for the \goldens. 
Dotted histogram represents the expectation from simulated Monte-Carlo QE events, 
while the hatched one represents the simulated background from non-QE events. 
Both contributions are summed in the filled histogram and normalized to the 
\goldens data. }
\label{fig:muon_var}
\end{center}
\end{figure}

Other variables embedding the reconstructed kinematics of the protons
are sensitive to genuine nuclear effects. In particular, 
we analyzed two of them: the acollinearity and the missing transverse momentum of the event.
The former is defined as
\begin{equation}
A \equiv \mathrm{acos} \left[ \frac{ p_{xp} \ p_{x\mu} + p_{yp} \ p_{y\mu} }
{\sqrt { (p_{xp}^2 + p_{yp}^2) (p_{x\mu}^2 + p_{y\mu}^2) } } \right]
\end{equation}
$p_{xp}$ and $p_{yp}$ being the transverse momentum components of the
proton and $p_{x\mu}$ and $p_{y\mu}$ the corresponding quantities for
the muon. For a purely QE scattering on a nucleon, the muon and the
proton ought to be back-to back in the transverse plane so that the
acollinearity is zero. Fig.~\ref{fig:event_var} (right) shows the
acollinearity distribution for the \goldens and the
expectation from simulations. For this particular variable, the inclusion of nuclear 
effects in the Monte-Carlo does not show a striking difference with respect to the case 
where no nuclear effects are taken into account. The selection cuts (in particular, 
the fact that the proton should be fully contained), the resolution NOMAD has for 
muon reconstruction and the contamination from non-QE events are the main reasons 
for the appearance of events with large acollinearity values. The distortion introduced 
in the tail of the acollinearity distribution by nuclear effects is much smaller than 
the one due to detector effects.
Hence this variable is not adequate to show how the event kinematics varies in the 
presence of nuclear matter.

The influence of nuclear effects on the event kinematics
is best seen when we consider the missing transverse momentum $P_{miss}^T$ 
(see left plot on Fig.~\ref{fig:event_var}).
%However, nuclear effects
%have been observed as perturbations in event kinematics, especially at higher values
%of acollinearity and missing transverse momentum (Fig.~\ref{fig:event_var}), and 
%therefore they must be included in the Monte-Carlo.
We observe that a naive approach that takes into account Fermi motion
and Pauli blocking disregarding nuclear effects (dashed histograms in 
Fig.~\ref{fig:event_var} left) does not reproduce the $P_{miss}^T$ data.
Once nuclear effects are added, simulated events show a good agreement with data.
The effect of background events is also relevant. 
Being the non-QE events more unbalanced than QE events in the transverse plane, 
they have the effect of shifting the expected theoretical distributions to 
higher values of $P_{miss}^T$, just like data do.

\begin{figure}[!ht]
\begin{center}
  \begin {tabular}{cc}
    \includegraphics[width=7.2cm,height=7.5cm]{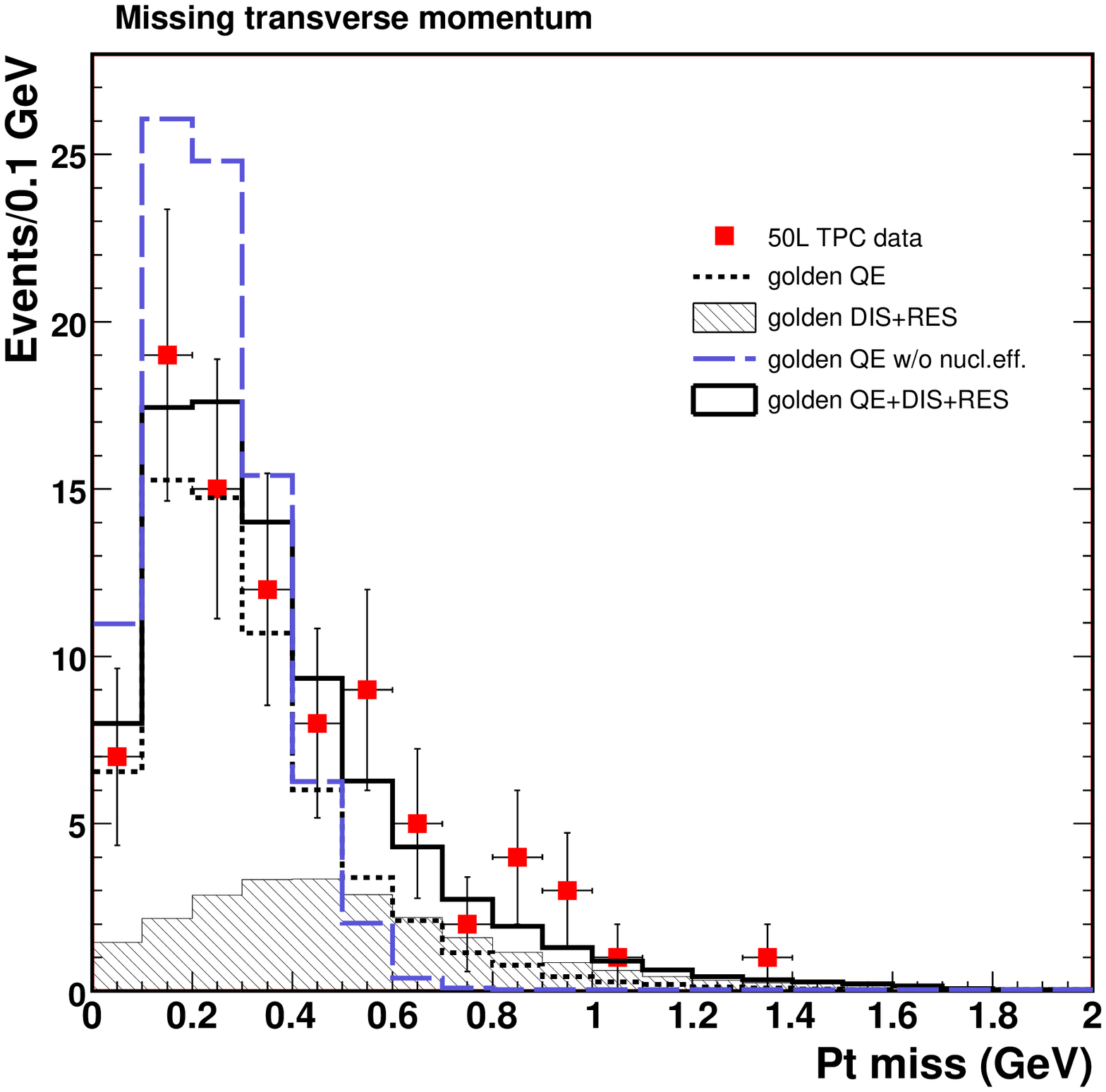} &
    \includegraphics[width=7.2cm,height=7.5cm]{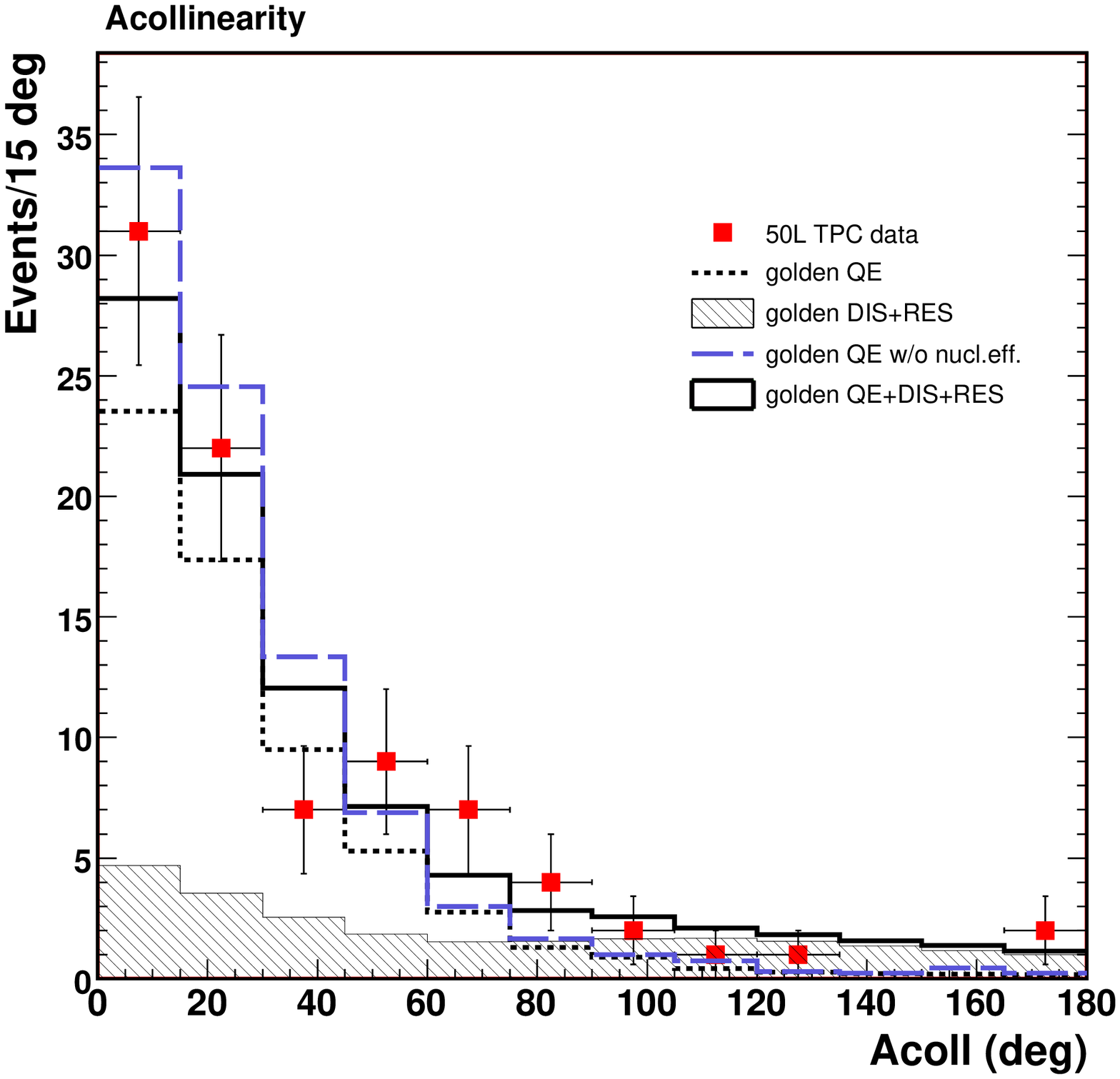} 
  \end{tabular}	
\caption[Distribution of the missing transverse momentum and acollinearity 
for the \goldens]
{Distribution of the missing transverse momentum (left) and
acollinearity (right) for the \goldens. 
Dotted histogram represents the expectation from simulated Monte-Carlo QE events, 
while the hatched one represents the simulated background from non-QE events. 
Both contributions are summed in the filled histogram and normalized to the \goldens data. 
The dashed histograms show the expected distributions in case no nuclear
effects are taken into account.}
\label{fig:event_var}
\end{center}
\end{figure}

In order to quantify the level of agreement between the theoretical expectations
and the data, we have performed a Kolmogorov test \cite{NUMERICAL_RECIPES,ROOT} 
for some of the transverse variables. Such a test is based on a comparison
between the cumulative distributions of the samples being analyzed 
(see Fig.~\ref{fig:event_var_Cum},
where cumulative distributions for $P_{miss}^T$ and the acollinearity are shown). 
Its goal is to give a statistical estimation about how compatible is the data 
sample with being a random sampling from a given theoretical distribution. 
In Tab.~\ref{table:Kol} we quote the results of such a test in terms of the Kolmogorov 
probability. There, we test experimental data against three theoretical approaches: 
\emph{golden} QE with no nuclear effects, \emph{golden} QE with nuclear 
effects included and, finally, against \emph{golden} QE plus the corresponding DIS 
and RES background\footnote{Theoretical distributions are always normalized to 
the \goldens data.}. Clearly, nuclear effects offer a much better description of 
data than naive approaches commented above, mainly for the case of 
$P_{miss}^T$ (see Tab.~\ref{table:Kol}). 
On the other hand, the effect of adding the background leads to
a better modeling of the missing transverse momentum and the acollinearity.

\begin{table*}[!ht]
 \begin{center}
  \begin{tabular}{|l|c|c|c|c|}
   \hline\hline
   \multirow{2}{*}{Variable}      &\multicolumn{3}{c|}{Kolmogorov probability}  \\ \cline{2-4}
                & QE w/o nucl. eff. & QE with nucl. eff & QE+DIS+RES with nucl. eff. \\ \hline
   $P_P^T$      & $0.35$            & $0.99$            &  $0.98$                  \\
   $P_{miss}^T$ & $0.01$            & $0.28$            &  $0.99$                  \\
   $Acoll$      & $0.38$            & $0.61$            &  $0.99$                  \\
   \hline\hline
   \end{tabular}
  \caption[Table of Kolmogorov probabilities for transverse variables]
{Table of Kolmogorov probabilities for transverse variables (see text).}
\label{table:Kol}
\end{center}
\end{table*}

%\begin{table*}[!ht]
% \begin{center}
%  \begin{tabular}{|l|c|c|c|c|}
%   \hline\hline
%   Variable      & Extended Kolmogorov probability \\ \hline
%   $T_P$         & $0.70$ \\
%   $P_P^T$       & $0.70$ \\
%   $T_\mu$       & $0.14$ \\
%   $P_\mu^T$     & $0.13$ \\
%   $P_{miss}^T$    & $0.56$ \\
%   $Acoll$       & $0.69$ \\ \hline
%   \hline
% \end{tabular}
%  \caption[Table of extended Kolmogorov probabilities]
%{Table of extended Kolmogorov probabilities}
%\label{table:ExtKol}
%\end{center}
%\end{table*}

\begin{figure}[!ht]
\begin{center}
  \begin {tabular}{cc}
  \includegraphics[width=6.5cm,height=6.5cm]{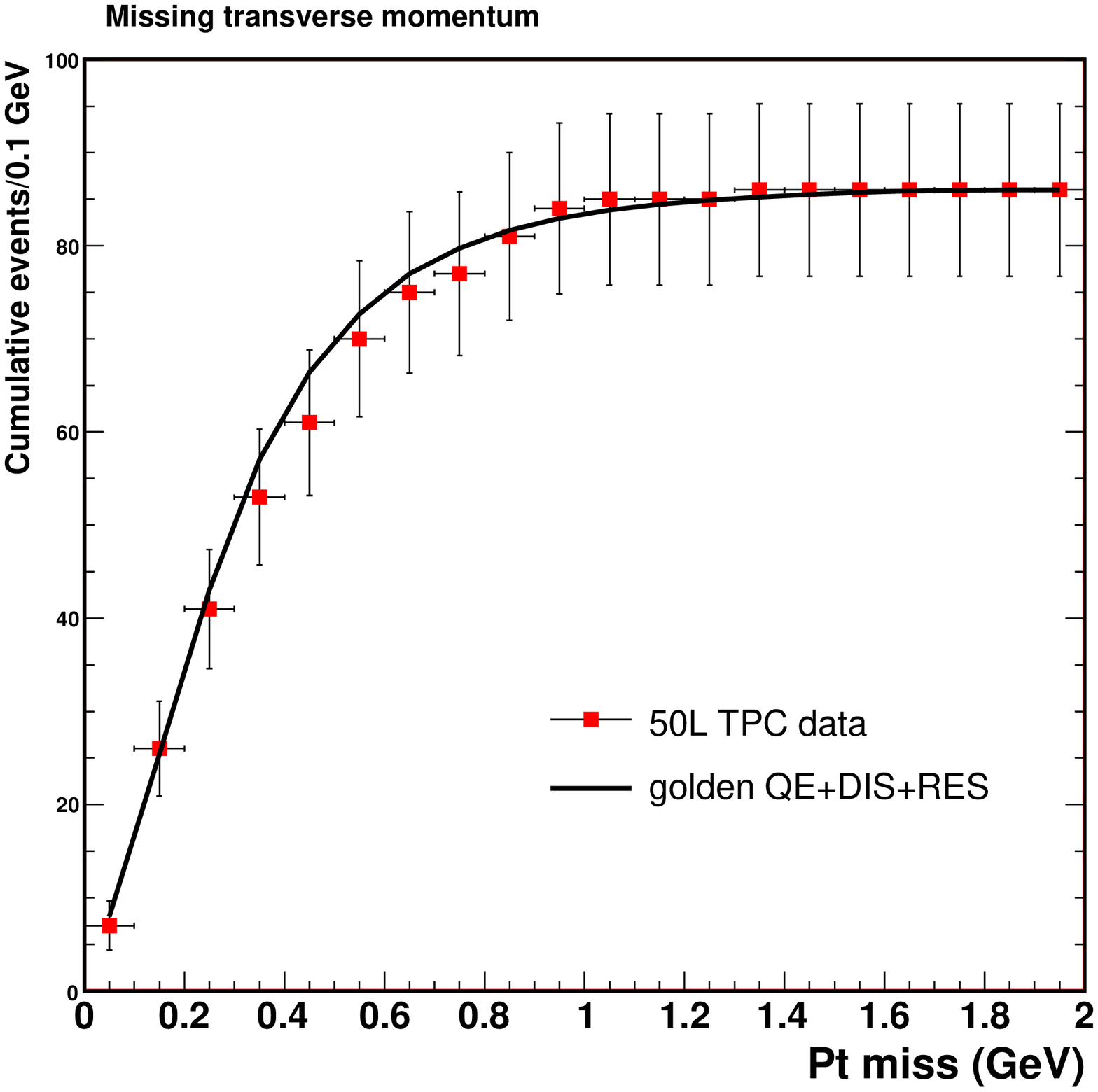} &
  \includegraphics[width=6.5cm,height=6.5cm]{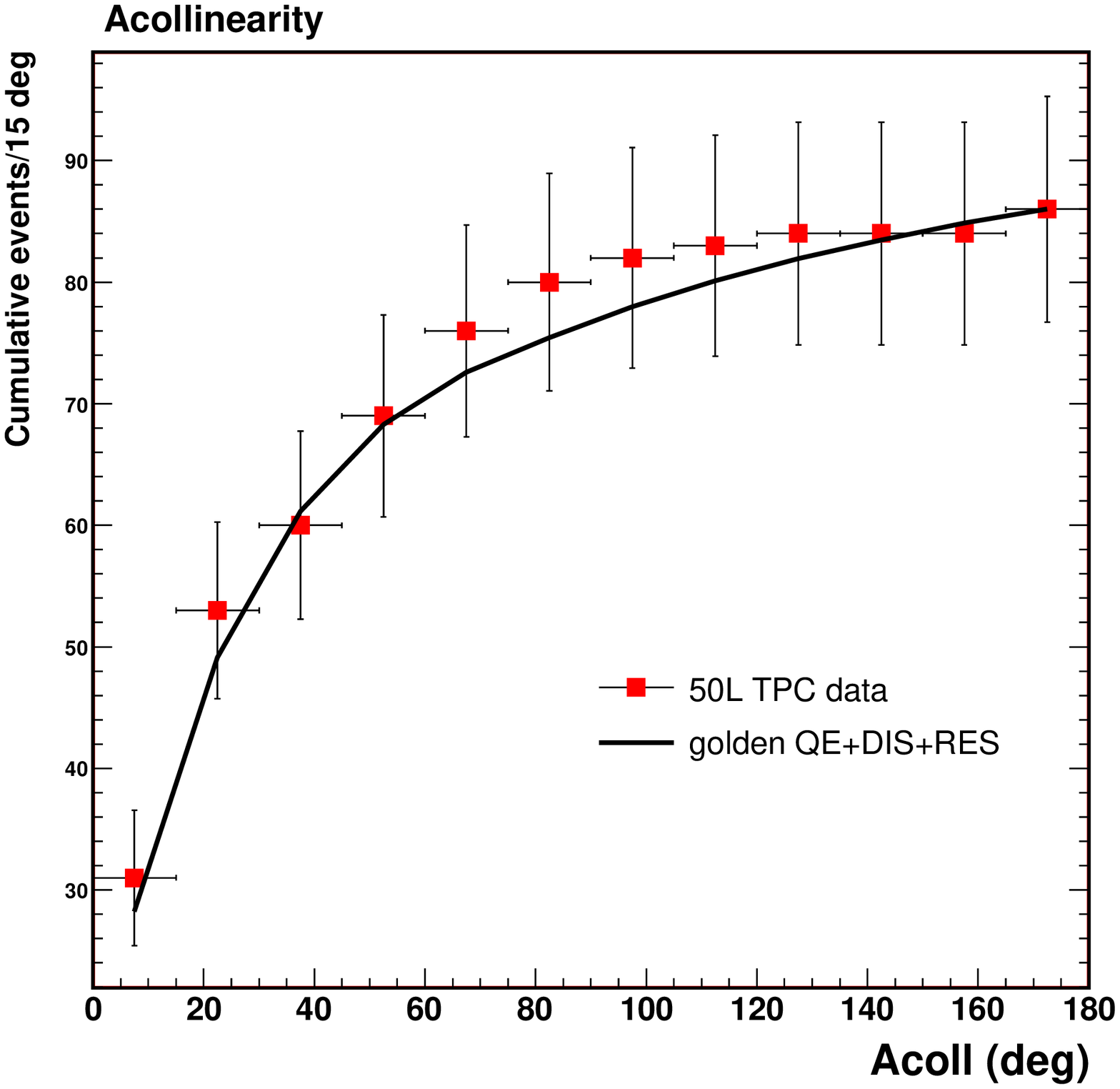}
  \end{tabular}
\caption[Cumulative distribution of the missing transverse momentum and the acollinearity for the \goldens]
{Cumulative distribution of the missing transverse momentum and the acollinearity 
for the \emph{golden sample}. Continuous line represents the expectation from simulated 
Monte-Carlo events (QE\-+\-DIS\-+\-RES).}
\label{fig:event_var_Cum}
\end{center}
\end{figure}

In summary, we have seen that nuclear effects are an important 
source of perturbation for the kinematics of quasi-elastic neutrino interactions, 
however the smallness of the accumulated statistics does not allow a systematic 
survey of these effects~\cite{nuint,nuint2}. Still, these results empirically demonstrate the 
effectiveness of LAr imaging in the characterization of low-multiplicity multi-GeV 
neutrino interactions~\cite{50Lqe}. In particular, as we discuss in Chap.~\ref{chap:tausearch},
a good description of transverse variables is fundamental since they are key to a 
kinematic-based approach for $\nu_\mu~\rightarrow~\nu_\tau(\nu_e)$ oscillation 
appearance~\cite{oscsearch}.

\section{Measurement of the quasi-elastic $\nu_\mu$ CC cross section}
\label{sec:xsection}
%%% Data Set %%%
The collected sample of \emph{golden} events gives us the opportunity of measuring
for first time the quasi-elastic $\nu_\mu$ CC cross section using a LAr TPC.
To this purpose a precise knowledge of the beam composition, the beam profile,
the exposition time, trigger efficiencies, acceptance cuts, event selection
rates and background ratios is needed. In previous sections 
most of the required quantities have been already obtained, specially in
Sec.~\ref{sec:Ev_rates} where an estimation of the expected number of QE events
was discussed. In fact, the calculation of the cross section is like
inverting the process we followed there, i.e. given the 
total number of measured \emph{golden} events extract the quasi-elastic cross section.
However, in order to give a competitive measurement instead of just a cross-check,
a careful treatment of systematic errors has been done.

\subsection*{Event counting}
As we wrote in Sec.~\ref{sec:Sel_QE}, the total number of \emph{golden events}
collected is $N=86$ events. 
The data taking period corresponds to a total exposure $E_{xp}=1.2\times 10^{19}$ 
protons on target, which must be corrected by the effective lifetime of the detectors 
$T_{live} = 75\%$ (see Sec.~\ref{sec:Ev_rates}).
The events contained in the \goldens  are pure QE interactions 
($\nu_\mu \ n \ \rightarrow \mu^- \ p$) plus an 
intrinsic and instrumental background (see Sec.~\ref{sec:background}).
To compute the QE cross section, this background must be clearly understood 
and removed. 

%%% Background estimation %%%
\subsection*{Background subtraction}
The \goldens is contaminated with a sizable fraction of 
deep-inelastic (DIS) and resonance (RES) events. To estimate this 
contamination, we use 
\begin{equation}
\mathcal{B}=\frac{N^{nQE}}{N^{QE}}=\frac{R^{nQE}}{R^{QE}}\times\frac{Acc^{nQE}}{Acc^{QE}}\times\frac{\mathcal{G}^{nQE}}{\mathcal{G}^{QE}}
\label{equ:backrate}
\end{equation}
$N^{QE}$ ($N^{nQE}$) is the number of true QE (non-QE) events
inside the \goldens ($N=N^{QE} + N^{nQE}$). $R$ stands for 
the fraction of a given event class 
(i.e., in our Monte-Carlo, QE events amount to $R^{QE}=2.3\%$ of the 
total. For non-QE events $R^{nQE}=97.7\%$). 
$Acc^{QE}$ ($Acc^{nQE}$) is the acceptance for QE (non-QE) events and 
according to our simulations their average 
values are $Acc^{QE}=0.96$ and $Acc^{nQE}=0.72$. 
Finally, $\mathcal{G}$ measures the fraction of events of a given class
that are classified as \emph{golden}. For QE events, we obtain
$\mathcal{G}^{QE}=0.166$. The fraction of non-QE events 
classified as \emph{golden} is $\mathcal{G}^{nQE}=0.00145$. 
Despite the smallness of this number, the final contamination in our
\goldens is not negligible since, at these high energies, the majority of 
collected events are of non-QE nature. In particular, given the
reduced size of our detector, most of the background comes from
undetected $\pi^0$ since, as computed in Sec.~\ref{sec:background} the probability to
miss the two photons is almost one third. We have estimated that 
the final contamination inside the \goldens due to 
non-QE events is $\mathcal{B}=0.28$. 
%---------------------------------------------------------------

%%% The measurement %%%
\subsection*{The measurement}
After background subtraction, the \goldens contains 
$N^{QE}=\frac{N}{1+\mathcal{B}}= 67\pm 8\ \mathrm{(stat)}$ 
genuine QE events. 
The beam simulation predicts a neutrino flux of 
$\Phi = 2.37\times 10^{-7} \ \nu_\mu$/cm$^2$/p.o.t. Knowing 
the values for the effective detector lifetime and the total 
exposure, we compute the total cross sections using
\begin{equation}
\sigma_{QE} \ = \ \frac{N^{QE}}{\Phi \times E_{xp} \times N_{fid} \times T_{live} \times Acc^{QE} \times \mathcal{G}^{QE}}
\label{equ:xsecexp}
\end{equation}
where $N_{fid}=2.167\times 10^{28}$ 
is the number of target neutrons contained in the fiducial
volume. The figures used for the rest of the variables have been
discussed above. Hence, for an average neutrino energy of 29.5~GeV, 
the total QE $\nu_\mu$ cross section amounts to
\begin{equation*}
\sigma_{QE} = (0.90 \pm~0.10 \ \mathrm{(stat)}) 
\times 10^{-38} cm^2 \,.
 \end{equation*}
Although a wide-band beam has been used, the
reduced statistics accumulated during the test, do not allow to obtain
a set of total cross sections measurements 
for several incoming neutrino energies.
%---------------------------------------------------------------

%%% Evaluation of systematics %%%
\subsection*{Evaluation of systematics}
The systematic error associated to our measurement was evaluated exploiting the
full Monte-Carlo simulation of the experimental set-up (Sec.~\ref{sec:montecarlo}).
To better identify the different contributions to the systematic 
uncertainty, it is much more convenient to express Eq.~\eqref{equ:xsecexp}, 
after a trivial change of variables, as:
\begin{equation}
\begin{split}
\sigma_{QE} = & \ \frac{N}{\Phi \times E_{xp} \times N_{fid} \times T_{live}} \times
              \frac{1}{Acc^{QE} \times \mathcal{G}^{QE}+ (R^{nQE}/R^{QE}) \times Acc^{nQE} \times \mathcal{G}^{nQE}}
\end{split}
\label{equ:cross}
\end{equation}

We identify three types of systematic errors: 
The first one involves the knowledge of the experimental setup, the neutrino flux 
and the detector performance (see Sec.~\ref{sec:Setup}). 
The second is related to the \emph{golden} event selection criteria 
(Sec.~\ref{sec:Sel_QE}) and the geometrical acceptance (Sec.~\ref{sec:acc}).
Finally, we have the systematics associated with the determination of the 
fraction of \emph{golden} events and the expected ratio between QE and non-QE events,
which strongly depends on the modeling of neutrino-nucleus interactions 
(see Sec.~\ref{sec:montecarlo}) and the experimental uncertainties in the cross 
sections at these energies.

In the following, we summarize and describe in detail the different sources of 
systematics entering in the cross section calculation through Eq.~\eqref{equ:cross}:
\begin{itemize}
\item The uncertainty in the WANF neutrino flux has been taken
from~\cite{flux_nomad} and it amounts to 8$\%$. 
The systematic error on the detector lifetime is $4\%$. 
It is dominated by the systematics associated to the 
NOMAD detector lifetime~\cite{NOMAD}. 

\item The uncertainty associated to the fiducial volume 
has been evaluated varying its dimensions by $\pm 1$ wire in
the wire coordinate and $\pm 10$ drift samples on the time
coordinate. The measured resolution on the time coordinate is much better,
however we have adopted a conservative approach when evaluating
this kind of systematic error. The discussed 
modification of the fiducial volume causes a 
variation on the number of \emph{golden} events that amounts to
$3\%$. We take this value as the systematic error associated to the fiducial cut. 

\item The criteria used for event selection are another source of 
systematics. The energy resolution for stopping hadrons is
estimated to be about 4$\%$ and for converted photons we take
$11\%/\sqrt{E}$. This translates into a final systematic error of 
$6\%$ due to the definition of a \emph{golden} candidate. 

\item The main source of systematics for the geometrical acceptance comes 
from the criteria used to match the reconstructed muons in NOMAD with
the ones that exit the 50 liter LAr TPC (see Sec.~\ref{sec:muon_rec}). 
Due to the small lever-arm of the TPC, the resolution on 
the measurement of the muon direction is, on average, $10$~mrad. 
To evaluate how this uncertainty affects the muon acceptance, 
we have modified the matching criteria described above 
by $\pm 10$~mrad. The error on the geometrical acceptance 
turns out to be $2\%$ for QE and non-QE samples. These errors
translate into a systematic contribution of $1.5\%$ ($0.5\%$) for QE
(non-QE) events.

\item The fraction of {\it golden} events inside the QE and non-QE 
samples is affected by nuclear effects, which modify the topology and kinematics 
of the final state. 
Following the discussion in~\cite{psala}, where
the acceptance for identification of QE CC events in a fine grained
detector is studied as a function of the proton detection threshold
and nuclear effects, we conservatively take that $\mathcal{G}^{QE}$ varies by 
$\pm 25 \%$ due to uncertainties in the modeling of 
nuclear effects inside the Monte-Carlo. 
With this assumption, the systematic error 
on the cross section due to the fraction of {\it golden} events 
inside the QE sample amounts to 16$\%$. This is the dominant 
error source. A similar error 
on $\mathcal{G}^{nQE}$ ($\pm 25 \%$) contributes to 
$5\%$ on the total systematic. 

\item Finally, the ratio of QE versus non-QE events is
assumed to be known at the level of $20\%$ from experimental uncertanties 
in the cross section~\cite{psala2}. Its contribution to the cross
section systematic error is $4\%$.

\end{itemize}
Table~\ref{table:syserror} summarizes the contributions from the different 
systematic uncertainties. 

All in all, the estimated total systematic error is $20.6\%$. 
This translates onto a measurement of the total QE 
$\nu_\mu$ charged-current cross section:
\begin{equation}
\sigma_{QE} = (0.90 \pm 0.10 \ \mathrm{(stat)} \pm 0.18
\ \mathrm{(sys)}) \times 10^{-38} cm^2 \,.
\end{equation}
This value agrees with previous
measurements of this reaction reported in the literature 
(see Fig.~\ref{fig:xsection}
and references \cite{anl,ggm,bnl,fnal,sepur,skat,wa25}), and it stands as the first
Physics measurement ever done with a LAr TPC using a neutrino beam 
from an accelerator.

\begin{figure}[!htb]
\begin{center}
\includegraphics[width=9cm]{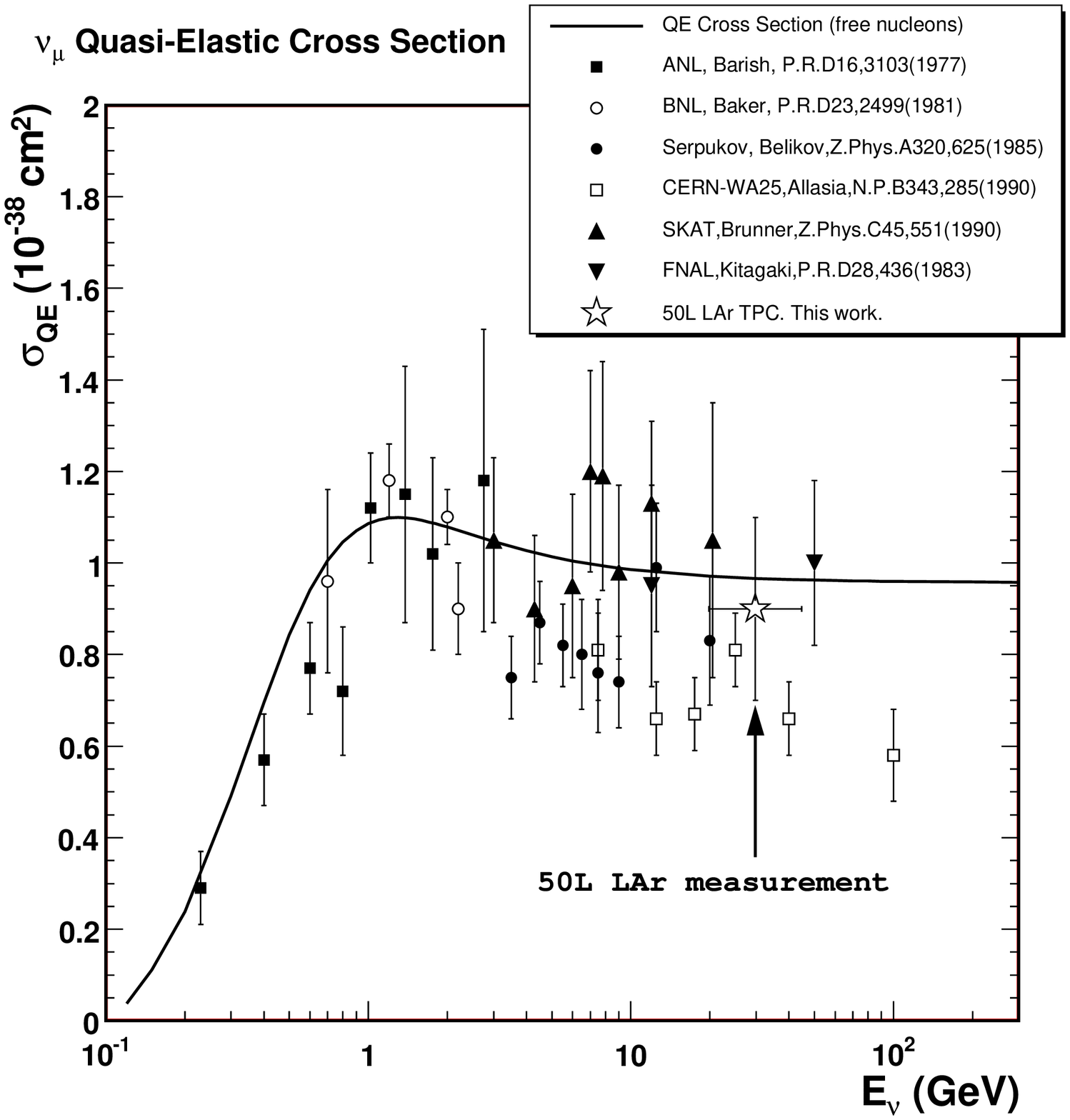}
\caption[Muon neutrino-nucleus quasi-elastic cross section data]
{Muon neutrino-nucleus quasi-elastic cross section data. Continuous line
is the theoretical behavior of the QE cross section in free nucleon 
approximation. The measurement 
reported in this work appears in the figure as a star. The shown
vertical error bar is obtained adding in quadrature the 
systematic and statistical uncertainties. The horizontal error bar
corresponds to the RMS of the simulated neutrino energy spectrum 
(it amounts to 18~GeV).}
\label{fig:xsection}
\end{center}
\end{figure}

\begin{table*}[!htb]
 \begin{center}
  \begin{tabular}{l l c}
   \hline
   \hline
Source                                                           &       & Systematic error (\%) \\
   \hline
Neutrino flux                                                     &                     & 8  \\
Detector lifetime                                                 &                     & 4  \\
Vertex detection (fiducial cut)                                   &                     & 3  \\
{\it Golden} event selection cuts                                   &                     & 6  \\
Acceptance for QE events, $Acc^{QE} (\pm2\%)$                      &                      & 1.5  \\
Acceptance for non-QE events, $Acc^{nQE} (\pm2\%)$                 &                      & 0.5  \\
Fraction of {\it golden} events in QE sample, $\mathcal{G}^{QE}$  ($\pm25\%$)  &            & 16 \\
Fraction of {\it golden} events in non-QE sample, $\mathcal{G}^{nQE}$  ($\pm25\%$)  &       & 5 \\
Fraction of non-QE to QE events, $\frac{R^{nQE}}{R^{QE}}$ ($\pm20\%$)                 &       & 4     \\\hline

Total                                                    &                                & 20.6 \\
   \hline
   \hline
  \end{tabular}
  \caption{Summary of systematic errors. The largest source of error
comes from nuclear effects affecting the expected fraction of 
{\it golden} events inside the genuine QE sample.}
  \label{table:syserror}
 \end{center}
\end{table*}
%---------------------------------------------------------------

%%%%%%%%%%%%%%%%%%%%%%%%%%%%%
%%%%%%%%%%%%%%%%%%%%%%%%%%%%%
                           %%
\chapter{Searching for $\nu_{\mu}\rightarrow\nu_{\tau}$ oscillations at CNGS} %%
                           %%
%%%%%%%%%%%%%%%%%%%%%%%%%%%%%
%%%%%%%%%%%%%%%%%%%%%%%%%%%%%
\label{chap:tausearch}
The Super-Kamiokande Collaboration has measured a first
evidence of the sinusoidal behaviour of neutrino disappearance as
dictated by neutrino oscillations \cite{SUPERK2}. However, although
the most favoured hypothesis for the observed $\nu_{\mu}$ disappearance
is that of $\nu_{\mu}\rightarrow\nu_{\tau}$ oscillations, no direct
evidence for $\nu_{\tau}$ appearance exists up to date. A long baseline
neutrino beam, optimized for the parameters favoured by atmospheric
oscillations, has been approved and is now running in Europe to look 
for explicit $\nu_{\tau}$ appearance: the CERN-Laboratori Nazionali del 
Gran Sasso (CNGS) beam \cite{BEAM}. 
The approved experimental program consists of two experiments
ICARUS \cite{ICARUS} and OPERA \cite{OPERA} that will search for
$\nu_{\mu}\rightarrow\nu_{\tau}$ oscillations using complementary
techniques.

Given the previous experimental efforts \cite{NOMAD,CHORUS} and present
interest in direct $\nu_{\tau}$ appearance, we assess in this Chapter
the performance of several statistical techniques applied to the search
for $\nu_{\tau}$ using kinematic techniques. Classic statistical
methods (like multi-dimensional likelihood and Fischer's discriminant
schemes) and \emph{Neural Networks} based ones (like multi-layer perceptron
and self-organized neural networks) have been applied in order to
find the approach that offers the best sensitivity.

The underlying idea consists on finding the kinematical features 
which characterize the signal respect to the dominant
background. To this purpose is obvious that optimal 
reconstruction capabilities are of utmost importance. 
In Chapter~\ref{chap:50Lexperiment} we have assessed the excellent performance 
of our reconstruction tools when they are applied to the analysis of the
kinematics of $\nu_\mu$~CC events. Furthermore, we have also seen that our 
Monte-Carlo accurately describes neutrino-nucleus interactions and nuclear effects. 
Thanks to these successfully tested capabilities, in this Chapter, we can use them 
to evaluate the performance of a LAr TPC when searching for direct appearance 
of $\nu_\tau$ at CNGS. 

\section{On the problem of $\nu_{\tau}$ appearance at CNGS}
In the CERN-NGS beam \cite{BEAM,BEAM2}, the expected $\nu_{e}$ and $\nu_{\tau}$
contamination are of the order of $10^{-2}$ and $10^{-7}$ respectively
compared with the dominant $\nu_{\mu}$ composition. These properties
allow to search for oscillations by looking at the appearance of $\nu_{e}$
and $\nu_{\tau}$ charged current events. In these cases, the detector
must be able to tag efficiently the interaction of $\nu_{e}$'s and
$\nu_{\tau}$'s out of the bulk of $\nu_{\mu}$ events.

In the case of $\nu_{\mu}\rightarrow\nu_{\tau}$ oscillations, the
golden channel to look for the $\nu_{\tau}$ appearance is the decay
of the tau into an electron and a pair neutrino anti-neutrino due
to: (a) the excellent electron identification capabilities of a LAr TPC; 
(b) the low background level, since the intrinsic $\nu_{e}$ and $\bar{\nu}_{e}$
charged current contamination of the beam is at the level of one per cent.

Kinematical identification of the $\tau$ decay, which follows the
$\nu_{\tau}$ CC interaction, requires excellent detector performance:
good calorimetric features together with tracking and topology reconstruction
capabilities. In order to separate $\nu_{\tau}$ events from the background,
two basic criteria can be used:

\begin{itemize}
\item An unbalanced total transverse momentum due to neutrinos produced
in the $\tau$ decay.
\item A kinematical isolation of hadronic prongs and missing momentum in
the transverse plane.
\end{itemize}

\subsection{Detector configuration and data simulation}
As a reference for this study, we consider a detector configuration
consisting of 5 identical T600 modules (T3000 detector)
\footnote{See Sec.~\ref{ICARUSproject} for a brief description of the
ICARUS detector or consult Ref.~\cite{ICARUS} for a much more detailed description.}. 
As the active Argon for one of these modules is 470 tons, the total mass amounts
to 2.35 ktons. We also assume five years running of the CNGS beam
in shared mode ($4.5\times10^{19}$ p.o.t. per year), which translates
in a total exposure of $5\times0.47\times5=11.75$ Kton$\times$year.

Events have been fully simulated in the whole of each T600 module
($\approx3.2\, h\times6\, w\times18\, l$ m\ensuremath{^3}), but only
$65\%$ have the primary interaction vertex inside the active part
of the LAr volume.

The study of the capabilities to reconstruct high-energy neutrino
events was done using fully simulated $\nu_{e}$CC events inside the
considered LAr detector \cite{T600-Geometry}. The neutrino cross
sections and the generation of neutrino interactions is based on the
NUX code \cite{nux}; final state particles are then tracked using
the FLUKA package \cite{FLUKA}. 
The angular and energy resolutions 
used in the simulation of final state electrons and individual hadrons are 
identical to those quoted in \cite{ICARUS}: $\sigma(E)/E$ equal to 
$20\%/\sqrt{E}\oplus5\%$ for hadrons and $3\%/\sqrt{E}\oplus1\%$ for
electromagnetic showers ($E$ in units of GeV). Concerning the reconstruction
of the particle direction, the assumed angular resolutions are $0.13/\sqrt{E}$
for electrons and $0.04$ for single hadrons. In the analyses, the particle
four vectors are smeared according to the previous parameters.

%In order to apply the most efficient kinematic selection, it is mandatory
%to reconstruct with the best possible resolution the energy and the
%angle of the hadronic jet and the prompt lepton, with particular attention
%to the tails of the distributions. Therefore, the energy flow algorithm
%have been designed with care, taking into account the
%needs of the tau search analyses.

%A detailed discussion of the algorithms and the expected resolution
%in reconstructing the event kinematics can be found in Sec.~2 of
%\cite{Taue}.

\paragraph{The fiducial volume\\ \\}
The ability to look for tau appearance events is limited by the containment
of high energy neutrino events. Energy leakage outside the active
image volume creates tails in the kinematical variables that fake
the presence of neutrinos in the final state. We therefore impose
fiducial cuts in order to guarantee that on average the events will
be sufficiently contained.

The fiducial volume is defined by looking at the profiles of the total
missing transverse momentum and of the total visible energy of the
events. The average value of these variables is a good estimator of
how much energy is leaking on average.

The profiles of missing $P_{T}$ and visible energy as a function
of the vertex positions for the T600 module are shown in 
Fig.~\ref{fig:ptmissprofile}. We have to leave 20 cm at the top and bottom of
the active volume (x coordinate), and 25 cm from the sides (y coordinate).
Longitudinally, we cut out 1.7 m from the total length (z coordinate).
This results in a total fiducial volume of $2.7\, h\times5\, w\times16.3\, l$
m\ensuremath{^3}. Thus with this conservative approach, we keep 65\%
of the total number of events occurring in the active LAr volume.
The total exposure in a 5 T600 setup with 5 years shared running CNGS,
taking in account the fiducial cuts, is 7.6 kton$\times$year. 
In what follows, tau selection and background rejection efficiencies are
normalized to this figure.

\begin{figure}[!ht]
\begin{center}
\includegraphics[width=10cm,keepaspectratio]{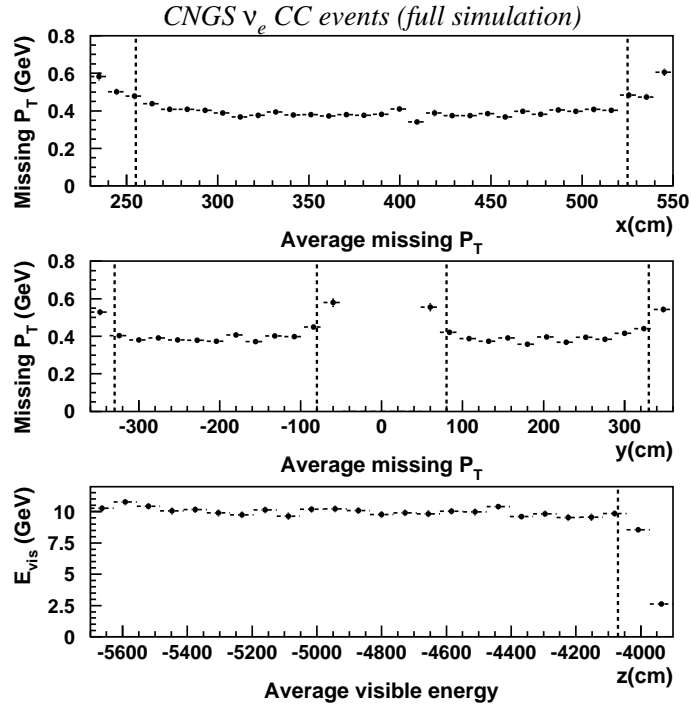}
\end{center}
\caption[Performance of the energy flow reconstruction in a T600 module]
{Performance of the energy flow reconstruction in a T600 module:
Transverse momentum versus x (top plot) and y (middle plot) position.
Average visible energy versus z position (bottom plot) (a cut $E_{\nu}<20\;$GeV
has been imposed).\label{fig:ptmissprofile}}
\end{figure}

\subsection{Expected rates and generated data}
Tab.~\ref{tab:nurates} summarizes the total event rates expected
for 5 T600 modules and 5 years CNGS running. It also shows, for different
values of $\Delta m^{2}$, the expected $\nu_{\tau}$ CC rates in
case oscillations take place. 

\begin{table}[!ht]
\begin{center}
\begin{tabular}{|c|c|c|}
\hline 
&
Expected events&
Expected events\tabularnewline
Process&
\textbf{Active LAr. }&
\textbf{Fiducial LAr. }\tabularnewline
&
(11.75 Kton$\times$year)&
(7.6 Kton$\times$year)\tabularnewline
\hline
$\nu_{\mu}$ CC &
 32600 &
21190\tabularnewline
 $\bar{\nu}_{\mu}$ CC &
 652 &
424\tabularnewline
 $\nu_{e}$ CC &
 252 &
163\tabularnewline
 $\bar{\nu}_{e}$ CC &
 17 &
11\tabularnewline
 $\nu$ NC &
 10600 &
6890\tabularnewline
 $\bar{\nu}$ NC &
 245 &
159\tabularnewline
\hline
\hline 
$\nu_{\tau}$ CC, $\Delta m^{2}$ (eV$^{2}$)&
\multicolumn{1}{c}{}&
\multicolumn{1}{c|}{}\tabularnewline
\hline
$1\times10^{-3}$&
 31 (6) &
20 (4)\tabularnewline
 $2\times10^{-3}$&
 125 (23) &
81 (15)\tabularnewline
 $3\times10^{-3}$&
 280 (50)&
182 (33)\tabularnewline
 $5\times10^{-3}$&
 750 (135)&
488 (88)\tabularnewline
\hline
\end{tabular}
\end{center}
\caption[Expected event rates for the 5 years of CNGS running and 5 T600 modules]
{Expected event rates in active and fiducial LAr for the 5 years of
CNGS running and 5 T600 modules. For standard processes, no oscillations
are assumed. For $\nu_{\tau}$CC, we take two neutrino $\nu_{\mu}\rightarrow\nu_{\tau}$
oscillations with $\sin^2 2\theta=1$. In brackets, the number of $\tau\rightarrow e$
events ($\tau$ decay rate is 18\%).}

\label{tab:nurates}
\end{table}

%\subsubsection{The generated data\label{sub:generated-data}}

The $\nu_{\tau}$ rate has a strong dependence on the exact value
of the $\Delta m^2$ in the parameter region favoured by atmospheric
data as we see in table \ref{tab:nurates}. For our central value
taken from atmospheric neutrino results ($\Delta m^{2}=3\times10^{-3}\; eV^{2}$),
the number of $\nu_{\tau}$CC with $\tau\rightarrow e$ is about 50;
the signal over background ratio is $50/252\simeq0.2$.

Tab.~\ref{tab:mcevents} summarizes the total amount of simulated
data. 
We note that $\nu_{\tau} \ (\nu_{e})$ CC sample, generated
in active LAr, is more than a factor 250 (50) larger than the expected
number of collected events after five years of CNGS running. 

\begin{table}[!ht]
\begin{center}\begin{tabular}{|c|c|c|}
\hline 
\textbf{Process}& \textbf{$\nu_{e}$CC} & \textbf{$\nu_{\tau}$CC }\tabularnewline
 & & \textbf{($\tau\rightarrow e$)}\tabularnewline \hline \hline 
\textbf{Total LAr}& 22500& 22100\tabularnewline
\hline 
\textbf{Active LAr}& 14200& 13900 \tabularnewline
\hline 
\textbf{Fiducial Vol.}& \textbf{\emph{9250}}& \textbf{\emph{9000}}\tabularnewline
\hline
\end{tabular}\par\end{center}
\caption[Amount of fully generated data]
{Amount of fully generated data. }

\label{tab:mcevents}
\end{table}

To save CPU power, only $\nu_{e}$CC events with an incoming neutrino
energy below 30 GeV have been generated, since oscillated events,
with a $\Delta m^{2}=3\times10^{-3}eV^{2}$, tend to accumulate in
the low energy region (only 3\% of the total $\nu_{\tau}$CC events
have an energy above 30 GeV). 
%In Fig.~\ref{fig:enu} we see the
%neutrino expected spectra for $\nu_{\tau}$ and $\nu_{e}$.

\section{Statistical pattern recognition applied to oscillation searches}
In the case of a $\nu_{\mu}\rightarrow\nu_{\tau}$ oscillation search with Liquid Argon, 
the golden channel to look for $\nu_{\tau}$ appearance is the decay
of the tau into an electron and a pair neutrino anti-neutrino due
to: (a) the excellent electron identification capabilities; (b) the
low background level, since the intrinsic $\nu_{e}$ and $\bar{\nu}_{e}$
charged current contamination of the beam is at the level of one per
cent.

Kinematic identification of the $\tau$ decay \cite{Astier:1999vc}, which follows the
$\nu_{\tau}$CC interaction, requires excellent detector performance:
good calorimetric features together with tracking and topology reconstruction
capabilities. In order to separate $\nu_{\tau}$ events from the background,
a basic criteria can be used: an unbalanced total transverse momentum due to neutrinos 
produced in the $\tau$ decay.

In Fig.~\ref{fig:mainvars} we illustrate the difference on kinematics
for signal and background events. We plot four of the most discriminating
variables: 
\begin{itemize} 
\item $E_{vis}$: Visible energy. 
\item $P_{T}^{miss}$: Missing momentum in the transverse plane with respect to the direction of the 
incident neutrino beam. 
\item $P_{T}^{lep}$: Transverse momentum of the prompt electron candidate. 
\item $\rho_{l}=\frac{{\textstyle P_{T}^{lep}}}{{\textstyle P_{T}^{lep}+P_{T}^{had}+P_{T}^{miss}}}$%
\end{itemize}
 Signal events tend to accumulate in low $E_{vis}$, low
$P_{T}^{lep}$, low $\rho_{l}$ and high $P_{T}^{miss}$ regions.

\begin{figure}[!ht]
\begin{center}
\includegraphics[width=10cm,keepaspectratio]{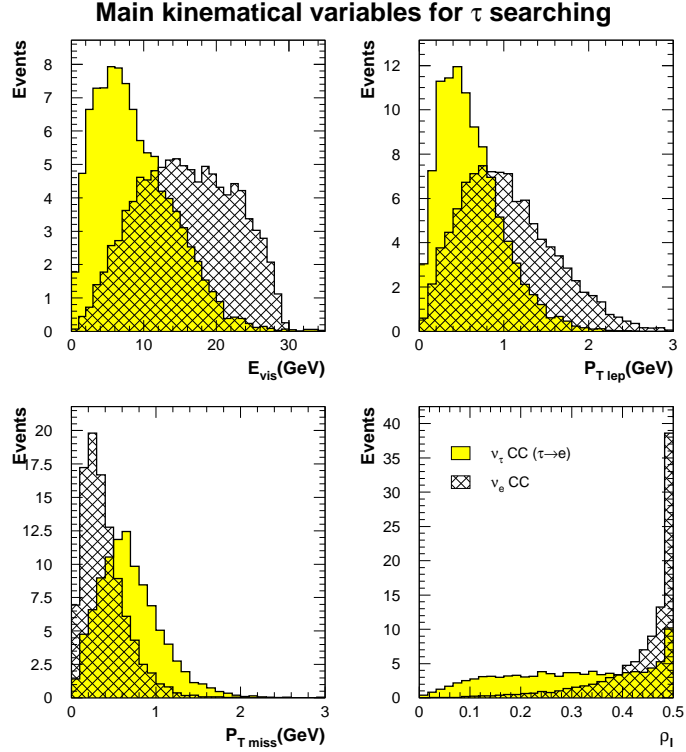}
\end{center}
\caption[Distributions of the main kinematical variables used for $\tau$ searching]
{Visible energy (top left), transverse missed momentum (bottom left),
transverse electron momentum (top right) and $\rho_{l}$ (bottom right).
Histograms have an arbitrary normalization.}

\label{fig:mainvars}
\end{figure}

Throughout this work, we take into account only the background due to electron neutrino 
charged current interactions. Due to the low content the beam has on $\bar{\nu}_e$, 
charged currents interactions of this type have been observed to give a negligible 
contribution to the total expected background. 
We are confident that neutral current background can be reduced to a negligible level
using LAr imaging capabilities and algorithms based on the different
energy deposition showed by electrons and $\pi^{0}$ (see for example~\cite{pi0}). 
Therefore it will not be further considered. The contamination due to charm production 
and $\nu_\mu$ CC events, where the prompt muon is not identified as such, was studied 
by the ICARUS Collaboration \cite{ICANOE} and showed to be less important 
than $\nu_{e}$CC background. 

\subsection{Direct Background rejection using Sequential Cuts}
The most straightforward way to look for $\nu_{\tau}$ appearance
is by means of a series of sequential cuts. This procedure does not
take into account possible correlations among the selected kinematical
variables. We take advantage of the soft energy spectrum for oscillated
$\nu_{\tau}$ induced events with respect to the visible energy spectrum
of $\nu_{e}$CC. Oscillated events tend to accumulate at relatively
low values of $P_{T}^{lep}$, as well, therefore an upper cut on $P_{T}^{lep}$
efficiently removes the background. Since $\nu_{e}$CC events are
well balanced and very few populate the high $P_{T}^{miss}$ region
(missing momentum in the plane perpendicular to the incoming neutrino
direction), a cut in this variable is also applied.

Tab.~\ref{tab:scuts} summarizes the list of sequential cuts applied
to reduce the $\nu_{e}$CC background. The expected number of signal
events, for $\Delta m^{2}=3\times10^{-3}$ eV\ensuremath{^2}, after
all cuts is $11.7\pm0.4$, for a total background of $2.2\pm0.3$
expected events%
\footnote{Errors in the number of expected events are of statistical nature.
The usual Poisson statistical error in probability density estimation
is rescaled according to the distribution normalization (5 years of
CNGS running). %
}. Fig.~\ref{fig:scuts} shows some kinematical variables for signal
and background events surviving the cuts. The signal to background
ratio (S/B) for this approach is around five. More sophisticated techniques,
taking into account correlations among discriminating variables, should
be used in order to improve the S/B value. 

\begin{table}[!ht]
\begin{center}\begin{tabular}{|l||c|c|c|}
\hline 
&
\textbf{$\nu_{\tau}$}CC($\tau\rightarrow e$) &
 \textbf{}&
 \textbf{$\nu_{\tau}$}CC ($\tau\rightarrow e$)\tabularnewline
Cuts &
Efficiency&
 \textbf{$\nu_{e}$}CC &
 \textbf{$\Delta m^{2}=$}\tabularnewline
&
 \textbf{}($\%$) &
&
 \textbf{$3\times10^{-3}$} eV$^{2}$\tabularnewline
\hline
Initial &
 100 &
 252 &
 50 \tabularnewline
\hline
Fiducial volume &
64&
 161&
 32\tabularnewline
\hline
$E_{visible}<18$ GeV &
 62 &
 $37.2\pm1.1$&
 $30.8\pm0.7$ \tabularnewline
\hline
$P_{T}^{lep}<0.9$ GeV &
 51&
 $20.0\pm0.8$&
 $25.6\pm0.6$\tabularnewline
\hline
$Q_{T}^{lep}>0.3$ GeV &
 43&
 $18.5\pm0.8$&
 $21.6\pm0.6$\tabularnewline
\hline
$P_{T}^{miss}>0.6$ GeV &
 \textbf{23}&
 \textbf{$\mathbf{2.2\pm0.3}$}&
 \textbf{$\mathbf{11.7\pm0.4}$}\tabularnewline
\hline
\end{tabular}\end{center}
\caption[Rejection of the $\nu_{e}$CC background in the $\tau\rightarrow e$analysis]
{Rejection of the $\nu_{e}$CC background in the $\tau\rightarrow e$
analysis. 5 years of CNGS running and 5 T600 modules exposure has
been considered. We quote statistical errors only.\label{tab:scuts}}
\end{table}

\begin{figure}[!ht]
\begin{center}
\includegraphics[width=10cm]{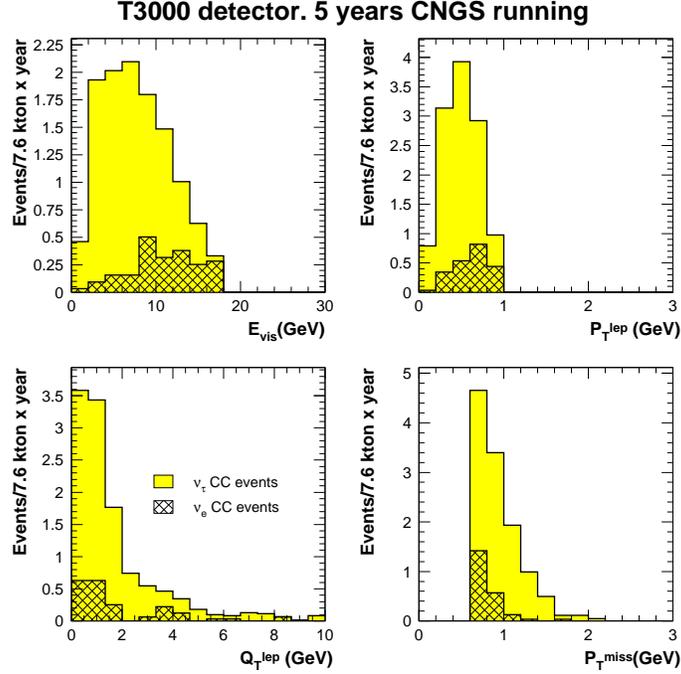}
\end{center}
\caption[Distribution of signal and background events after sequential cut]
{Sequential cuts. Distributions of $\nu_{\tau}$CC events (solid histograms)
and $\nu_{e}$CC events (hatched histograms) after sequential cuts
are imposed. }

\label{fig:scuts}
\end{figure}

\subsection{Kinematical search using classic statistical methods}

\subsubsection{A multi-dimensional likelihood\label{sub:likelihood}}

The second method adopted for the $\tau\rightarrow e$ analysis is
the construction of a multi-dimensional likelihood function (see for
example \cite{CowanStat}), which is used as the unique discriminant between
signal and background. This approach is, a priori, an optimal discrimination 
tool since it takes into account correlations between the chosen variables.

A complete likelihood function should contain five variables (three
providing information of the plane normal to the incident neutrino
direction and two more providing longitudinal information). However,
in a first approximation, we limit ourselves to the discrimination information
provided by the three following variables: $E_{visible}$, $P_{T}^{miss}$ and $\rho_l$.

As we will see later, all the discrimination power is contained in
these variables, therefore we can largely reduce the complexity of
the problem without affecting the sensitivity of the search. Two likelihood
functions were built, one for $\tau$ signal ($\mathcal{L_{S}}$)
and another for background events ($\mathcal{L_{B}}$). The discrimination
was obtained by taking the ratio of the two likelihoods: 

\begin{equation}
ln(\lambda)\equiv\mathcal{L}([E_{visible},P_{T}^{miss},\rho_{l}])=\frac{\mathcal{L_{S}}([E_{visible},P_{T}^{miss},\rho_{l}])}{\mathcal{L_{B}}([E_{visible},P_{T}^{miss},\rho_{l}])}\label{eq:likelihood}\end{equation}

In order to avoid a bias in our estimation, 
half of the generated data was used to build 
the likelihood functions and the other 
half was used to evaluate overall efficiencies. 
Full details about the multi-dimensional likelihood algorithm can
be found elsewhere \cite{pi0}. However, we want to point out here some
important features of the method.

A partition of the hyperspace of input variables is required: The
multi-dimensional likelihood will be, in principle, defined over a
lattice of bins. The number of bins to be filled when constructing
likelihood tables grows like $n^{d}$ where $n$ is the number of
bins per variable and $d$ the number of these variables. This leads
to a ``dimensionality'' problem when we increment the number of
variables, since the amount of data required to have a well defined
value for ln$\lambda$ in each bin of the lattice will grow exponentially. 

In order to avoid regions populated with very few events, input variables
must be redefined to have the signal uniformly distributed in the
whole input hyperspace, hence $E_{visible}$, $P_{T}^{miss}$ and
$\rho_{l}$ are replaced by ``flat'' variables (see Fig.~\ref{fig:flat}).
In general, a variable $g(a)$ defined as
\begin{equation}
g(a) = f(x(a))\Bigl\lvert\frac{dx}{da}\Bigr\rvert
\end{equation}
is uniformly distributed (``flat'') between 0 and 1 by choosing 
$a(x)=\int^{x}_{-\infty} f(x^\prime)dx^\prime$, where $x$ is a variable
distributed according to a p.d.f. $f(x)$. With this approach, more bins
are automatically taken in those regions of faster variation of the signal.
The combination of the ``flat'' variables into a single one did not give
rise to an uniform three dimensional distribution, due to the presence
of correlations between the variables.

Besides, an adequate smoothing algorithm is needed in order to alleviate
fluctuations in the distributions in the hyperspace and also, to provide
a continuous map from the input variables to the multi-dimensional
likelihood one (ln$\lambda$).

\begin{figure}[!ht]
\begin{center}
\includegraphics[width=10cm,keepaspectratio]
{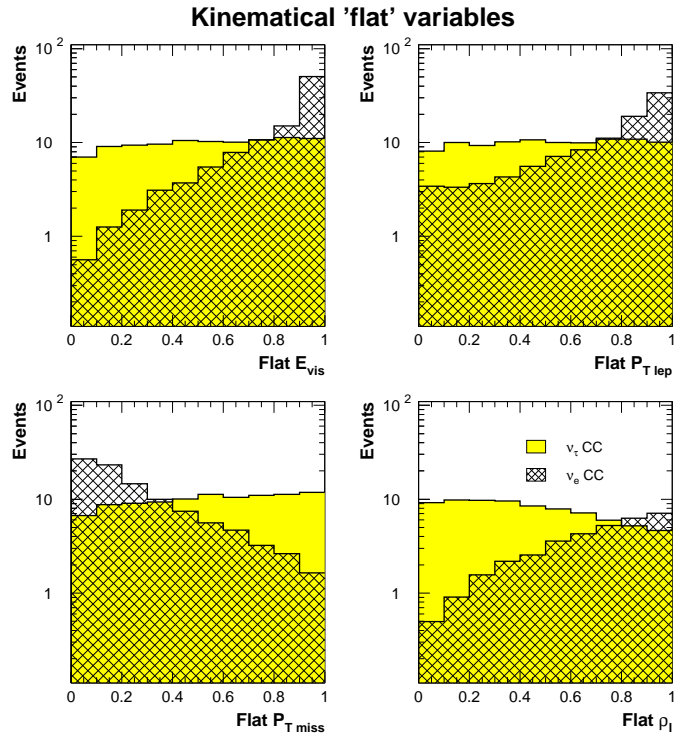}
\end{center}
\caption[Distribution of ``flat'' variables]
{Comparison for {}``flat'' $E_{visible}$, $P_{T}^{lep}$, $\rho_{l}$
and $P_{T}^{miss}$ variables between $\tau$ signal and $\nu_{e}$
CC events. Arbitrary normalization has been taken into account when
plotting background events.\label{fig:flat}}
\end{figure}

Ten bins per variable were used, giving rise to a total of $10^{3}$
bins. Fig.~\ref{fig:likes} shows the likelihood distributions for
background and tau events assuming five years running of CNGS (total
exposure of 7.6 kton $\times$ year for events occurring inside the
fiducial volume). %
\begin{figure}[!ht]
\begin{center}
\includegraphics[width=10cm,height=10cm]
{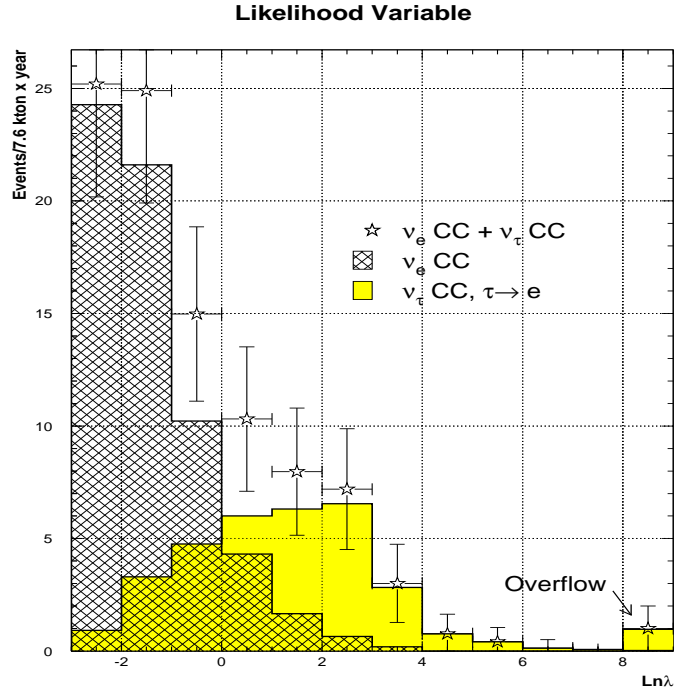}
\end{center}
\caption[The Multi-dimensional likelihood variable]
{Multi-dimensional likelihood distributions for $\nu_{e}$ CC and
$\tau\rightarrow e$ events. The last bin in signal includes the event
overflow. Error bars in $\nu_{e}$CC + $\nu_{\tau}$CC sample represent
statistical fluctuations in the expected profile measurements after
5 years of data taking with shared running CNGS and a 5 T600 detector
configuration.\label{fig:likes}}
\end{figure}

\begin{figure}[!ht]
\begin{center}\includegraphics[width=10cm,height=10cm]{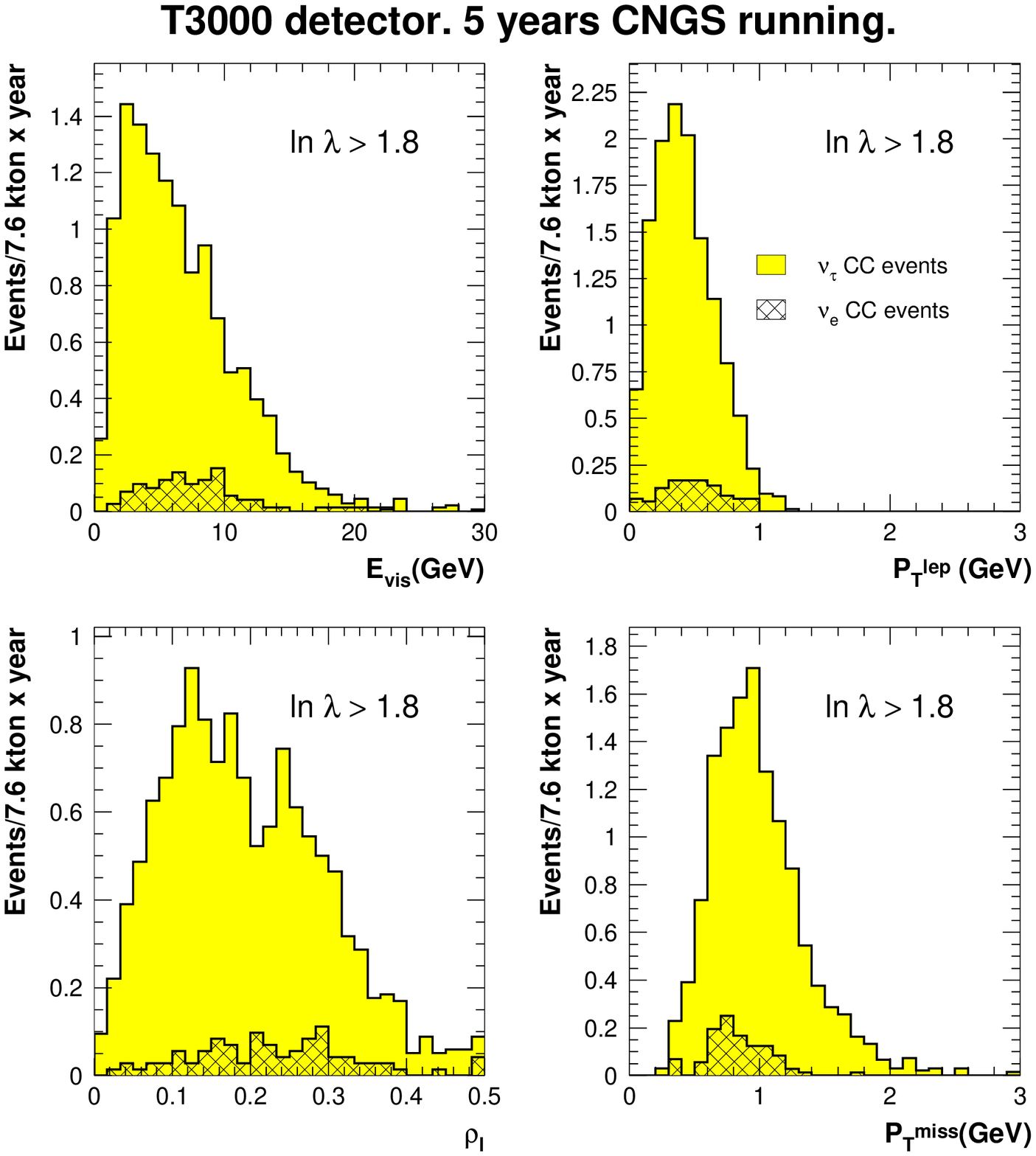}
\end{center}
\caption[Distribution of signal and background events after \emph{likelihood} cut]
{Distribution of signal and background events in $E_{visible}$, $P_{T}^{lep}$,
$\rho_{l}$ and $P_{T}^{miss}$ after the cut in ln $\lambda$. We
see how signal appearance is considerably enhanced respect to background
in regions defined by the ln$\lambda>1.8$ cut.\label{fig:lkl_cutted_var}}
\end{figure}

Tab.~\ref{tab:lkl} shows, for different cuts of $\ln\lambda$, the
expected number of tau and $\nu_{e}$ CC background events. As reference
for future comparisons, we focus our attention in the cut ln$\lambda>1.8$.
It gives a signal selection efficiency around 25\% (normalized to
the total number of $\tau$ events in active LAr). This $\tau$ efficiency
corresponds to 13 signal events. For this cut, we expect $1.1\pm0.2$
background events.  After cuts are imposed, this approach predicts
a S/B ratio similar to 13. The background rejection power is enhanced with respect
to the sequential cuts, since correlations among variables are taken
into account. In Fig.~\ref{fig:lkl_cutted_var}, we show the distribution
of the surviving signal and background events for ln$\lambda>1.8$.
The region selected by the multidimensional likelihood approach, in
the three dimensional space of the chosen variables, does not present
sharp edges like in the case of sequential cuts: The topology of the
selected region is more complex thanks to the use of variable correlations.

\begin{table}[!ht]
\begin{center}\begin{tabular}{|l|c|c|c|}
\hline 
Cuts &
$\nu_{\tau}$CC $(\tau\rightarrow e)$ &
&
 $\nu_{\tau}$ CC $(\tau\rightarrow e)$ \tabularnewline
&
Efficiency &
$\nu_{e}$CC&
 $\Delta m^{2}=$\tabularnewline
&
 ($\%$) &
&
 $3\times10^{-3}$ eV$^{2}$\tabularnewline
\hline
Initial &
 100 &
 252 &
 50 \tabularnewline
\hline
Fiducial volume &
 65 &
 164 &
 33 \tabularnewline
\hline
$\ln\lambda>0.0$&
 48 &
 $6.8\pm0.5$ &
 $24.0\pm0.6$\tabularnewline
 $\ln\lambda>0.5$&
 42 &
$3.6\pm0.3$&
 $20.8\pm0.6$\tabularnewline
 $\ln\lambda>1.0$&
 36 &
$2.5\pm0.3$&
$18.0\pm0.6$\tabularnewline
 $\ln\lambda>1.5$&
 30 &
 $1.7\pm0.2$ &
$15.2\pm0.5$\tabularnewline
\textbf{ln $\mathbf{\lambda>1.8}$}&
\textbf{25}&
\textbf{$\mathbf{1.1\pm0.2}$}&
\textbf{$\mathbf{12.9\pm0.5}$}\tabularnewline
 ln $\lambda>2.0$&
 23&
$0.86\pm0.16$&
$11.7\pm0.5$\tabularnewline
 $\ln\lambda>2.5$&
 16 &
 $0.40\pm0.12$ &
 $8.1\pm0.4$ \tabularnewline
 $\ln\lambda>3.0$&
 10 &
$0.22\pm0.08$&
 $5.2\pm0.3$ \tabularnewline
 $\ln\lambda>3.5$&
 7 &
$0.12\pm0.06$&
 $3.3\pm0.2$ \tabularnewline
\hline
\end{tabular}\end{center}
\caption[Multi-dimensional likelihood: Expected number of signal and background events]
{Expected number $\nu_{e}$CC background and signal events in the
$\tau\rightarrow e$ analysis. A multi-dimensional likelihood function
is used as the unique discriminant. Numbers are normalized to 5 years
running of CNGS. Errors in the number of expected events are given
by Poisson statistics.\label{tab:lkl}}
\end{table}

\subsubsection{The Fisher discriminant method}

The Fisher discriminant method \cite{CowanStat} is a standard statistical
procedure that, starting from a large number of input variables, allows
us to obtain a single variable that will efficiently distinguish among
different hypotheses. Like in the likelihood method, the Fisher discriminant
will contain all the discrimination information. 

The Fisher approach  tries to find a linear combination of the following kind
\[
t(\{ x_{j}\})=a_{0}+\sum_{i=1}^{n}a_{i}x_{i}\]
of an initial set of variables $\{ x_{j}\}$ which maximizes

\begin{equation}
J(\{ a_{j}\})=\frac{(\bar{t}_{sig}-\bar{t}_{bkg})^{2}}{(\sigma_{sig}^{2}-\sigma_{bkg}^{2})}\label{eq:fisher}\end{equation}
where $\bar{t}$ is the mean of the $t$ variable and $\sigma$ its
variance. This last expression is nothing but a measurement, for the variable
$t$, of how well separated signal and background are. Thus, by maximizing
Eq.~\eqref{eq:fisher} we find the optimal linear combination of initial
variables that best discriminates signal from background. The parameters
$a_{j}$ which maximize Eq.~\eqref{eq:fisher} can be obtained analytically
by (see \cite{CowanStat})

\begin{equation}
a_{i}=W_{ij}^{-1}(\mu_{j}^{sig}-\mu_{j}^{bkg})
\label{eq:fisher_coef}
\end{equation}
where $\mu_{j}^{sig}$ and $\mu_{j}^{bkg}$ are the mean
in the variable $x_{j}$ for signal and background respectively, and
$W=V_{sig}+V_{bkg}$. $V_{sig, \ back}$, the covariance matrices for the two event 
classes considered, are defined as: 
\begin{equation}
V^k_{ij}=\Sigma (x-\mu^k)_i (x-\mu^k)_j/N
\end{equation}
where k stands for signal or background classes, $(x_1,...,x_n)$ is a vector of data 
with means $(\mu_1,...,\mu_n)$ and N is the total number of events. 

\paragraph{A Fisher function for $\tau$ search\\ \\}
From the distributions of kinematical variables for $\nu_{\tau}$CC
and $\nu_{e}$CC, we can immediately construct the Fisher function
for a given set of variables. Initially we selected the same set of
variables we used for the likelihood approach, namely: $E_{visible}$,
$P_{T}^{miss}$ and $\rho_{l}$. We need only the vector of means
and covariance matrices in order to calculate the optimum Fisher variable
(Eq. \eqref{eq:fisher_coef}). Distributions are shown in
Fig.~\ref{fig:Fischer-discriminant-variable.}, where the usual normalization
has been assumed. In Tab.~\ref{tab:Fisher-tab} values for the expected
number of signal and background events are shown as a function of
the cut on the Fisher discriminant. Since correlations among variables
are taken into account, it is not surprising that the result for the
Fisher discriminant method is similar to the one obtained using a
multidimensional likelihood.

\begin{figure}[!ht]
\begin{center}
\includegraphics[width=10cm,height=10cm]{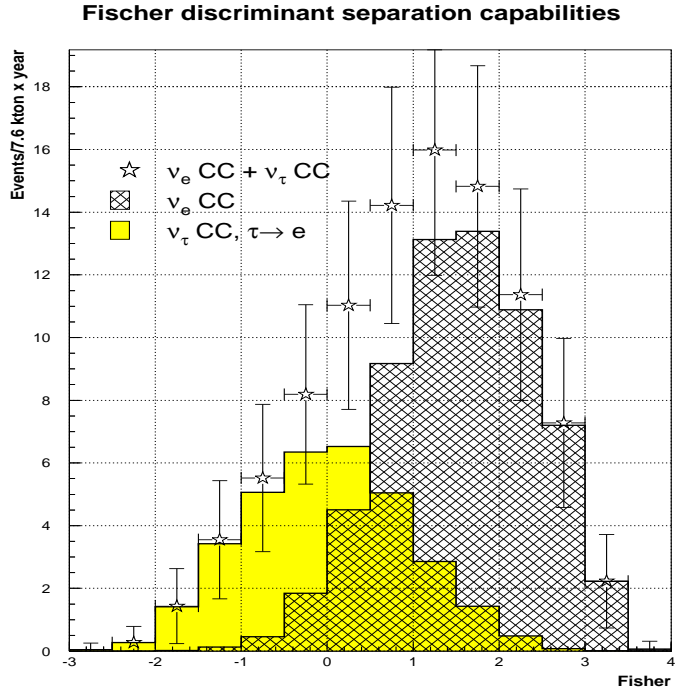}
\end{center}
\caption[The Fischer discriminant variable]
{The Fischer discriminant variable. Error bars in $\nu_{e}$CC + $\nu_{\tau}$CC
sample represent statistical fluctuations in the expected profile
measurements after 5 years of data taking with shared running CNGS
and a 5 T600 detector configuration\label{fig:Fischer-discriminant-variable.}}
\end{figure}

\begin{table}[!ht]
\begin{center}\begin{tabular}{|l|c|c|c|}
\hline 
Cuts &
 $\nu_{\tau}$CC ($\tau\rightarrow e$) &
&
 $\nu_{\tau}$ CC ($\tau\rightarrow e$)\tabularnewline
&
Efficiency&
 $\nu_{e}$CC&
 $\Delta m^{2}=$\tabularnewline
&
(\%)&
&
 $3\times10^{-3}$ eV$^{2}$\tabularnewline
\hline
Initial &
 100 &
 252 &
 50 \tabularnewline
\hline
Fiducial volume &
 65 &
 164 &
 33 \tabularnewline
\hline
 Fisher$>0.5$&
 46 &
$6.9\pm0.3$&
$23.1\pm0.4$\tabularnewline
 Fisher$>0.0$&
 33 &
 $2.4\pm0.2$ &
$16.6\pm0.4$\tabularnewline
\textbf{Fisher}$\mathbf{>-0.27}$&
\textbf{25}&
$\mathbf{1.15\pm0.13}$&
$\mathbf{12.9\pm0.3}$\tabularnewline
 Fisher$>-0.5$&
 20&
$0.60\pm0.10$&
$10.2\pm0.3$\tabularnewline
 Fisher$>-1.0$&
 10 &
 $0.14\pm0.05$ &
 $5.2\pm0.2$ \tabularnewline
\hline
\end{tabular}\end{center}
\caption[Fisher method: Expected number of signal and background events]
{Expected number $\nu_{e}$CC background and signal events in the
$\tau\rightarrow e$ analysis. A Fisher variable is used as the unique
discriminant. Numbers are normalized to 5 years running of CNGS. Errors
in the number of expected events are given by Poisson statistics.\label{tab:Fisher-tab}}
\end{table}

\paragraph{Increasing the number of initial kinematical variables\\ \\}
Contrary to what happens with a multi-dimensional likelihood (where the increase in 
the number of discriminating variables demands more Monte-Carlo data and therefore it 
is an extreme CPU-consuming process), the application of the Fisher method to a 
larger number of kinematic variables is straightforward, since the main characteristic 
of the Fisher method is that the final discriminant
can be obtained algebraically from the initial distributions of kinematic
variables. For instance, a Fisher discriminant built out of 9 kinematic variables
($E_{vis}$, $P_{T}^{miss}$, $\rho_{l}$, $P_{T}^{lep}$, $E_{lep}$,
$\rho_{m}$, $Q_{T}$, $m_{T}$, $Q_{lep}$)\footnote{see~\cite{Astier:1999vc} 
for a detailed explanation of the 
variables} predicts for $12.9\pm0.3$
taus a background of $1.17\pm0.14$ $\nu_{e}$ CC events. We conclude
that, for the Fisher method, increasing the number of variables does not improve the discrimination
power we got with the set $E_{vis}$, $P_{T}^{miss}$, $\rho_{l}$ and therefore these three 
variables are enough to perform an efficient $\tau$ appearance search.

\subsection{Kinematical search using neural networks}

In the context of signal vs background discrimination, neural networks
arise as one of the most powerful tools. The crucial point that makes
these algorithms so good is their ability to adapt themselves to the
data by means of non-linear functions.

Artificial Neural Networks techniques have become a promising approach
to many computational applications. It is a mature and well founded
computational technique able to \emph{learn} the natural behaviour
of a given data set, in order to give future predictions or take decisions
about the system that the data represent (see \cite{Bishop} and \cite{Haykin}
for a complete introduction to neural networks). During the last decade,
neural networks have been widely used to solve High Energy Physics
problems (see \cite{CERN_NN} for a introduction to neural networks
techniques and applications to HEP). Multilayer Perceptrons efficiently
recognize signal features from, an a priori, dominant background environment
(\cite{Ametller}, \cite{Higgs-Chiapetta}).

We have evaluated neural network performance when looking for $\nu_{\mu}\rightarrow\nu_{\tau}$
oscillations. As in the case of a multidimensional likelihood, a single
valued function will be the unique discriminant. This is obtained
adjusting the free parameters of our neural network model by means
of a \emph{training period}. During this process, the neural network
is taught to distinguish signal from background using a \emph{learning}
data sample.

Two different neural networks models have been studied: the \emph{multi-layer
perceptron} and the \emph{learning vector quantization} self-organized
network. In the following, their foundations and the results obtained
with both methods are discussed.

\subsubsection{The Multi-layer Perceptron\label{sub:The-Multi-layer-Perceptron}}

The multi-layer perceptron (MLP) function has a topology based on
different layers of neurons which connect input variables (the variables
that define the problem, also called \emph{feature} variables) with
the output unit (see Fig.~\ref{fig:mlp}). The value (or ``state'')
a neuron has, is a non-linear function of a weighted sum over the
values of all neurons in the previous layer plus a constant called
bias:

\begin{equation}
s_{i}^{l}=g\,(\sum_{j}\omega_{ij}^{l}s_{j}^{l-1}+b_{i}^{l})\label{eq:mlp_neurons_output}\end{equation}
where $s_{i}^{l}$ is the value of the neuron $i$ in layer $l$;
$\omega_{ij}^{l}$ is the weight associated to the link between neuron
$i$ in layer $l$ and neuron $j$ in the previous layer $(l-1)$;
$b_{i}^{l}$ is a bias defined in each neuron and $g(x)$ is a non-linear
function called \emph{transfer function}. The transfer function is
used to regularize the neuron's output to a bounded value between
0 and 1 (or -1,1). An usual choice for the transfer function is the
\emph{sigmoid function} (see Fig.~\ref{fig:sigmoid}),

\begin{equation}
g_{T}(x)=\frac{1}{1+e^{-\frac{x}{T}}}=\frac{1}{2}[1+tanh(\frac{x}{2T})]\label{eq:mlp_sigmoid}\end{equation}

This function tries to mimic the response of biological neurons by
putting the neuron in a high ($\sim1$) or low ($\sim0$) state depending
on the output of the previous neurons and how the links (weights)
are set. The \emph{temperature} parameter $T$ is not determinant
in principle and is always set to 1.

These kind of structures for a Neural Network, where the information
is propagated from the input units to the output ones passing through
different ordered layers, are called \emph{feed forward} neural networks. 

\begin{figure}[!ht]
\begin{center}
\includegraphics[width=10cm,keepaspectratio]{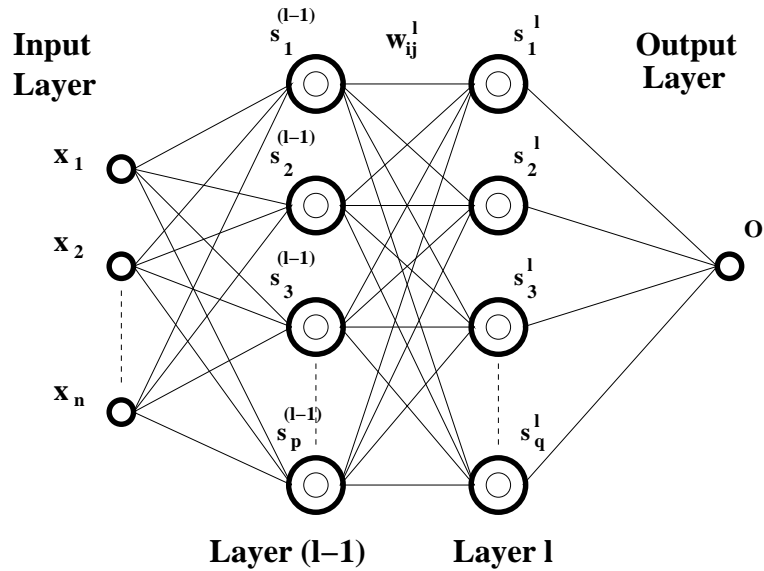}
\end{center}
\caption[General Multi-Layer Perceptron diagram]
{A general Multi-Layer Perceptron diagram. The optimal non-linear
function of input variables ($x_{i}$) is constructed using a set
of basic units called \emph{neurons}. Each neuron has two free parameters
that must be adjusted minimizing an error function.\label{fig:mlp}}
\end{figure}
\begin{figure}[!ht]
\begin{center}
\includegraphics[width=8cm,keepaspectratio]{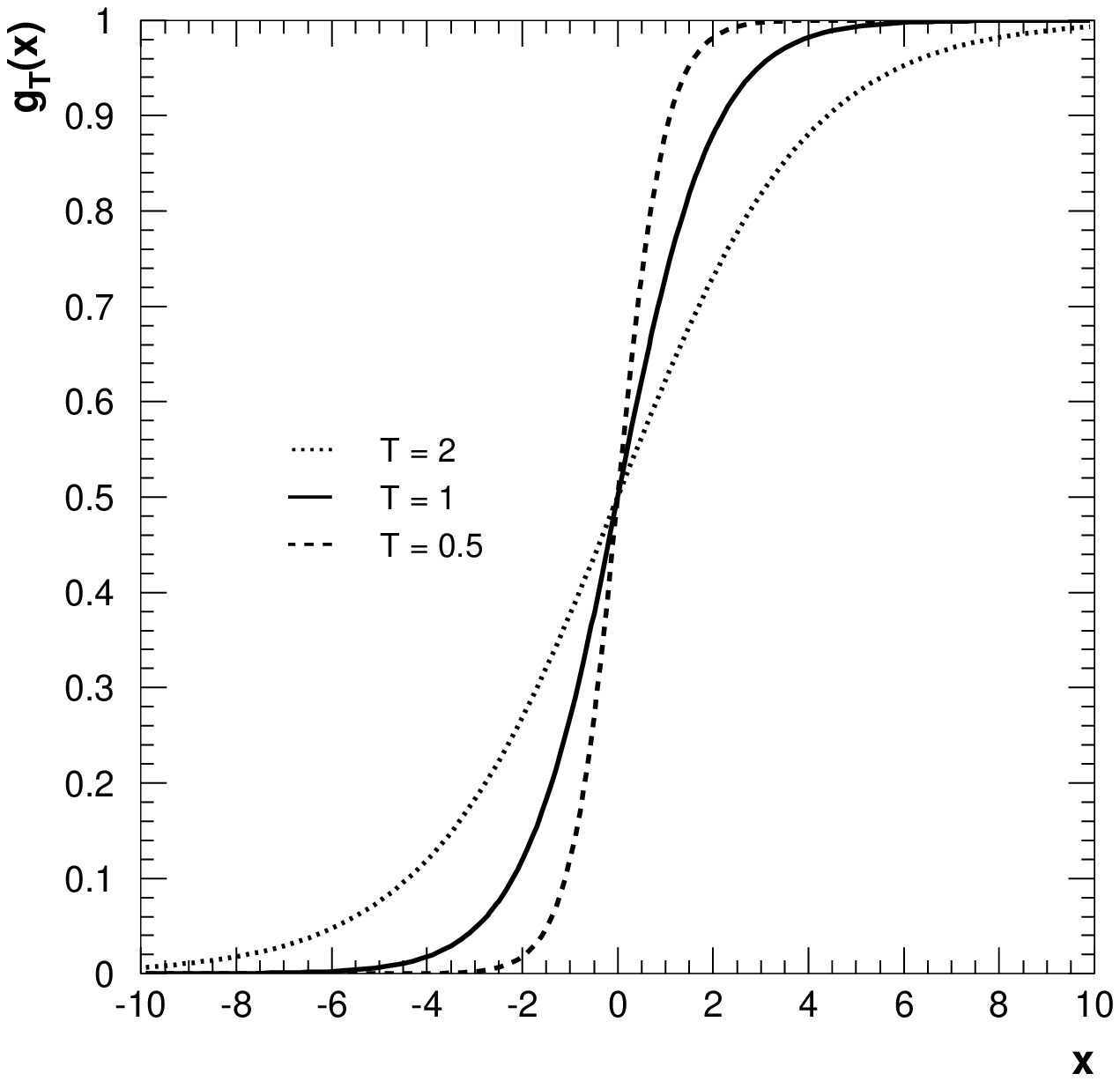}
\end{center}

\caption[Sigmoid function]
{Sigmoid function. This is the usual choice for the transfer function
of neurons in the MLP model. It has the ability of regularizing the
initial input variables to a value between 0 and 1, imitating, from
the point of view of biological models, the response of real neurons.
\label{fig:sigmoid}}
\end{figure}

\paragraph{Training the MLP\\ \\}
In a multi-layer perceptron a non-linear function is used to  obtain
the discriminating variable. The MLP is fed with a set of input variables
which define the problem. The way the MLP sets its free parameters
is through minimization of an error function. This error function
is defined as:
\begin{equation}
E=\sum_{e \epsilon Sample}E_{e}=\sum_{e \epsilon Sample}(o_{e}-d_{e})^{2}
\label{eq:mlp_error}
\end{equation}
where $o_{e}$ represents the MLP output for a given event belonging
to the training sample, and $d_{e}$ is the ``desired output''.
The previous expression is nothing but a distance between the current
MLP performance and the ideal case. In pattern recognition tasks,
we would like to have some kind of discrimination function which,
in an ideal case, always gives the correct answer, i.e., 1 for signal
events and 0 for background events. 

Once a topology is chosen for the MLP (number of layers and neurons
per layer), we need a first sample of signal and background events
for the training stage. The weights and the biases are then adapted,
minimizing the error function \eqref{eq:mlp_error}, following the standard
\emph{back-propagation algorithm} (see \cite{Bishop} or \cite{Haykin}),
which is the general procedure for a general feed-forward neural network
and an arbitrary differentiable activation function. The process of
adapting the parameters is repeated over the training sample during
a number of ``epochs'' until some condition is fulfilled (usually
when a stable minimum is reached). 

At the end of the training stage, the network is made, and final weights
and biases of the network are the ones which provide the best response
for the training sample. In order to determine the real discrimination
capabilities of the MLP, we need to test the neural network performance
over an statistically independent sample (to avoid biases).

\paragraph{Freedom of choice\\ \\}
The construction of a MLP implies that several choices must be made
a priori:

\begin{itemize}
\item \textbf{Number of input variables:} The MLP can be fed, in principle,
with as many variables as we consider fit in order to fully define
our pattern recognition problem. We need to know a priori which variables
(or combination of them) represent the main features of the classes
in our pattern selection task. Redundant variables contribute to a
worse performance of the MLP, hence is better to reduce the number
of input variables to avoid a loss of discrimination power.
\item \textbf{Number of hidden layers:} It depends on the specific task
one wants the network to perform, but a general statement is that
no more than two hidden layers are needed. In pattern classification
tasks, regions in input space must be created depending on the class
they belongs. It could be shown (see \cite{Bishop}) that a MLP with
no hidden layers only can made decision boundaries consisting in a
hyperplane, one hidden layer can perform a single convex region consisting
in hyperplanes segments, finally, two hidden layers could generate
arbitrary decision regions. Hence, if your classes are very mixed
in the variable space, two hidden layers will be able to construct
more complex region borders.
\item \textbf{Number of neurons per layer:} The view of pattern classification
as a ``cut out'' procedure gives a clue to the minimum number of
hidden units needed for a specific classification task. In the ``hard''
case, where the first hidden layer has to define the border of a closed
volume in the input space, at least (N+1) units (where N is the number
of input variables) are needed. The second hidden layer usually performs
some kind of logical function on the units belonging to the first
hidden layer, therefore, a small number of neurons (1 or 2) is enough%
\footnote{Logical function as AND or OR can be exactly performed using a perceptron
with no more than one hidden layer with two neurons. See \cite{CERN_NN}
for an example.%
}.
\item \textbf{Number of events in each sample:} The size of the simulated
data set is crucial in order to optimize the training algorithm performance.
If the training sample is small, it is likely for the MLP to adjust
itself extremely well to this particular data set, thus losing generalization
power (when this occurs the MLP is \emph{over-learning} the data).
Over-learning can be tested on-line by defining a statistical independent
sample to compute the error function in each step of the learning
procedure. 
\item \textbf{Number of epochs}: The number of iterations in the training
procedure depends on how low and stable is the minimum we get. In
addition over-learning must be avoided: by definition, the error curve
in the learning sample is always \emph{macroscopically} decreasing,
however the error curve in test sample could start to rise. At this
moment the MLP loses generalization power. Hence, we must stop our
learning phase before over-learning starts.
\end{itemize}

\paragraph{Multi-layer Perceptron setup\\ \\}
As already mentioned in \ref{sub:likelihood}, five variables (three
in the transverse plane and two in longitudinal direction) utterly
describe the event kinematics, provided that we ignore the jet structure;
therefore, using five independent variables we fully define our tagging
problem. Initially we build a MLP that contains only three input variables,
and in a latter step we incorporate more variables to see how the
discrimination power varies. The three chosen variables are $E_{visible}$,
$P_{T}^{miss}$ and $\rho_{l}$. Our election is similar to the one
used for the multidimensional likelihood approach, this allows to
make a direct comparison of the sensitivities provided by the two
methods.

\begin{figure}[!ht]
\begin{center}
\includegraphics[width=10cm,keepaspectratio]{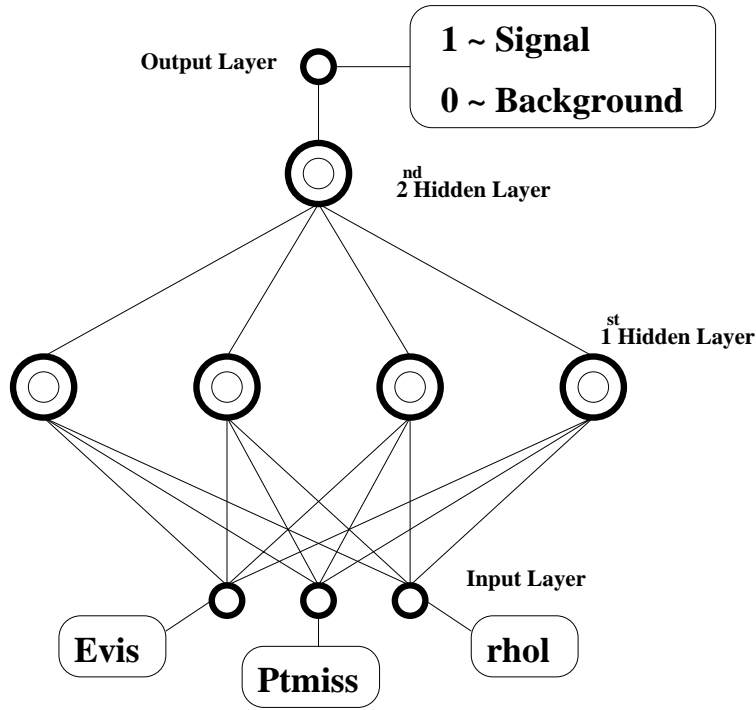}
\end{center}
\caption[The Multi-Layer Perceptron setup]
{The MLP setup. Chosen topology for the MLP. We feed a two layered
MLP (4 neurons in first layer and 1 in second) with input variables:
$E_{visible}$, $P_{T}^{miss}$ and $\rho_{l}$. \label{fig:mlp_setup}}
\end{figure}

There are plenty of packages available on the market which implement
a multi-layer perceptron. The simplest choice for us was to use the
MLPfit package \cite{PAW_mlp}, interfaced in PAW.

We have opted for a neural network topology consisting on two hidden
layers with four neurons in the first hidden layer and one in the
second (see Fig.~\ref{fig:mlp_setup}).

Simulated data was divided in three, statistically independent, subsets
of 5000 events each (2500 signal events and a similar amount of background).

The MLP was trained with a first ``training'' data sample. Likewise,
the second ``test'' data set was used as a training sample to check
that over-learning does not occur. Once the MLP is set, the evaluation
of final efficiencies is done using the third independent data sample
(namely, a factor 40 (75) larger than what is expected for background
(signal) after five years of CNGS running with a T3000 detector).

Error curves (Eq.~\eqref{eq:mlp_error}) during learning are shown in
Fig.~\ref{fig:learning_curves} for training and test samples. We
see that even after 450 epochs, over-learning does not take place.
Final distributions in the multi-layer perceptron discriminating variable
can be seen in Fig.~\ref{fig:MLP-output.}.

\begin{figure}[!ht]
\begin{center}
\includegraphics[width=10cm,height=10cm]{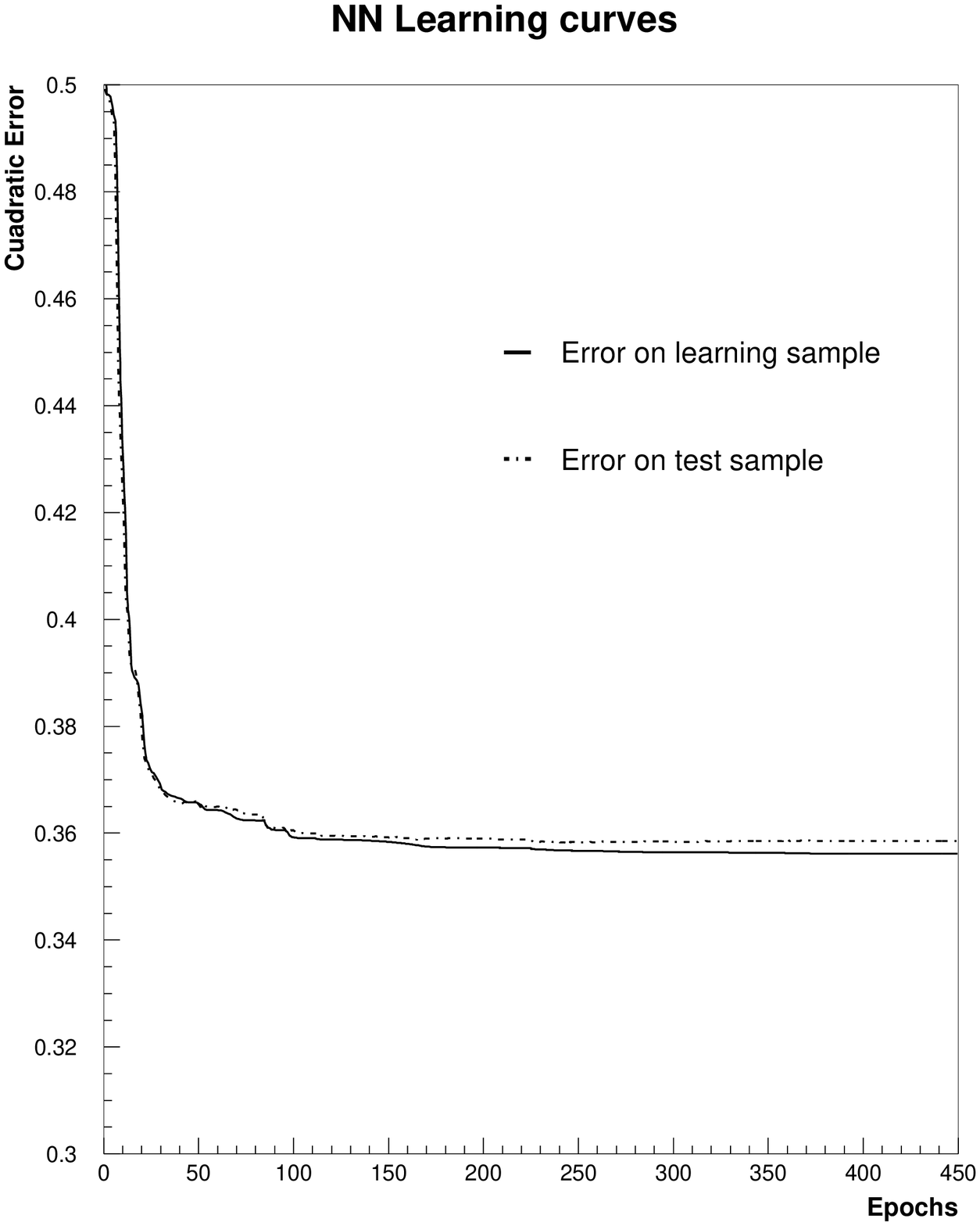}
\end{center}
\caption[Learning curves for the MLP]
{Learning curves for the MLP. The neural network is trained for 450
epochs in order to reach a stable minimum. The solid line represents
the error on training sample, the dashed line is the error on the
test sample. Both lines run almost parallel: no over-learning occurs.\label{fig:learning_curves}}
\end{figure}

\begin{figure}[!ht]
\begin{center}
\includegraphics[width=10cm,height=10cm]{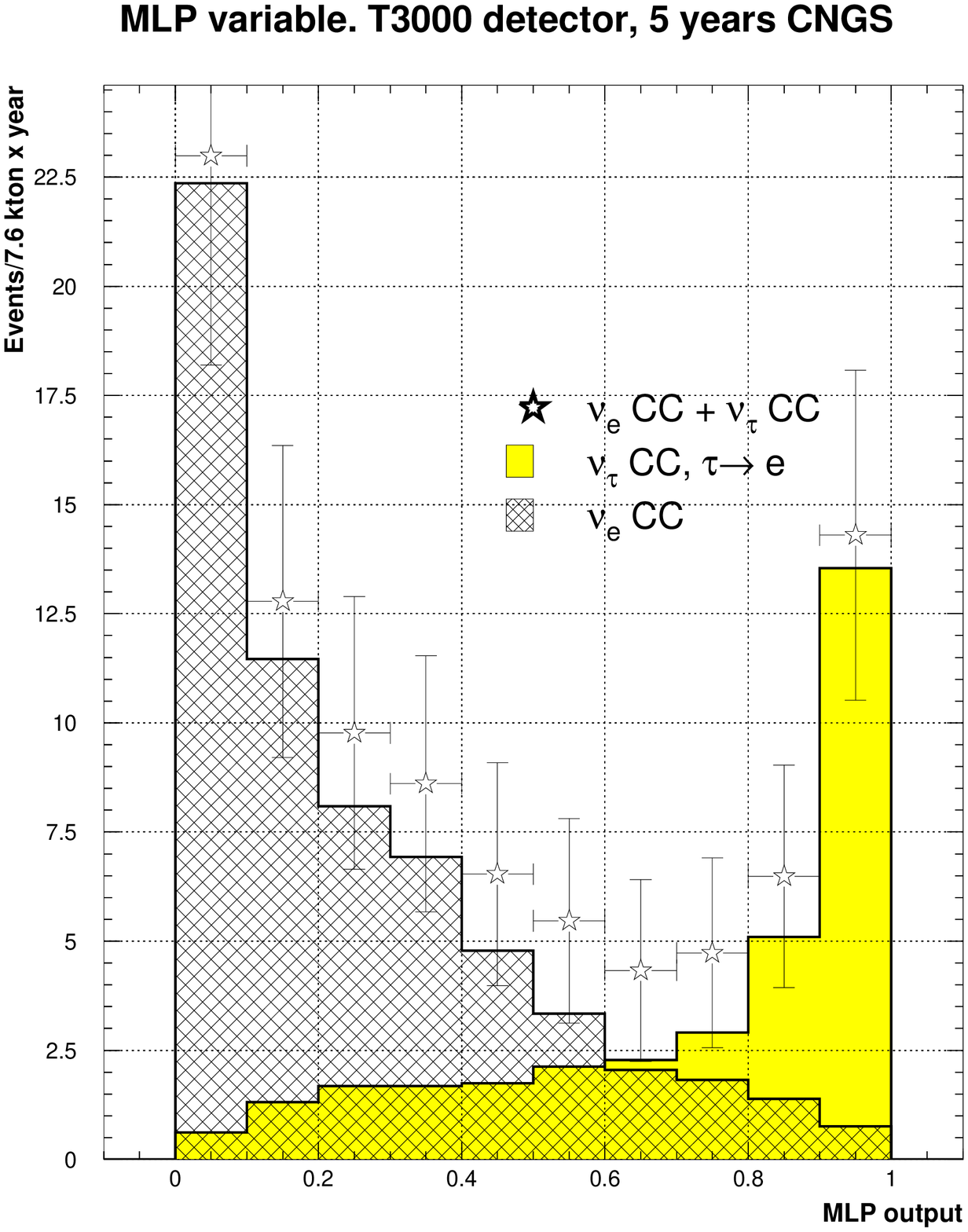}
\end{center}
\caption[The Multi-layer perceptron output]
{Multi-layer perceptron output for $\nu_{\tau}$CC ($\tau\rightarrow e$)
and $\nu_{e}$CC events. We see how signal events accumulate around
1 while background peaks at 0.\label{fig:MLP-output.}}
\end{figure}

\paragraph{Results of MLP analysis\\ \\}
Fig.~\ref{fig:mlp_sig_bac_cut} shows the number of signal and background
expected after 5 years of data taking as a function of the cut in
the MLP variable. In Fig.~\ref{fig:mlp_eff} we represent the probability
of an event, falling in a region of the input space characterized
by \emph{MLP output}~$>$~\emph{cut} to be a signal event (top plot), and the
statistical significance as a function of the MLP cut (bottom plot).
Like in the case of the likelihood, a cut based on the MLP output
variable can select regions of complicated topology in the kinematical
hyperspace (see Fig.~\ref{fig:mlp_cutted_vars}).

\begin{figure}[!ht]
\begin{center}
\includegraphics[width=10cm,height=10cm]{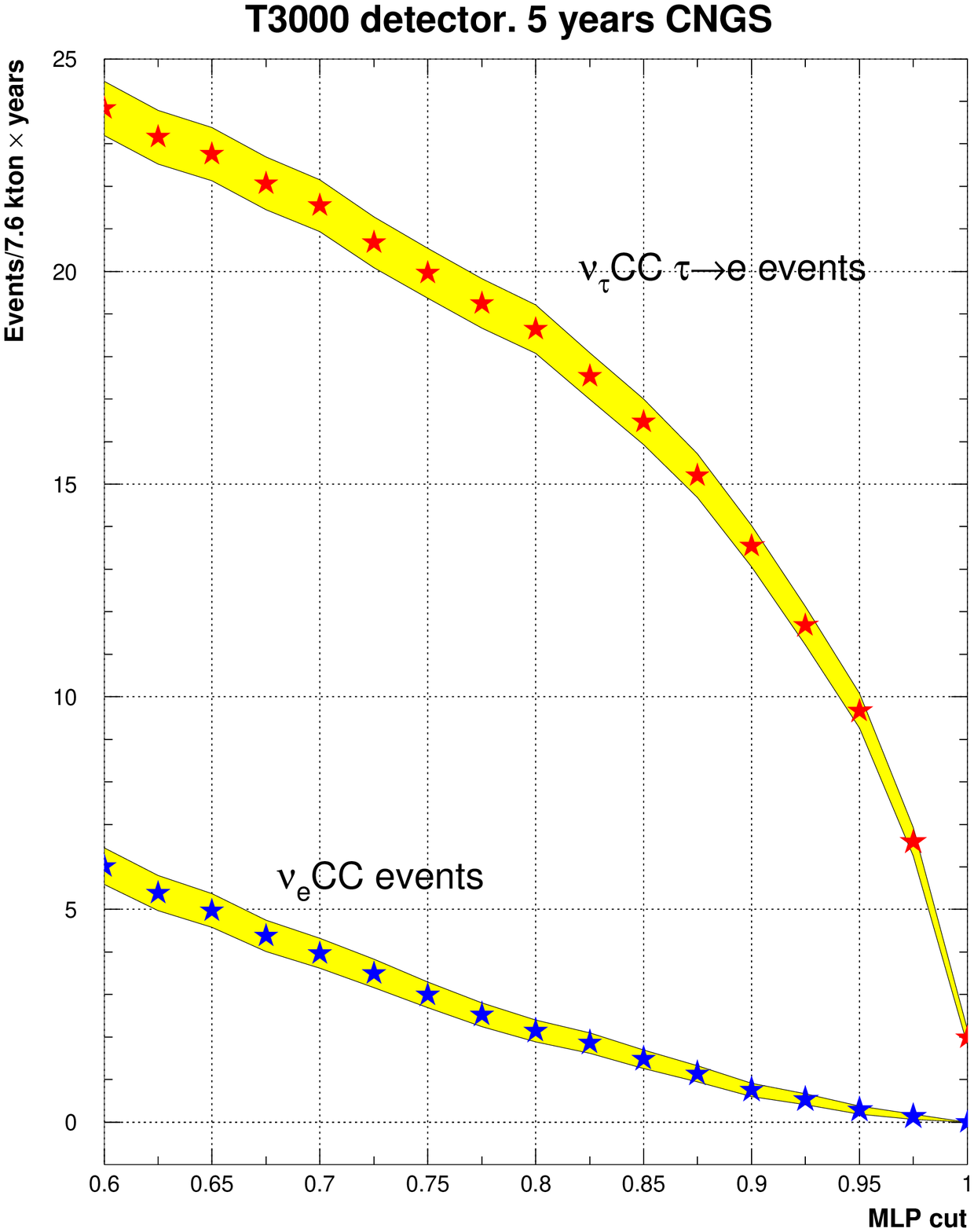}
\end{center}
\caption[The Multi-Layer perceptron: Expected number of signal and background events]
{Number of signal and background events after 5 years of running CNGS
as a function of the MLP cut. Shadowed zones correspond to statistical
errors.\label{fig:mlp_sig_bac_cut}}
\end{figure}

\begin{figure}[!ht]
\begin{center}
\includegraphics[width=10cm,height=10cm]{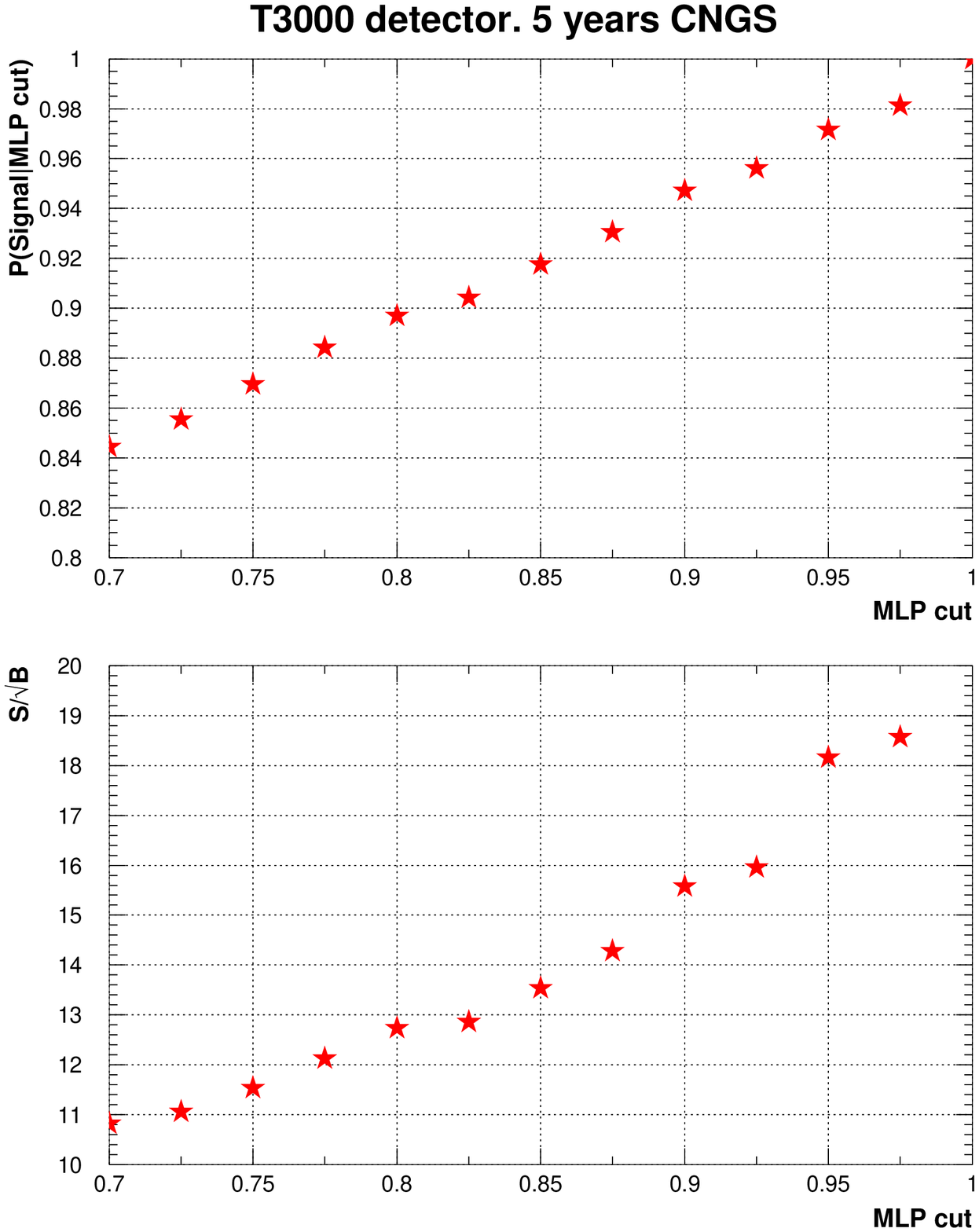}
\end{center}
\caption[The Multi-Layer perceptron: Statistical significance]
{(Top) probability of an event belonging to a region in input variable
space characterized by $MLP>cut$ of being a signal event. (Bottom)
Statistical significance of signal events as a function of the cut
in MLP.\label{fig:mlp_eff}}
\end{figure}
Selecting $MLP>0.91$ (overall $\tau$ selection efficiency
$=25$\%), the probability that an event falling in this region is
signal amounts to $\sim0.95$. For 5 years of running CNGS and a T3000
detector, we expect a total amount of $12.9\pm0.5$ $\nu_{\tau}$CC
($\tau\rightarrow e$) events and $0.66\pm0.14$ $\nu_{e}$CC events.
Tab.~\ref{tab:mlp} summarizes as a function of the applied MLP cut
the expected number of signal and background events.

 If we compare the outcome of this approach with the one
obtained in section \ref{sub:likelihood}, we see that for the same
$\tau$ selection efficiency, the multidimensional likelihood expects
$1.1\pm0.2$ background events. Therefore, for this particular cut,
the MLP achieves a 60\% reduction in the number of expected $\nu_{e}$
CC background events.

\begin{table}[!ht]
\begin{center}\begin{tabular}{|l|c|c|c|}
\hline 
Cuts &
 $\nu_{\tau}$CC ($\tau\rightarrow e$) &
&
 $\nu_{\tau}$ CC ($\tau\rightarrow e$) \tabularnewline
&
Efficiency&
$\nu_{e}$CC&
 $\Delta m^{2}=$\tabularnewline
&
(\%) &
&
 $3\times10^{-3}$ eV$^{2}$\tabularnewline
\hline
Initial &
 100 &
 252 &
 50 \tabularnewline
\hline
Fiducial volume &
 65 &
 164 &
 33 \tabularnewline
\hline
$MLP>0.70$&
 42 &
 $4.0\pm0.4$ &
 $21.4\pm0.6$\tabularnewline
$MLP>0.75$&
 40 &
$3.0\pm0.3$&
 $19.9\pm0.6$\tabularnewline
$MLP>0.80$&
 37 &
$2.1\pm0.3$&
$18.6\pm0.5$\tabularnewline
$MLP>0.85$&
 33 &
 $1.5\pm0.2$ &
$16.4\pm0.5$\tabularnewline
\textbf{$MLP>0.90$}&
 27&
$0.76\pm0.15$&
$13.5\pm0.5$\tabularnewline
$\mathbf{MLP>0.91}$&
\textbf{25}&
$\mathbf{0.66\pm0.14}$&
$\mathbf{12.9\pm0.5}$\tabularnewline
$MLP>0.95$&
 19 &
 $0.28\pm0.09$ &
 $9.6\pm0.4$ \tabularnewline
$MLP>0.98$&
12&
$0.09\pm0.05$&
$5.8\pm0.3$\tabularnewline
\hline
\end{tabular}\end{center}
\caption[The Multi-Layer Perceptron method: Expected number of signal and background 
events]
{Expected number of background and signal events when a multi-layer
perceptron function is used as the unique discriminant. Numbers are
normalized to 5 years running of CNGS. Errors in the number of events
expected are of statistical nature.\label{tab:mlp}}
\end{table}

\begin{figure}[!ht]
\begin{center}
\includegraphics[width=12cm]{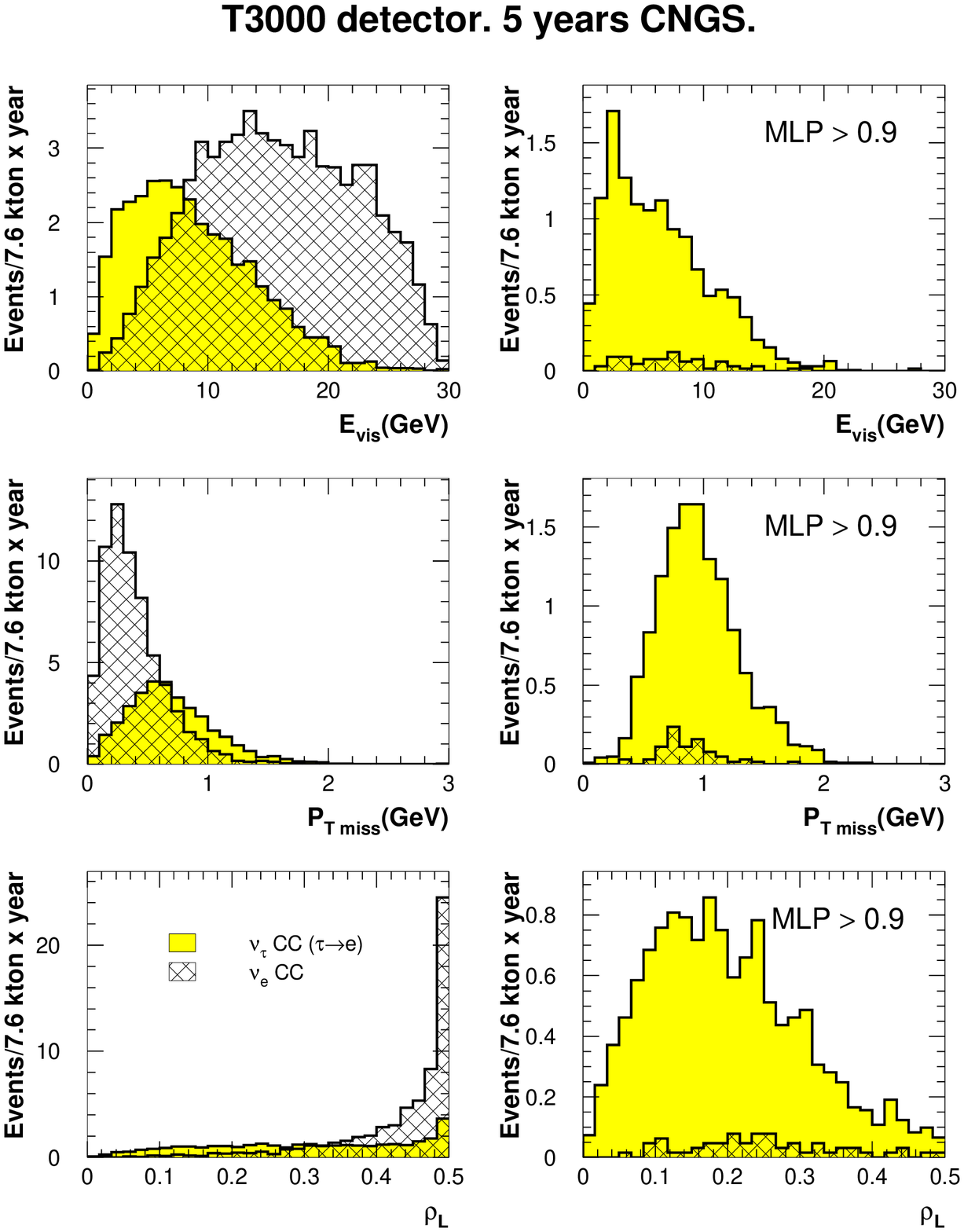}
\end{center}
\caption[Distribution of signal and background events before and after the MLP cut]
{Kinematical variables before (left histograms) and after (right histograms)
cut are applied based on the MLP output. We see how the MLP has learnt
that signal events favour low $E_{visible}$, high $P_{T}^{miss}$
and low $\rho_{l}$values.\label{fig:mlp_cutted_vars}}
\end{figure}

\paragraph{ Increasing the number of \emph{feature} variables\\ \\}
Contrary to the likelihood approach, where the lack of very
large Monte-Carlo samples, increases the complexity of the problem
as the number of input variables grows (even for three input variables,
smoothing algorithms have to be applied in order to get smooth distributions
throughout the hyper-space), the MLP approach shows no technical difficulties
in what concerns an increase of the input (feature) variables. 

Given that five variables define the problem under study,
we expect, a priori, an enhancement of the sensitivity when we move
from three to five feature variables. Notwithstanding the increase
on the number of inputs, we observed that the change in the final
sensitivity is negligible. The discrimination power is provided by
$E_{visible}$, $P_{T}^{miss}$ and $\rho_{l}$. The surviving background
can not be further reduced by increasing the dimensionality of the
problem. The application of statistical techniques able to find complex
correlations among the input variables is the only way to enhance
background rejection capabilities. In this respect, neural networks
are an optimal approach.

\paragraph{ Extended fisher discriminant method\\ \\}
Given that an increase on the number of input variables
does not improve the discrimination power of the Multi-layer perceptron,
we tried to enhance signal efficiency following a different approach:
optimizing the set of input variables by finding new linear combinations
of the original ones (or functions of them like squares, cubes, etc). 

 To this purpose, using the fast computation capabilities
of the Fisher method, we can operate in a systematic way in order
to find the most relevant feature variables. Starting from an initial
set of input variables, the algorithm described in \cite{Roe} tries
to gather all the discriminant information in an smaller set of optimized
variables. These last variables are nothing but successive Fisher
functions of different combinations of the original ones (see \cite{Roe}).
In order to allow not only linear transformations, we can add non-linear
functions of the kinematical variables like independent elements of
the initial set.

 We performed the analysis described in \cite{Roe} with
5 initial kinematical variables ($E_{vis}$, $P_{T}^{miss}$, $\rho_{l}$,
$P_{T}^{lep}$ and $E_{lep}$) plus their cubes and their exponentials
(in total 15 initial variables). At the end, we chose a smaller subset
of six optimized Fisher functions that we use like input features
variables for a new Multi-layer perceptron. 

 The MLP analysis with six Fisher variables does not enhance
the oscillation search sensitivity that we got with the three usual
variables $E_{vis}$, $P_{T}^{miss}$and $\rho_{l}$. We therefore
conclude that neither the increase on the number of features variables
nor the use of optimized linear combinations of kinematical variables
as input, enhances the sensitivity provided by the MLP.

\subsubsection{ Self organizing neural networks: LVQ network}
A \emph{self-organizing} (SO) network operates in a different
way than a multilayer perceptron does. These networks have the ability
to organize themselves according to the ``natural structure'' of
the data. They can learn to detect regularities and correlations in
their input and adapt their future response to that input accordingly.
A SO network usually has, besides the input, only one layer of neurons
that is called \emph{competitive layer}.

\paragraph{ Competitive self-organization\\ \\}
Neurons in the competitive layer are able to learn the structure
of the data following a simple scheme called \emph{competitive self-organization}
(see \cite{CERN_NN}), which ``moves'' the basic units (neurons)
in the competitive layer in such a way that they imitate the natural
structure of the data. It works as follows:

Suppose we have a set of $N$ input variables $\vec{x}^{(p)}=(x_{1}^{(p)},x_{2}^{(p)},...,x_{N}^{(p)})$
taking values over the total training sample (superscript ($p$) denotes
one element of the sample). These input nodes are connected with each
of the $M$ neurons in the competitive layer by a vector of weights
$\vec{w}_{i}=(w_{i1},w_{i2},...,w_{iN})$ (see Fig.~\ref{fig:Self-organized-network}).
From an initial random distribution of the weights, competitive self-organization
operates computing the following quantity in each neuron:
\[h_{i}=|\vec{w}_{i}-\vec{x}^{(p)}|\]
where $i$ is an index over the number of neurons in the
competitive layer. Once the $h_{i}$ are computed, a ``winner neuron''
is selected:
\[h_{m}=min(h_{j})\]

 A winner neuron is the one which has its vector of weights
closest to the pattern. The output of the network is a vector $\vec{o}=(o_{1},o_{2},...,o_{M})$
which returns 1 in the $m$ component corresponding to the winner
neuron and 0 for all the others. The weight vector $\vec{w}_{m}$
belonging to the winner neuron is then moved closer to $\vec{x}^{(p)}$
according to

\begin{equation}
\Delta\vec{w}_{m}=\eta(\vec{x}^{(p)}-\vec{w}_{m})
\label{eq:competitive rule}
\end{equation}
where $\eta$ is the learning rate. Thus, the neuron whose
weight vector is closest to the input vector is promoted to be even
closer. The result is that the winner neuron is more likely to win
the competition the next time a similar vector is presented, and less
likely to win when a different input vector is presented. As more
and more inputs are presented, each neuron in the layer closest to
a group of vectors soon adjusts its weights vector toward those input
vectors. Eventually, if there are enough neurons, every cluster of
similar input vectors will have a neuron that outputs 1 when a vector
in the cluster is presented, while outputting a 0 at all other times.
Thus, the competitive network learns to categorize the input vector
it sees.

\begin{figure}[!ht]
\begin{center}
\includegraphics[width=10cm,keepaspectratio]{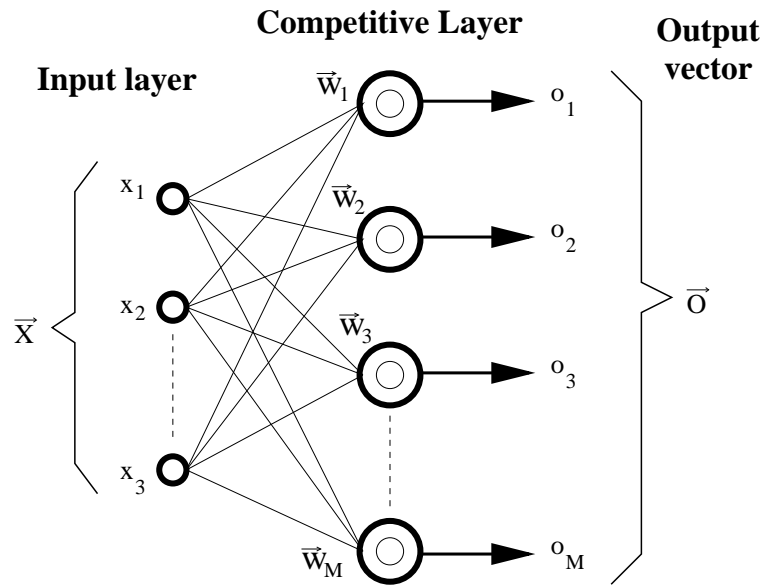}
\end{center}
\caption[Schematic diagram of the general topology for a \emph{self-organized}
neural network]
{Schematic diagram of the general topology for a \emph{self-organized}
neural network. Neurons in the competitive layer are connected with
each one of the input nodes. As output, it returns a vector indicating
the winner neuron.\label{fig:Self-organized-network}}
\end{figure}

\paragraph{ Learning vector quantization networks\\ \\}
Although competitive self-organization is an unsupervised
learning algorithm, for classification purposes one can improve the
algorithm with supervised learning in order to fine tune final positions
of the neurons in the competitive layer. This is called \emph{learning
vector quantization} (LVQ). 

 The modifications respect to the previous scheme are quite
simple: First we have to associate neurons in the competitive layer
with the different classes we try to classify, i.e. if we use 8 neurons,
4 can be associated to signal class and 4 to the background class.
In addition to the input vector $\vec{x}^{(p)}$ we need to provide
information about the class it belongs, i.e. signal or background
event. Now, for a given input vector belonging to a certain class,
a winner neuron is selected that could belong to the same class than
the input vector or not. If the classes match, the vector weight of
the winner neuron is then adapted according to Eq.~\eqref{eq:competitive rule}; 
if they mismatch the modified adapting rule is simple
\[ \Delta\vec{w}_{m}=-\eta(\vec{x}^{(p)}-\vec{w}_{m})\]

 As a result, neurons belonging to a given class are favoured
to adapt themselves to the regions where the patterns of their parent
class lie, and all the others tend to go away and trying to find their
corresponding regions, as well.

 At the end of training period, all the vector weights are
placed in an optimized way around the input variable space. Afterward,
we can test the neural network performance with a statistical independent
sample. When the LVQ is fed with new patterns, it selects the closest
neuron to each pattern and, depending on the class label of that neuron,
the pattern is classified. An important difference with respect to
the multi-layer perceptron approach is that in LVQ we always get a
discrete classification, namely, an event is always classified in
one of the classes. The only thing you can estimate is the degree
of belief in the LVQ choice.

\paragraph{ A LVQ network for $\tau$ search\\ \\}
Once again we use $E_{visible}$, $P_{T}^{miss}$ and $\rho_{l}$
as discriminating variables. A LVQ network with 10 neurons has been
trained with samples of 2500 events for both signal and background.
Given that, before any cuts, a larger background sample is expected,
we have chosen an asymmetric configuration for the competitive layer.
Out of 10 neurons, 6 were assigned to recognize background events,
and the rest were associated to the signal class. After the neurons
are placed by the training procedure, the LVQ network is fed with
a larger data sample (statistically independent) of 6000 signal and
background events. The output provided by the network is plotted in
Fig.~\ref{fig:LVQ-neural-network}. We see how events are classified
in two independent classes: \emph{signal like} events (labeled with
2) and \emph{background like} ones (labeled with 1). 68\% of $\nu_{\tau}$CC
($\tau\rightarrow e$) events and 10\% of $\nu_{e}$CC events, occurring
in fiducial volume, are classified to be a \emph{signal like} event.
This means a $\tau$ efficiency around 45\% with respect to the tau
events generated in active LAr. For the same $\tau$ efficiency, the
multi-layer perceptron only misclassifies around 8\% of $\nu_{e}$CC
events.

 Several additional tests have been performed with LVQ networks,
by increasing the number of input variables and/or the number of neurons
in the competitive layer. No improvement on the separation capabilities
have been seen. For instance, a topology with 16 feature neurons in
the competitive layer and 4 input variables (we add the transverse
lepton momentum) leads to exactly the same result. 

 The simple geometrical interpretation of this kind of neural
networks supports our statement that the addition of new variables
to the original set \{$E_{visible}$, $P_{T}^{miss}$, $\rho_{l}$\}
does not enhance the discrimination power: the bulk of signal and
background events are not better separated when we increment the dimensionality
of the input space.

\begin{figure}[!ht]
\begin{center}
\includegraphics[width=10cm,height=10cm]{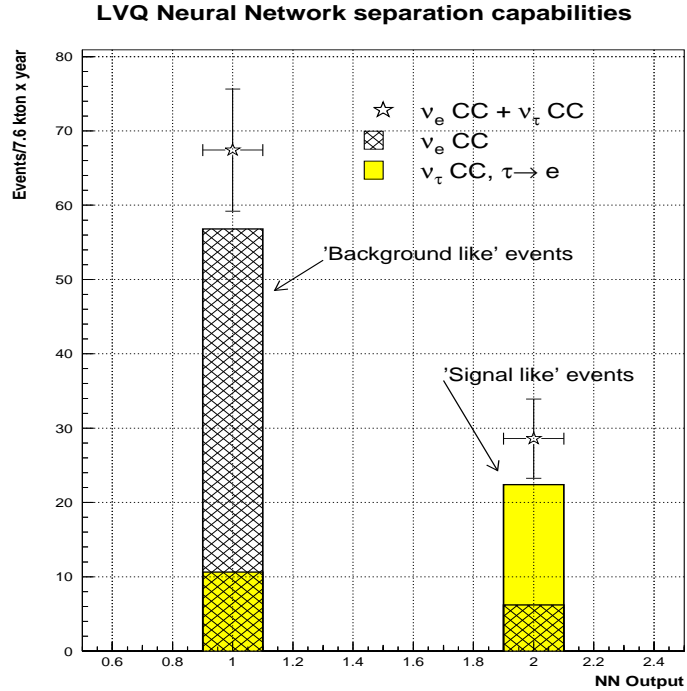}
\end{center}
\caption[LVQ neural network separation capabilities]
{LVQ neural network separation capabilities. In competitive self-organized
networks a discrete decision is always issued: \emph{signal like}
events are labeled with 2 and \emph{background like} with 1.\label{fig:LVQ-neural-network}}
\end{figure}

\paragraph{ Combining MLP with LVQ\\ \\}
We have seen that LVQ networks returns a discrete output.
The whole event sample is classified by the LVQ in two classes: \emph{signal-like}
and \emph{background-like.} We can use the classification of a LVQ
as a \emph{pre-}classification for the MLP. A priori, it seems reasonable
to expect an increase on the oscillation search sensitivity if we
combine the LVQ and MLP approaches. The aim is to evaluate how much
additional background rejection, from the \emph{contamination inside
the signal-like} sample, can be obtained by means of a MLP. 

 We present in Fig.~\ref{fig:LVQ-and-MLP} the MLP output
for events classified as \emph{signal-like} by the LVQ network (see
Fig.~\ref{fig:LVQ-neural-network}). Applying a cut on the MLP output
such that we get $12.9\pm0.5$ signal events (our usual reference
point of 25\% $\tau$ selection efficiency), we get $0.82\pm0.19$
background events, similar to what was obtained with the MLP approach
alone. This outcome conclusively shows that, contrary to our a priori
expectations, an event pre-classification, by means of a learning
vector quantization neural network, does not help improving the discrimination
capabilities of a multi-layer perceptron.

\begin{figure}[!ht]
\begin{center}
\includegraphics[width=10cm,height=10cm]{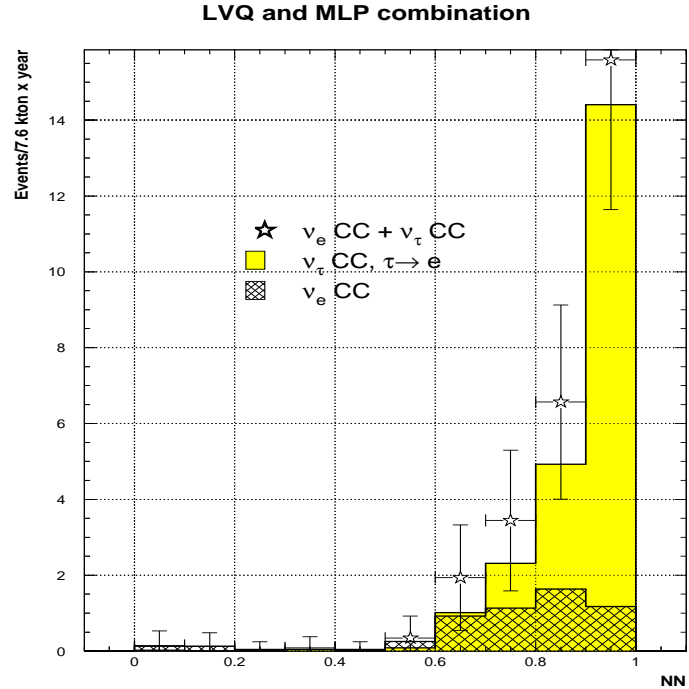}
\end{center}
\caption[Performance of LVQ and MLP networks combined]
{LVQ and MLP networks combined. Distributions are given in the continuous
MLP variable. Only events labeled by LVQ with 2 (\emph{signal like})
have been used for the analysis.\label{fig:LVQ-and-MLP}}
\end{figure}

\newpage
\section{$\nu_{\tau}$ appearance discovery potential}

We have studied several pattern recognition techniques applied to
the particular problem of searching for $\nu_{\mu}\rightarrow\nu_{\tau}$
oscillations. Based on \emph{discovery criteria}, similar to the ones
proposed in \cite{discovery} for statistical studies of prospective
nature, we try to quantify how much the discovery potential varies
depending on the statistical method used.

We define $\mu_{S}$ and $\mu_{B}$ as the average number of expected
signal and background events, respectively. With this notation, the
discovery criteria are defined as:

\begin{enumerate}
\item We require that the probability for a background fluctuation, giving
a number of events equal or larger than $\mu_{S}+\mu_{B}$, be smaller
than $\epsilon$ (where $\epsilon$ is $5.733\times10^{-7}$, the
usual $5\sigma$ criteria applied for Gaussian distributions). 
\item We also set at which confidence level $(1-\delta)$, the distribution
of the total number of events with mean value $\mu_{S}+\mu_{B}$ fulfills
the background fluctuation criteria stated above. 
\end{enumerate}
For instance, if $\delta$ is $0.10$ and $\epsilon$ is $5.733\times10^{-7}$,
we are imposing that $90\%$ of the times we repeat this experiment,
we will observe a number of events which is $5\sigma$ or more above
the background expectation.

For all the statistical techniques used, we fix $\delta$=$0.10$
and $\epsilon$=$5.733\times10^{-7}$. In this way we can compute
the minimum number of events needed to establish that, in our particular
example, a direct $\nu_{\mu}\rightarrow\nu_{\tau}$ oscillation has
been discovered at the CNGS beam.

In Tab.~\ref{tab:discovery} we compare the number of signal and
background events obtained for the multi-layer perceptron and the
multi-dimensional likelihood approaches after 5 years of data taking
with a 3 kton detector. We also compare the minimum exposure needed
in order to claim a discovery. The minimum exposure is expressed in
terms of a scale factor $\alpha$, where $\alpha=1$ means a total
exposure of $11.75$~kton$\times$year. For the multi-layer perceptron
approach ($\alpha=0.86$),  a statistically significant signal can be 
obtained after a bit more than four years of data taking.
On the other hand, the multi-dimensional likelihood approach requires 
5 full years of data taking to claim a discovery. 
 Therefore, when applied to 
the physics quest for neutrino oscillations, neural 
network techniques are more performing than classic statistical methods.
\begin{table}[!ht]
\begin{center}\begin{tabular}{|l|c|c|}\cline{2-3}
\multicolumn{1}{l|}{}&
 \textbf{Multi-layer}&
 \textbf{Multi-dimensional}\tabularnewline
\multicolumn{1}{l|}{}&
 \textbf{Perceptron}&
 \textbf{Likelihood}\tabularnewline
\hline
\multicolumn{1}{|l|}{\textbf{\# Signal}}&
 12.9&
 12.9\tabularnewline
\hline
\textbf{\# Background}&
 0.66&
 1.1\tabularnewline
\hline
\textbf{$\alpha$~factor}&
 \textbf{0.86}&
 \textbf{1.01} \tabularnewline
\hline
\end{tabular}\end{center}
\caption[MLP vs LKL: Expected rates]
{Number of signal and background events for the multi-layer perceptron
and the multi-dimensional likelihood approaches. Numbers are normalized
to 5 years of data taking in shared CNGS running mode and a 3 kton
detector configuration. The last row displays the scale factor $\alpha$
needed to compute the minimum exposure fulfilling the discovery criteria
described in the text.\label{tab:discovery}}
\end{table}

\subsection*{Conclusions}
We have considered the general problem of $\nu_\mu\to\nu_{\tau}$ 
oscillation search based on kinematic criteria to assess the performance 
of several statistical pattern recognition methods.

Two are the main conclusions of this study: 
\begin{itemize}
\item An optimal discrimination power 
is obtained using only the following variables:
$E_{visible}$, $P_{T}^{miss}$ and $\rho_{l}$ and their correlations.
Increasing the number of variables (or combinations of variables)
only increases the complexity of the problem, but does not result
in a sensible change of the expected sensitivity. 
\item Among the set of statistical methods considered, the multi-layer
perceptron offers the best performance. 
\end{itemize}

As an example, we have considered the case of the CNGS beam and 
$\nu_\tau$ appearance search (for the $\tau\to e$ decay channel) using a 
very massive (3 kton) Liquid Argon TPC detector. Fig.~\ref{fig:mlpvslkl}
compares the discrimination capabilities of multi-dimensional likelihood
and multi-layer perceptron approaches. We see that, for the low background
region, the multi-layer perceptron gives the best sensitivity. For
instance, choosing a $\tau$ selection efficiency of 25\% as a reference
value, we expect a total of $12.9\pm0.5$ $\nu_{\tau}$CC ($\tau\rightarrow e$)
signal and $0.66\pm0.14$ $\nu_{e}$CC background. Compared to multi-dimensional
likelihood predictions, this means a 60\% reduction on the number
of expected background events. Hence, using a multi-layer perceptron, 
four years of data taking will suffice to get a statistically significant 
signal, while five years are needed when the search approach is based on 
a multi-dimensional likelihood.

\begin{figure}[!ht]
\begin{center}
\includegraphics[width=10cm,height=10cm]{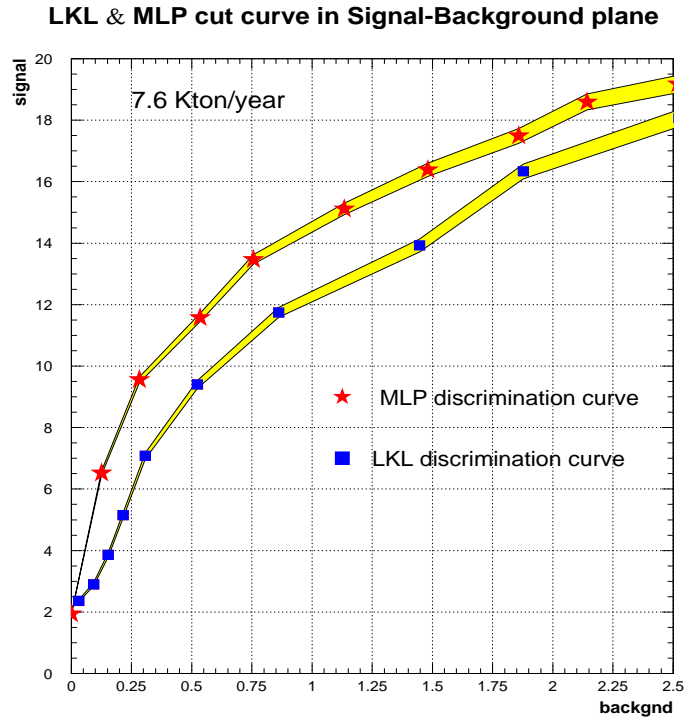}
\end{center}
\caption[Signal selection efficiency of Multi-layer perceptron vs multi-dimensional 
likelihood]
{Multi-layer perceptron vs multi-dimensional likelihood. We assume
5 T600 module configuration and 5 years of CNGS running (7.6 Kton/year
exposure). The shadowed area shows the statistical error.\label{fig:mlpvslkl}}
\end{figure}

\typeout{}
\chapter*{Conclusions}
\markboth{Conclusions}{Conclusions}
\addcontentsline{toc}{chapter}{Conclusions}

The results of this dissertation involve different aspects of Neutrino Physics.
After a brief review of the main theoretical topics on neutrino interactions 
and the problem of their masses, a phenomenological study of the possibilities 
offered by future lepton colliders for the exploration of the limits on 
the mixing of light and new heavy neutrinos is presented. 
Afterwards we have focused our interest on experimental Neutrino Physics with 
LAr detectors, which is the main issue of this thesis.
A sound and innovative work has been carried out in the development of the LAr 
technique and the necessary tools to perform Physics analyses, thus establishing 
the high potentiality of these detectors for the study of Neutrino Physics.

Assuming the possibility of the existence of heavy neutrinos with masses of 
few hundreds of GeV, we have analysed the contribution of these neutrinos to 
electroweak process in future lepton colliders, exploring the limits on heavy 
neutrino mixing and masses.
%Heavy neutrinos with masses near the electroweak scale and large mixing angles
%$\sim 0.1 - 0.01$ with the SM leptons are observable at ILC if they exist. 
We have studied the ILC potential for their detection in the process
$e^+ e^- \to \ell W \nu$, taking into account the SM background and the effects
of ISR and breamstrahlung, paying special attention to the relevance of the final
state lepton flavour and initial beam polarization. Using a parton simulation it
has been shown that \textbf{it is possible to observe a heavy neutrino signal in this
final state if it has a mixing with the electron $V_{eN} \gtrsim 10^{-2}$,
and a mass below $400$ GeV}.
Although a mixing with the electron of this size is necessary to observe a heavy
neutrino at ILC, \textbf{the signal may be more visible in the muon or tau channel if it
also has a relatively large coupling to them}. The production cross sections and
then the discovery limits do not depend on the Dirac or Majorana nature of the
heavy neutrino. 
These non-decoupled heavy neutrinos do not explain the observed light neutrino
masses nor the baryon asymmetry of the universe. In this sense this search for
heavy neutrinos at large lepton colliders is complementary to the joint
experimental effort for determining the light neutrino properties, and in
particular the neutrino mixing matrix (Ref~\cite{paper1}).

%Thereafter, the Thesis have been dedicated to the experimentation on neutrino Physics 
%with LAr detectors. A sound and innovative work have been carried out in the
%development of the LAr technique and its reconstrution tools, which arise as 
%very competitive detectors for the study of current and future experiments on 
%neutrino physics.

%Within the framework of the minimal extension of the Standard Model to include
%massive neutrinos
Thereafter, in the bulk of the dissertation, we have been mainly concentrated
on the study of the Physics potential of LAr TPCs.
\textbf{We have developed a set of tools and procedures which makes possible the calorimetric 
and 3D reconstruction of ionizing events occurring within the chamber}. 
These tools have shown an excellent performance working on events of low-multiplicity 
composed of several well defined tracks.

The technical operation of a 50 liter LAr TPC prototype, exposed to a multi-GeV $\nu_\mu$ 
neutrino beam, was rather good: 
The LAr purity was kept stable at a high-level during the whole data taking period and 
the electronic read-out configuration supplied an unbeatable signal-background 
ratio with no saturation of the electronic response, allowing an excellent 
imaging of the ionizing events.

\textbf{The 50L TPC offered the opportunity to measure and reconstruct for first time 
beam neutrino interactions using the LAr technique}. 
Due to the smallness of the prototype, a joint operation with the NOMAD detector 
acting as a muon spectrometer downstream of the TPC has been necessary in order to 
measure the high energy muons resulting from the neutrino interactions.
Before the data analyses, a careful calibration of the TPC was done.
All the needed parameters (the electron drift velocity, the alignment angles 
with respect to the beam, the electron lifetime and the recombination factors) 
were successfully obtained.

During its operation, the 50L TPC gathered a substantial sample of neutrino 
interactions in the multi-GeV region, with a sizable amount of low-multiplicity 
events. From the whole sample of $\nu_\mu$ CC events, we have selected a set of 
events called the \emph{golden sample}. They are events having a clear two prong topology 
with a muon leaving the chamber and an identified contained proton.
\textbf{This sample contains 86 successfully reconstructed events and, in spite of 
its limited size, they have made possible to study the accuracy of our
current Monte-Carlo models for the interactions of neutrinos with 
nuclear matter}. We expect that most of these events come from quasi-elastic 
neutrino interactions, but there is also a sizable background coming from 
deep-inelastic and resonant events.
Thanks to the superb reconstruction capabilities of the
LAr TPC, the kinematical distributions of the events have been obtained
with high resolution. This allows to discriminate between models with 
a full description of nuclear effects and naive approaches that only 
include Fermi motion and Pauli blocking (Ref.~\cite{50Lqe}).

On the other hand, the collected sample of \emph{golden} events gives us 
the opportunity of \textbf{measuring for first time the quasi-elastic $\nu_\mu$ CC 
cross section using a LAr TPC}.
To this purpose a precise knowledge of the beam composition, the beam profile,
the exposition time, trigger efficiencies, acceptance cuts, event selection
rates and background ratios was needed. In addition, a careful evaluation
of the systematic errors entering in the computation has been done exploiting
a full Monte-Carlo simulation of the neutrino beam, the nuclear effects and 
the experimental set-up.
Our measurement of the total QE 
$\nu_\mu$ charged-current cross section~(Ref.~\cite{50Lxsec}) is 
\begin{equation*}
\boxed{
\sigma_{QE} = (0.90 \pm 0.10 \ \mathrm{(stat)} \pm 0.18
\ \mathrm{(sys)}) \times 10^{-38}\,\text{cm}^2
}
\end{equation*}
This value agrees with previous
measurements of this reaction reported in the literature, and it stands as the 
first Physics measurement ever done with a LAr TPC using a neutrino beam from
an accelerator.

Finally, given the excellent reconstruction capabilities of the LAr 
TPC technique and the accuracy in the reconstruction of the kinematic
variables in the plane transverse to the incoming neutrino, 
\textbf{we have evaluated our capability to perform a 
$\nu_{\mu}\rightarrow\nu_{\tau}$ oscillation search based on kinematical criteria}. 
A systematic assessment of the performance of several statistical pattern 
recognition methods shows that an optimal discrimination power is obtained 
using only three variables and their correlations: 
The visible energy ($E_{vis}$), the transverse missed momentum ($P_{T}^{miss}$) 
and the fraction of transverse momentum carried by the prong lepton ($\rho_{l}$).
Moreover, among the set of statistical methods considered, 
the multi-layer perceptron offers the best performance.
\textbf{Considering the case of the $\nu_{\tau}$ appearance search at the CNGS $\nu_\mu$
neutrino beam using a very massive (3 kton) Liquid Argon TPC detector, 
we expect a total of $12.9\pm0.5$ $\nu_{\tau}$CC ($\tau\rightarrow e$) signal 
and $0.66\pm0.14$ $\nu_{e}$CC background after 5 years running}. 
This means a 60\% reduction of the number of expected background events with 
respect to other broadly use methods in Particle Physics like for example, 
multi-dimensional likelihoods (Ref.~\cite{oscsearch}).

%This study have been done using a Monte-Carlo simulation of the neutrino interactions 
%we would expect within a large LAr detector of 2.35 ktons located at the Gran Sasso 
%laboratory.

%%%%%%%%%%%%%%%%%%%
%\renewcommand{\chaptermark}[1]{         % Lower Case Chapter marker style
%\markboth{\appendixname\ \thechapter.\ #1}{}} %
%%%%%%%%%%%%%%%%%%%

\appendix
%\begin{fmffile}{myFeynDiag}
%%%%%%%%%%%%%%%%%%%%%%%%%%%%%%
\chapter{Electroweak interactions}
\label{sec:WeakInt}
%%%%%%%%%%%%%%%%%%%%%%%%%%%%%%

The standard electroweak model is based on the gauge symmetry group 
$SU(2)_L\times U(1)_Y$, and the corresponding gauge bosons $W^i_\mu,\,i=1,2,3$, 
and $B_\mu$.
After spontaneus symmetry breaking and the usual redefinition of the gauge 
fields~\cite{PDG}, the Lagrangian which describes the electroweak gauge interactions 
for fermions reads:

\begin{equation}
\begin{split}
 \mathcal{L}^{int}_F 
& = -\frac{g}{2\sqrt{2}}\sum_i\overline{\psi_i}\gamma^\mu(1-\gamma^5)(T^+W^+_\mu + T^-W^-_\mu)\psi_i \\
& \quad -\frac{g}{2\cos{\theta_W}}\sum_i\overline{\psi_i}\gamma^\mu(g^i_V-g^i_A\gamma^5)Z_\mu\psi_i  \\
& \quad -e\sum_i Q_i\overline{\psi_i}\gamma^\mu A_\mu\psi_i \,,
\end{split}
\label{eq:EWlag}
\end{equation}
where $\psi_i = \Bigl( \begin{smallmatrix} \psi_i^\uparrow \\ 
\psi_i^\downarrow \end{smallmatrix}\Bigr) $ is the usual $SU(2)$ doublet of the 
$i^{th}$ fermion family, being $T^+$ and $T^-$ the raising and lowering operators
of the same $SU(2)$ representation. The first line of Eq.~\eqref{eq:EWlag} describes 
charge current (CC) interactions, while the second one rules the neutral currents (NC);
the electromagnetic interactions are governed by the third term.

%%%%%%%%%%%%%%%%%%%%%%%%%%%%%%%%%%%%%%%%%%%%
\section{Charged Current Interactions}
\label{sec:CCint}
%%%%%%%%%%%%%%%%%%%%%%%%%%%%%%%%%%%%%%%%%%%%
In the SM left-handed fermions are put together in three different families 
of $SU(2)_L$ doublets, while the right-handed components are singlets.
Thus, the emission of a charged vector boson $W^{\pm}$ causes a transition between the 
left-handed fermions within the weak isospin $\nicefrac{1}{2}$ doublet 
$\Bigl( \begin{smallmatrix} \psi_i^\uparrow \\ 
\psi_i^\downarrow \end{smallmatrix}\Bigr)$.

These processes $ {\psi_i}^{\downarrow} \rightarrow {\psi_i}^{\uparrow} + W^-$ and 
${\psi_i}^{\uparrow} \rightarrow {\psi_i}^{\downarrow} + W^+$ are governed by the 
Lagrangian
\begin{equation}
\mathcal{L}^{\text{cc}} = -\frac{g}{\sqrt{2}}\,({j^{\mu}_{\text{cc}}} W^+_{\mu} 
                          + {j^{\mu}_{\text{cc}}}^\dagger W^-_{\mu}) 
\end{equation}
with the vector field operators\footnote{$W^+_{\mu}$ is the operator that 
annihilates a $W^+$ or creates a $W^-$.} $W^{\pm}_{\mu}$ and the fermion current 
${j^{\mu}_{\text{cc}}}$ summed over the 3 lepton and quark doublets:
\begin{equation}
{j^{\mu}_{\text{cc}}} = \sum_{i=1}^3 \overline{\nu_i}\gamma^{\mu}\frac{1}{2}{(1-\gamma^5)}{l_i} 
+ \sum_{i=1}^3 \overline{u_i}\gamma^{\mu}\frac{1}{2}{(1-\gamma^5)}{d_i}
\end{equation}
Here, the field operators $\psi_{i}$ have been replaced by the particle symbols 
($\nu_i, l_i \leftrightarrow \text{leptons}$, $u_i, d_i \leftrightarrow \text{quarks}$) 
for simplicity. 
The projection operator $\frac{1}{2}{(1-\gamma^5)}$ assures that only leptons with 
left-handed chirality take part in the process
\begin{equation}
\overline{\nu_i}\gamma^{\mu}\frac{1-\gamma^5}{2}{l_i}
=\overline{{\nu_i}_L}\gamma^{\mu}{l_i}_L \,.
\end{equation}

For massive fermions the weak isospin eigenstates are generally a mixture of the mass 
eigenstates. The basic mechanism by means of which fermions adquire mass in the SM
preserving the gauge symmetry is through Yukawa couplings with the Higgs boson~\cite{CottSM}. 
It is always possible to choose a fermion basis where just
the fermions of one of the isospin eigenstates are rotated respect to their mass 
eigenstates after symmetry breaking. Thus, without loss of generality, only one
unitary $(3\times3)$-matrix operating in the $\psi^{\downarrow}$ is needed in order to 
describe the relation between interaction and mass eigenstates:
\begin{equation}
\begin{pmatrix} \psi^{\downarrow}_1\\ \psi^{\downarrow}_2\\ \psi^{\downarrow}_3 \end{pmatrix}
=  U \begin{pmatrix} \psi^{\downarrow}_{m\,1}\\  \psi^{\downarrow}_{m\,2}\\ \psi^{\downarrow}_{m\,3} \end{pmatrix}
\label{eqn:mixingGeneral}
\end{equation}
It could have been equivalently done with the $\psi^\uparrow_i$ fermions.
Therefore, quarks and lepton basis are usually choosen so that charge leptons 
(e~$\mu$~$\tau$) and \emph{up} quarks ($u$~$c$~$t$) coincide with their mass 
eigenstates, while their opposite weak isospin partners are rotated respect to their 
mass eigenstates.
The charged fermion current can now be rewritten as a sum over the observed fermions
\begin{equation}
j^{\mu}_{\text{cc}} = (\overline{\nu_{m\,1}}~\overline{\nu_{m\,2}}~\overline{\nu_{m\,3}}) U_l^\dagger
 \gamma^{\mu}\frac{1}{2}{(1-\gamma^5)}
 \begin{pmatrix} e \\ \mu \\ \tau \end{pmatrix}  +
(\overline{u}~\overline{c}~\overline{t})
 \gamma^{\mu}\frac{1}{2}{(1-\gamma^5)} U_q
 \begin{pmatrix} d \\s \\ b \end{pmatrix} 
\label{eq:lepcurrent}
\end{equation}
The quarks and charged leptons correspond to the mass eigenstates shown in 
Tab.~\ref{table:fermions}. Eq.~\eqref{eq:lepcurrent}
essentially implies that charged current weak interactions cause transitions 
between quarks or leptons of the three generations.

%%%%%%%%%%%%%%%%%%%%%%%%%%%%%%%%%%%%%%%
\subsection{Charged Leptonic Current}
\label{sec:lepCC}
%%%%%%%%%%%%%%%%%%%%%%%%%%%%%%%%%%%%%%%

The recently found neutrino oscillations imply that the neutrinos $\nu_{e}$, 
$\nu_{\mu}$ and $\nu_{\tau}$ are not the mass eigenstates $\nu_{m\,i}$. 
This means that the mixing matrix $U_l$ is not diagonal \cite{PDG}. 
In fact, current limits indicate mixing angles which correspond to almost maximal 
mixing for $\nu_e \leftrightarrow \nu_\mu$ and $\nu_\mu \leftrightarrow \nu_\tau$. 
Also, there might be more than three neutrinos taking part in the mixing 
(see Sec~\ref{sec:neumass}). 
Because of the small neutrino masses and since they interact only weakly, 
the mass eigenstates are not yet directly observable, but have to be determined by 
means of oscillation rates and cosmological measurements. 
The experimental evidence in favor of oscillations of solar (KamLAND) and 
atmospheric (K2K) neutrinos can be accommodated in the SM with three massive neutrinos.
The charged leptonic current is often also written in terms of the weak eigenstates:
\begin{equation}
{j^{\mu}_{\text{cc},l}} =  (\overline{\nu_e}~\overline{\nu_\mu}~\overline{\nu_\tau})
 \gamma^{\mu}\frac{1}{2}{(1-\gamma^5)}
 \begin{pmatrix} e \\ \mu \\ \tau \end{pmatrix}\,.
\end{equation}

%%%%%%%%%%%%%%%%%%%%%%%%%%%%%%
\subsection{Charged Quark Current}
\label{sec:quarkCC}
%%%%%%%%%%%%%%%%%%%%%%%%%%%%%%
The mixing in the quark sector is described by the Cabibbo-Kobayashi-Maskawa 
(CKM) matrix:
\begin{equation}
U_q =
\left(
\begin{array}{ccc}
U_{ud} & U_{us} & U_{ub} \\
U_{cd} & U_{cs} & U_{cb} \\
U_{td} & U_{ts} & U_{tb} \\
\end{array}
\right).
\end{equation}
The squared matrix elements $|U_{ij}|^2$ give the probability of the quark $u_{i}$ 
to change into the quark $d_{j}$. Conservation of probability requires the 
CKM matrix to be unitary. 
In principle, the individual matrix elements can all be measured in weak decays of 
the relevant quarks or in deep inelastic neutrino-nucleon scattering. 
The \unit[90]{\%} confidence limits on the magnitude of the matrix elements, 
as given in \cite{PDG}, are
\begin{equation}
\left(
\begin{array}{ccc}
0.9739-0.9751 & 0.221-0.227   & 0.0029-0.0045 \\
0.221-0.227   & 0.9730-0.9744 & 0.039-0.044   \\
0.0048-0.014  & 0.037-0.043   & 0.9990-0.9992 \\
\end{array}
\right).
\end{equation}

In contrast with the leptonic sector, the mixing in the quark sector is easily 
observable, since the mass differences are bigger and the quarks interact also 
via the strong force.

%%%%%%%%%%%%%%%%%%%%%%%%%%%%%%%%%%%%%%%%%%% 
\section{Neutral Current Interactions}
%%%%%%%%%%%%%%%%%%%%%%%%%%%%%%%%%%%%%%%%%%%
%The fundamental process of the weak neutral current interactions is the emission 
%of a $Z$ boson, $f \rightarrow f + Z$. 
The $Z$ boson, just as the photon, 
is a mixture of the neutral gauge field of the $SU(2)_L$ symmetry group (the other two 
weak bosons being $W^+$ and $W^-$) and the gauge field of the $U(1)$ symmetry group. 
The form of the weak neutral current, therefore, also depends on whether or not the 
fermions carry charge. The corresponding Lagrangian is given by
\begin{equation}
\mathcal{L}^{nc} = -\frac{g}{2\cos{\theta_W}}\, j_{\text{nc}}^{\mu} Z_{\mu}
\end{equation}
where $\theta_W$ is the \emph{Weinberg angle} defined by $\cos\theta_W={m_W}/{m_Z}$. 
The neutral current is summed over all fermions
\begin{equation}
\begin{split}
j_{\text{nc}}^{\mu} 
& = \sum_i \overline{\psi}_i \gamma^{\mu}(g^i_V-g^i_A\gamma^5) \psi_i \\
& = \sum_i \overline{\psi}_i \gamma^{\mu}({T^3}^i (1-\gamma^5)-2 Q^i\sin^2\theta_W) \psi_i 
\end{split}
\end{equation}
and
\begin{equation}
\begin{split}
g^i_V & = {T^3}^i - 2 Q^i\sin^2\theta_W \,,\\
g^i_A & = {T^3}^i \,,
\end{split}
\end{equation}
where $T^3$ is the weak isospin of the left-handed fermion, i.\ e.\ 
$\pm\nicefrac{1}{2}$, and $Q^i$ is the charge of the fermion in units 
of $e$.

%\include{Theory/nNScatt}
%%%%%%%%%%%%%%%%%%%%%%%%%%%%%%
\chapter{Calculation of the $e^+ e^- \to \ell^- W^+ \nu$ process with heavy neutrinos}
\label{sec:NuMassCalc}
%%%%%%%%%%%%%%%%%%%%%%%%%%%%%%

Following the notation that we introduced in Sec.~\ref{sec:HeavyNu} for the 
$6\times6$ neutrino mixing matrix (Eq.~\eqref{eq:mixmassext}), 
we write down the extended (MNS) matrix \cite{MNS,Pont} 
as ${\cal V} = ( U ~ V )$. This matrix  parameterises the charged and neutral 
current gauge interactions. On the other hand, the neutrino mass eigenstates have been
renamed as in that Section in order to make explicit the distintion between 
the light ($\nu_L$) and the heavy ($N_L$) sectors. The electroweak Lagrangian is then 
written as
\begin{eqnarray}
\label{ec:lagrW}
\mathcal{L}^W_{CC} & = & - \frac{g}{\sqrt{2}}
\; \bar l_L \gamma^\mu \; {\cal V} 
\left( \!\! \begin{array}{c} \nu_L \\ N_L \end{array} \!\! \right)
\; W_\mu^- + \mathrm{H.c.} \,, \\
\label{ec:lagrZ}
\mathcal{L}^Z_{NC}\ & = & -
\frac {g}{2 \cos \theta_W} 
\left(\bar \nu_L \; \bar N_L \right) \gamma^\mu  \; {\cal V}^\dagger {\cal V} 
\left( \!\! \begin{array}{c} \nu_L \\ N_L \end{array} \!\! \right)
\;  Z_\mu  \,. 
\end{eqnarray}
In the evaluation of $e^+ e^- \to \ell^- W^+ \nu$, with $W^+ \to q \bar q'$, 
we will only consider the contributions from the diagrams in Fig.~\ref{diag}, 
neglecting diagrams with four fermions $e^- q \bar q' \nu$ in the final state 
but with $q \bar q'$ not resulting from a $W$ decay. At any rate, we have
checked  that the corresponding contributions are negligible in the phase space
region  of interest.
The discovery of a new heavy neutrino in $e^+ e^- \to \ell W \nu$ requires its
observation as a peak in the invariant $\ell W$ mass distribution, otherwise
the irreducible SM background is overwhelming. This requires to reconstruct the
$W$, what justifies to consider $\ell W \nu$ production (instead of general
four fermion production), with $W$ decaying hadronically.

Let us discuss the contributions and sizes of the different diagrams for  
$e^+ e^- \to \ell^- W^+ \nu$ in Fig. \ref{diag} and what  we can learn from
this type of processes at ILC. 
The first four diagrams are SM contributions. Diagrams 
5, $7-9$ are present within the SM, mediated by a light neutrino, but
they can also involve a heavy one. Diagrams 6 and 10 are exclusive to Majorana
neutrino exchange. The SM contribution has a substantial part from resonant $W^+
W^-$ production, diagrams 4 and 8, especially for final states with $\ell = \mu,
\tau$. The heavy neutrino signal is dominated by diagrams 5 and 6
with $N$ produced on its mass shell, because the $\Gamma_N$ enhancement of the
amplitude partially cancels the mixing angle factor in the decay vertex,
yielding the corresponding branching ratio. It must be remarked that the
$s$-channel $N$ production diagram 7 is negligible (few per mille) when
compared to the  $t$-, $u$-channel diagrams 5 and 6. This behaviour is general,
because the $s$-channel propagator is fixed by the large collider energy and
suppresses the contribution of this diagram, whereas the $t$- and $u$-channel
propagators do not have such suppression.
Since both diagrams 5 and 6 involve an $eNW$
vertex to produce a heavy neutrino, only in the presence of this interaction the
signal is observable. Once the heavy neutrino is produced, it can decay
to $\ell W$ with $\ell = e,\mu,\tau$, being the corresponding branching 
ratios in the ratio $|V_{eN}|^2\, :\, |V_{\mu N}|^2\, :\, |V_{\tau N}|^2$.

\begin{figure}[t]
\begin{center}
\includegraphics[width=13cm]{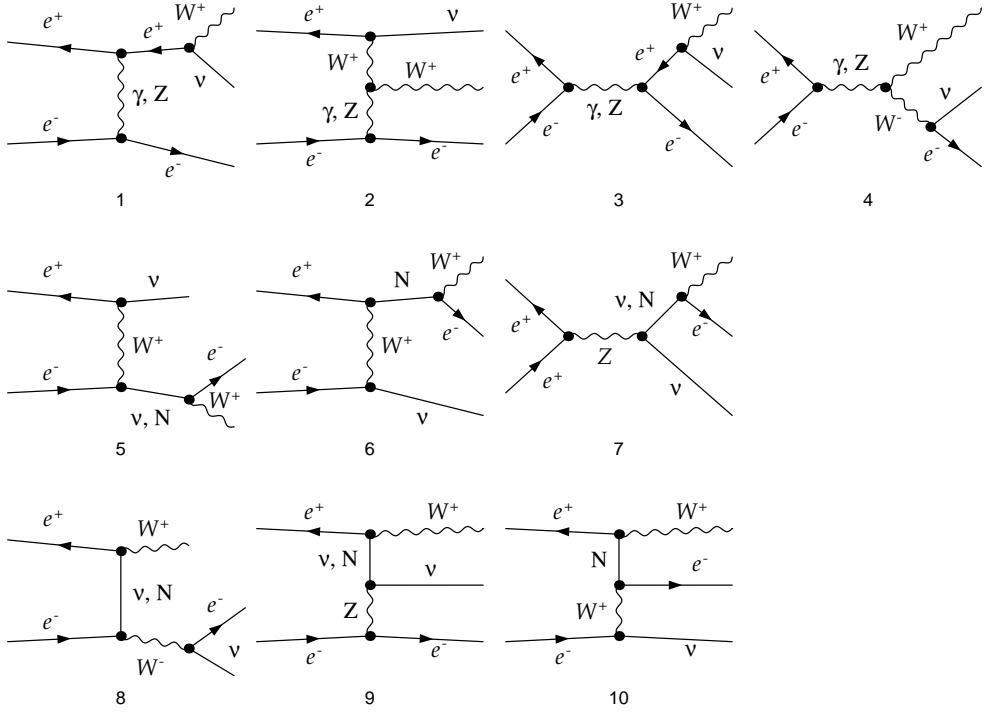}
\end{center}
\caption[Diagrams contributing to $e^+e^- \to e^- W^+ \nu$]
{Diagrams contributing to $e^+e^- \to e^- W^+ \nu$. For $\ell =
\mu,\tau$ only diagrams 3-8,10 contribute.}
\label{diag}
\end{figure}

In case that the signal is dominated by $t$ and $u$ channel on-shell $N$
production (the only situation in which it is observable), negative electron
polarisation and positive positron polarisation increase its statistical
significance. For the signal contributions alone we have 
$\sigma_{e_R^+ e_L^-} \,:\, \sigma_{e_L^+ e_R^-} = 1200 \,:\, 1$ (for $m_N =
300$ GeV, $V_{eN} = 0.073$), with
$\sigma_{e_R^+ e_R^-} = \sigma_{e_L^+ e_L^-} = 0$. For the SM process,
$\sigma_{e_R^+ e_L^-} \,:\, \sigma_{e_R^+ e_R^-} \,:\,
\sigma_{e_L^+ e_R^-} = 150 \,:\, 7 \,:\, 1$, $\sigma_{e_L^+ e_L^-} = 0$.
In the limit of perfect beam polarisations $P_{e^-} = -1$, $P_{e^+}=1$,
the signal is enhanced with respect to the background by a factor of 1.05 and,
what is more important, the ratio $S/\sqrt B$ increases by a factor of two.
Using right-handed electrons and left-handed positrons decreases the $S/B$
ratio by a factor of 8, and $S/\sqrt B$ by a factor of 50.
On the other hand, if the neutrino does not mix with the electron but mixes with
the muon or tau, the behaviour is the opposite. Since the only contribution
comes from diagram 7 the use of left-handed positrons and
right-handed electrons actually increases the signal, while reducing the SM
cross section for this process \cite{Meloni}. This case is of limited practical
interest, since for $V_{eN} = 0$ the signal is barely observable. 

We finally point out that the signal cross section exhibits little dependence on
the heavy neutrino mass, except close to the kinematical limit \cite{GZ1,GZ2}, and
the final results are almost independent of $m_N$ within the range 200-400 GeV
\cite{paper2}. For our calculations we take
$m_N = 300$ GeV. In contrast with what has been claimed in the literature
\cite{ACMV}, we find equal production cross sections for Dirac and Majorana
neutrinos to a very good approximation. The reason is easy to understand:
while in the present case the signal is
strongly dominated by diagrams 5 and 6 (which give equal contributions to the
cross section and do not interfere because light neutrino masses can be safely 
neglected), for a Dirac neutrino only diagram 5 is
present. On the other hand, the width of a Dirac neutrino is one half of the
width of a Majorana neutrino with the same mixing angles \cite{paper2}.

\subsection*{Generation of signals}
\label{sec:SigGen}

The matrix elements for $e^+ e^- \to \ell^- W^+ \nu \to \ell^- q \bar q' \nu$
are calculated using {\tt HELAS} \cite{helas}, including all spin correlations
and finite width effects. We sum SM and heavy neutrino-mediated diagrams at
the amplitude level. The charge conjugate process is included in all our
results unless otherwise noted.
We assume a CM energy of 500 GeV, with electron polarisation $P_{e^-} = -0.8$
and positron polarisation $P_{e^+} = 0.6$.
The luminosity is taken as 345 fb$^{-1}$ per year \cite{lum}.
In our calculations we take into account the effects of
ISR \cite{isr} and beamstrahlung \cite{peskin,BS2}. For the design luminosity
at 500 GeV we use the parameters $\Upsilon = 0.05$, $N = 1.56$ \cite{lum}.
The actual expressions for ISR and beamstrahlung used in our calculation are
collected  in Ref.~\cite{npb}. We also include a beam energy spread of 1\%.

In final states with $\tau$ leptons, we select $\tau$ decays to $\pi$, $\rho$
and $a_1$ mesons (with a combined branching fraction of 55\% \cite{PDG}), in
which a single
$\nu_\tau$ is produced, discarding other hadronic and leptonic decays.
We simulate the $\tau$ decay assuming that the meson and $\tau$ momenta are
collinear (what is a good approximation for high $\tau$ energies) and assigning
a random fraction $x$ of the $\tau$ momentum to the meson, according to the
probability distributions \cite{taudecays}
\begin{equation}
P(x) = 2 (1-x)
\end{equation}
for pions, and
\begin{equation}
P(x) = \frac{2}{2 \zeta^3-4 \zeta^2+1} \left[ (1-2 \zeta^2)-(1-2 \zeta) x
\right]
\end{equation}
for $\rho$ and $a_1$ mesons, where $\zeta = m_{\rho,a_1}^2/m_\tau^2$. We assume
a $\tau$ jet tagging efficiency of 50\%.

We simulate the calorimeter and tracking resolution of the detector by
performing a Gaussian smearing of the energies of electrons, muons 
and jets, using the specifications in Ref.~\cite{tesla2},
\begin{equation}
\frac{\Delta E^e}{E^e} = \frac{10\%}{\sqrt{E^e}} \oplus 1 \% \;, \quad
\frac{\Delta E^\mu}{E^\mu} = 0.02 \% \, E^\mu \;, \quad
\frac{\Delta E^j}{E^j} = \frac{50\%}{\sqrt{E^j}} \oplus 4 \% \;,
\end{equation}
respectively, where the two terms are added in quadrature and the energies 
are in GeV.
We apply kinematical cuts on transverse momenta, $p_T \geq 10$ GeV, and
pseudorapidities $|\eta| \leq 2.5$, the latter corresponding to polar angles
$10^\circ \leq \theta \leq 170^\circ$. To ensure high $\tau$ momenta (so that
the meson resulting from its decay is effectively collinear) we require $p_T
\geq 30$ GeV for $\tau$ jets. We reject events in which the
leptons or jets are not isolated, requiring a ``lego-plot'' separation
$\Delta R = \sqrt{\Delta \eta^2+\Delta \phi^2} \geq 0.4$.
For the Monte Carlo integration in 6-body phase space we use
{\tt RAMBO} \cite{rambo}.

In final states with electrons and muons the light neutrino momentum $p_\nu$ is
determined from the missing transverse and longitudinal momentum of the event
and the requirement that $p_\nu^2 = 0$ 
(despite ISR and beamstrahlung, the missing longitudinal momentum approximates
with a reasonable accuracy the original neutrino momentum).
In final states with $\tau$
leptons, the reconstruction is more involved, due to the additional neutrino
from the $\tau$ decay. We determine the ``first'' neutrino momentum and the
fraction $x$ of the $\tau$ momentum retained by the $\tau$ jet using the
kinematical constraints
\begin{align}
E_W + E_\nu + \frac{1}{x} E_j & = \sqrt s \,, \nonumber \\
\vec p_W + \vec p_\nu + \frac{1}{x}  \vec p_j & = 0 \,, \nonumber \\
p_\nu^2 & = 0 \,,
\end{align}
in obvious notation.
These constraints only hold if ISR and beamstrahlung are ignored, and in the
limit of perfect detector resolution. When solving them for the generated Monte
Carlo events we sometimes obtain $x > 1$ or $x < 0$. In the first case we
arbitrarily set $x=1$, and in the second case we set $x = 0.55$, which is the
average momentum fraction of the $\tau$ jets. With the procedure outlined here,
the reconstructed $\tau$ momentum reproduces with a fair accuracy the original
one, while the obtained $p_\nu$ is often quite different from its original 
value.

%\end{fmffile}

\backmatter
\listoffigures
\addcontentsline{toc}{chapter}{List of figures}
\listoftables
\addcontentsline{toc}{chapter}{List of tables}

\bibliographystyle{JHEP}
\bibliography{biblio}
\addcontentsline{toc}{chapter}{Bibliography}

\end{document}